\documentclass[apj]{emulateapj}

\usepackage{graphicx}
\usepackage{amssymb}
\usepackage{amsmath}
\usepackage{hyperref}
\usepackage{natbib}
\usepackage{longtable}

\begin{document}

%%%%% TITLE %%%%%
\title{The Ages of Early-Type Stars: Str\"{o}mgren Photometric Methods Calibrated, Validated, Tested, and Applied to Hosts and Prospective Hosts of Directly Imaged Exoplanets}
\author{Trevor J. David\altaffilmark{1,2} and Lynne A. Hillenbrand\altaffilmark{1}}
\affiliation{$^1$Department of Astronomy; MC 249-17; California Institute of Technology; Pasadena, CA 91125, USA; \href{mailto:tjd,lah@astro.caltech.edu}{tjd,lah@astro.caltech.edu}}
\altaffiltext{2}{NSF Graduate Research Fellow}

%%%%% ABSTRACT %%%%%
\begin{abstract} 
Age determination is undertaken for nearby early-type (BAF) stars, which constitute attractive targets for high-contrast debris disk and planet imaging surveys.  Our analysis sequence consists of: acquisition of $uvby\beta$ photometry from catalogs, correction for the effects of extinction, interpolation of the photometry onto model atmosphere grids from which atmospheric parameters are determined, and finally, comparison to the theoretical isochrones from pre-main sequence through post-main sequence stellar evolution models, accounting for the effects of stellar rotation. We calibrate and validate our methods at the atmospheric parameter stage by comparing our results to fundamentally determined $T_\text{eff}$ and $\log g$ values.  We validate and test our methods at the evolutionary model stage by comparing our results on ages to the accepted ages of several benchmark open clusters (IC 2602, $\alpha$ Persei, Pleiades, Hyades). Finally, we apply our methods to estimate stellar ages for 3493 field stars, including several with directly imaged exoplanet candidates.
\end{abstract}

\maketitle

%%%%% INTRODUCTION %%%%%
\section{Introduction}
\label{sec:intro}

In contrast to other fundamental stellar parameters such as mass, radius, and angular momentum -- that for certain well-studied stars and stellar systems can be anchored firmly in observables and simple physics -- stellar ages for stars other than the Sun have no firm basis.  Ages are critical, however, for many investigations  involving time scales including formation and evolution of planetary systems, evolution of debris disks, and interpretation of low mass stars, brown dwarfs, and so-called planetary mass objects that are now being detected routinely as faint point sources near bright stars in high contrast imaging surveys.

\subsection{The Era of Direct Imaging of Exoplanets}
\label{subsec:directimaging}

Intermediate-mass stars ($1.5-3.0\ M_{\odot}$) have proven themselves attractive targets for planet search work. Hints of their importance first arose during initial data return from IRAS in the early 1980s, when several A-type stars (notably Vega but also $\beta$ Pic and Fomalhaut) as well K-star Eps Eri -- collectively known as ``the fab four'' -- distinguished themselves by showing mid-infrared excess emission due to optically thin dust in Kuiper-Belt-like locations. Debris disks are signposts of planets, which dynamically stir small bodies resulting in dust production. Spitzer results in the late 2000s solidified the spectral type dependence of debris disk presence (e.g. \citealt{carpenter2006, wyatt2008}) for stars of common age. For a random sample of field stars, however, the primary variable determining the likelihood of debris is stellar age \citep{kains2011}. 

The correlation in radial velocity studies of giant planet frequency with stellar mass \citep{fischer2005, gaidos2013} is another line of evidence connecting planet formation efficiency to stellar mass.  The claim is that while $\sim$14\% of A stars have one or more $>1 M_\mathrm{Jupiter}$ companions at $<$5 AU, only $\sim$2\% of M stars do (\citealt{johnson2010}, c.f. \citealt{lloyd2013, schlaufman2013}).

Consistently interpreted as indicators of hidden planets, debris disks finally had their long-awaited observational connection to planets with the watershed discovery of {\it directly imaged} planetary mass companions. These were -- like the debris disks before them -- found first around intermediate-mass A-type stars, rather than the solar-mass FGK-type stars that had been the subject of much observational work at high contrast during the 2000s. HR 8799 \citep{marois2008, marois2010} followed by Fomalhaut \citep{kalas2008} and $\beta$ Pic \citep{lagrange2009, lagrange2010} have had their planets {\it and indeed one planetary system}, digitally captured by ground-based and/or space-based high contrast imaging techniques.  Of the known {\it bona fide} planetary mass ($< 10 M_\text{Jup}$) companions that have been directly imaged, six of the nine are located around the three A-type host stars mentioned above, with the others associated with lower mass stars including the even younger 5-10 Myr old star 1RXS 1609-2105 \citep{lafreniere2008, ireland2011} and brown dwarf 2MASS 1207-3933 \citep{chauvin2004} and the probably older GJ 504 \citep{kuzuhara2013}. Note that to date these directly imaged objects are all ``super-giant planets" and not solar system giant planet analogs (e.g. Jupiter mass or below).

Based on the early results, the major direct imaging planet searches have attempted to optimize success by preferentially observing intermediate-mass, early-type stars.  The highest masses are avoided due to the limits of contrast.  Recent campaigns include those with all the major large aperture telescopes: Keck/NIRC2, VLT/NACO, Gemini/NICI, and Subaru/HiCAO. Current and near-future campaigns include Project 1640 (P1640; Hinkley et al. 2011) at Palomar Observatory, Gemini Planet Imager (GPI), operating on the Gemini South telescope, VLT/SPHERE, and Subaru/CHARIS. The next-generation TMT and E-ELT telescopes both feature high contrast instruments.

\citet{mawet2012} compares instrumental contrast curves in their Figure 1. Despite the technological developments over the past decade, given the as-built contrast realities, only the largest, hottest, brightest, and therefore the youngest planets, i.e. those less than a few to a few hundred Myr in age, are still self-luminous enough to be amenable to direct imaging detection. 
Moving from the 3-10 $M_\text{Jupiter}$ detections at several tens of AU that are possible today/soon, to detection of lower mass, more Earth-like planets located at smaller, more terrestrial zone, separations, will require pushing to higher contrast from future space-based platforms. The targets of future surveys, whether ground or space, are however not likely to be substantially different from the samples targeted in today's ground-based surveys.

The most important parameter really is age, since the brightness of planets decreases so sharply with increasing age due to the rapid gravitational contraction and cooling \citep{fortney2008, burrows2004}. 
There is thus a premium on identifying the closest, youngest stars.

\subsection{The Age Challenge}
\label{subsec:agechallenge}

Unlike the other fundamental parameters of stellar mass (unambiguously determined from measurements of double-line eclipsing binaries and application of Kepler's laws) and stellar radius (unambiguously measured from interferometric measurements of angular diameters and parallax measurements of distances), there are no directly interpretable observations leading to stellar age.

Solar-type stars ($\sim 0.7-1.4 M_{\odot}$, spectral types F6-K5) were the early targets of radial velocity planet searches and later debris disk searches that can imply the presence of planets.  For these objects, although more work remains to be done, there are established activity-rotation-age diagnostics that are driven by the presence of convective outer layers and can serve as proxies for stellar age \citep[e.g.][]{mamajek2008}.  

For stars significantly different from our Sun, however, and in particular the intermediate-mass stars ($\sim 1.5-3.0 M_{\odot}$, spectral types A0-F5 near the main sequence) of interest here, empirical age-dating techniques have not been sufficiently established or calibrated. Ages have been investigated recently for specific samples of several tens of stars using color-magnitude diagrams by \citet{nielsen2013, vigan2012, moor2006, su2006, rhee2007, lowrance2000}.

Perhaps the most robust ages for young BAF stars come from clusters and moving groups, which contain not only the early-type stars of interest, but also lower mass stars to which the techniques mentioned above can be applied. These groups are typically dated using a combination of stellar kinematics, lithium abundances, rotation-activity indicators, and placement along theoretical isochrones in a color-magnitude diagram.
The statistics of these coeval stellar populations greatly reduce the uncertainty in derived ages. However, only four such groups exist within $\sim$ 60 pc of the Sun and the number of early-type members is small.

Field BAF stars having late-type companions at wide separation could have ages estimated using the methods valid for F6-K5 age dating.  However, these systems are not only rare in the solar neighborhood, but considerable effort is required in establishing companionship e.g. \cite{stauffer1995, barrado1997, song2000}. Attempts to derive fractional main sequence ages for A-stars based on the evolution of rotational velocities are ongoing \citep{zorec2012}, but this method is undeveloped and a bimodal distribution in $v \sin i$ for early-type A-stars may inhibit its utility. Another method, asteroseismology, which detects low-order oscillations in stellar interiors to determine the central density and hence age, is a heavily model-dependent method, observationally expensive, and best suited for older stars with denser cores.

The most general and quantitative way to age-date A0-F5 field stars is through isochrone placement.  As intermediate-mass stars evolve quickly along the H-R diagram, they are better suited for age-dating via isochrone placement relative to their low-mass counterparts which remain nearly stationary on the main sequence for many Gyr \citep{soderblom2010}. Indeed, the mere presence of an early-type star on the main sequence suggests moderate youth, since the hydrogen burning phase is relatively short-lived. However, isochronal ages are obviously model-dependent and they do require precise placement of the stars on an H-R diagram implying a parallax.  The major uncertainties arise from lack of information regarding metallicity \citep{nielsen2013}, rotation \citep{collinssmith1985} and multiplicity \citep{derosa2014}.

\subsection{Our Approach}

\begin{figure*}
\centering
\includegraphics[width=0.95\textwidth]{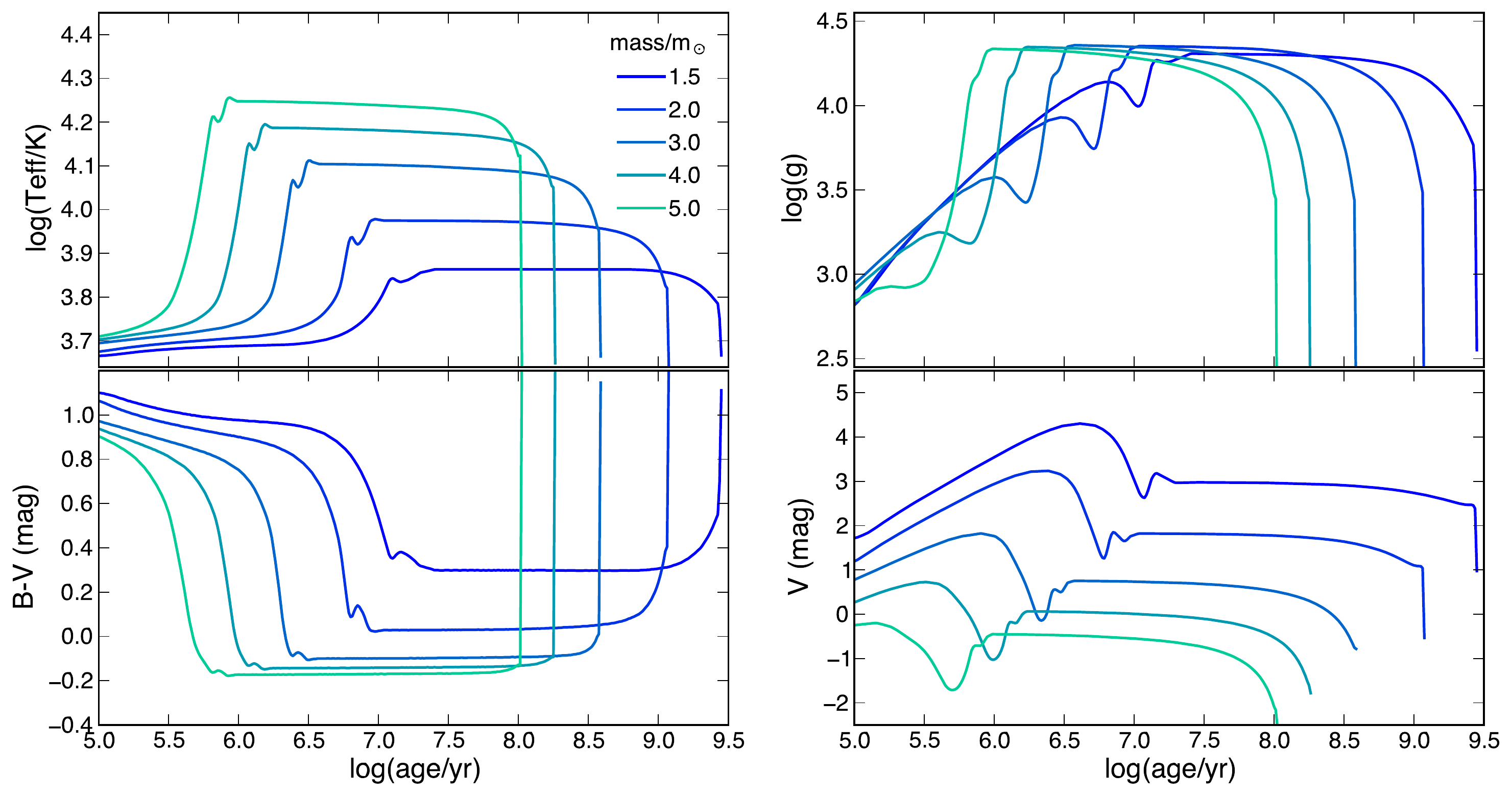}      
\caption{\emph{Top panels:} Evolution of $\log{T_\mathrm{eff}}$ and $\log{g}$ with age for intermediate-mass stars, as predicted by PARSEC evolutionary models \citep{bressan2012}. \emph{Bottom panels:} Same evolutionary trends for $B-V$ (close to $b-y$) and $M_V$ mag, as might be used to discern ages from color-magnitude diagram evolution \citep[e.g.][]{nielsen2013}. While the color and temperature trends reflect one another, the absolute magnitude trends are not as strong as the surface gravity trends when the stars are evolving from the main sequence after a few hundred Myr. The PARSEC models predict the precision in $\log g$ needed to distinguish a $1.5 M_\odot$ star and a $2.0 M_\odot$ star evolves from 0.0397 dex at $\sim$ 30 Myr to 0.0242 dex at 100 Myr to 0.0378 dex at $\sim$ 300 Myr. The precision in $\log g$ needed to distinguish a $1.5 M_\odot$ star and a $3.0 M_\odot$ evolves from 0.0085 dex at $\sim$ 30 Myr to 0.0694 dex at 100 Myr to 0.5159 dex at $\sim$ 300 Myr. The precision in $\log g$ needed to distinguish a $2.0 M_\odot$ star and a $3.0 M_\odot$ evolves from 0.0312 dex at $\sim$ 30 Myr to 0.0936 dex at 100 Myr to 0.4781 dex at 300 Myr.}
\label{fig:evolutiont}
\end{figure*}

Despite that many nearby BAF stars are well-studied, historically, there is no modern data set leading to a set of consistently derived stellar ages for this population of stars. Here we apply Str\"{o}mgren photometric techniques, and by combining modern stellar atmospheres and modern stellar evolutionary codes, we develop the methods for robust age determination for stars more massive than the Sun. The technique uses specific filters, careful calibration, definition of photometric indices, correction for any reddening, interpolation from index plots of physical atmospheric parameters, correction for rotation, and finally Bayesian estimation of stellar ages from evolutionary models that predict the atmospheric parameters as a function of mass and age. 

Specifically, our work uses high-precision archival $uvby\beta$ photometry and model atmospheres so as to determine the fundamental stellar atmospheric parameters $T_\text{eff}$ and $\log g$. Placing stars accurately in an $\log T_\mathrm{eff}$ vs. $\log g$ diagram leads to derivation of their ages and masses.  We consider \cite{bressan2012} evolutionary models that include pre-main sequence evolutionary times (2 Myr at 3 $M_{\odot}$ and 17 Myr at 1.5 $M_{\odot}$), which are a significant fraction of any intermediate mass star's absolute age, as well as \cite{ekstrom2012} evolutionary models that self-consistently account for stellar rotation, which has non-negligible effects on the inferred stellar parameters of rapidly rotating early-type stars. Figure ~\ref{fig:evolutiont} shows model predictions for the evolution of both physical and observational parameters.

The primary sample to which our technique is applied in this work consists of 3499 BAF field stars within 100 pc and with $uvby\beta$ photometry available in the \cite{hauck1998} catalog, hereafter HM98. The robustness of our method is tested at different stages with several control samples. To assess the uncertainties in our atmospheric parameters we consider (1) 69 $T_\mathrm{eff}$ standard stars from \cite{boyajian2013} or \cite{napiwotzki1993}; (2) 39 double-lined eclipsing binaries with standard $\log{g}$ from \cite{torres2010}; (3) 16 other stars from \cite{napiwotzki1993}, also for examining $\log{g}$. To examine isochrone systematics, stars in four open clusters are studied (31 members of IC 2602, 51 members of $\alpha$ Per, 47 members of the Pleiades, and 47 members of the Hyades). Some stars belonging to sample (1) above are also contained in the large primary sample of field stars.

%%%%% METHODOLOGY %%%%%

\section{The Str\"{o}mgren Photometric System}
\label{sec:uvby}

\begin{figure}[!h]
\centering
\includegraphics[width=0.45\textwidth]{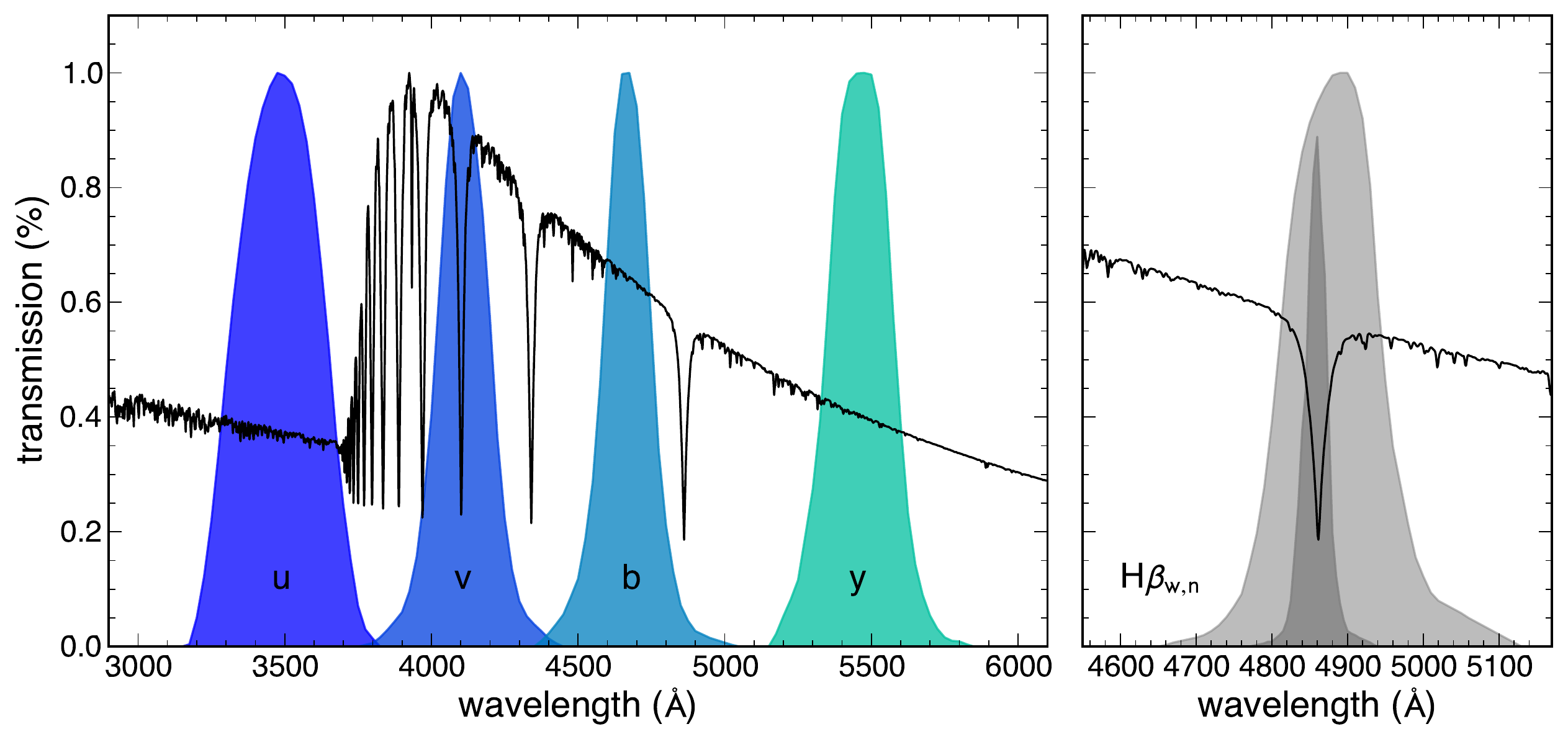}
\caption{The $u, v, b, y,$ H$\beta_\text{wide}$ and H$\beta_\text{narrow}$ passbands. Overplotted on an arbitrary scale is the synthetic spectrum of an A0V star generated by \cite{munari2005} from an ATLAS9 model atmosphere. The $uvby$ filter profiles are those of \cite{bessell2011}, while the H$\beta$ filter profiles are those originally described in \cite{crawford1966} and the throughput curves are taken from \cite{castelli2006}.}
\label{fig:filters}
\end{figure}

Historical use of Str\"{o}mgren photometry methods indeed has been for the purpose of determining stellar parameters for early-type stars. Recent applications include work by \citet{nieva2013,dallemese2012,onehag2009,allende1999}. An advantage over more traditional color-magnitude diagram techniques \citep{nielsen2013,derosa2014} is that distance knowledge is not required, so the distance-age degeneracy is removed. Also, metallicity effects are relatively minor (as addressed in an Appendix) and rotation effects are well-modelled and can be corrected for (\S~\ref{subsec:vsinicorrection}).

% The uvby-Beta Photometric System %
\subsection{Description of the Photometric System}
\label{subsec:uvbydescription}

The $uvby\beta$ photometric system is comprised of four intermediate-band filters ($uvby$) first advanced by \cite{stromgren1966} plus the H$\beta$ narrow and wide filters developed by \cite{crawford1958}; see Figure~\ref{fig:filters}. Together, the two filter sets form a well-calibrated system that was specifically designed for studying earlier-type BAF stars, for which the hydrogen line strengths and continuum slopes in the Balmer region rapidly change with temperature and gravity.

From the fluxes contained in the six passbands, five $uvby\beta$ indices are defined. The color indices, ($b-y$) and ($u-b$), and the $\beta$-index,

\begin{equation}
\beta = \mathrm{H}\beta_\text{narrow} - \mathrm{H}\beta_\text{wide},
\end{equation}

are all sensitive to temperature and weakly dependent on surface gravity for late A- and F-type stars. The Balmer discontinuity index, 

\begin{equation}
c_1 = (u-v) - (v-b),
\end{equation}

is sensitive to temperature for early type (OB) stars and surface gravity for intermediate (AF) spectral types. Finally, the metal line index,  

\begin{equation}
m_1 = (v-b) - (b-y),
\end{equation}

is sensitive to the metallicity $[M/H]$.

For each index, there is a corresponding intrinsic, dereddened index denoted by a naught subscript with e.g $c_0, (b-y)_0,$ and $(u-b)_0$, referring to the intrinsic, dereddened equivalents of the indices $c_1, (b-y),$ and $(u-b)$, respectively.  Furthermore, although reddening is expected to be negligible for the nearby sources of primary interest to us, automated classification schemes that divide a large sample of stars for analysis into groups corresponding to earlier than, around, and later than the Balmer maximum will sometimes rely on the reddening-independent indices defined by \cite{crawfordmandwewala1976} for A-type dwarfs:

    \begin{align}
    [c_1] &= c_1 - 0.19 (b-y) \\
    [m_1] &= m_1 + 0.34 (b-y) \\
    [u-b] &= [c_1] + 2 [m_1].
    \end{align}

Finally, two additional indices useful for early A-type stars, $a_0$ and $r^*$, are defined as follows:

    \begin{align}
    a_0 &= 1.36(b-y)_0 + 0.36m_0 + 0.18c_0 - 0.2448 \\
        &= (b-y)_0 + 0.18[(u-b)_0 - 1.36], \\
    r^* &= 0.35c_1 - 0.07(b-y)-(\beta-2.565).
    \end{align}

Note that $r^*$ is a reddening free parameter, and thus indifferent to the use of reddened or unreddened photometric indices.

% Extinction correction
\subsection{Extinction Correction}
\label{subsec:reddening}

Though the sample of nearby stars to which we applying the Str\"{o}mgren methodology are assumed to be unextincted or only lightly extincted, interstellar reddening is significant for the more distant stars including those in the open clusters used in \S ~\ref{subsec:openclustertests} to test the accuracy of the ages derived using our $uvby\beta$ methodology. In the cases where extinction is thought to be significant, corrections are performed using the \texttt{UVBYBETA}\footnote{\url{http://idlastro.gsfc.nasa.gov/ftp/pro/astro/uvbybeta.pro}} and \texttt{DEREDD}\footnote{\url{http://idlastro.gsfc.nasa.gov/ftp/pro/astro/deredd.pro}} programs for IDL.

These IDL routines take as input $(b-y), m_1, c_1, \beta$, and a class value (between 1-8) that is used to roughly identify what region of the H-R diagram an individual star resides in. For our sample, stars belong to only four of the eight possible classes. These classes are summarized as follows: (1) B0-A0,  III-V, $2.59 < \beta < 2.88$, $-0.20 < c_0 < 1.00$, (5) A0-A3, III-V, $2.87 < \beta < 2.93$, $-0.01 < (b-y)_0 < 0.06$, (6) A3-F0, III-V, $2.72 < \beta < 2.88$, $0.05 < (b-y)_0 < 0.22$, and (7) F1-G2, III-V, $2.60 < \beta < 2.72$, $0.22 < (b-y)_0 < 0.39$. The class values in this work were assigned to individual stars based on their known spectral types (provided in the XHIP catalog \citep{anderson2011}), and $\beta$ values where needed. In some instances, A0-A3 stars assigned to class (5) with values of $\beta < 2.87$, the dereddening procedure was unable to proceed. For these cases, stars were either assigned to class (1) if they were spectral type A0-A1, or to class (6), if they were spectral type A2-A3.  

Depending on the class of an individual star, the program then calculates the dereddened indices $(b-y)_0, m_0, c_0$, the color excess $E(b-y)$, $\delta m_0$, the absolute V magnitude, $M_V$, the stellar radius and effective temperature. Notably, the $\beta$ index is unaffected by reddening as it is the flux difference between two narrow band filters with essentially the same central wavelength. Thus, no corrections are performed on $\beta$ and this index can be used robustly in coarse classification schemes.

To transform $E(b-y)$ to $A_V$, we use the extinction measurements of \cite{schlegel1998} and to propagate the effects of reddening through to the various $uvby\beta$ indices we use the calibrations of \cite{crawfordmandwewala1976}:

\begin{align}
E(m_1) &= -0.33 E(b-y) \\
E(c_1) &= 0.20 E(b-y) \\
E(u-b) &= 1.54 E(b-y).
\end{align}

From these relations, given the intrinsic color index $(b-y)_0$, the reddening free parameters $m_0, c_0, (u-b)_0,$ and $a_0$ can be computed.

In \S ~\ref{subsec:tefflogguncertainties} we quantify the effects of extinction and extinction uncertainty on the final atmospheric parameter estimation, $T_\mathrm{eff}, \log g$.

\subsection{Utility of the Photometric System}
\label{subsec:uvbyutility}

From the four basic Str\"{o}mgren indices  -- $b-y$ color, $\beta$, $c_1$, and $m_1$ -- accurate determinations of the stellar atmospheric parameters $T_\text{eff}, \log g$, and $[M/H]$ are possible for B, A, and F stars. Necessary are either empirical \citep[e.g.][]{crawford1979, lester1986, olsen1988, smalley1993, smalley1995, clem2004}, or theoretical \citep[e.g.][]{balona1984, moon1985, napiwotzki1993, balona1994, lejeune1999, castelli2006, castelli2004, onehag2009} calibrations. Uncertainties of 0.10 dex in $\log g$ and 260 K in $T_\text{eff}$ are claimed as achievable and we reassess these uncertainties ourselves \S ~\ref{subsec:tefflogguncertainties}.

% Determination of Atmospheric Parameters %

\section{Determination of Atmospheric Parameters $T_\mathrm{eff}, \log g$}
\label{sec:atmosphericparameters}
\subsection{Procedure}
\label{subsec:atmosphericparameters}    

\begin{figure*}
\centering
\includegraphics[width=0.99\textwidth]{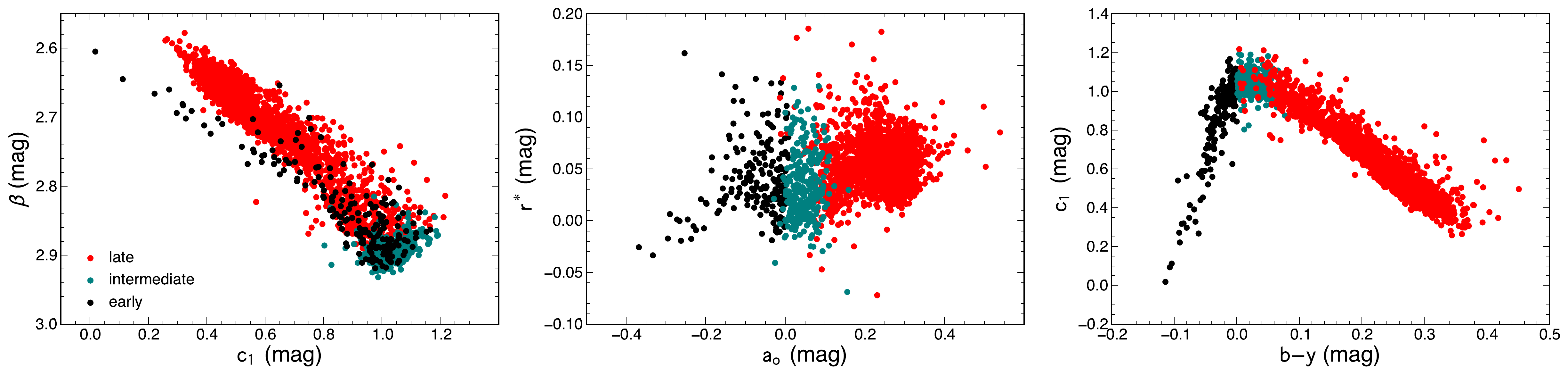}
\includegraphics[width=0.99\textwidth]{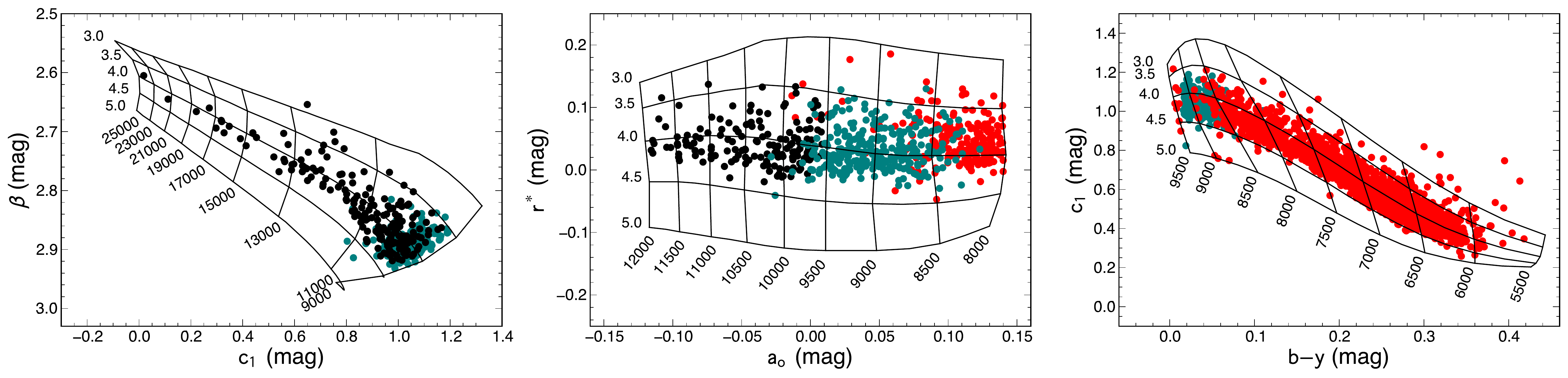}
\caption{\emph{Top:} Three relevant $uvby\beta$ spaces for atmospheric parameter determination of our sample of BAF stars with $uvby\beta$ photometry in the HM98 catalog, and located within 100 pc of the Sun. Two stars were excluded from these figures for favorable scaling: Castor, which is an outlier in all three planes ($\beta <$2.4, $a_0 >$1, $b-y >$0.6), and HD 17300, a poorly studied F3V star with $b-y >$0.6. \emph{Bottom:} The same plots as above, with the model color grids of \cite{castelli2006, castelli2004} overlaid in the relevant regions of parameter space. The lines of constant $T_\mathrm{eff}$ (largely vertical) and of constant $\log{g}$ (largely horizontal) are annotated with their corresponding values. Some outliers have been pruned, and irrelevant groups of stars eliminated, for clarity in this second plot.}
\label{fig:uvbygrids}
\end{figure*}

Once equipped with $uvby\beta$ colors and indices and understanding the effects of extinction, arriving at the fundamental parameters $T_\text{eff}$ and $\log g$ for program stars, proceeds by interpolation among theoretical color grids (generated by convolving filter sensitivity curves with model atmospheres) or explicit formulae (often polynomials) that can be derived empirically or using the theoretical color grids. In both cases, calibration to a sample of stars with atmospheric parameters that have been independently determined through fundamental physics is required. See e.g. \cite{figueras1991} for further description.

Numerous calibrations, both theoretical and empirical, of the $uvby\beta$ photometric system exist. 
For this work we use the \cite{castelli2006, castelli2004} color grids generated from solar metallicity (Z=0.017, in this case) ATLAS9 model atmospheres using a microturbulent velocity parameter of $\xi = $ 0 km s$^{-1}$ and the new ODF. We do not use the alpha-enhanced color grids. The grids are readily available from F. Castelli\footnote{\url{http://wwwuser.oats.inaf.it/castelli}} or R. Kurucz\footnote{\url{http://kurucz.harvard.edu/grids/gridP00ODFNEW/uvbyp00k0odfnew.dat}}.   

Prior to assigning atmospheric parameters to our program stars directly from the model grids, we first investigated the accuracy of the models on samples of BAF stars with fundamentally determined $T_\mathrm{eff}$ (through interferometric measurements of the angular diameter and estimations of the total integrated flux) and $\log g$ (from measurements of the masses and radii of double lined eclipsing binaries).  We describe these validation procedures in \S ~\ref{subsec:teffvalidation} and \S ~\ref{subsec:loggvalidation}.
   
Atmospheric parameter determination occurs in three different observational Str\"{o}mgren planes depending on the temperature regime (see Figure ~\ref{fig:uvbygrids}); this is in order to avoid the degeneracies that are present in all single observational planes when mapped onto the physical parameter space of $\log T_\text{eff}$ and $\log g$.  

Building off of the original work of e.g. \cite{stromgren1951, stromgren1966},  \cite{moon1985}, and later \cite{napiwotzki1993}, suggested assigning physical parameters in the following three regimes:  for cool stars ($T_\text{eff} \leq$ 8500 K), $\beta$ or $(b-y)$ can be used as a temperature indicator and $c_0$ is a surface gravity indicator;  for intermediate temperature stars (8500 K $\leq T_\text{eff} \leq$ 11000 K), the temperature indicator is $a_0$ and surface gravity indicator $r^*$;  finally, for hot stars ($T_\text{eff} \gtrsim$ 11000 K), the $c_0$ or the $[u-b]$ indices can be used as a temperature indicator while $\beta$ is a gravity indicator (note that the role of $\beta$ is reversed for hot stars compared to its role for cool stars). We adopt here $c_1$ vs. $\beta$ for the hottest stars, $a_0$ vs. $r^*$ for the intermediate temperatures, and $(b-y)$ vs. $c_1$ for the cooler stars.

Choosing the appropriate plane for parameter determination effectively means establishing a crude temperature sequence prior to fine parameter determination; in this, the $\beta$ index is critical.  Because the $\beta$ index switches from being a temperature indicator to a gravity indicator in the temperature range of interest to us (spectral type B0-F5, luminosity class IV/V stars), atmospheric parameter determination proceeds depending on the temperature regime.  For the $T_\mathrm{eff}$ and $\log g$ calibrations described below, temperature information existed for all of the calibration stars, though this is not the case for our program stars. In the general case we must rely on photometric classification to assign stars to the late, intermediate, and early groups, and then proceed to determine atmospheric parameters in the relevant $uvby\beta$ planes.

\cite{ttmoon1985} provides a scheme, present in the \texttt{UVBYBETA} IDL routine, for roughly identifying the region of the H-R diagram in which a star resides. However, because our primary sample of field stars are assumed to be unextincted, and because the \texttt{UVBYBETA} program relies on user-inputted class values based on unverified spectral types from the literature, we opt for a classification scheme based solely on the $uvby\beta$ photometry.

\cite{monguio2014}, hereafter M14, designed a sophisticated classification scheme, based on the work of \cite{stromgren1966}. The M14 scheme places stars into early (B0-A0), intermediate (A0-A3), and late (later than A3) groups based solely on $\beta$, the reddened color $(b-y)$, and the reddening-free parameters $[c_1], [m_1], [u-b]$. The M14 scheme improves upon the previous method of \cite{figueras1991} by imposing two new conditions (see their Figure 2 for the complete scheme) intended to prevent the erroneous classification of some stars. For our sample of 3499 field stars (see \S ~\ref{subsec:fieldstars}), there are 699 stars lacking $\beta$ photometry, all but three of which cannot be classified by the M14 scheme. For such cases, we rely on supplementary spectral type information and manually assign these unclassified stars to the late group. Using the M14 scheme, the final makeup of our field star sample is 85.9\% late, 8.4\% intermediate, and 5.7\% early.

\subsection{Sample and Numerical Methods}

For all stars in this work, $uvby\beta$ photometry is acquired from the \cite{hauck1998} compilation (hereafter HM98), unless otherwise noted. HM98 provides the most extensive compilation of $uvby\beta$ photometric measurements, taken from the literature and complete to the end of 1996
(the photometric system has seen less frequent usage/publication in more modern times). The HM98 compilation includes 105,873 individual photometric measurements for 63,313 different stars, culled from 533 distinct sources, and are presented both as individual measurements and weighted means of the literature values. 

The HM98 catalog provides $(b-y), m_1, c_1,$ and $\beta$ and the associated errors in each parameter if available. From these indices $a_0$ and $r^*$ are computed according to Equations (7), (8) \& (9). The ATLAS9 $uvby\beta$ grids provide a means of translating from ($b-y, m_1, c_1, \beta, a_0, r^*$) to a precise combination of ($T_\mathrm{eff}, \log g$). Interpolation within the model grids is performed on the appropriate grid: ($(b-y)$ vs. $c_1$ for the late group, $a_0$ vs. $r^*$ for the intermediate group, and $c_1$ vs. $\beta$ for the early group). 

The interpolation is linear and performed using the SciPy routine \texttt{griddata}. Importantly, the model $\log g$ values are first converted into linear space so that $g$ is determined from the linear interpolation procedure before being brought back into log space. The model grids used in this work are spaced by 250 K  in $T_\mathrm{eff}$ and 0.5 dex in $\log g$. To improve the precision of our method of atmospheric parameter determination in the future, it would be favorable to use model color grids that have been calculated at finer resolutions, particularly in $\log g$, directly from model atmospheres. However, the grid spacings stated above are fairly standardized among extant $uvby\beta$ grids.

% Rotational velocity correction
\subsection{Rotational Velocity Correction}
\label{subsec:vsinicorrection}

\begin{figure}
\centering
\includegraphics[width=0.49\textwidth]{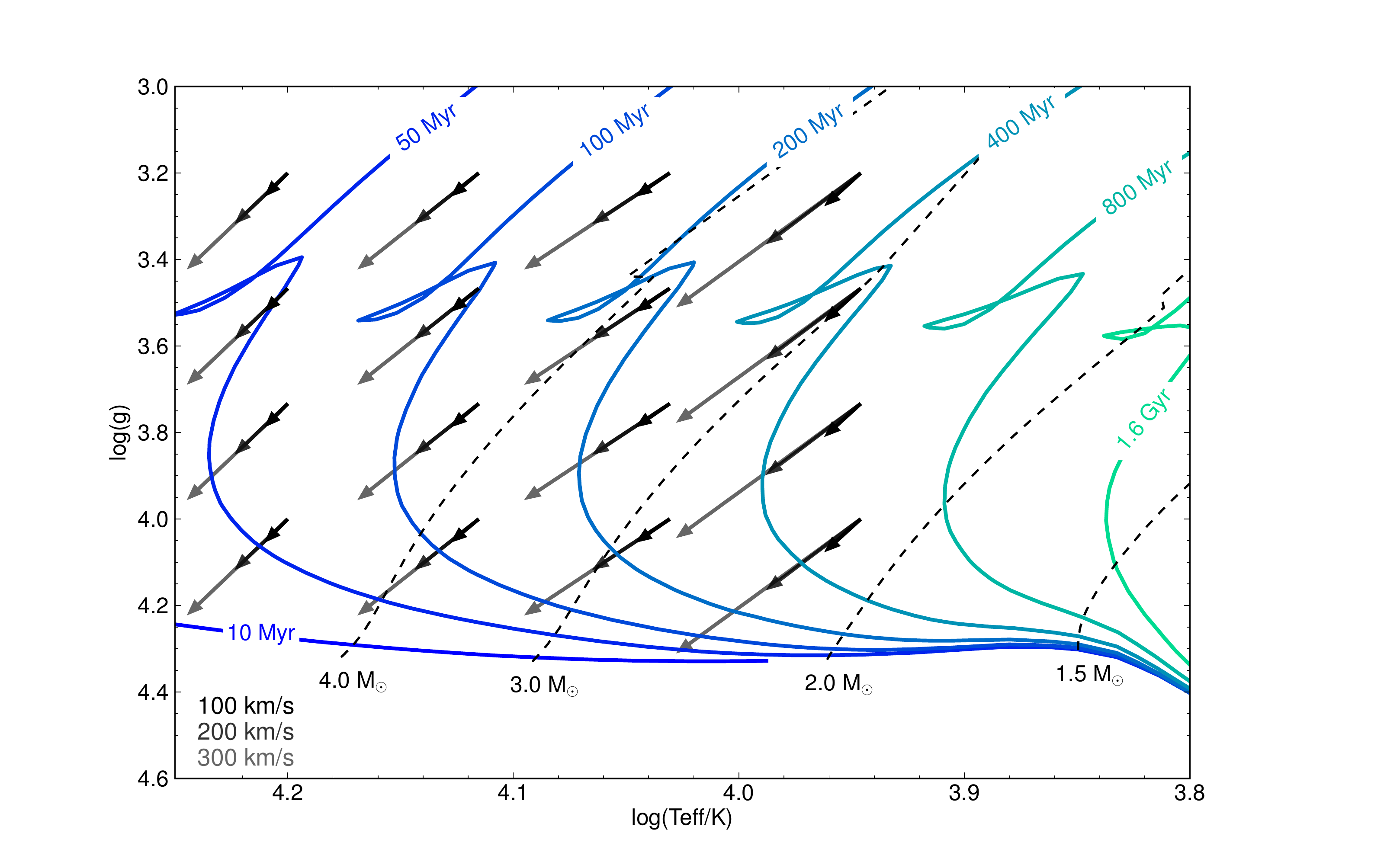}
\caption{Vectors showing the magnitude and direction of the rotational velocity corrections at 100 (black), 200, and 300 (light grey) km s$^{-1}$ for a grid of points in log(Teff)-log$g$ space, with PARSEC isochrones overlaid for reference.  While typical A-type stars rotate at about 150 km s$^{-1}$, high-contrast imaging targets are sometimes selected for slow rotation and hence favorable inclinations, typically $v \sin i <$50 km s$^{-1}$ or within the darkest black vectors. For rapid rotators, a 100$\%$ increase in the inferred age due to rotational effects is not uncommon.}
\label{fig:rotation-vectors}
\end{figure}

Early-type stars are rapid rotators, with rotational velocities of $v \sin i \gtrsim 150$ km s$^{-1}$ being typical. For a rotating star, both surface gravity and effective temperature decrease from the poles to the equator, changing the mean gravity and temperature of a rapid rotator relative to a slower rotator \citep{sweetroy1953}. Vega, rotating with an inferred equatorial velocity of $v_\mathrm{eq}  \sim 270$ km s$^{-1}$ at a nearly pole-on inclination, has measured pole-to-equator gradients in $T_\mathrm{eff}$ and $\log{g}$ that are $\sim$ 2400 K and $\sim$ 0.5 dex,  respectively \citep{peterson2006}. The apparent luminosity change due to rotation depends on the inclination: a pole-on ($i=0^\circ$) rapid rotator appears more luminous than a nonrotating star of the same mass, while an edge-on ($i=90^\circ$) rapid rotator appears less luminous than a nonrotating star of the same mass. \cite{sweetroy1953} found that a $(v \sin i)^2$ correction factor could describe the changes in luminosity, gravity, and temperature.
  
The net effect of stellar rotation on inferred age is to make a rapid rotator appear cooler, more luminous, and hence older when compared to a nonrotating star of the same mass (or more massive when compared to a nonrotating star of the same age).  Optical colors can be  affected since the spectral lines of early type stars are strong and broad. \citet{kraftwrubel1965} demonstrated specifically in the Str{\"o}mgren system that the effects are predominantly in the gravity indicators ($c_1$, which then also affects the other gravity indicator $r^*$) and less so in the temperature indicators ($b-y$, which then affects $a_0$).

\cite{figueras1998}, hereafter FB98, used Monte-Carlo simulations to investigate the effect of rapid rotation on the measured $uvby\beta$ indices, derived atmospheric parameters, and hence isochronal ages of early-type stars. Those authors concluded that stellar rotation conspires to artificially enhance isochronal ages derived through $uvby\beta$ photometric methods by 30-50\% on average.

To mitigate the effect of stellar rotation on the parameters $T_\text{eff}$ and $\log(g)$, FB98 presented the following corrective formulae for stars with $T_\mathrm{eff} > 11000$ K:

\begin{align}
\Delta T_\mathrm{eff} &= 0.0167 (v \sin i)^2 + 218, \\ 
\Delta \log g &= 2.10 \times 10^{-6} (v \sin i)^2 + 0.034.
\end{align}

For stars with $8500 \mathrm{K} \leq T_\mathrm{eff} \leq 11000 \mathrm{K}$, the analogous formulae are:

\begin{align}
\Delta T_\mathrm{eff} &= 0.0187 (v \sin i)^2 + 150, \\ 
\Delta \log g &= 2.92 \times 10^{-6} (v \sin i)^2 + 0.048.
\end{align}

In both cases, $\Delta T_\mathrm{eff}$ and $\Delta \log g$ are \emph{added} to the $T_\mathrm{eff}$ and $\log g$ values derived from $uvby\beta$ photometry.

Notably, the rotational velocity correction is dependent on whether the star belongs to the early, intermediate, or late group.  Specifically, FB98 define three regimes: $T_\mathrm{eff}<$8830 K (no correction), 8830 K$<$Teff$<$9700 K (correction for intermediate A0-A3 stars), Teff$>$9700 K (correction for stars earlier than A3). 

\cite{song2001}, who performed a similar isochronal age analysis of A-type stars using $uvby\beta$ photometry, extended the FB98 rotation corrections to stars earlier and later than B7 and A4, respectively. In the present work, a more conservative approach is taken and the rotation correction is applied only to stars in the early or intermediate groups, as determined by the classification scheme discussed in \S ~\ref{subsec:atmosphericparameters}. This decision was partly justified by the abundance of late-type stars that fall below the ZAMS in the open cluster tests (\S ~\ref{subsec:openclustertests}), for which the rotation correction would have a small (due to the lower rotational velocities of late-type stars) but exacerbating effect on these stars whose surface gravities are already thought to be overestimated.

We include these corrections and, as illustrated in Figure~\ref{fig:rotation-vectors}, emphasize that in their absence we would err on the side of over-estimating the age of a star, meaning conservatively overestimating rather than underestimating companion masses based on assumed ages. As an example, for a star with $T_\mathrm{eff} \approx$ 13,275 K and log$g$ $\approx$ 4.1, assumed to be rotating edge-on at 300 km s$^{-1}$, neglecting to apply the rotation correction would result in an age of $\sim$ 100 Myr. Applying the rotation correction to this star results in an age of $\sim$ 10 Myr.

Of note, the FB98 corrections were derived for atmospheric parameters determined using the synthetic $uvby\beta$ color grids of \cite{moon1985}. It is estimated that any differences in derived atmospheric parameters resulting from the use of color grids other than those of \cite{moon1985} are less than the typical measurement errors in those parameters. In \S ~\ref{subsec:tefflogguncertainties} we quantify the effects of rotation and rotation correction uncertainty on the final atmospheric parameter estimation, $T_\mathrm{eff}, \log g$.

\section{Calibration and Validation Using the HM98 Catalog}
In this section we assess the effective temperatures and surface gravities derived from atmospheric models and $uvby\beta$ color grids relative to fundamentally determined temperatures (\S ~\ref{subsec:teffvalidation}) and surface gravities (\S ~\ref{subsec:loggvalidation}).

\subsection{Effective Temperature}
\label{subsec:teffvalidation}

A fundamental determination of $T_\mathrm{eff}$ is possible through an interferometric measurement of the stellar angular diameter and an estimate of the total integrated flux. We gathered 69 stars (listed in Table ~\ref{table:teffcal}) with fundamental $T_\mathrm{eff}$ measurements from the literature and determine photometric temperatures for these objects from interpolation of $uvby\beta$ photometry in ATLAS9 model grids.

Fundamental $T_\mathrm{eff}$ values were sourced from \cite{boyajian2013}, hereafter B13, and \cite{napiwotzki1993}, hereafter N93. Several stars have multiple interferometric measurements of the stellar radius, and hence multiple fundamental $T_\mathrm{eff}$ determinations. For these stars, identified as those objects with multiple radius references in Table ~\ref{table:teffcal}, the mean $T_\mathrm{eff}$ and standard deviation were taken as the fundamental measurement and standard error. Among the 16 stars with multiple fundamental $T_\mathrm{eff}$ determinations by between 2 and 5 authors, there is a scatter of typically several percent (with 0.1-4\% range).

Additional characteristics of the $T_\mathrm{eff}$ ``standard'' stars are summarized as follows: spectral types B0-F9, luminosity classes III-V, 2 km s$^{-1}$ $\leq v \sin i \leq$ 316 km s$^{-1}$, mean and median $v \sin i$ of 58 and 26 km s$^{-1}$, respectively, 2.6 pc $\leq d \leq$ 493 pc, and a mean and median [Fe/H] of -0.08 and -0.06 dex, respectively. Line-of-sight rotational velocities were acquired from the \cite{glebocki2005} compilation and [Fe/H] values were taken from SIMBAD. Variability and multiplicity were considered, and our sample is believed to be free of any possible contamination due to either of these effects.

From the HM98 compilation we retrieved $uvby\beta$ photometry for these ``effective temperature standards.'' The effect of reddening was considered for the hotter, statistically more distant stars in the N93 sample. Comparing mean $uvby\beta$ photometry from HM98 with the dereddened photometry presented in N93 revealed that nearly all of these stars have negligible reddening ($E(b-y) \leq$ 0.001 mag). The exceptions are HD 82328, HD 97603, HD 102870, and HD 126660 with color excesses of $E(b-y)=$ 0.010, 0.003, 0.011, and 0.022 mag, respectively. Inspection of Table ~\ref{table:teffcal} indicates that despite the use of the reddened HM98 photometry the $T_\mathrm{eff}$ determinations for three of these four stars are still of high accuracy. For HD 97603, there is a discrepancy of $>$ 300 K between the fundamental and photometric temperatures. However, the $uvby\beta$ $T_\mathrm{eff}$ using reddened photometry for this star is actually hotter than the fundamental $T_\mathrm{eff}$. Notably, the author-to-author dispersion in multiple fundamental $T_\mathrm{eff}$ determinations for HD 97603 is also rather large. As such, the HM98 photometry was deemed suitable for all of the ``effective temperature standards.''

\begin{figure}
\includegraphics[width=0.45\textwidth]{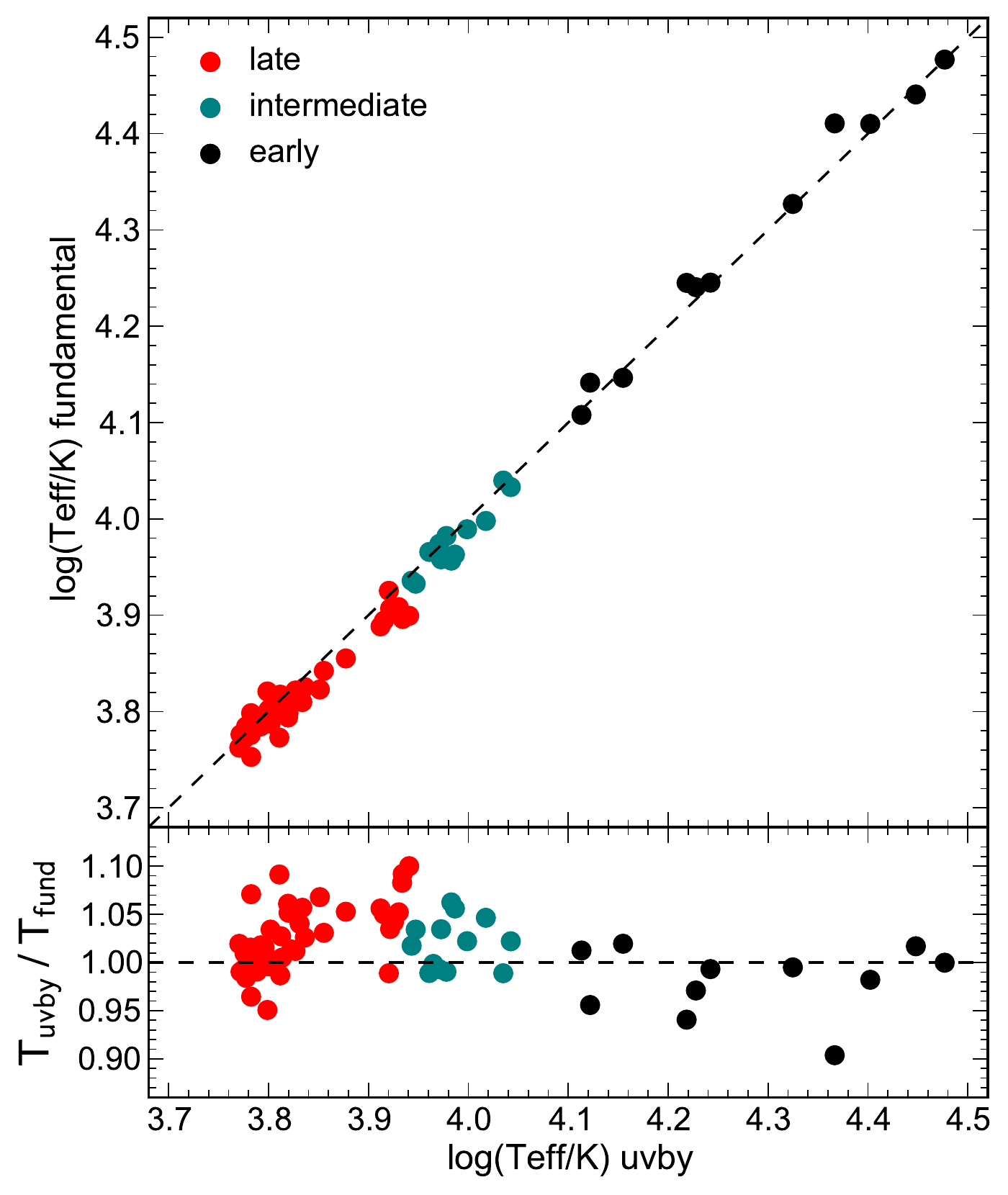}
\caption{\emph{Top:} Comparison of the temperatures derived from the ATLAS9 $uvby\beta$ color grids (T$_{uvby}$) and the fundamental effective temperatures ($\mathrm{T_{fund}}$) taken from B13 and N93. \emph{Bottom:} Ratio of $uvby\beta$ temperature to fundamental temperature, as a function of $T_\mathrm{uvby}$. For the majority of stars, the $uvby\beta$ grids can predict $\mathrm{T_{eff}}$ to within $\sim 5 \%$ without any additional correction factors.
}
\label{fig:teff-cal-3}
\end{figure}

For the sake of completeness, different model color grids were investigated, including those of \cite{fitzpatrick2005}, which were recently calibrated for early group stars, and those of \cite{onehag2009}, which were calibrated from MARCS model atmospheres for stars cooler than 7000 K. We found the grids that best matched the fundamental effective temperatures were the ATLAS9 grids of solar metallicity with no alpha-enhancement, microturbulent velocity of 0 km s$^{-1}$, and using the new opacity distribution function (ODF). The ATLAS9 grids with microturbulent velocity of 2 km s$^{-1}$ were also tested, but were found to worsen both the fractional $T_\mathrm{eff}$ error and scatter, though only nominally (by a few tenths of a percent).

For the early group stars, temperature determinations were attempted in both the $c_1-\beta$ and [$u-b$]-$\beta$ planes. The $c_1$ index was found to be a far better temperature indicator in this regime, with the [$u-b$] index underestimating $T_\mathrm{eff}$ relative to the fundamental values $>$10\% on average. Temperature determinations in the $c_1-\beta$ plane, however, were only $\approx$ 1.9\% cooler than the fundamental values, regardless of whether $c_1$ or the dereddened index $c_0$ was used. This is not surprising as the $c_1-\beta$ plane is not particularly susceptible to reddening. 

At intermediate temperatures, the $a_0-r^*$ plane is used. In this regime, the ATLAS9 grids were found to overestimate $T_\mathrm{eff}$ by $\approx$ 2.0\% relative to the fundamental values. 

Finally, for the late group stars temperature determinations were attempted in the $(b-y)-c_1$ and $\beta-c_1$ planes. In this regime, $(b-y)$ was found to be a superior temperature indicator, improving the mean fractional error marginally and reducing the RMS scatter by more than 1\%. In this group, the model grids overpredict $T_\mathrm{eff}$ by $\approx$ 2.4\% on average, regardless of whether the reddened or dereddened indices are used.

Figure \ref{fig:teff-cal-3} shows a comparison of the temperatures derived from the ATLAS9 $uvby\beta$ color grids and the fundamental effective temperatures given in B13 and N93. For the majority of stars the color grids can predict the effective temperature to within about 5 $\%$. A slight systematic trend is noted in Figure~\ref{fig:teff-cal-3}, such that the model color grids overpredict $T_\mathrm{eff}$ at low temperatures and underpredict $T_\mathrm{eff}$ at high temperatures. We attempt to correct for this systematic effect by applying $T_\mathrm{eff}$ offsets in three regimes according to the mean behavior of each group: late and intermediate group stars were shifted to cooler temperatures by 2.4\% and 2.0\%, respectively, and early group stars were shifted by 1.9\% toward hotter temperatures. After offsets were applied, the remaining RMS error in temperature determinations for these ``standard'' stars was 3.3\%, 2.5\%, and 3.5\% for the late, intermediate, and early groups, respectively, or 3.1\% overall. 

Taking the uncertainties or dispersions in the fundamental $T_\mathrm{eff}$ determinations as the standard error, there is typically a 5-6 $\sigma$ discrepancy between the fundamental and photometric $T_\mathrm{eff}$ determinations. However, given the large author-to-author dispersion observed for stars with multiple fundamental $T_\mathrm{eff}$ determinations, it is likely that the formal errors on these measurements are underestimated. Notably, N93 does not publish errors for the fundamental $T_\mathrm{eff}$ values, which are literature means. However, those authors did find fractional errors in their photometric $T_\mathrm{eff}$ ranging from 2.5-4\% for BA stars. 

In \S ~\ref{subsec:openclustertests}, we opted not to apply systematic offsets, instead assigning $T_\mathrm{eff}$ uncertainties in three regimes according to the average fractional uncertainties noted in each group. In our final $T_\mathrm{eff}$ determinations for our field star sample (\S ~\ref{subsec:fieldstars}) we attempted to correct for the slight temperature systematics and applied offsets, using the magnitude of the remaining RMS error (for all groups considered collectively) as the dominant source of uncertainty in our $T_\mathrm{eff}$ measurement (see \S ~\ref{subsec:tefflogguncertainties}).

\begin{figure}
\centering
\includegraphics[width=0.48\textwidth]{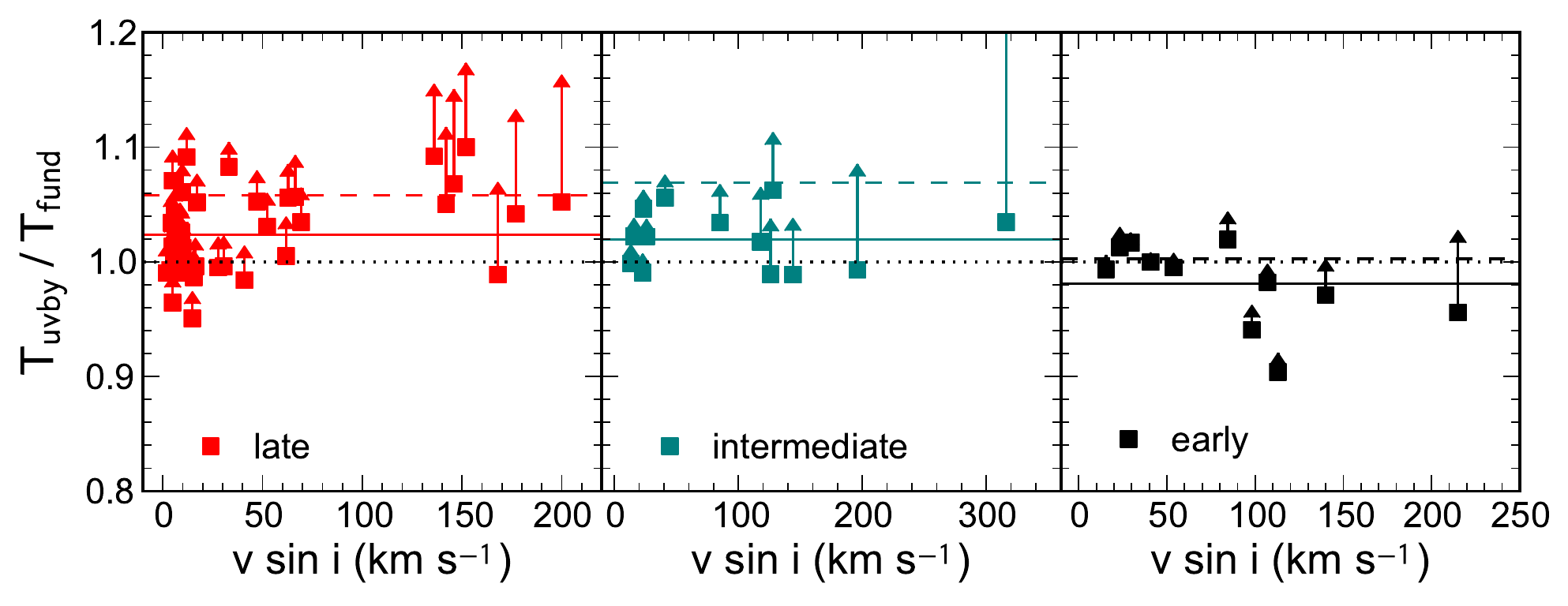}
\caption{Ratio of the $uvby\beta$ temperature to fundamental temperature as a function of $v \sin i$, for the late (left), intermediate (middle), and early (right), group stars. The solid horizontal colored lines indicate the mean ratios in each case. The arrows reperesent both the magnitude and direction of change to the ratio $T_{uvby}/T_\mathrm{fund}$ after applying the FB98 rotation corrections. The dashed horizontal colored lines indicate the mean ratios after application of the rotation correction. The rotation correction appears to improve temperature estimates for early group stars, but worsen estimates for the late and intermediate groups. Notably, however, the vast majority of $T_\mathrm{eff}$ standards are slowly rotating ($v\sin{i}<150$ km s$^{-1}$). Note one rapidly rotating intermediate group star extends beyond the scale of the figure, with a rotation corrected $T_{uvby}/T_\mathrm{fund}$ ratio of $\approx 1.26$.}
\label{fig:teffcal-vsini}
\end{figure}

As demonstrated in Figure ~\ref{fig:teffcal-vsini}, rotational effects on our temperature determinations for the $T_\mathrm{eff}$ standards were investigated. Notably, the FB98 $v\sin{i}$ corrections appear to enhance the discrepancy between our temperature determinations and the fundamental temperatures for the late and intermediate groups, while moderately improving the accuracy for the early group. For the late group this is expected, as the correction formulae were originally derived for intermediate and early group stars. Notably, however, only two stars in the calibration sample exhibit projected rotational velocities $>200$ km s$^{-1}$. We examine the utility of the $v\sin{i}$ correction further in \S ~\ref{subsec:loggvalidation} \& \S ~\ref{subsec:openclustertests}. 

\begin{figure}
\centering
\includegraphics[width=0.3\textwidth]{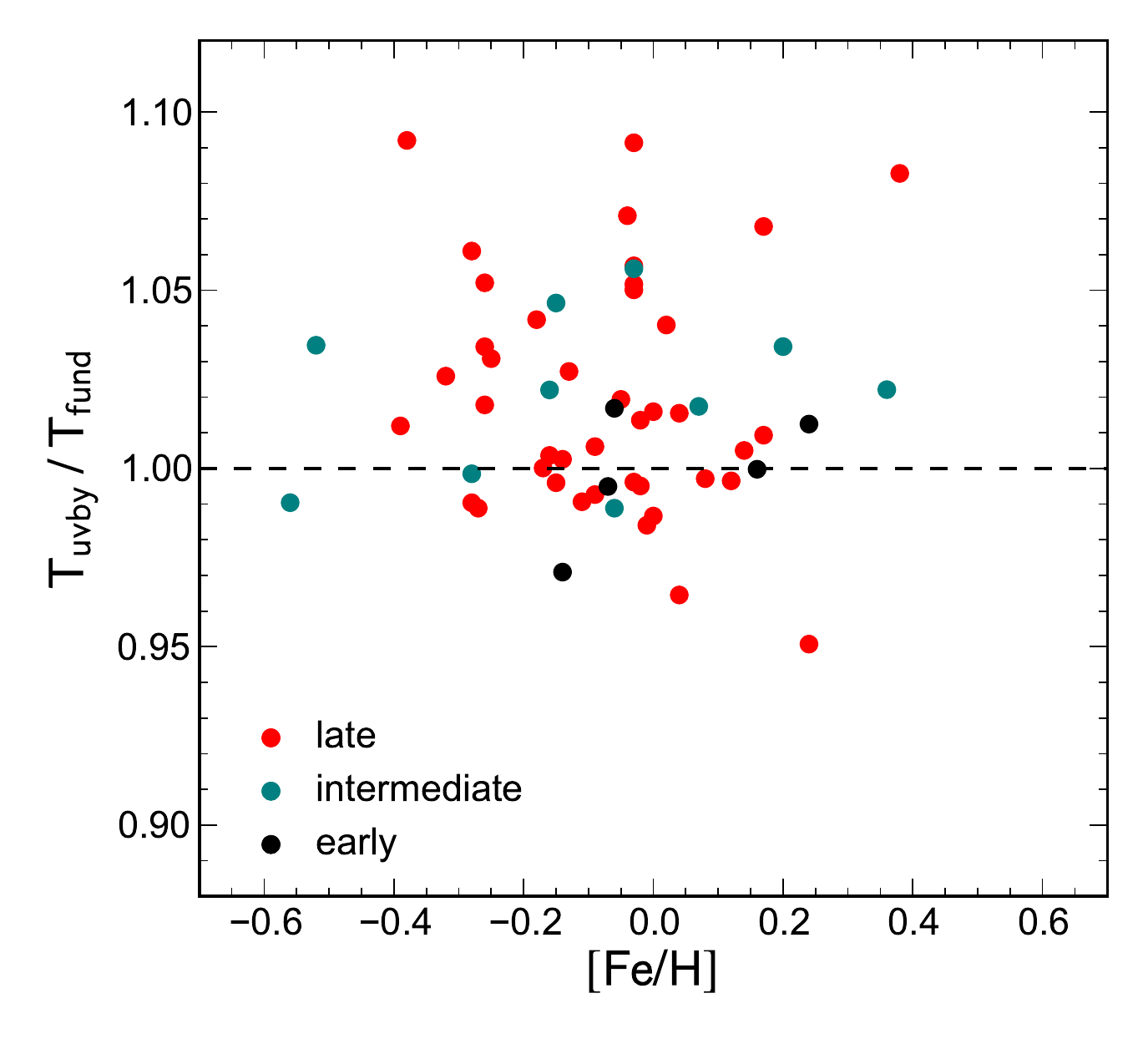}
\caption{Ratio of the $uvby\beta$ temperature to fundamental temperature as a function of [Fe/H]. There is no indication that the grids systematically overestimate or underestimate $T_\mathrm{eff}$ for different values of [Fe/H].}
\label{fig:teff-cal-2}
\end{figure}

The effect of metallicity on the determination of $T_\mathrm{eff}$ from the $uvby\beta$ grids is investigated in Figure \ref{fig:teff-cal-2} showing the ratio of the grid-determined temperature to the fundamental temperature as a function of [Fe/H]. The sample of temperature standards spans a large range in metallicity, yet there is no indication of any systematic effect with [Fe/H], justifying our choice to assume solar metallicity throughout this work (see further discussion of metallicity effects in the Appendix).

The effect of reddening on our temperature determinations was considered but since the vast majority of sources with fundamental effective temperatures are nearby, no significant reddening was expected. Indeed, no indication of a systematic trend of the temperature residuals as a function of distance was noted.

In summary our findings that the ATLAS9 predicted $T_\mathrm{eff}$ values are $\sim 2 \%$ hotter than fundamental values for AF stars are consistent with the results of \citet{bertone2004}, who found 4-8\% shifts warmer in $T_\mathrm{eff}$ from fits of ATLAS9 models to spectrophotometry relative to $T_\mathrm{eff}$ values determined from the infrared flux method (IRFM). We attempt systematic corrections with offsets of magnitude $\sim 2\%$ according to group, and the remaining RMS error between $uvby\beta$ temperatures and fundamental values is $\sim 3\%$.

\begin{deluxetable*}{cccccccccccc}
\tabletypesize{\footnotesize}
\tablecaption{Stars with fundamental determinations of $T_\mathrm{eff}$ through Interferometry\label{table:teffcal}}
\tablehead{
\colhead{HD} & 
\colhead{Sp. Type} & 
\colhead{$T_\mathrm{fund}$} & 
\colhead{Radius Ref.$^{a}$} & 
\colhead{$T_{uvby\beta}$} &
\colhead{$\log{g_{uvby\beta}}$} &
\colhead{[Fe/H]} & 
\colhead{$v \sin i$} & 
\colhead{$(b-y)$} & 
\colhead{$m_1$} & 
\colhead{$c_1$} & 
\colhead{$\beta$} 
\\
 & & (K) & & (K) & (dex) & (dex) & (km s$^{-1}$) & (mag) & (mag) & (mag) & (mag)
} 
\startdata

4614 & F9V & 5973 $\pm$ 8 & 3 & 5915 & 4.442 & -0.28 & 1.8 & 0.372 & 0.185 & 0.275 & 2.588 \\
5015 & F8V & 5965 $\pm$ 35 & 3 & 6057 & 3.699 & 0.04 & 8.6 & 0.349 & 0.174 & 0.423 & 2.613 \\
5448 & A5V & 8070  & 18 & 8350 & 3.964 & \textemdash & 69.3 & 0.068 & 0.189 & 1.058 & 2.866 \\
6210 & F6Vb & 6089 $\pm$ 35 & 1 & 5992 & 3.343 & -0.01 & 40.9 & 0.356 & 0.183 & 0.475 & 2.615 \\
9826 & F8V & 6102 $\pm$ 75 & 2,4 & 6084 & 3.786 & 0.08 & 8.7 & 0.346 & 0.176 & 0.415 & 2.629 \\
16765 & F7V & 6356 $\pm$ 46 & 1 & 6330 & 4.408 & -0.15 & 30.5 & 0.318 & 0.160 & 0.355 & 2.647 \\
16895 & F7V & 6153 $\pm$ 25 & 3 & 6251 & 4.118 & 0.00 & 8.6 & 0.325 & 0.160 & 0.392 & 2.625 \\
17081 & B7V & 12820  & 18 & 12979 & 3.749 & 0.24 & 23.3 & -0.057 & 0.104 & 0.605 & 2.717 \\
19994 & F8.5V & 5916 $\pm$ 98 & 2 & 5971 & 3.529 & 0.17 & 7.2 & 0.361 & 0.185 & 0.422 & 2.631 \\
22484 & F9IV-V & 5998 $\pm$ 39 & 3 & 5954 & 3.807 & -0.09 & 3.7 & 0.367 & 0.173 & 0.376 & 2.615 \\
30652 & F6IV-V & 6570 $\pm$ 131 & 3,6 & 6482 & 4.308 & 0.00 & 15.5 & 0.298 & 0.163 & 0.415 & 2.652 \\
32630 & B3V & 17580  & 18 & 16536 & 4.068 & \textemdash & 98.2 & -0.085 & 0.104 & 0.318 & 2.684 \\
34816 & B0.5IV & 27580  & 18 & 28045 & 4.286 & -0.06 & 29.5 & -0.119 & 0.073 & -0.061 & 2.602 \\
35468 & B2III & 21230  & 18 & 21122 & 3.724 & -0.07 & 53.8 & -0.103 & 0.076 & 0.109 & 2.613 \\
38899 & B9IV & 10790  & 18 & 11027 & 3.978 & -0.16 & 25.9 & -0.032 & 0.141 & 0.906 & 2.825 \\
47105 & AOIV & 9240  & 18 & 9226 & 3.537 & -0.28 & 13.3 & 0.007 & 0.149 & 1.186 & 2.865 \\
48737 & F5IV-V & 6478 $\pm$ 21 & 3 & 6510 & 3.784 & 0.14 & 61.8 & 0.287 & 0.169 & 0.549 & 2.669 \\
48915 & A0mA1Va & 9755 $\pm$ 47 & 7,8,9,10,11 & 9971 & 4.316 & 0.36 & 15.8 & -0.005 & 0.162 & 0.980 & 2.907 \\
49933 & F2Vb & 6635 $\pm$ 90 & 12 & 6714 & 4.378 & -0.39 & 9.9 & 0.270 & 0.127 & 0.460 & 2.662 \\
56537 & A3Vb & 7932 $\pm$ 62 & 3 & 8725 & 4.000 & \textemdash & 152 & 0.047 & 0.198 & 1.054 & 2.875 \\
58946 & F0Vb & 6954 $\pm$ 216 & 3,18 & 7168 & 4.319 & -0.25 & 52.3 & 0.215 & 0.155 & 0.615 & 2.713 \\
61421 & F5IV-V & 6563 $\pm$ 24 & 11,13,14,15,18 & 6651 & 3.983 & -0.02 & 4.7 & 0.272 & 0.167 & 0.532 & 2.671 \\
63922 & BOIII & 29980  & 18 & 29973 & 4.252 & 0.16 & 40.7 & -0.122 & 0.043 & -0.092 & 2.590 \\
69897 & F6V & 6130 $\pm$ 58 & 1 & 6339 & 4.290 & -0.26 & 4.3 & 0.315 & 0.149 & 0.384 & 2.635 \\
76644 & A7IV & 7840  & 18 & 8232 & 4.428 & -0.03 & 142 & 0.104 & 0.216 & 0.856 & 2.843 \\
80007 & A2IV & 9240  & 18 & 9139 & 3.240 & \textemdash & 126 & 0.004 & 0.140 & 1.273 & 2.836 \\
81937 & F0IVb & 6651 $\pm$ 27 & 3 & 7102 & 3.840 & 0.17 & 146 & 0.211 & 0.180 & 0.752 & 2.733 \\
82328 & F5.5IV-V & 6299 $\pm$ 61 & 3,18 & 6322 & 3.873 & -0.16 & 7.1 & 0.314 & 0.153 & 0.463 & 2.646 \\
90839 & F8V & 6203 $\pm$ 56 & 3 & 6145 & 4.330 & -0.11 & 8.6 & 0.341 & 0.171 & 0.333 & 2.618 \\
90994 & B6V & 14010  & 18 & 14282 & 4.219 & \textemdash & 84.5 & -0.066 & 0.111 & 0.466 & 2.730 \\
95418 & A1IV & 9181 $\pm$ 11 & 3,18 & 9695 & 3.899 & -0.03 & 40.8 & -0.006 & 0.158 & 1.088 & 2.880 \\
97603 & A5IV(n) & 8086 $\pm$ 169 & 3,6,18 & 8423 & 4.000 & -0.18 & 177 & 0.067 & 0.195 & 1.037 & 2.869 \\
102647 & A3Va & 8625 $\pm$ 175 & 5,6,18 & 8775 & 4.188 & 0.07 & 118 & 0.043 & 0.211 & 0.973 & 2.899 \\
102870 & F8.5IV-V & 6047 $\pm$ 7 & 3,18 & 6026 & 3.689 & 0.12 & 5.4 & 0.354 & 0.187 & 0.416 & 2.628 \\
118098 & A2Van & 8097 $\pm$ 43 & 3 & 8518 & 4.163 & -0.26 & 200 & 0.065 & 0.183 & 1.006 & 2.875 \\
118716 & B1III & 25740  & 18 & 23262 & 3.886 & \textemdash & 113 & -0.112 & 0.058 & 0.040 & 2.608 \\
120136 & F7IV-V & 6620 $\pm$ 67 & 2 & 6293 & 3.933 & 0.24 & 14.8 & 0.318 & 0.177 & 0.439 & 2.656 \\
122408 & A3V & 8420  & 18 & 8326 & 3.500 & -0.27 & 168 & 0.062 & 0.164 & 1.177 & 2.843 \\
126660 & F7V & 6202 $\pm$ 35 & 3,6,18 & 6171 & 3.881 & -0.02 & 27.7 & 0.334 & 0.156 & 0.418 & 2.644 \\
128167 & F4VkF2mF1 & 6687 $\pm$ 252 & 3,18 & 6860 & 4.439 & -0.32 & 9.3 & 0.254 & 0.134 & 0.480 & 2.679 \\
130948 & F9IV-V & 5787 $\pm$ 57 & 1 & 5899 & 4.065 & -0.05 & 6.3 & 0.374 & 0.191 & 0.321 & 2.625 \\
136202 & F8IV & 5661 $\pm$ 87 & 1 & 6062 & 3.683 & -0.04 & 4.9 & 0.348 & 0.170 & 0.427 & 2.620 \\
141795 & kA2hA5mA7V & 7928 $\pm$ 88 & 3 & 8584 & 4.346 & 0.38 & 33.1 & 0.066 & 0.224 & 0.950 & 2.885 \\
142860 & F6V & 6295 $\pm$ 74 & 3,6 & 6295 & 4.130 & -0.17 & 9.9 & 0.319 & 0.150 & 0.401 & 2.633 \\
144470 & BlV & 25710  & 18 & 25249 & 4.352 & \textemdash & 107 & -0.112 & 0.043 & -0.005 & 2.621 \\
162003 & F5IV-V & 5928 $\pm$ 81 & 3 & 6469 & 3.916 & -0.03 & 11.9 & 0.294 & 0.147 & 0.497 & 2.661 \\
164259 & F2V & 6454 $\pm$ 113 & 3 & 6820 & 4.121 & -0.03 & 66.4 & 0.253 & 0.153 & 0.560 & 2.690 \\
168151 & F5Vb & 6221 $\pm$ 39 & 1 & 6600 & 4.203 & -0.28 & 9.7 & 0.281 & 0.143 & 0.472 & 2.653 \\
169022 & B9.5III & 9420  & 18 & 9354 & 3.117 & \textemdash & 196 & 0.016 & 0.102 & 1.176 & 2.778 \\
172167 & AOVa & 9600  & 18 & 9507 & 3.977 & -0.56 & 22.8 & 0.003 & 0.157 & 1.088 & 2.903 \\
173667 & F5.5IV-V & 6333 $\pm$ 37 & 3,18 & 6308 & 3.777 & -0.03 & 16.3 & 0.314 & 0.150 & 0.484 & 2.652 \\
177724 & A0IV-Vnn & 9078 $\pm$ 86 & 3 & 9391 & 3.870 & -0.52 & 316 & 0.013 & 0.146 & 1.080 & 2.875 \\
181420 & F2V & 6283 $\pm$ 106 & 16 & 6607 & 4.187 & -0.03 & 17.1 & 0.280 & 0.157 & 0.477 & 2.657 \\
185395 & F3+V & 6516 $\pm$ 203 & 3,4 & 6778 & 4.296 & 0.02 & 5.8 & 0.261 & 0.157 & 0.502 & 2.688 \\
187637 & F5V & 6155 $\pm$ 85 & 16 & 6192 & 4.103 & -0.09 & 5.4 & 0.333 & 0.151 & 0.380 & 2.631 \\
190993 & B3V & 17400  & 18 & 16894 & 4.195 & -0.14 & 140 & -0.083 & 0.100 & 0.295 & 2.686 \\
193432 & B9.5V & 9950  & 18 & 10411 & 3.928 & -0.15 & 23.4 & -0.021 & 0.134 & 1.015 & 2.852 \\
193924 & B2IV & 17590  & 18 & 17469 & 3.928 & \textemdash & 15.5 & -0.092 & 0.087 & 0.271 & 2.662 \\
196867 & B9IV & 10960  & 18 & 10837 & 3.861 & -0.06 & 144 & -0.019 & 0.125 & 0.889 & 2.796 \\
209952 & B7IV & 13850  & 18 & 13238 & 3.913 & \textemdash & 215 & -0.061 & 0.105 & 0.576 & 2.728 \\
210027 & F5V & 6324 $\pm$ 139 & 6 & 6496 & 4.187 & -0.13 & 8.6 & 0.294 & 0.161 & 0.446 & 2.664 \\
210418 & A2Vb & 7872 $\pm$ 82 & 3 & 8596 & 3.966 & -0.38 & 136 & 0.047 & 0.161 & 1.091 & 2.886 \\
213558 & A1Vb & 9050 $\pm$ 157 & 3 & 9614 & 4.175 & \textemdash & 128 & 0.002 & 0.170 & 1.032 & 2.908 \\
215648 & F6V & 6090 $\pm$ 22 & 3 & 6198 & 3.950 & -0.26 & 7.7 & 0.331 & 0.147 & 0.407 & 2.626 \\
216956 & A4V & 8564 $\pm$ 105 & 5,18 & 8857 & 4.198 & 0.20 & 85.1 & 0.037 & 0.206 & 0.990 & 2.906 \\
218396 & F0+($\lambda$ Boo) & 7163 $\pm$ 84 & 17 & 7540 & 4.435 & \textemdash & 47.2 & 0.178 & 0.146 & 0.678 & 2.739 \\
219623 & F8V & 6285 $\pm$ 94 & 1 & 6061 & 3.85 & 0.04 & 4.9 & 0.351 & 0.169 & 0.395 & 2.624 \\
222368 & F7V & 6192 $\pm$ 26 & 3 & 6207 & 3.988 & -0.14 & 6.1 & 0.330 & 0.163 & 0.399 & 2.625 \\
222603 & A7V & 7734 $\pm$ 80 & 1 & 8167 & 4.318 & \textemdash & 62.8 & 0.105 & 0.203 & 0.891 & 2.826

\enddata
\tablenotetext{a}{Interferometric radii references: (1) \cite{boyajian2013}, (2) \cite{baines2008}, (3) \cite{boyajian2012}, (4) \cite{ligi2012}, (5) \cite{difolco2004}, (6) \cite{vanbelle2009}, (7) \cite{davis2011}, (8) \cite{hanburybrown1974}, (9) \cite{davis1986}, (10) \cite{kervella2003a}, (11) \cite{mozurkewich2003}, (12) \cite{bigot2011}, (13) \cite{chiavassa2012}, (14) \cite{nordgren2001}, (15) \cite{kervella2004}, (16) \cite{huber2012}, (17) \cite{baines2012}, (18) \cite{napiwotzki1993}. Note, that (18) simply provides means of the $T_\mathrm{eff}$ values published by \cite{code1976, beeckmans1977, malagnini1986}, all three of which used the radii of (9).}
\end{deluxetable*}

%%%%%%%%%%%%%%%%%%%%%%%%%%%%%%%
% SURFACE GRAVITY CALIBRATION %
%%%%%%%%%%%%%%%%%%%%%%%%%%%%%%%
\subsection{Surface Gravity}
\label{subsec:loggvalidation}

To assess the surface gravities derived from the $uvby\beta$ grids, we compare to results on both double-lined eclipsing binary and spectroscopic samples.

\subsubsection{Comparison with Double-Line Eclipsing Binaries}

\cite{torres2010} compiled an extensive catalog of 95 double-lined eclipsing binaries with fundamentally determined surface gravities for all 180 individual stars. Eclipsing binary systems allow for dynamical determinations of the component masses and geometrical determinations of the component radii. From the mass and radius of an individual component, the Newtonian surface gravity, $g=GM/R^2$ can be calculated.

From these systems, 39 of the primary components have $uvby\beta$ photometry available for determining surface gravities using our methodology. The spectral type range for these systems is O8-F2, with luminosity classes of IV, V. The mass ratio (primary/secondary) for these systems ranges from $\approx$ 1.00-1.79, and the orbital periods of the primaries range from $\approx$ 1.57-8.44 days. In the cases of low mass ratios, the primary and secondary components should have nearly identical fundamental parameters, assuming they are coeval. In the cases of high mass ratios, given that the individual components are presumably unresolved, we assume that the primary dominates the $uvby\beta$ photometry. For both cases (of low and high mass ratios), we assume that the photometry allows for accurate surface gravity determinations for the primary components and so we only consider the primaries from the \cite{torres2010} sample.

It is important to note that the eclipsing binary systems used for the surface gravity calibration are more distant than the stars for which we can interferometrically determine angular diameters and effective temperatures for. Thus, for the surface gravity calibration it was necessary to compute the dereddened indices $(b-y)_0, m_0, c_0$ in order to obtain the highest accuracy possible for the intermediate-group stars, which rely on $a_0$ (an index using dereddened colors) as a temperature indicator. Notably, however, we found that the dereddened photometry actually worsened $\log{g}$ determinations for the early and late groups. Dereddened colors were computed using the IDL routine \texttt{UVBYBETA}.

\begin{figure*}
\centering
\includegraphics[width=0.99\textwidth]{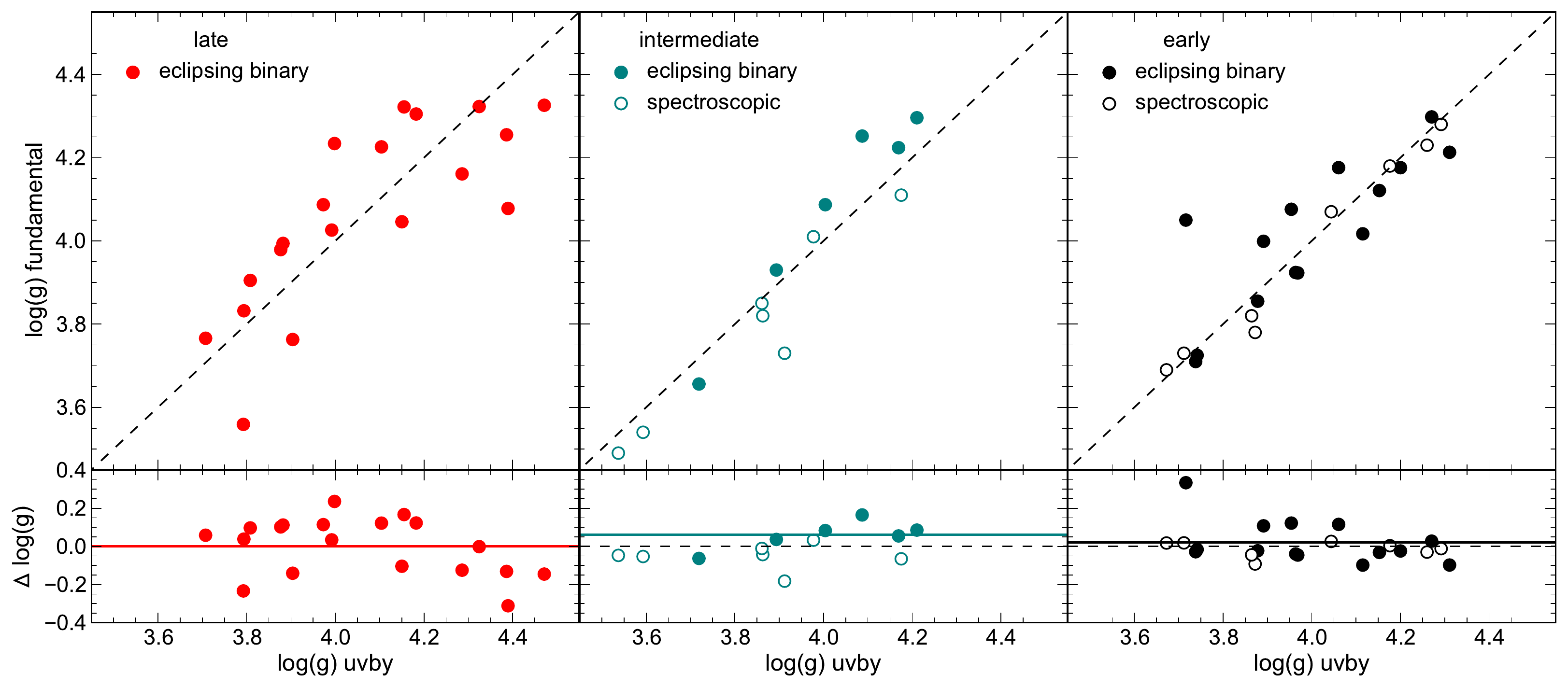}
\caption{Comparison of the $uvby\beta$ derived $\log g$ values with fundamental values for the primary components of the double lined eclipsing binaries compiled in \cite{torres2010}. Red, teal, and black points represent late, intermediate, and early group stars, respectively. In each case the solid colored line represents the mean of the residuals, $\Delta \log g$ (in the sense of fundamental-$uvby\beta$). As can be seen, the mean offsets for the late and early groups is negligible. For the intermediate group, however, while only five stars were used for calibration, the $uvby\beta$ $\log g$ values are about 0.13 dex lower than the fundamental values on average.}
\label{fig:logg-cal-fig1}
\end{figure*}

The results of the $\log g$ calibration are presented in Table~\ref{table:loggcal} and Figure ~\ref{fig:logg-cal-fig1}.  As described above, for the late group stars. ($T_\mathrm{eff} <$8500 K), $\log g$ is determined in the $(b-y)-c_1$ plane. The mean and median of the $\log g$ residuals (in the sense of grid-fundamental) are -0.001 dex and -0.038 dex, respectively, and the RMS error 0.145 dex. As in \S ~\ref{subsec:teffvalidation}, we found that the $\beta-c_1$  plane produced less accurate atmospheric parameters, relative to fundamental determinations, for late group stars.

For the intermediate group stars (8500 K $\leq T_\mathrm{eff} \leq$ 11000 K), $\log g$ is determined in the $a_0-r^*$ plane. The mean and median of the $\log g$ residuals are -0.060 dex and -0.069 dex, respectively with RMS error 0.091 dex. For the early group stars ($T_\mathrm{eff} >$ 11000 K), $\log g$ is determined in the $c_1-\beta$ plane. The mean and median of the $\log g$ residuals are -0.0215 dex and 0.024 dex, respectively, with RMS error 0.113 dex. The $[u-b]-\beta$ plane was also investigated for early group stars, but was found to produce $\log{g}$ values of lower accuracy relative to the fundamental determinations.

When considered collectively, the mean and median of the $\log g$ residuals for all stars are -0.017 dex and -0.034 dex, and the RMS error 0.127 dex.
The uncertainties in our surface gravities that arise from propagating the photometric errors through our atmospheric parameter determination routines are of the order $\sim$ 0.02 dex, significantly lower than the uncertainties demonstrated by the comparison to fundamental values of $\log g$.

As stated above, the main concern with using double-lined eclipsing binaries as surface gravity calibrators for our photometric technique is contamination from the unresolved secondary components. The $\log g$ residuals were examined as a function of both mass ratio and orbital period. While the amplitude of the scatter is marginally larger for low mass ratio or short period systems, in all cases our $\log g$ determinations are within 0.2 dex of the fundamental values $\approx 85\%$ of the time.

\begin{figure}
\centering
\includegraphics[width=0.45\textwidth]{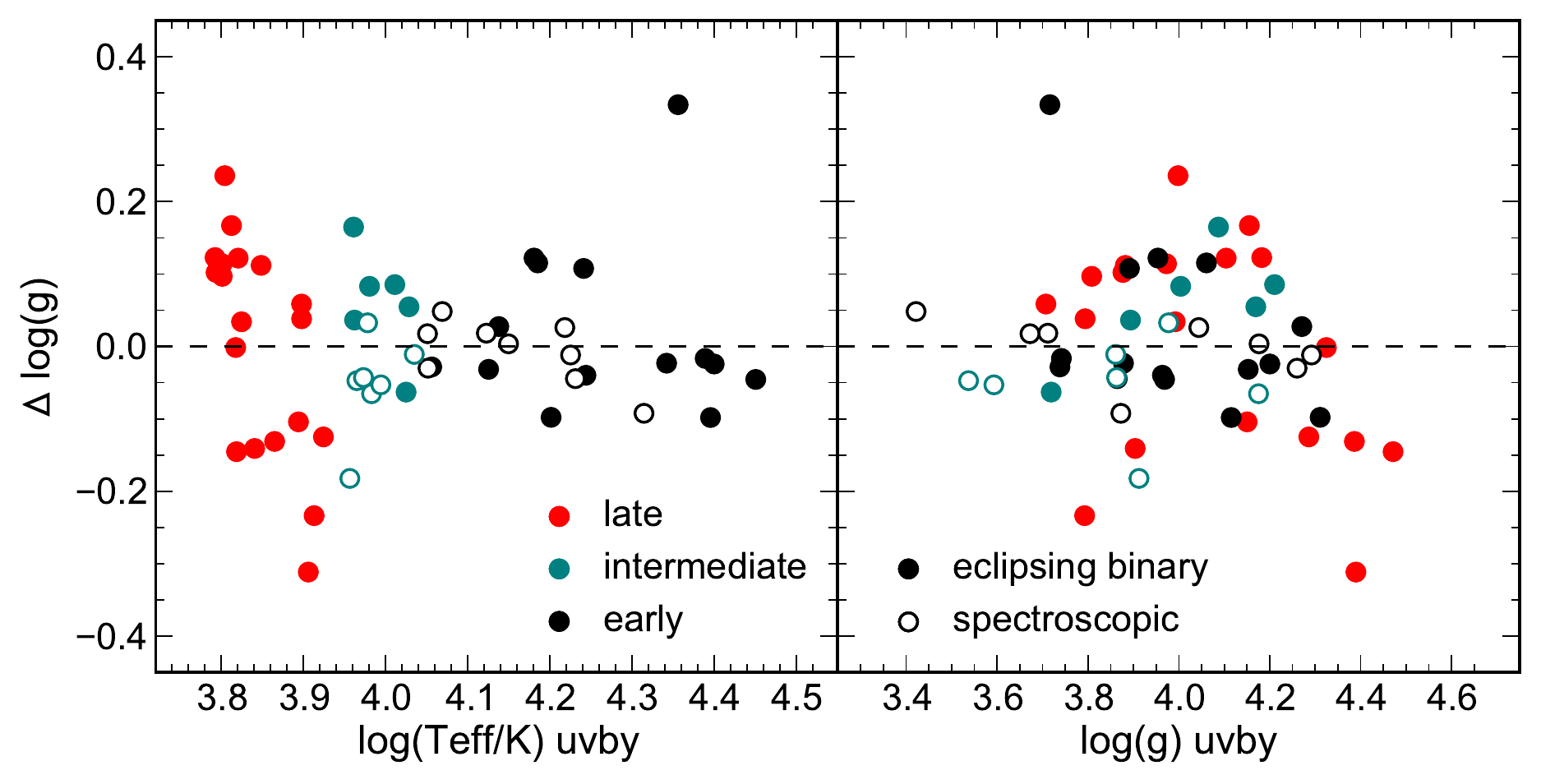}
\caption{Surface gravity residuals, $\Delta \log g$ (in the sense of fundamental-$uvby\beta$), as a function of $uvby\beta$-determined $\log{(T_\mathrm{eff})}$ (left) and $\log{g}$ (right). Solid points represent eclipsing binary primaries from \cite{torres2010} and open circles are stars with spectroscopic $\log{g}$ determinations in N93. Of the 39 eclipsing binaries, only six have residuals greater than 0.2 dex in magnitude. This implies that the $uvby\beta$ grids determine $\log g$ to within 0.2 dex of fundamental values $\sim 85\%$ of the time. Surface gravity residuals are largest for the cooler stars. Photometric surface gravity measurements are in better agreement with spectroscopic determinations than the eclipsing binary sample. There is no indication for a global systematic offset in $uvby\beta$-determined $\log g$ values as a function of either $T_\mathrm{eff}$ or $\log g$.
}
\label{fig:logg-cal-fig3}
\end{figure}

To assess any potential systematic inaccuracies of the grids themselves, the surface gravity residuals were examined as a function of $T_\mathrm{eff}$ and the grid-determined $\log g$. Figures \ref{fig:logg-cal-fig3} show the $\log g$ residuals as a function of $T_\mathrm{eff}$ and $\log g$, respectively. No considerable systematic effects as a function of either effective temperature or $\log g$ were found in the $uvby\beta$ determinations of $\log g$.

\begin{figure}
\centering
\includegraphics[width=0.45\textwidth]{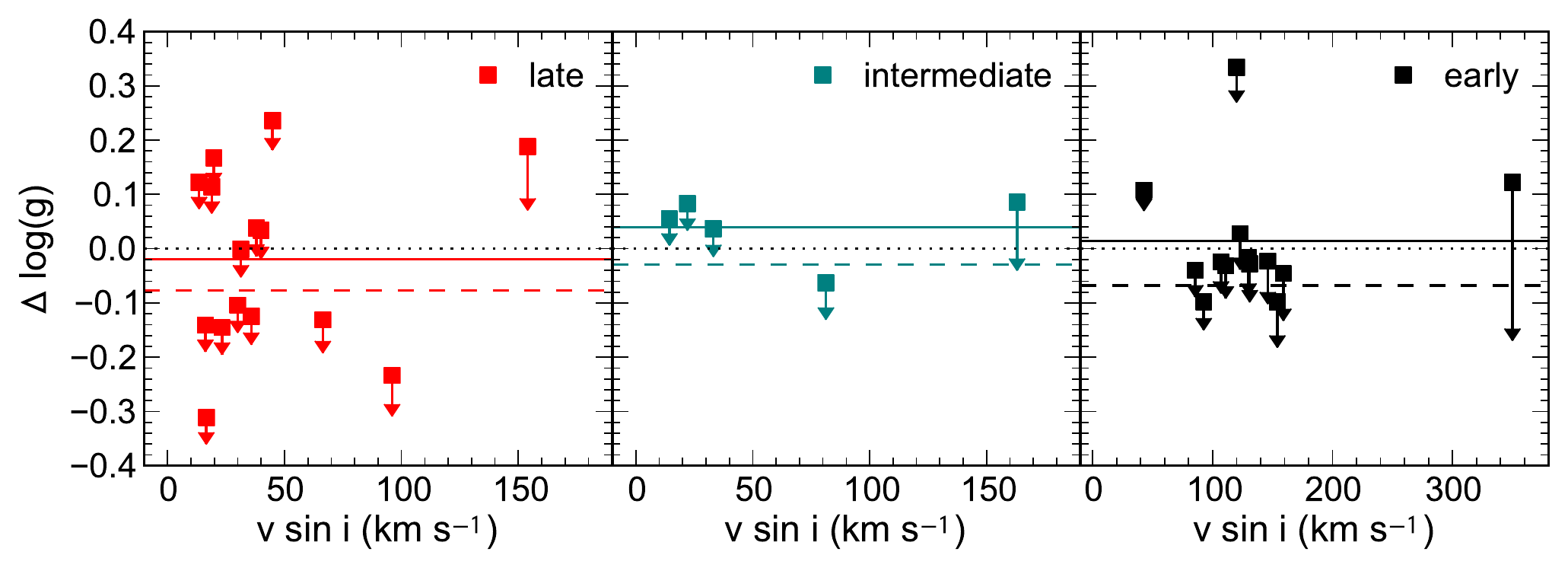}
\caption{Surface gravity residuals, $\Delta \log g$ (in the sense of fundamental-$uvby\beta$), of eclipsing binary primaries as a function of $v \sin i$. Arrows indicate the locations of points after application of the \cite{figueras1998} $v\sin{i}$ correction, where in this case late group stars received the same correction as the intermediate group.}
\label{fig:dlogg-vsini}
\end{figure}

The effect of rotational velocity on our $\log g$ determinations was considered. As before, $v \sin i$ data for the surface gravity calibrators was collected from \cite{glebocki2005}. As seen in Figure \ref{fig:dlogg-vsini}, the majority of the $\log g$ calibrators are somewhat slowly rotating ($v \sin i \leq$ 150 km s$^{-1}$). While the $v \sin i$ correction increases the accuracy of our $\log g$ determinations for the early-group stars in most cases, the correction appears to worsen our determinations for the intermediate group, which appear systematically high to begin with.

\begin{figure}
\centering
\includegraphics[width=0.45\textwidth]{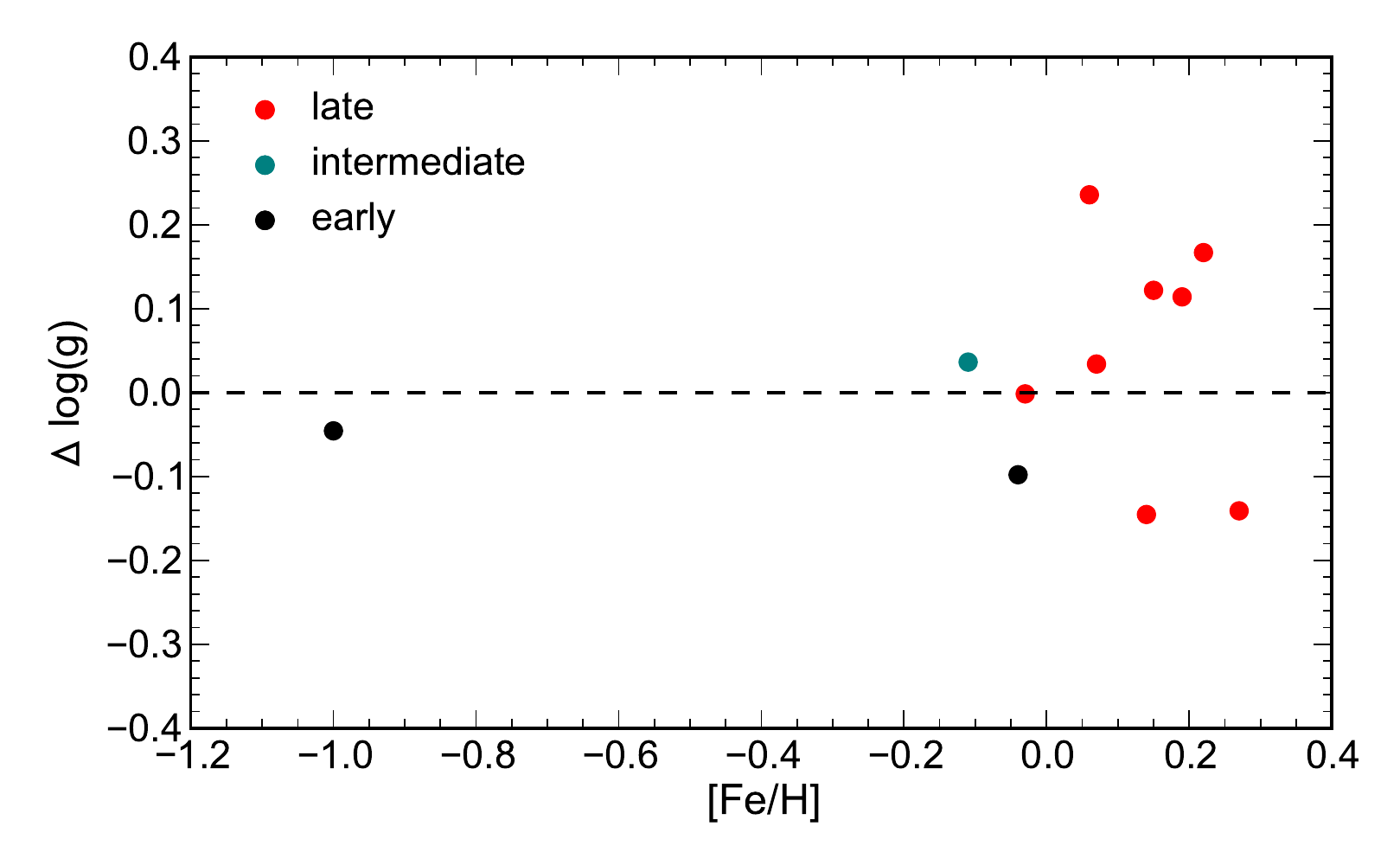}
\caption{Surface gravity residuals, $\Delta \log g$ (in the sense of fundamental-$uvby\beta$), as a function of [Fe/H]. The metallicity values have been taken primarily from \cite{ammons2006}, with additional values coming from \cite{anderson2012}. While metallicities seem to exist for very few of the surface gravity calibrators used here, there does not appear to be a systematic trend in the residuals with [Fe/H]. There is a larger amount of scatter for the more metal-rich late-type stars, however the scatter is confined to a relatively small range in [Fe/H] and it is not clear that this effect is due to metallicity effects.
}
\label{fig:logg-cal-fig8}
\end{figure}

The potential systematic effect of metallicity on our $\log g$ determinations is considered in Figure \ref{fig:logg-cal-fig8}, showing the surface gravity residuals as a function of [Fe/H]. Metallicity measurements were available for very few of these stars, and were primarily taken from \cite{ammons2006, anderson2012}. Nevertheless, there does not appear to be a global systematic trend in the surface gravity residuals with metallicity. There is a larger scatter in $\log g$ determinations for the more metal-rich, late-type stars, however it is not clear that this effect is strictly due to metallicity.

In summary, for the open cluster tests we assign $\log g$ uncertainties in three regimes: $\pm$ 0.145 dex for stars belonging to the late group, $\pm$ 0.091 dex for the intermediate group, and $\pm$ 0.113 dex for the early group. 

For our sample of nearby field stars we opt to assign a uniform systematic uncertainty of $\pm$ 0.116 dex for all stars. We do not attempt to correct for any systematic effects by applying offsets in $\log g$, as we did with $T_\mathrm{eff}$.  As noted in discussion of the $T_\mathrm{eff}$ calibration, we do apply the $v \sin i$ correction to both intermediate and early group stars, as these corrections permit us to better reproduce open cluster ages (as presented in \S ~\ref{subsec:openclustertests}).

\begin{deluxetable*}{ccccccccccccc}
\tabletypesize{\footnotesize}
\tablewidth{0.99\textwidth}
\tablecaption{Primary Components of Double-lined Eclipsing Binaries with fundamental determinations of $\log g$.}
\tablehead{
\colhead{Star} & 
\colhead{Sp. Type} & 
\colhead{$T_\mathrm{eff}$} & 
\colhead{$T_{uvby}$} & 
\colhead{$\log g_\mathrm{EB}$} & 
\colhead{$\log g_{uvby}$} & 
\colhead{$v \sin i$} &
\colhead{[Fe/H]} &
\colhead{$(b-y)$} & 
\colhead{$m_1$} & 
\colhead{$c_1$} &
\colhead{$\beta$} 
\\
 & & (K) & (K) & (dex) & (dex) & (km s$^{-1}$) & (dex) & (mag) & (mag) & (mag) & (mag)
}
\startdata

EM Car     & O8V     & 34000 $\pm$ 2000 & 21987 & 3.855 $\pm$ 0.016 & 3.878 & 146.0 & 	\textemdash & 0.279 & -0.042 & 0.083 & 2.617 \\
V1034 Sco  & O9V     & 33200 $\pm$ 900 & 28228 & 3.923 $\pm$ 0.008 & 3.969 & 159.0 & -1.0 & 0.190 & -0.024 & -0.068 & 2.587 \\
AH Cep     & B0.5Vn  & 29900 $\pm$ 1000 & 24867 & 4.017 $\pm$ 0.009 & 4.115 & 154.0 & 	\textemdash & 0.290 & -0.064 & 0.003 & 2.611 \\
V578 Mon   & B1V     & 30000 $\pm$ 740 & 25122 & 4.176 $\pm$ 0.015 & 4.200 & 107.0 & 	\textemdash & 0.206 & -0.024 & -0.003 & 2.613 \\
V453 Cyg   & B0.4IV  & 27800 $\pm$ 400 & 24496 & 3.725 $\pm$ 0.006 & 3.742 & 130.0 & 	\textemdash & 0.212 & -0.004 & -0.004 & 2.590 \\
CW Cep     & B0.5V   & 28300 $\pm$ 1000 & 22707 & 4.050 $\pm$ 0.019 & 3.716 & 120.0 & 	\textemdash & 0.355 & -0.077 & 0.050 & 2.601 \\
V539 Ara   & B3V     & 18100 $\pm$ 500 & 17537 & 3.924 $\pm$ 0.016 & 3.964 & 85.6 & 	\textemdash & -0.033 & 0.089 & 0.268 & 2.665 \\
CV Vel     & B2.5V   & 18100 $\pm$ 500 & 17424 & 3.999 $\pm$ 0.008 & 3.891 & 42.8 & 	\textemdash & -0.057 & 0.083 & 0.273 & 2.659 \\
AG Per     & B3.4V   & 18200 $\pm$ 800 & 15905 & 4.213 $\pm$ 0.020 & 4.311 & 92.6 & -0.04 & 0.048 & 0.079 & 0.346 & 2.708 \\
U Oph      & B5V     & 16440 $\pm$ 250 & 15161 & 4.076 $\pm$ 0.004 & 3.954 & 350.0 & 	\textemdash & 0.081 & 0.050 & 0.404 & 2.695 \\
V760 Sco   & B4V     & 16900 $\pm$ 500 & 15318 & 4.176 $\pm$ 0.019 & 4.061 & 	\textemdash & 	\textemdash & 0.169 & 0.023 & 0.392 & 2.701 \\
GG Lup     & B7V     & 14750 $\pm$ 450 & 13735 & 4.298 $\pm$ 0.009 & 4.271 & 123.0 & 	\textemdash & -0.049 & 0.115 & 0.514 & 2.747 \\
$\zeta$ Phe   & B6V     & 14400 $\pm$ 800 & 13348 & 4.121 $\pm$ 0.004 & 4.153 & 111.0 & 	\textemdash & -0.039 & 0.118 & 0.559 & 2.747 \\
$\chi^2$ Hya   & B8V     & 11750 $\pm$ 190 & 11382 & 3.710 $\pm$ 0.007 & 3.738 & 131.0 & 	\textemdash & -0.020 & 0.110 & 0.841 & 2.769 \\
V906 Sco   & B9V     & 10400 $\pm$ 500 & 10592 & 3.656 $\pm$ 0.012 & 3.719 & 81.3 & 	\textemdash & 0.039 & 0.101 & 0.996 & 2.805 \\
TZ Men     & A0V     & 10400 $\pm$ 500 & 10679 & 4.224 $\pm$ 0.009 & 4.169 & 14.4 & 	\textemdash & 0.000 & 0.142 & 0.918 & 2.850 \\
V1031 Ori  & A6V     & 7850 $\pm$ 500 & 8184 & 3.559 $\pm$ 0.007 & 3.793 & 96.0 & 	\textemdash & 0.076 & 0.174 & 1.106 & 2.848 \\
$\beta$ Aur   & A1m     & 9350 $\pm$ 200 & 9167 & 3.930 $\pm$ 0.005 & 3.894 & 33.2 & -0.11 & 0.017 & 0.173 & 1.091 & 2.889 \\
V364 Lac   & A4m:    & 8250 $\pm$ 150 & 7901 & 3.766 $\pm$ 0.005 & 3.707 & 	\textemdash & 	\textemdash & 0.107 & 0.168 & 1.061 & 2.875 \\
V624 Her   & A3m     & 8150 $\pm$ 150 & 7902 & 3.832 $\pm$ 0.014 & 3.794 & 38.0 & 	\textemdash & 0.111 & 0.230 & 1.025 & 2.870 \\
V1647 Sgr  & A1V     & 9600 $\pm$ 300 & 9142 & 4.252 $\pm$ 0.008 & 4.087 & 	\textemdash & 	\textemdash & 0.040 & 0.174 & 1.020 & 2.899 \\
VV Pyx     & A1V     & 9500 $\pm$ 200 & 9560 & 4.087 $\pm$ 0.008 & 4.004 & 22.1 & 	\textemdash & 0.028 & 0.161 & 1.013 & 2.881 \\
KW Hya     & A5m     & 8000 $\pm$ 200 & 8053 & 4.078 $\pm$ 0.006 & 4.390 & 16.6 & 	\textemdash & 0.122 & 0.232 & 0.832 & 2.827 \\
WW Aur     & A5m     & 7960 $\pm$ 420 & 8401 & 4.161 $\pm$ 0.005 & 4.286 & 35.8 & 	\textemdash & 0.081 & 0.231 & 0.944 & 2.862 \\
V392 Car   & A2V     & 8850 $\pm$ 200 & 10263 & 4.296 $\pm$ 0.011 & 4.211 & 163.0 & 	\textemdash & 0.097 & 0.108 & 1.019 & 2.889 \\
RS Cha     & A8V     & 8050 $\pm$ 200 & 7833 & 4.046 $\pm$ 0.022 & 4.150 & 30.0 & 	\textemdash & 0.136 & 0.186 & 0.866 & 2.791 \\
MY Cyg     & F0m     & 7050 $\pm$ 200 & 7054 & 3.994 $\pm$ 0.019 & 3.882 & 	\textemdash & 	\textemdash & 0.219 & 0.226 & 0.709 & 2.756 \\
EI Cep     & F3V     & 6750 $\pm$ 100 & 6928 & 3.763 $\pm$ 0.014 & 3.904 & 16.2 & 0.27 & 0.234 & 0.199 & 0.658 & 2.712 \\
FS Mon     & F2V     & 6715 $\pm$ 100 & 6677 & 4.026 $\pm$ 0.005 & 3.992 & 40.0 & 0.07 & 0.266 & 0.148 & 0.594 & 2.688 \\
PV Pup     & A8V     & 6920 $\pm$ 300 & 7327 & 4.255 $\pm$ 0.009 & 4.386 & 66.4 & 	\textemdash & 0.200 & 0.169 & 0.636 & 2.722 \\
HD 71636   & F2V     & 6950 $\pm$ 140 & 6615 & 4.226 $\pm$ 0.014 & 4.104 & 13.5 & 0.15 & 0.278 & 0.157 & 0.496 & \textemdash \\
RZ Cha     & F5V     & 6450 $\pm$ 150 & 6326 & 3.905 $\pm$ 0.006 & 3.808 & 	\textemdash & 0.02 & 0.312 & 0.155 & 0.482 & \textemdash \\
BW Aqr     & F7V     & 6350 $\pm$ 100 & 6217 & 3.979 $\pm$ 0.018 & 3.877 & 	\textemdash & 	\textemdash & 0.328 & 0.165 & 0.432 & 2.650 \\
V570 Per   & F3V     & 6842 $\pm$ 50 & 6371 & 4.234 $\pm$ 0.019 & 3.998 & 44.9 & 0.06 & 0.308 & 0.165 & 0.441 & \textemdash \\
CD Tau     & F6V     & 6200 $\pm$ 50 & 6325 & 4.087 $\pm$ 0.007 & 3.973 & 18.9 & 0.19 & 0.314 & 0.178 & 0.436 & \textemdash \\
V1143 Cyg  & F5V     & 6450 $\pm$ 100 & 6492 & 4.322 $\pm$ 0.015 & 4.155 & 19.8 & 0.22 & 0.294 & 0.165 & 0.451 & 2.663 \\
VZ Hya     & F3V     & 6645 $\pm$ 150 & 6199 & 4.305 $\pm$ 0.003 & 4.182 & 	\textemdash & -0.22 & 0.333 & 0.145 & 0.370 & 2.629 \\
V505 Per   & F5V     & 6510 $\pm$ 50 & 6569 & 4.323 $\pm$ 0.016 & 4.325 & 31.4 & -0.03 & 0.287 & 0.142 & 0.435 & 2.654 \\
HS Hya     & F4V     & 6500 $\pm$ 50 & 6585 & 4.326 $\pm$ 0.005 & 4.471 & 23.3 & 0.14 & 0.287 & 0.160 & 0.397 & 2.648

\enddata
\tablenotetext{}{Spectral type, temperature, and fundamental $\log{g}$ information originate from \cite{torres2010}. The $uvby\beta \log{g}$ values are from this work. Projected rotational velocities are from \cite{glebocki2005}, [Fe/H] from \cite{anderson2012, ammons2006}, and the $uvby\beta$ photometry are from HM98. The surface gravities in the column $\log{g_{uvby}}$ are derived together with $T_{uvby}$, and not for the $T_\mathrm{eff}$ values given by \cite{torres2010}.}
\label{table:loggcal}
\end{deluxetable*}

%%%%%%%%%%%%%%%%%%%%%%%%%%%%%%%%
% COMPARISON WITH SPECTROSCOPY %
%%%%%%%%%%%%%%%%%%%%%%%%%%%%%%%%

\subsubsection{Comparison with Spectroscopic Measurements}

The Balmer lines are a sensitive surface gravity indicator for stars hotter than $T_\mathrm{eff} \gtrsim$ 9000 K and can be used as a semi-fundamental surface gravity calibration for the early- and intermediate-group stars. The reason why surface gravities derived using this method are considered semi-fundamental and not fundamental is because the method still relies on model atmospheres for fitting the observed line profiles. Nevertheless, surface gravities determined through this method are considered of high fidelity and so we performed an additional consistency check, comparing our $uvby\beta$ values of $\log g$ to those with well-determined spectroscopic $\log g$ measurements.

N93 fit theoretical profiles of hydrogen Balmer lines from \cite{kurucz1979} to high resolution spectrograms of the H$\beta$ and H$\gamma$ lines for a sample of 16 stars with $uvby\beta$ photometry. The sample of 16 stars was mostly drawn from the list of photometric $\beta$ standards of \cite{crawford1966}. We compared the $\log g$ values we determined through interpolation in the $uvby\beta$ color grids to the semi-fundamental spectroscopic values determined by N93. The results of this comparison are presented in Table ~\ref{table:loggspec}.

Though N93 provide dereddened photometry for the spectroscopic sample, we found using the raw HM98 photometry produced significantly better results (yielding an RMS error that was three times lower). For the early group stars, the atmospheric parameters were determined in both the $c_0-\beta$ plane and the $[u-b]-\beta$ plane. In both cases, $\beta$ is the gravity indicator, but we found that the $\log g$ values calculated when using $c_0$ as a temperature indicator for hot stars better matched the semi-fundamental spectroscopic $\log g$ values. This result is consistent with the result from the effective temperature calibration that suggests $c_0$ better predicted the effective temperatures of hot stars than $[u-b]$.  As before, $\log{g}$ for intermediate group stars is determined in the $a_0-r^*$ plane.

We tested $uvby\beta$ color grids of different metallicity, alpha-enhancement, and microturbulent velocity and determined that the non-alpha-enhanced, solar metallicity grids with microturbulent velocity $v_\mathrm{turb}$ = 0 km s$^{-1}$ best reproduced the spectroscopic surface gravities for the sample of 16 early- and intermediate-group stars measured by N93.

The $\log g$ residuals, in the sense of (spectroscopic -- grid), as a function of the grid-calculated effective temperatures are plotted in Figure ~\ref{fig:logg-cal-fig3}. There is no evidence for a significant systematic offset in the residuals as a function of either the $uvby\beta$-determined $T_\mathrm{eff}$ or $\log{g}$. For the early group, the mean and median surface gravity residuals are -0.007 dex and 0.004 dex, respectively, with RMS 0.041 dex.  For the intermediate group, the mean and median surface gravity residuals are -0.053 dex and -0.047 dex, respectively, with RMS 0.081 dex.  Considering both early and intermediate group stars collectively, the mean and median surface gravity residuals are -0.027 dex and -0.021 dex, and the RMS 0.062 dex.

One issue that may cause statistically larger errors in the $\log g$ determinations compared to the $T_\mathrm{eff}$ determinations is the linear interpolation in a low resolution logarithmic space (the $uvby\beta$ colors are calculated at steps of 0.5 dex in $\log g$). In order to mitigate this effect one requires either more finely gridded models or an interpolation scheme that takes the logarithmic gridding into account.

\begin{deluxetable*}{ccccccccccc}
\tabletypesize{\footnotesize}
\tablewidth{0.99\textwidth}
\tablecaption{Stars with semi-fundamental determinations of $\log g$ through Balmer-line fitting.}
\tablehead{
\colhead{HR} & 
\colhead{Sp. Type} & 
\colhead{$T_\mathrm{eff}$} &
\colhead{$T_{uvby}$} &
\colhead{$\log g_\mathrm{spec}$} & 
\colhead{$\log g_{uvby}$} &
\colhead{$(b-y)$} &
\colhead{$m_1$} &
\colhead{$c_1$} &
\colhead{$[u-b]$} &
\colhead{$\beta$}
\\
 & & (K) & (K) & (dex) & (dex) & (mag) & (mag) & (mag) & (mag) & (mag)
}
\startdata

63 & A2V & 8970 & 9047 & 3.73 & 3.912 & 0.026 & 0.181 & 1.050 & 1.425 & 2.881 \\
153 & B2IV & 20930 & 20635 & 3.78 & 3.872 & -0.090 & 0.087 & 0.134 & 0.264 & 2.627 \\
1641 & B3V & 16890 & 16528 & 4.07 & 4.044 & -0.085 & 0.104 & 0.319 & 0.485 & 2.683 \\
2421 & AOIV & 9180 & 9226 & 3.49 & 3.537 & 0.007 & 0.149 & 1.186 & 1.487 & 2.865 \\
4119 & B6V & 14570 & 14116 & 4.18 & 4.176 & -0.062 & 0.111 & 0.481 & 0.673 & 2.730 \\
4554 & AOVe & 9360 & 9398 & 3.82 & 3.863 & 0.006 & 0.155 & 1.112 & 1.425 & 2.885 \\
5191 & B3V & 17320 & 16797 & 4.28 & 4.292 & -0.080 & 0.106 & 0.297 & 0.470 & 2.694 \\
6588 & B3IV & 17480 & 17025 & 3.82 & 3.864 & -0.065 & 0.079 & 0.292 & 0.418 & 2.661 \\
7001 & AOVa & 9540 & 9508 & 4.01 & 3.977 & 0.003 & 0.157 & 1.088 & 1.403 & 2.903 \\
7447 & B5III & 13520 & 13265 & 3.73 & 3.712 & -0.016 & 0.088 & 0.575 & 0.743 & 2.707 \\
7906 & B9IV & 10950 & 10838 & 3.85 & 3.861 & -0.019 & 0.125 & 0.889 & 1.130 & 2.796 \\
8585 & A1V & 9530 & 9615 & 4.11 & 4.175 & 0.002 & 0.170 & 1.032 & 1.373 & 2.908 \\
8634 & B8V & 11330 & 11247 & 3.69 & 3.672 & -0.035 & 0.113 & 0.868 & 1.077 & 2.768 \\
8781 & B9V & 9810 & 9868 & 3.54 & 3.593 & -0.011 & 0.128 & 1.129 & 1.380 & 2.838 \\
8965 & B8V & 11850 & 11721 & 3.47 & 3.422 & -0.031 & 0.100 & 0.784 & 0.969 & 2.725 \\
8976 & B9IVn & 11310 & 11263 & 4.23 & 4.260 & -0.035 & 0.131 & 0.831 & 1.076 & 2.833 \\

\enddata
\tablenotetext{}{Spectral type, $T_\mathrm{eff}$, and spectroscopic $\log{g}$ originate from N93. The $uvby\beta$ $T_\mathrm{eff}$ and $\log{g}$ values are from this work. Though N93 does not provide formal errors on the atmospheric parameters, those authors estimate uncertainties of $\sim$ 0.03 dex in their spectroscopically determined $\log{g}$. The fractional errors in their photometrically derived $T_\mathrm{eff}$ range from 2.5\% for stars cooler than $\approx$ 11000 K to 4\% for stars hotter than $\approx$ 20000 K. The photometry is from HM98.}
\label{table:loggspec}
\end{deluxetable*}

%%%%%%%%%%%%%%%%%%%%%%%%%%%%%%%%%%%%%%%
% ATMOSPHERIC PARAMETER UNCERTAINTIES %
%%%%%%%%%%%%%%%%%%%%%%%%%%%%%%%%%%%%%%%
\subsection{Summary of Atmospheric Parameter Uncertainties}
\label{subsec:tefflogguncertainties}

Precise and accurate stellar ages are the ultimate goal of this work. The accuracy of our ages is determined by both the accuracy with which we can determine atmospheric parameters and any systematic uncertainties associated with the stellar evolutionary models and our assumptions in applying them. The precision, on the other hand, is determined almost entirely by the precision with which we determine atmospheric parameters and, because there are some practical limits to how well we may ever determine $T_\mathrm{eff}$ and $\log{g}$, the location of the star in the H-R diagram (e.g. stars closer to the main sequence will always have more imprecise ages using this method).

It is thus important to provide a detailed accounting of the uncertainties involved in our atmospheric parameter determinations, as the final uncertainties quoted in our ages will arise purely from the values of the $\sigma_{T_\mathrm{eff}}, \sigma_{\log{g}}$ used in our $\chi^2$ calculations. Below we consider the contribution of the systematics already discussed, as well as the contributions from errors in interpolation, photometry, metallicity, extinction, rotational velocity, multiplicity, and spectral peculiarity.

{\bf Systematics:} The dominant source of uncertainty in our atmospheric parameter determinations are the systematics quantified in \S~\ref{subsec:teffvalidation} and \S~\ref{subsec:loggvalidation}. All systematic effects inherent to the $uvby\beta$ method, and the particular model color grids chosen, which we will call $\sigma_\mathrm{sys}$, are embedded in the comparisons to the stars with fundamentally or semi-fundamentally determined parameters, summarized as approximately $\sim 3.1\%$ in $T_\mathrm{eff}$ and $\sim 0.116$ dex in $\log{g}$.  We also found that for stars with available [Fe/H] measurements, the accuracy with which we can determine atmospheric parameters using $uvby\beta$ photometry does not vary systematically with metallicity, though we further address metallicity issues both below and in an Appendix.

{\bf Interpolation Precision:} 
To estimate the errors in atmospheric parameters due to the numerical precision of the interpolation procedures employed here, we generated 1000 random points in each of the three relevant $uvby\beta$ planes. For each point, we obtained ten independent $T_\mathrm{eff}, \log{g}$ determinations to test the repeatability of the interpolation routine. The scatter in independent determinations of the atmospheric parameters were found to be $<10^{-10}$ K, dex, and thus numerical errors are assumed zero.

%Propagation of Photometric Uncertainties
%========================================
{\bf Photometric Errors:} Considering the most basic element of our approach, there are uncertainties due to the propagation of photometric errors through our atmospheric parameter determination pipeline. As discussed in \S~\ref{subsec:fieldstars}, the photometric errors are generally small ($\sim 0.005$ mag in a given index).  Translating the model grid points in the rectangular regions defined by the magnitude of the mean photometric error in a given index, and then interpolating to find the associated atmospheric parameters of the perturbed point, we take the maximum and minimum values for $T_\mathrm{eff}$ and $\log{g}$ to calculate the error due to photometric measurement error.

To simplify the propagation of photometric errors for individual stars, we performed simulations with randomly generated data to ascertain the mean uncertainty in $T_\mathrm{eff}$, $\log{g}$ that results from typical errors in each of the $uvby\beta$ indices. 

We begin with the HM98 photometry and associated measurement errors for our sample (3499 stars within 100 pc, B0-F5, luminosity classes IV-V). Since the HM98 compilation does not provide $a_0$ or $r^*$, as these quantities are calculated from the four fundamental indices, we calculate the uncertainties in these parameters using the crude approximation that none of the $uvby\beta$ indices are correlated. Under this assumption, the uncertainties associated with $a_0$ and $r^*$ are as follows:

\begin{align}
\sigma_{a_0} &= \sqrt{1.36^2\sigma_{b-y}^2+0.36^2\sigma_{m_1}^2+0.18^2\sigma_{c_1}^2} \\
\sigma_{r^*} &= \sqrt{0.07^2\sigma_{b-y}^2+0.35^2\sigma_{c_1}^2+\sigma_{\beta}^2}.
\end{align}

A model for the empirical probability distribution function (hereafter PDF) for the error in a given $uvby\beta$ index is created through a normalized histogram with 25 bins. From this empirical PDF, one can randomly draw values for the error in a given index. For each $uvby\beta$ plane, 1,000 random points in the appropriate range of parameter space were generated with photometric errors drawn as described above. The eight ($T_\mathrm{eff}$, $\log{g}$) values corresponding to the corners and midpoints of the ``standard error rectangle'' centered on the original random data point are then evaluated. The maximally discrepant ($T_\mathrm{eff}$, $\log{g}$) values are saved and the overall distributions of $\Delta T_\mathrm{eff}/T_\mathrm{eff}$ and $\Delta \log{g}$ are then analyzed to assess the mean uncertainties in the atmospheric parameters derived in a given $uvby\beta$ plane due to the propagation of typical photometric errors.

For the late group, points were generated in the range of $(b-y)-c_1$ parameter space bounded by 6500 K $\leq T_\mathrm{eff} \leq$ 9000 K and $3.0 \leq \log{g} \leq 5.0$. In this group, typical photometric uncertainties of $\left \langle \sigma_\mathrm{b-y} \right \rangle$ = 0.003 mag and $\left \langle \sigma_\mathrm{c_1} \right \rangle$ = 0.005 mag lead to average uncertainties of 0.6 \% in $T_\mathrm{eff}$ and 0.055 dex in $\log{g}$. For the intermediate group, points were generated in the range of $a_0-r^*$ parameter space bounded by 8500 K $\leq T_\mathrm{eff} \leq$ 11000 K and $3.0 \leq \log{g} \leq 5.0$. In this group, typical photometric uncertainties of $\left \langle \sigma_\mathrm{a_0} \right \rangle$ = 0.005 mag and $\left \langle \sigma_\mathrm{r^*} \right \rangle$ = 0.005 mag lead to average uncertainties of 0.8 \% in $T_\mathrm{eff}$ and 0.046 dex in $\log{g}$. For the early group, points were generated in the range of $c_1-\beta$ parameter space bounded by 10000 K $\leq T_\mathrm{eff} \leq$ 30000 K and $3.0 \leq \log{g} \leq 5.0$. In this group, typical photometric uncertainties of $\left \langle \sigma_\mathrm{c_1} \right \rangle$ = 0.005 mag and $\left \langle \sigma_\mathrm{\beta} \right \rangle$ = 0.004 mag lead to average uncertainties of 1.1 \% in $T_\mathrm{eff}$ and 0.078 dex in $\log{g}$. Across all three groups, the mean uncertainty due to photometric errors is $\approx 0.9\%$ in $T_\mathrm{eff}$ and $\approx 0.060$ dex in $\log{g}$.

{\bf Metallicity Effects:} 
For simplicity and homogeneity, our method assumes solar composition throughout. However, our sample can more accurately be represented as a Gaussian centered at -0.109 dex with $\sigma \approx$ 0.201 dex. Metallicity is a small, but non-negligible, effect and allowing [M/H] to change by $\pm$ 0.5 dex can lead to differences in the assumed $T_\mathrm{eff}$ of $\sim$ 1-2$\%$  for late-, intermediate-, and some early-group stars, or differences of up to $6\%$ for stars hotter than $\sim$ 17000 K (of which there are few in our sample). In $\log{g}$, shifts of $\pm$0.5 dex in [M/H] can lead to differences of $\sim 0.1$ dex in the assumed $\log{g}$ for late- or early-group stars, or $\sim 0.05$ dex in the narrow region occupied by intermediate-group stars.

Here, we estimate the uncertainty the metallicity approximation introduces to the fundamental stellar parameters derived in this work. We begin with the actual $uvby\beta$ data for our sample, and [Fe/H] measurements from the XHIP catalog \citep{anderson2012}, which exist for approximately 68\% of our sample. Those authors collected photometric and spectroscopic metallicity determinations of Hipparcos stars from a large number of sources, calibrated the values to the high-resolution catalog of \cite{wu2011} in an attempt to homogenize the various databases, and published weighted means for each star. The calibration process is described in detail in \S 5 of \cite{anderson2012}. 

\begin{figure}
\centering
\includegraphics[width=0.49\textwidth]{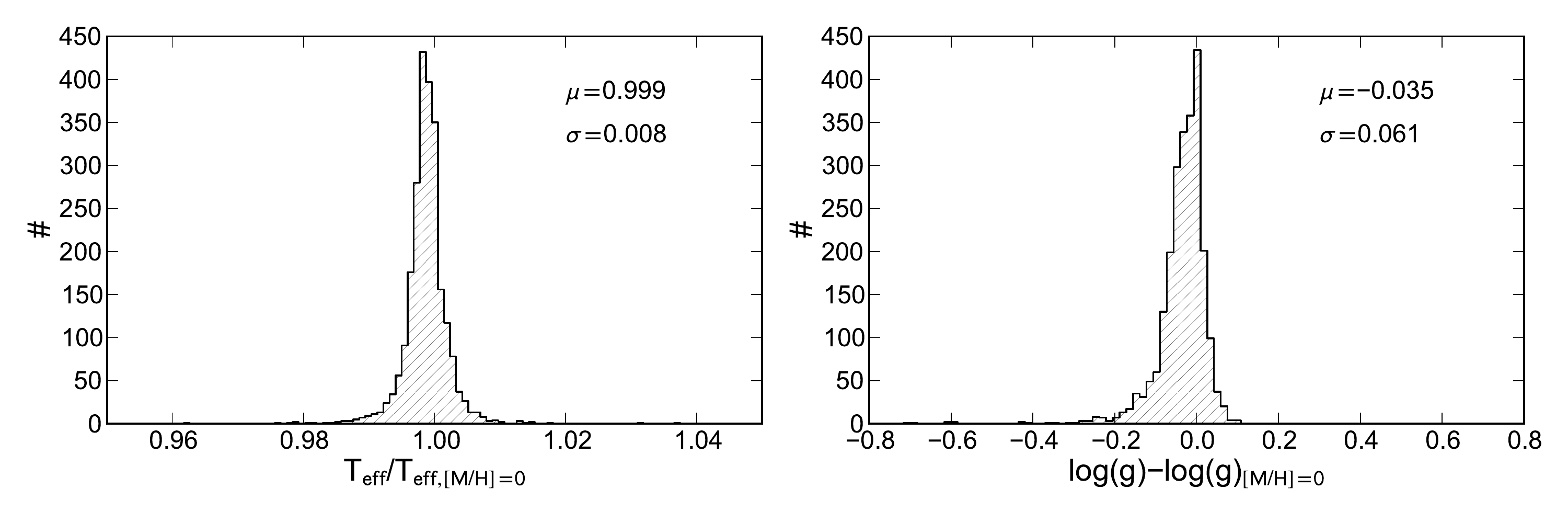}
\caption{Distributions of the true variations in $T_\mathrm{eff}$ (left) and $\log{g}$ (right) caused by our assumption of solar metallicity. The ``true'' $T_\mathrm{eff}$ and $\log{g}$ values are determined for the $\sim 68\%$ of our field star sample with [Fe/H] measurements in XHIP and from linear interpolation between the set of atmospheric parameters determined in eight ATLAS9  grids \citep{castelli2006, castelli2004} that vary from -2.5 to 0.5 dex in [M/H].}
\label{fig:metallicity-error}
\end{figure}

For each of the stars with available [Fe/H] in our field star sample, we derive $T_\mathrm{eff}, \log{g}$ in the appropriate $uvby\beta$ plane for the eight cases of [M/H]=-2.5,-2.0,-1.5,-1.0,-0.5,0.0,0.2,0.5. Then, given the measured [Fe/H], and making the approximation that [M/H]=[Fe/H], we perform a linear interpolation to find the most accurate values of $T_\mathrm{eff}, \log{g}$ given the color grids available. We also store the atmospheric parameters a given star would be assigned assuming [M/H]=0.0. Figure~\ref{fig:metallicity-error} shows the histograms of $T_\mathrm{eff}/T_\mathrm{eff, [M/H]=0}$ and $\log{g}-\log{g}_\mathrm{[M/H]=0}$. We take the standard deviations in these distributions to reflect the typical error introduced by the solar metallicity approximation. For $T_\mathrm{eff}$, there is a 0.8\% uncertainty introduced by the true dispersion of metallicities in our sample, and for $\log{g}$, the uncertainty is 0.06 dex. These uncertainties in the atmospheric parameters are naturally propagated into uncertainties in the age and mass of a star through the likelihood calculations outlined in \S ~\ref{subsubsec:formalism}.

\begin{figure}
\centering
\includegraphics[width=0.49\textwidth]{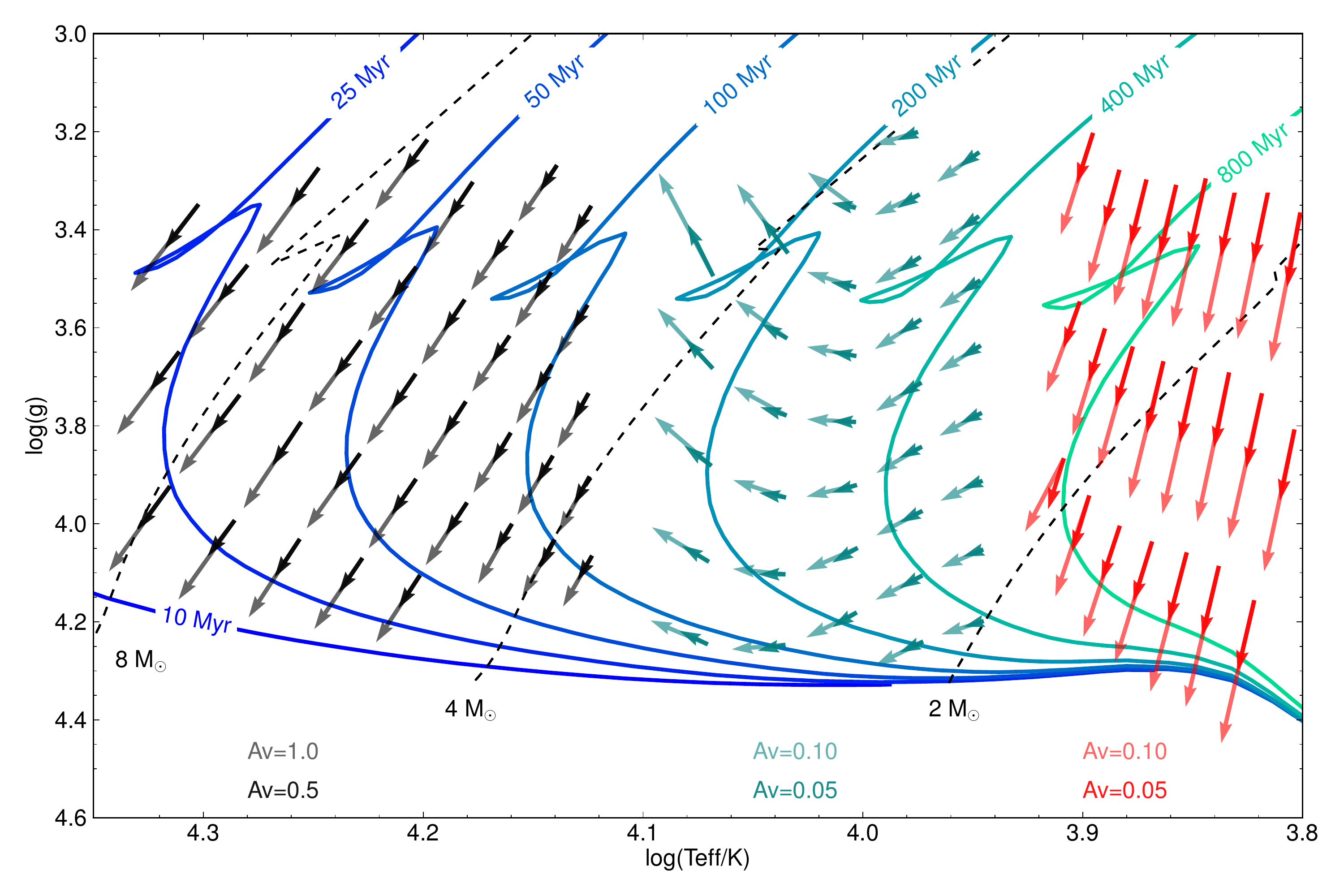}
\caption{The effect of interstellar reddening on atmospheric parameters derived from $uvby\beta$ photometry. The isochrones and mass tracks plotted are those of \cite{bressan2012}. The tail of each vector represents a given point in a specific photometric plane ($(b-y)-c_1$ for the late group stars in red, $a_0-r^*$ for the intermediate group stars in teal, and $c_1-\beta$ for the early group stars in black) and its corresponding value in [$T_\mathrm{eff}, \log{g}$]. The tip of the vector points to the new value of [$T_\mathrm{eff}, \log{g}$] after each point in photometric space has been ``dereddened'' assuming arbitrary values of $A_V$. The shifts in $uvby\beta$ space have been computed according to the extinction measurements of \cite{schlegel1998} and \cite{crawfordmandwewala1976}, assuming $A_V \simeq 4.237 E(b-y)$. The magnitudes of $A_V$ chosen for this figure represent the extremes of values expected for our sample of nearby stars and are meant to illustrate the directionality of the effects of reddening as propagated through the $uvby\beta$ planes. Finally, note for the early group (black vectors), the $A_V$ values are an order of magnitude larger and much higher than expected for our sample. Again, this is to illustrate the directionality of the reddening effect, which is particularly small for the early group which rely on $c_1$, the Balmer discontinuity index, for temperature, and $\beta$, a color between two narrow-band filters with nearly the same central wavelength, for $\log{g}$.}
\label{fig:redvectors}
\end{figure}

{\bf Reddening Effects:} 
For the program stars studied here, interstellar reddening is assumed negligible. Performing the reddening corrections (described in \S~\ref{subsec:reddening}) on our presumably unreddened sample of stars within 100 pc, we find for the $\sim 80\%$ of stars for which dereddening proved possible, that the distribution of $A_V$ values in our sample is approximately Gaussian with a mean and standard deviation of $\mu=0.007, \sigma=0.125$ mag, respectively (see Figure~\ref{fig:sample-specs}). Of course, negative $A_V$ values are unphysical, but applying the reddening corrections to our $uvby\beta$ photometry and deriving the atmospheric parameters for each star in both the corrected and uncorrected cases gives us an estimate of the uncertainties in those parameters due to our assumption of negligible reddening out to 100 pc. The resulting distributions of $T_\mathrm{eff,0}/T_\mathrm{eff}$ and $\log{g}_0-\log{g}$, where the naught subscripts indicate the dereddened values, are sharply peaked at 1 and 0, respectively. The FWHM of these distributions indicate an uncertainty $<0.2\%$ in $T_\mathrm{eff}$ and $\sim 0.004$ dex in $\log{g}$. For the general case of sources at larger distances that may suffer more significant reddening, the systematic effects of under-correcting for extinction are illustrated in Figure~\ref{fig:redvectors}.

{\bf Uncertainties in Projected Rotational Velocities:} 
The \cite{glebocki2005} compilation contains mean $v\sin{i}$ measurements, as well as individual measurements from multiple authors. Of the 3499 stars in our sub-sample of the HM98 catalog, 2547 stars have $v\sin{i}$ values based on 4893 individual $v\sin{i}$ measurements, 1849 of which have an accompanying measurement error. Of these measurements, 646 are for intermediate or early groups, for which rotation corrections are performed in our method. The mean fractional error in $v\sin{i}$ for this subset of measurements is $\sim 13\%$. Caclulating the atmospheric parameters for these stars, then performing the FB98 $v\sin{i}$ corrections using $v_\mathrm{rot}$ and $v_\mathrm{rot} \pm \sigma_{v_\mathrm{rot}}$ allows us to estimate the magnitude of the uncertainty in $T_\mathrm{eff}, \log{g}$ due to the uncertainties in $v\sin{i}$ measurements. The resulting RMS errors in $T_\mathrm{eff}, \log{g}$ are 0.7\% and 0.01 dex, respectively.  When $v\sin{i}$ measurements are not available, an average value based on the spectral type can be assumed, resulting in a somewhat larger error.
The systematic effects of under-correcting for rotation are illustrated in Figure~\ref{fig:rotation-vectors}.

{\bf Influence of Multiplicity:} In a large study such as this one, a high fraction of stars are binaries or higher multiples. Slightly more than 30\% of our sample stars are known as members of multiple systems. We choose not to treat these stars differently, given the unknown multiplicity status of much of the sample, and caution our readers to use due care regarding this issue.

{\bf Influence of Spectral peculiarities:} Finally, early-type stars possess several peculiar subclasses (e.g. Ap, Bp, Am, etc. stars) for which anomalous behavior has been reported in the $uvby\beta$ system with respect to their ``normal-type'' counterparts. Some of these peculiarities have been linked to rotation, which we do account for.  We note that peculiar subclasses constitute $\sim 4\%$ of our sample and these stars could suffer unquantified errors in the determination of fundamental parameters when employing a broad methodology based on calibrations derived from mostly normal-type stars (see Tables~\ref{table:teffcal} \&~\ref{table:loggcal} for a complete accounting of the spectral types used for calibrations). As these subclasses were included in the atmospheric parameter validation stage (\S ~\ref{sec:atmosphericparameters}), and satisfactory accuracies were still obtained, we chose not to adjust our approach for these stars and estimate the uncertainties introduced by their inclusion is negligible.

{\bf Final Assessment:}
Our final atmospheric parameter uncertainties are dominated by the systematic effects quantified in \S~\ref{subsec:teffvalidation} and \S~\ref{subsec:loggvalidation}, with the additional effects outlined above contributing very little to the total uncertainty. The largest additional contributor comes from the photometric error. Adding in quadrature the sources $\sigma_\mathrm{sys}, \sigma_\mathrm{num}, \sigma_\mathrm{phot}, \sigma_\mathrm{[Fe/H]}, \sigma_{v\sin{i}}$ and $\sigma_\mathrm{A_V}$ results in final error estimates of 3.4\% in $T_\mathrm{eff}$ and 0.14 dex in $\log{g}$.

The use of $uvby\beta$ photometry to determine fundamental stellar parameters is estimated in previous literature to lead to uncertainties of just 2.5\% in $T_\text{eff}$ and 0.1 dex in $\log g$ \citep{asiain1997}, with our assessment of the errors somewhat higher. 

The uncertainties that we derive in our Str\"{o}mgren method work can be compared with those given by other methods. The Geneva photometry system ($U,B1,B2, V1, G$ filters), like the  Str\"{o}mgren system, has been used to derive $T_\mathrm{eff}, \log{g}$, and [M/H] values based on atmospheric grids \citep{kobinorth1990,kunzli1997}, with \cite{kunzli1997} finding 150-250 K (few percent) errors in $\log{T_\mathrm{eff}}$ and 0.1-0.15 dex errors in log g, comparable to our values.  From stellar model atmosphere fitting to high dispersion spectra, errors of 1-5\% in $T_\mathrm{eff}$ and 0.05-0.15 dex (typically 0.1 dex) in $\log g$ are quoted for early type stars \citep[e.g.][]{nieva2011}, though systematic effects in log g on the order of an additional 0.1 dex may be present. Wu et al. (2011) tabulate the dispersions in atmospheric parameters among many different studies, finding author-to-author values that differ for OBA stars by 300-5000 K in $T_\text{eff}$ (3-12\%) and 0.2-0.6 dex in $\log{g}$ (cm/s$^2$), and for FGK stars 40-100 K in $T_\text{eff}$ and 0.1-0.3 dex in $\log{g}$ (cm/s$^2$).

%%%%%%%%%%%%%%%%%%%%%%%%%%%%
% METHOD OF AGE ESTIMATION %
%%%%%%%%%%%%%%%%%%%%%%%%%%%%

\section{Age Estimation from Isochrones}
\label{sec:ageestimation}
\subsection{Selection of Evolutionary Models}

Once $T_\text{eff}$ and $\log g$ have been established, ages are determined through a Bayesian grid search of the fundamental parameter space encompassed by the evolutionary models.  In this section we discuss the selection of evolution models, the Bayesian approach, numerical methods, and resulting age/mass uncertainties.

\begin{figure} 
\centering
\includegraphics[width=0.49\textwidth]{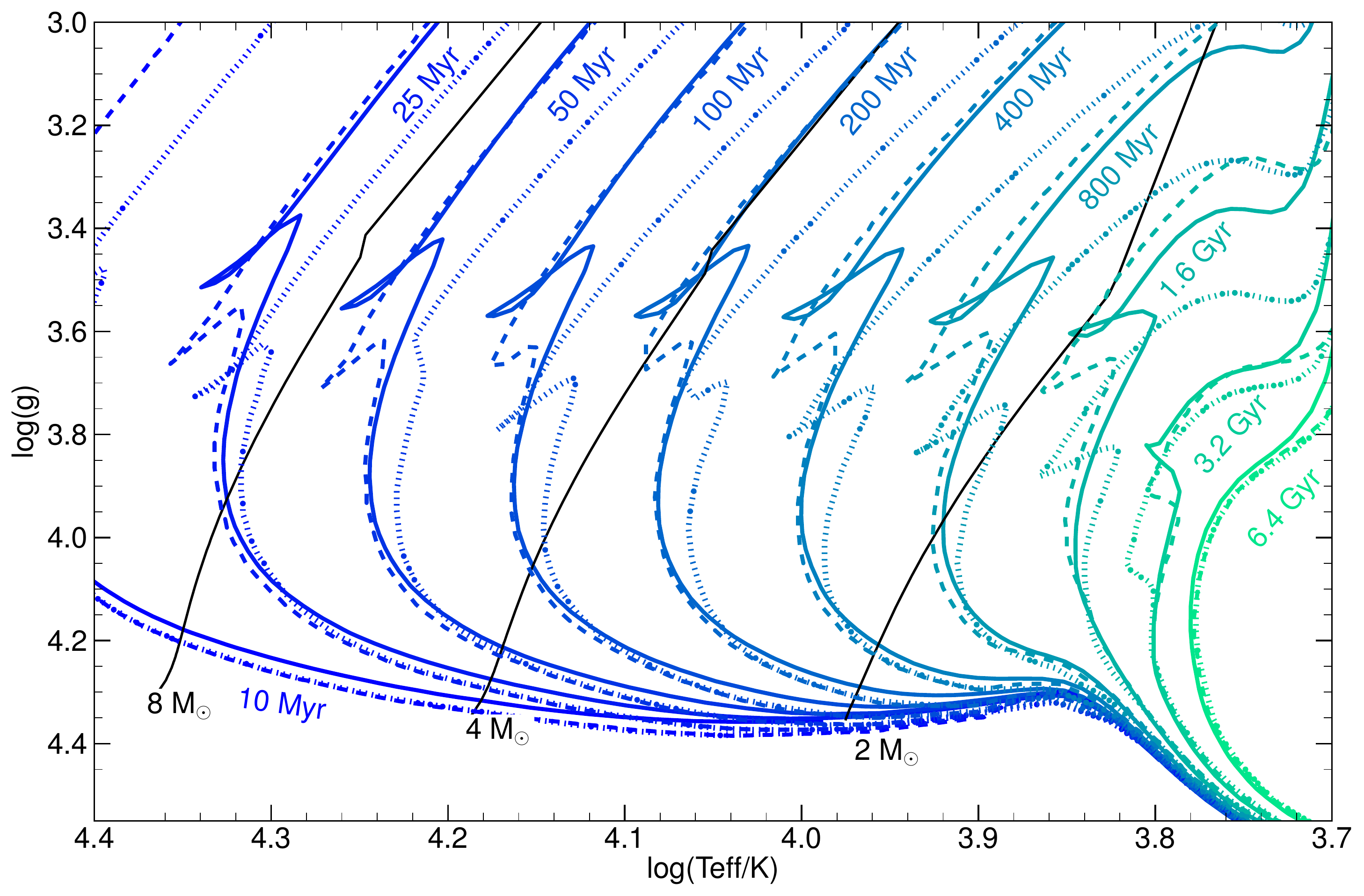}  
\caption{Comparison of PARSEC isochrones (solid lines), Ekstr\"{o}m isochrones in the rotating case (dashed lines), and Ekstr\"{o}m isochrones in the non-rotating case (dotted lines). The solid black lines are evolutionary tracks for stars of intermediate-mass, from the PARSEC models. All evolutionary tracks plotted are for solar metallicity.} %(Z=0.01524 in PARSEC and 0.014 in Ekstrom)
\label{fig:isochrone-compare}
\end{figure}

Two sets of isochrones are considered in this work. The model families are compared in Figure~\ref{fig:isochrone-compare}. The PARSEC solar-metallicity isochrones of \cite{bressan2012}, hereafter B12, take into account in a self-consistent manner the pre-main-sequence phase of evolution. The PARSEC models are the most recent iteration of the Padova evolutionary models, with significant revisions to the major input physics such as the equation of state, opacities, nuclear reaction rates and networks, and the inclusion of microscopic diffusion. The models are also based on the new reference solar composition, $Z = 0.01524$ from \cite{caffau2011}, but can be generated for a wide range of metallicities. The B12 models cover the mass range $0.1-12 M_\odot$.
    
PARSEC isochrones are attractive because early-type dwarfs have relatively rapid evolution with the pre-main-sequence evolution constituting a significant fraction of their lifetimes, i.e. $\tau_\text{PMS} / \tau_\text{MS}$ is larger compared to stars of later types.  For stars with effective temperatures in the range 6500 K - 25000 K (approximately spectral types B0-F5), the B12 models predict pre-main sequence lifetimes ranging from $\sim$ 0.2-40 Myr, main-sequence lifetimes from $\sim$ 14 Myr - 2.2 Gyr, and the ratio $\tau_\mathrm{PMS}/\tau_\mathrm{MS} \sim 1.6-2.4 \%$. A star of given initial mass thus can be followed consistently through the pre-MS, MS, and post-MS evolutionary stages.  As a consequence, most points in $T_\text{eff}-\log{g}$ space will have both pre-ZAMS and post-ZAMS ages as possible solutions. Figure ~\ref{fig:evolutiont} illustrates the evolution of atmospheric and corresponding photometric properties according to the PARSEC models. 
 
The solar-metallicity isochrones of \cite{ekstrom2012}, hereafter E12, also use updated opacities and nuclear reaction rates, and are the first to take into account the effects of rotation on global stellar properties at intermediate masses.  They are available for both non-rotating stars and stars that commence their lives on the ZAMS with a rotational velocity of 40$\%$ their critical rotational velocity ($v_\text{rot,i}/v_\text{crit}$ = 0.4); however, the \cite{ekstrom2012} models do not take the pre-main sequence phase into account. The E12 models currently exist only for solar metallicity (Z=0.014 is used), but cover a wider range of masses ($0.8-120 M_\odot$). 

The E12 models are attractive because they explictly account for rotation, though at a fixed percentage of breakup velocity. All output of stellar evolutionary models (e.g. lifetimes, evolution scenarios, nucleosynthesis) are affected by axial stellar rotation which for massive stars enhances the MS lifetime by about 30$\%$ and may increase isochronal age estimates by about 25$\%$ \citep{meynet2000}.  In terms of atmospheres, for A-type stars, stellar rotation increases the strength of the Balmer discontinuity relative to a non-rotating star with the same color index \citep{maeder1970}.
In the E12 models, the convective overshoot parameter was selected to reproduce the observed main sequence width at intermediate masses, which is important for our aim of distinguishing the ages of many field stars clustered on the main sequence with relatively large uncertainties in their surface gravities. Figure ~\ref{fig:isochrone-compare} shows, however, that there is close agreement between the B12 and the rotating E12 models. Thus, there is not a significant difference between the two models in regards to the predicted width of the MS band.

It should be noted that the $uvby\beta$ grids of \cite{castelli2006, castelli2004} were generated assuming a solar metallicity value of Z=0.017.  As discussed elsewhere, metallicity effects are not the dominant uncertainty in our methods and we are thus not concerned about the very small metallicty differences between the two model isochrone sets nor the third metallicity assumption in the model atmospheres.

In matching data to evolutionary model grids, a general issue is that nearly any given point in an H-R diagram (or equivalently in $T_\mathrm{eff}$-$\log g$ space), can be reproduced by multiple combinations of stellar age and mass. Bayesian inference can be used to determine the relative likelihoods of these combinations, incorporating prior knowledge about the distributions of the stellar parameters being estimated.

%%%%%%%%%%%%%%%%%%%%%%%%%%%%%%%%%%%%
% AGE ESTIMATION IN THE HR DIAGRAM %
%%%%%%%%%%%%%%%%%%%%%%%%%%%%%%%%%%%%
\subsection{Bayesian Age Estimation}

A simplistic method for determining the theoretical age and mass for a star on the Hertzsprung-Russell (H-R) diagram is interpolation between isochrones or evolutionary models. Some problems with this approach, as pointed out by \cite{gtakeda2007, pont2004}, is that interpolation between isochrones neither accounts for the nonlinear mapping of time onto the H-R diagram nor the non-uniform distribution of stellar masses observed in the galaxy. As a consequence, straightforward interpolation between isochrones results in an age distribution for field stars that is biased towards older ages compared to the distribution predicted by stellar evolutionary theory.

Bayesian inference of stellar age and mass aims to eliminate such a bias by accounting for observationally and/or theoretically motivated distribution functions for the physical parameters of interest. As an example, for a given point with error bars on the H-R diagram, a lower stellar mass should be considered more likely due to the initial mass function. Likewise, due to the longer main-sequence timescales for lower mass stars, a star that is observed to have evolved off the main sequence should have a probability distribution in mass that is skewed towards higher masses, i.e. because higher mass stars spend a more significant fraction of their entire lifetime in the post-MS stage.

%%%%%%%%%%%%%%%%%%%%%%%%%%%%%%%%%%%%%%%%%%%%%%%%
% THEORETICAL BASIS OF BAYESIAN AGE ESTIMATION %
%%%%%%%%%%%%%%%%%%%%%%%%%%%%%%%%%%%%%%%%%%%%%%%% 
\subsubsection{Bayes Formalism}
\label{subsubsec:formalism}

Bayesian estimation of the physical parameters can proceed from comparison of the data with a selection of models. Bayes' Theorem states:

\begin{equation}
P(\mathrm{model|data}) \propto P(\mathrm{data|model}) \times P(\mathrm{model})
\end{equation}

The probability of a model given a set of data is proportional to the product of the probability of the data given the model and the probability of the model itself. In the language of Bayesian statistics, this is expressed as:

\begin{equation}
\mathrm{posterior} \propto \mathrm{likelihood} \times \mathrm{prior}.
\end{equation}

Our model is the set of stellar parameters, age ($\tau$) and mass ($M_*$), and our data are the measured effective temperature, $T_\mathrm{eff}$, and surface gravity, $\log g$, for a given star. At any given combination of age and mass, the predicted $T_\mathrm{eff}$ and $\log g$ are provided by stellar evolutionary models. The $\chi^2$ statistic for an individual model can be computed as follows:

\begin{align}
\chi^2 (\tau, M_*) &= \sum \frac{(O-E)^2}{\sigma^2} \\ 
 &=  \frac{[(T_\mathrm{eff})_O-(T_\mathrm{eff})_E]^2}{\sigma_{T_\mathrm{eff}}^2} + \frac{[(\log{g})_O-(\log{g})_E]^2}{\sigma_{\log{g}}^2},
\end{align}

where the subscripts O and E refer to the observed and expected (or model) quantities, respectively, and $\sigma$ is the measurement error in the relevant quantity.

Assuming Gaussian statistics, the relative likelihood of a specific combination of ($T_\mathrm{eff}, \log g$) is:

\begin{align}
P(\mathrm{data|model}) &= P(T_\mathrm{eff, obs}, \log{g}_\mathrm{obs} | \tau, M_*) \\
&\propto \exp\left [{-\frac{1}{2}\chi^2(\tau, M_*}) \right ]. 
\end{align}

Finally, the joint posterior probability distribution for a model with age $\tau$ and mass $M_*$, is given by:

\begin{align}
P(\mathrm{model|data}) &= P(\tau, M_*|T_\mathrm{eff, obs}, \log{g}_\mathrm{obs}) \\
&\propto \exp\left [{-\frac{1}{2}\chi^2(\tau, M_*}) \right ] P(\tau)P(M_*),
\end{align}

where $P(\tau)$ and $P(M_*)$ are the prior probability distributions in age and mass, respectively. The prior probabilities of age and mass are assumed to be independent such that $P(\tau,M_*)=P(\tau)P(M_*)$.

%%%%%%%%%%%%%
% AGE PRIOR %
%%%%%%%%%%%%%

\subsubsection{Age and Mass Prior Probability Distribution Functions}
\label{subsubsec:priors}

Standard practice in the Bayesian estimation of stellar ages is to assume an age prior that is uniform in linear age, e.g. \cite{pont2004, jorgensen2005, gtakeda2007, nielsen2013}. There are two main justifications for choosing a uniform age prior: 1) it is the least restrictive choice of prior and 2) at this stage the assumption is consistent with observations that suggest a fairly constant star formation rate in solar neighborhood over the past 2 Gyr \citep{cignoni2006}.

Since the evolutionary models are logarithmically gridded in age, the relative probability of age bin $i$ is given by the bin width in linear age divided by the total range in linear age:

\begin{equation}
P(\log(\tau_{i}) \leq \log(\tau) < \log(\tau_{i+1})) = \frac{\tau_{i+1}-\tau_i}{\tau_n - \tau_0},
\end{equation}

where $\tau_n$ and $\tau_0$ are the largest and smallest allowed ages, respectively. This weighting scheme gives a uniform probability distribution in linear age.

As noted by \cite{gtakeda2007}, it is important to understand that assuming a flat prior in linear age corresponds to a highly non-uniform prior in the measured quantities of $\log T_\mathrm{eff}$ and $\log g$. This is due to the non-linear mapping between these measurable quantities and the physical quantities of mass and age in evolutionary models. Indeed, the ability of the Bayesian approach to implicitly account for this effect is considered one of its main strengths.

%%%%%%%%%%%%%%
% MASS PRIOR %
%%%%%%%%%%%%%%

%\subsubsection{Mass Prior Probability Distribution Function}
As is standard in the Bayesian estimation of stellar masses, an initial mass function (IMF) is assumed for the prior probability distribution of all possible stellar masses. 
Several authors point out that Bayesian estimates of physical parameters are relatively insensitive to the mass prior (i.e. the precise form of the IMF assumed), especially in the case of parameter determination over a small or moderate range in mass space.
For this work considering BAF stars, the power law IMF of \cite{salpeter1955} is assumed for the mass prior, so that the relative probability of mass bin $i$ is given by the following expression:

\begin{equation}
P(M_i \leq M < M_{i+1}) \propto M_i^{-2.35}.
\end{equation}

%%%%%%%%%%%%%%%%%%%%%%%%%%%%%%%%%%%%%%%%%%%%%%%%
% NUMERICAL METHODS OF AGE AND MASS ESTIMATION %
%%%%%%%%%%%%%%%%%%%%%%%%%%%%%%%%%%%%%%%%%%%%%%%%
\subsubsection{Numerical Methods}
\label{subsubsec:numericalmethods}

As \cite{gtakeda2007} point out, in Bayesian age estimation interpolation should be performed only along isochrones and not between them. To avoid biasing our derived physical parameters from interpolating between isochrones, we generated a dense grid of PARSEC models. The evolutionary models were acquired with a spacing of 0.0125 dex in log(age/yr) and 0.0001 $M_\odot$ in mass. All probabilities were then computed on a 321 $\times$ 321 grid ranging from log(age/yr)=6 to 10 and from 1-10 $M_\odot$.

%%%%%%%%%%%%%%%%%%%%%%%%
% CONFIDENCE INTERVALS %
%%%%%%%%%%%%%%%%%%%%%%%%
\subsubsection{Age and Mass Uncertainties}

Confidence intervals in age and mass are determined from the one-dimensional marginalized posterior probability distributions for each parameter. Since the marginalized probability distributions can often be assymetric, the region chosen for determining confidence intervals is that of the Highest Posterior Density (HPD). This method selects the smallest range in a parameter that encompasses $N \%$ of the probability. The HPD method is discussed in more detail in the appendix.

Notably, uncertainties in the ages depend on where in the $\log g$ and $\log T_\text{eff}$ parameter space the star is located, and whether a pre-main sequence or a post-zero-age-main sequence age is more appropriate. In the pre-main sequence phase, both atmospheric parameters are important in age determination.  For post-ZAMS stars, however, the relative importance of the two parameters changes. When stars are just bouncing back from the ZAMS and are starting to evolve through the MS phase, $\log g$ must be known precisely (within the range of $\sim$4.3 to 4.45) in order to derive a good age estimate.  The age at which this bounce occurs will be a function of mass (earlier for more massive stars). Otherwise, once late B, A, and early F stars are comfortably settled on the MS, their evolution is at roughly constant temperature (see Figure ~\ref{fig:isochrone-compare}) and so the gravity precision becomes far less important, with temperature precision now critical.

%%%%%%%%%%%%%%%%%%%%%%%%%%%%%%%%%%%
% AGE ESTIMATES AND UNCERTAINTIES %
%%%%%%%%%%%%%%%%%%%%%%%%%%%%%%%%%%%

\section{The Methodology Tested on Open Clusters}
\label{subsec:openclustertests}

%%%%%%%%%%%%%%%%%%%%%%%%%%%%
% OPEN CLUSTER HR DIAGRAMS %
%%%%%%%%%%%%%%%%%%%%%%%%%%%%

% CLUSTERS AV PLOT
\begin{figure}
\centering
\includegraphics[width=0.45\textwidth]{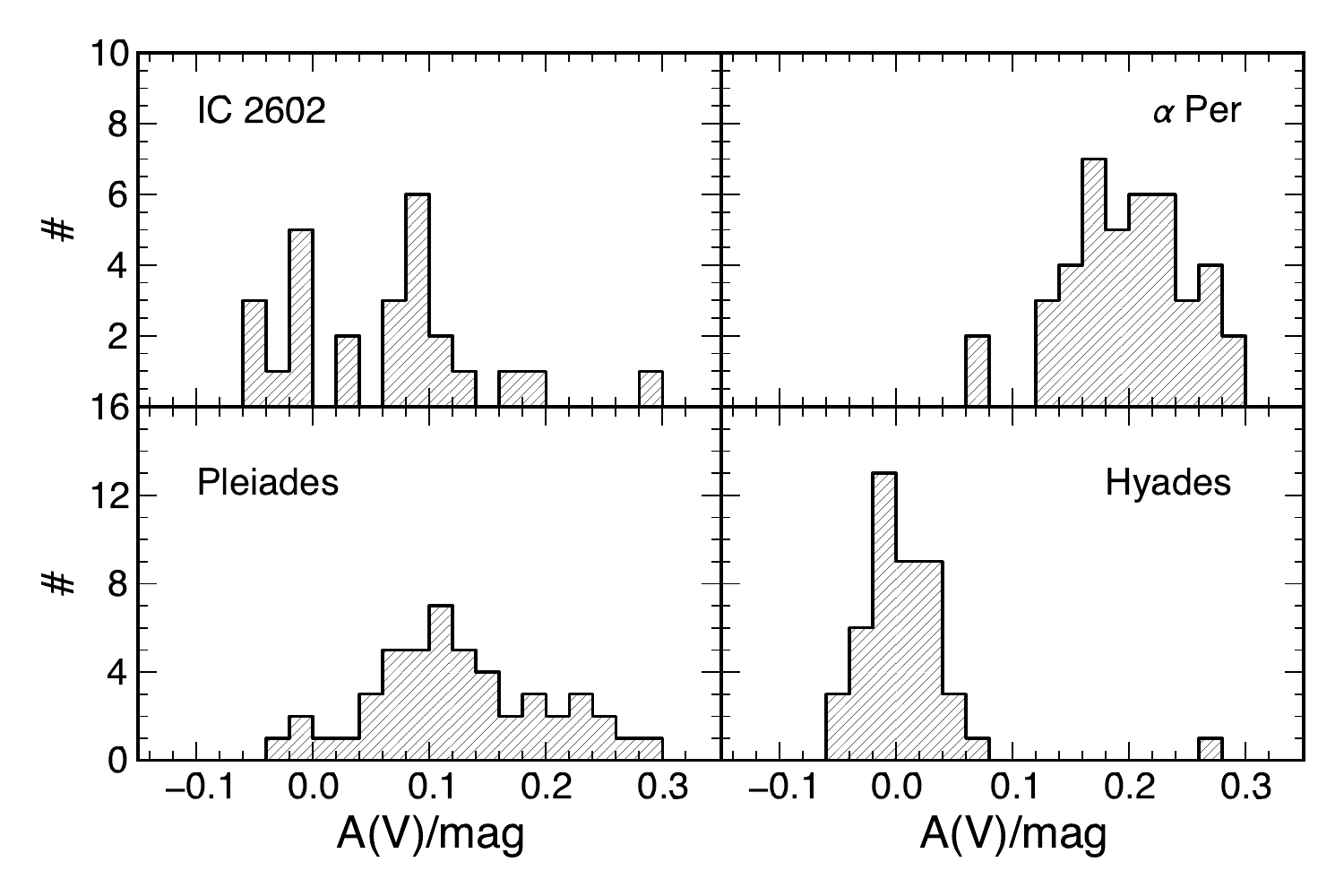}
\caption{Histograms of the visual extinction, $A_V$, in magnitudes for individual members of the four open clusters considered here. The extinction values are calculated using the relation $A_V=4.237E(b-y)$, with the $(b-y)$ color excesses computed as described in \S~\ref{subsec:reddening}.}
\label{fig:clusters-av}
\end{figure}

\begin{figure*}[!h] 
\centering
\includegraphics[width=0.49\textwidth]{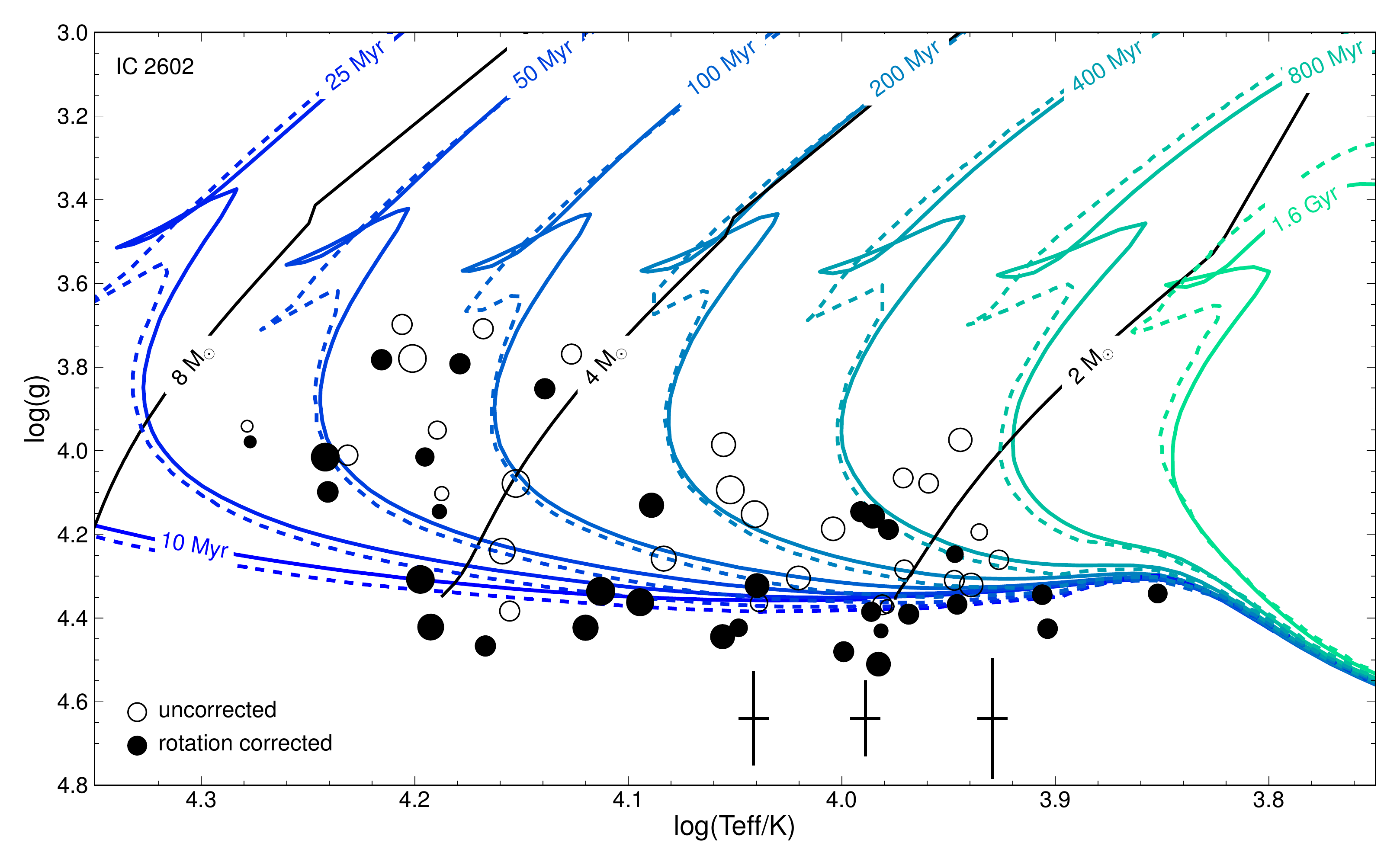}     
\includegraphics[width=0.49\textwidth]{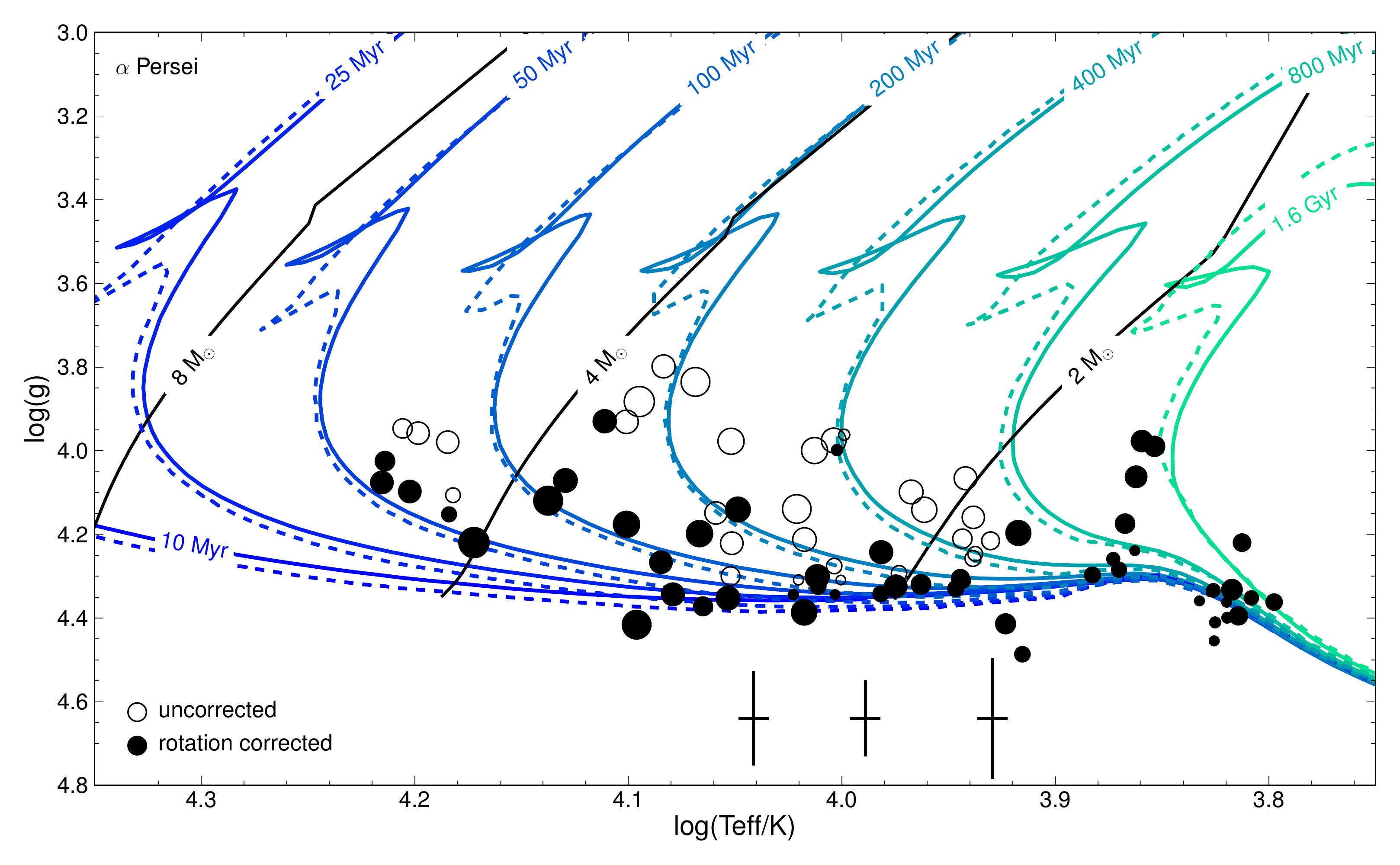} 
\includegraphics[width=0.49\textwidth]{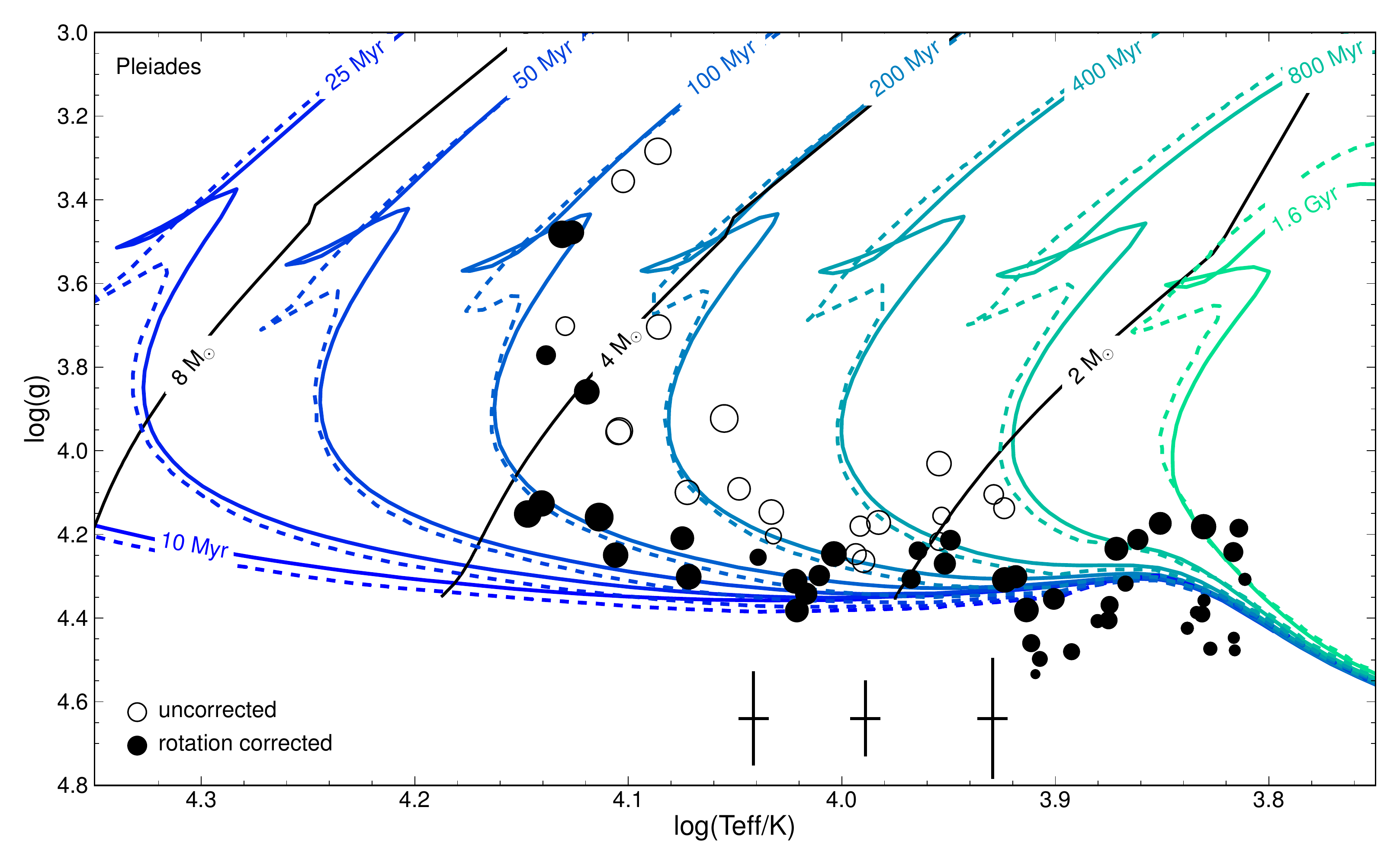}
\includegraphics[width=0.49\textwidth]{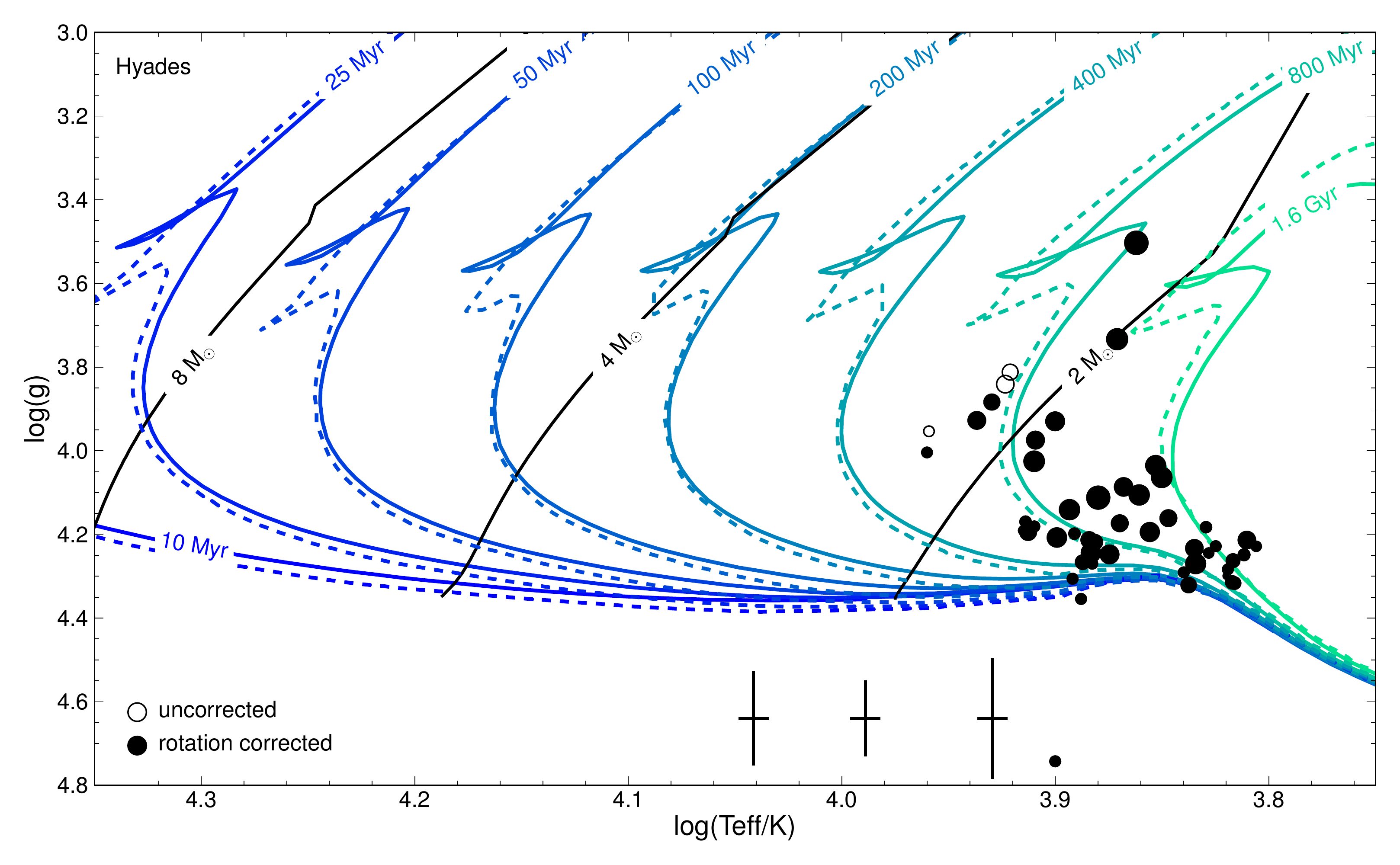}
\caption{PARSEC isochrones and mass tracks \citep{bressan2012} in $\log T_\mathrm{eff}$-$\log{g}$ space and the isochrones of \cite{ekstrom2012} (including rotation, plotted as dashed lines) with our $uvby\beta$ photometric determinations of the atmospheric parameters. For early and intermediate group stars, the black filled circles represent the $v \sin i$ corrected atmospheric parameters (using the FB98 formulae), while the open circles represent the uncorrected parameters. Note that the late-group stars do not receive a $v\sin{i}$ correction but are still plotted as filled circles. In both cases the point sizes are $\propto v \sin i$. The typical uncertainties in our $\log T_\mathrm{eff}$ and $\log{g}$ determinations are represented by the error bars at the bottom of the figure. These uncertainties correspond to 1.6$\%$ or $\approx 0.007$ dex in $\log T_\mathrm{eff}$ and 0.091 dex (intermediate), and 0.145 dex (late) in $\log g$, corresponding to the RMS errors as determined in the effective temperature and surface gravity calibrations. \emph{Top left:} IC 2602 members; the currently accepted age of IC 2602 is $\tau = 46^{+6}_{-5}$ Myr \citep{dobbie2010}.  \emph{Top right:} Members of the $\alpha$ Persei cluster, which has a currently accepted age of $\tau = 90 \pm 10$ Myr \citep{stauffer1999}. \emph{Bottom left}: Pleiades members where the currently accepted age of the Pleiades is $\tau =  125 \pm 8$ Myr \cite{stauffer1998}.  Of the $\sim$ 20 Pleiads that sit below the zero age main sequence, 5 are known pulsators of the $\delta$ Scu or $\gamma$ Dor variety. Additionally, there is an excess of slow rotators sitting below the ZAMS. Possible reasons for this observed behavior include systematics of the atmospheric models (several authors have noted problems with the treatment of convection in ATLAS9 models at this mass range), failure of the evolutionary models to predict the true width of the main sequence (though this effect is unlikely to be as large as the scatter seen here), and overaggressive dereddening procedures. \emph{Bottom right}: Hyades cluster members where the currently accepted age of the Hyades is $\tau =  625 \pm 50$ Myr \cite{perryman1998}. Note the far left outlier, HD 27962, is a known blue straggler \citep{abt1985, eggen1995} and was excluded by \cite{perryman1998} in their isochrone-fitting analysis. The outlier far below the ZAMS, HD 27268, is a spectroscopic binary \citep{debernardi2000}.}
\label{fig:cluster-hrds}
\end{figure*}

An important test of our methods is to assess the ages derived from our combination of $uvby\beta$ photometry, atmospheric parameter placement, and comparison to evolutionary models relative to the accepted ages for members of well-studied open clusters. We investigate four such clusters with rigorous age assessment in previous literature: IC 2602, $\alpha$ Persei, the Pleiades, and the Hyades.

The youngest ($\lesssim 20-30$ Myr) open clusters may be age-dated kinematically, by tracing the space motions of individual members back to the time when the stars were in closest proximity to one another \citep{soderblom2010}. After $\lesssim$ 1 galactic rotation period, however, individual member motions are randomized to the extent of limiting the utility of the kinematic method. Beyond $\sim 20-30$ Myr, the most precise open cluster ages come from the lithium depletion boundary (LDB) technique. This method uses the lithium abundances, which diminish predictably with time, of the lowest mass cluster members to converge on precise ($\sim$10\%) ages. LDB ages are available for IC 2602: $\tau = 46^{+6}_{-5}$ Myr  \citep{dobbie2010},  $\alpha$ Per: $\tau = 90 \pm 10$ Myr \citep{stauffer1999}, and the Pleiades: $\tau = 125 \pm 8$ Myr \citep{stauffer1998}. The LDB technique does not work past $\sim$ 250 Myr, so the Hyades is dated based on isochrone fitting in the H-R diagram using stars with high precision distance measurements, with currently accepted age $625 \pm 50$ Myr \citep{perryman1998}.

\subsection{Process}

Membership probabilities, $uvby\beta$ photometry, and projected rotational velocities are obtained for members of these open clusters via the WEBDA open cluster database\footnote{\url{http://www.univie.ac.at/webda/}}. For the Pleiades, membership information was augmented and cross-referenced with \cite{stauffer2007}. Both individual $uvby\beta$ measurements and calculations of the mean and scatter from the literature measurements are available from WEBDA in each of the photometric indices. As the methodology requires accurate classification of the stars according to regions of the H-R diagram, we inspected the spectral types and $\beta$ indices and considered only spectral types B0-F5 and luminosity classes III-V for our open cluster tests. 

In contrast to the field stars studied in the next section, the open clusters studied here are distant enough for interstellar reddening to significantly affect the derived stellar parameters.  The photometry is thus dereddened as described in \S~\ref{subsec:reddening}. Figure~\ref{fig:clusters-av} shows the histograms of the visual extinction $A_V$ for each cluster, with the impact of extinction on the atmospheric parameter determination illustrated above in Figure~\ref{fig:redvectors}.

In many cases, individual cluster stars have multiple measurements of $v \sin i$ in the WEBDA database and we select the measurement from whichever reference is the most inclusive of early-type members. In very few cases does a cluster member have no rotational velocity measurement present in the database; for these stars we assume the mean $v \sin i$ according to the $T_\mathrm{eff}-v\sin{i}$ relation presented in Appendix B of \cite{gray2005book}.

Atmospheric parameters are determined for each cluster member, as described in \S~\ref{sec:atmosphericparameters}.  Adopting our knowledge from the comparison to fundamental and semi-fundamental atmospheric parameters (\S~\ref{subsec:teffvalidation} \& \S~\ref{subsec:loggvalidation}), a uniform 1.6$\%$ shift towards cooler $T_\mathrm{eff}$ was applied to all temperatures derived from the model color grids to account for systematic effects in those grids. The FB98 $v \sin i$ corrections were then applied to the atmospheric parameters. The $v\sin{i}$ corrections prove to be a crucial step in achieving accurate ages for the open clusters (particularly for the Pleiades). 

\subsection{Results}

The results of applying our procedures to open cluster samples appear in Figure~\ref{fig:cluster-hrds}. While the exact cause(s) of the remaining scatter observed in the empirical isochrones for each cluster is not known, possible contributors may be systematic or astrophysical in nature, or due to incorrect membership information. Multiplicity, variability, and spectral peculiarities were among the causes investigated for this scatter, but the exclusion of objects on the basis of these criteria did not improve age estimation for any individual cluster. The number of stars falling below the theoretical ZAMS, particulary for stars with $\log{T_\mathrm{eff}} \lesssim 3.9$, is possibly systematic and may be due to an incomplete treatment of convection by the ATLAS9 models. This source of uncertainty is discussed in further detail in \S~\ref{subsec:belowzams}.

For each cluster, we publish the individual stars considered, along with relevant parameters, in Tables ~\ref{table:ic2602}, ~\ref{table:alphaper}, ~\ref{table:pleiades}, \& ~\ref{table:hyades}. In each table, the spectral types and $v\sin{i}$ measurements are from WEBDA, while the dereddened $uvby\beta$ photometry and atmospheric parameters are from this work.

\subsubsection{Ages from Bayesian Inference}

Once atmospheric parameters have been determined, age determination proceeds as outlined in \S~\ref{sec:ageestimation}. For each individual cluster member, the $\chi^2$, likelihood, and posterior probability distribution are calculated for each point on a grid ranging from log(age/yr)=6.5 to 10, with masses restricted to $1 \leq M/M_\odot \leq 10$. The resolution of the grid is 0.0175 dex in log(age/yr) and 0.045 $M_\odot$ in mass. The 1-D marginalized posterior PDFs for each individual cluster member are normalized and then summed to obtain an overall posterior PDF in age for the cluster as a whole. This composite posterior PDF is also normalized prior to the determination of statistical measures (mean, median, confidence intervals). Additionally, the posterior PDFs in log(age) for each member are multiplied to obtain the total probability in each log(age) bin that all members have a single age. While the summed PDF depicts better the behavior of individual stars or groups of stars, the multiplied PDF is best for assigning a single age to the cluster and evaluating any potential systematics of the isochrones themselves. 

\begin{figure*}
\centering
\includegraphics[width=0.9\textwidth]{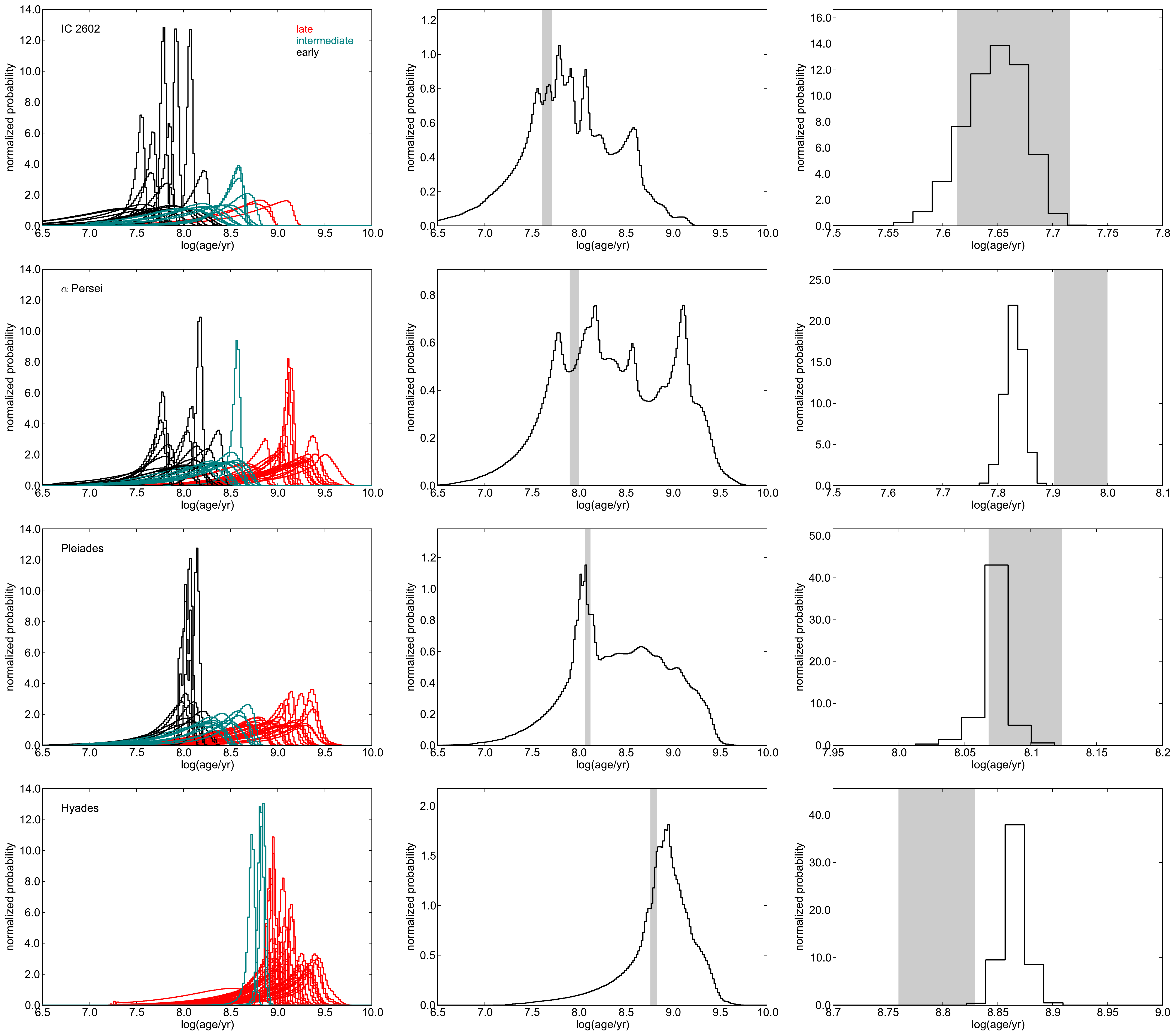}
\caption{\emph{Left panels}: 1D marginalized, normalized posterior PDFs in age, calculated from \cite{bressan2012} evolutionary models, for individual open cluster members. Black, teal, and red histograms represent early, intermediate, and late group stars, respectively. \emph{Middle panels}: Sums of the individual PDFs depicted on the left. This figure shows the total probability associated with the 200 age bins between log(age/yr)=6.5 to 10. The grey shaded regions indicate the currently accepted ages of IC 2602 (46$^{+6}_{-5}$ Myr), $\alpha$ Per (90$\pm$10 Myr), the Pleiades (125$\pm$8 Myr), and the Hyades (625$\pm$50 Myr). \emph{Right panels}: Products of the individual PDFs depicted in the left panels. The grey shaded regions again depict the accepted literature age ranges of each cluster.}
\label{fig:cluster-hist}
\end{figure*}

As shown in Figure ~\ref{fig:cluster-hist}, the summed age PDFs for each cluster generally follow the same behavior: (1) the peaks are largely determined by the early group (B-type) stars which have well-defined ages due to their unambiguous locations in the $T_\mathrm{eff}-\log{g}$ diagram; (2) examining the age posteriors for individual stars, the intermediate group stars tend to overpredict the cluster age relative to the early group stars, and the same is true for the late group stars with respect to the intermediate group stars, resulting in a large tail at older ages for each of the summed PDFs due to the relatively numerous and broad PDFs of the later group stars. For IC 2602 and the Pleiades, the multiplied PDFs have median ages and uncertainties that are in close agreement with the literature ages. Notably, the results of the open cluster tests favor an age for the Hyades that is older ($\sim$ 800 Myr) than the accepted value, though not quite as old as the recent estimate of ~950$\pm$100 Myr from \cite{brandt2015}. The Bayesian age analysis also favors an age for $\alpha$ Per that is younger ($\sim$ 70 Myr) than the accepted value based on lithium depletion, but older than the canonical 50 Myr from the Upper Main Sequence Turnoff \cite{mermilliod1981}. In an appendix, we perform the same analysis for the open clusters on p($\tau$) rather than p($\log{\tau}$), yielding similar results.

%old open clusters figure
%\begin{figure*}
%\centering
%\includegraphics[width=0.9\textwidth]{f21.pdf}
%\includegraphics[width=0.9\textwidth]{f22.pdf}
%\includegraphics[width=0.9\textwidth]{f23.pdf}
%\includegraphics[width=0.9\textwidth]{f24.pdf}
%\caption{\emph{Left panels}: 1D marginalized, normalized posterior PDFs in age, calculated from \cite{bressan2012} evolutionary models, for individual open cluster members. Black, teal, and red histograms represent early, intermediate, and late group stars, respectively. \emph{Right panels}: Sums of the individual PDFs depicted on the left. This figure shows the total probability associated with the 200 age bins between log(age/yr)=6.5 to 10. The grey hatched regions indicate the currently accepted ages of IC 2602 (46$^{+6}_{-5}$ Myr), $\alpha$ Per (90$\pm$10 Myr), the Pleiades (125$\pm$8 Myr), and the Hyades (625$\pm$50 Myr). The scatter point at the top represents the expected value of the summed PDF below, while the solid and dashed lines represent the 68$\%$ and 95$\%$ confidence intervals of the summed pdf, respectively.}
%\label{fig:cluster-hist}
%\end{figure*}

The results of the open cluster test are presented in Table ~\ref{table:clusters}. It is noted that all statistical measures of the marginalized age PDFs quoted hereafter are from PDFs normalized in log(age), as opposed to converting to linear age and then normalizing. This choice was made due to the facts that 1) the isochrones are provided in uniform logarithmic age bins, and 2) the marginalized PDFs of individual stars are more symmetric (and thus better characterized by traditional statistical measures) in log(age) than in linear age. Notably, the median age is equivalent regardless of whether one chooses to analyze prob($\log \tau$) or prob($\tau$). This issue is discussed further in an appendix. In general, there is very close agreement in the Bayesian method ages between B12 and rotating E12 models. For IC 2602 and the Pleiades, our analysis yields median cluster ages (as determined from the multiplied PDFs) that are within 1-$\sigma$ of accepted values, regardless of the evolutionary models considered. The Bayesian analysis performed with the PARSEC models favor an age for $\alpha$ Persei that is $\sim$ 20\% younger than the currently accepted value, or $\sim$ 20\% older for the Hyades.

\begin{deluxetable*}{cccccccccccc}
\tabletypesize{\footnotesize}
\tablecolumns{8}
\tablewidth{0.99\textwidth}
\tablecaption{ Open Cluster Ages \label{table:clusters}}
\tablehead{
&
& 
&
\colhead{Summed PDF} & 
\colhead{Summed PDF} & 
\colhead{Multiplied PDF} & 
\colhead{Multiplied PDF} & 
& 
\\
\colhead{Cluster}  &
\colhead{Lit. Age}  & 
\colhead{Models} &
\colhead{Median}  &
\colhead{68$\%$ C.I.} &
\colhead{Median}  &
\colhead{68$\%$ C.I.} &
\colhead{$\chi^2_\mathrm{min}$}  \\
&
\colhead{(Myr)} & 
&
\colhead{(Myr)} & 
\colhead{(Myr)} & 
\colhead{(Myr)} & 
\colhead{(Myr)} & 
\colhead{(Myr)} &
}

\startdata

IC 2602 & 46$^{+6}_{-5}$ & \cite{ekstrom2012} & 80 & 32-344 & 42 & 41-46 & 39 \\
 &  & \cite{bressan2012} & 79 & 27-284 & 46 & 44-50 & 37 \\

$\alpha$ Persei & 90$^{+10}_{-10}$ & \cite{ekstrom2012} & 234 & 83-1618 & 71 & 68-74 & 50 \\
 &  & \cite{bressan2012} & 226 & 74-1500 & 70 & 69-74 & 48 \\

Pleiades & 125$^{+8}_{-8}$ & \cite{ekstrom2012} & 277 & 81-899 & 128 & 126-130 & 126 \\
 &  & \cite{bressan2012} & 271 & 85-948 & 123 & 121-126 & 115 \\

Hyades & 625$^{+50}_{-50}$ & \cite{ekstrom2012} & 872 & 518-1940 & 827 & 812-837 &  631 \\
 &  & \cite{bressan2012} & 844 & 487-1804 & 764 & 747-780 & 501

\enddata
\tablenotetext{}{Literature ages (column 2) come from the sources referenced in \S~\ref{subsec:openclustertests}. For each set of evolutionary models, the median and 68\% confidence interval are computed for both the summed PDF (columns 4,5) and multiplied PDF (columns 6,7). The final column indicates the best-fit isochrone found through $\chi^2$-minimization of all cluster members in $\log{(T_\mathrm{eff})}-\log{g}$ space. Note, the Hyades analysis includes the blue straggler HD 27962 and the spectroscopic binary HD 27268. Excluding these outliers results in a median and 68\% confidence interval of 871 Myr [517-1839 Myr] of the summed PDF or 832 Myr [812-871 Myr] of the multiplied PDF, using the B12 models.}
\end{deluxetable*}

\subsubsection{Ages from Isochrone Fitting}

\begin{figure*}
\centering
\includegraphics[width=0.45\textwidth]{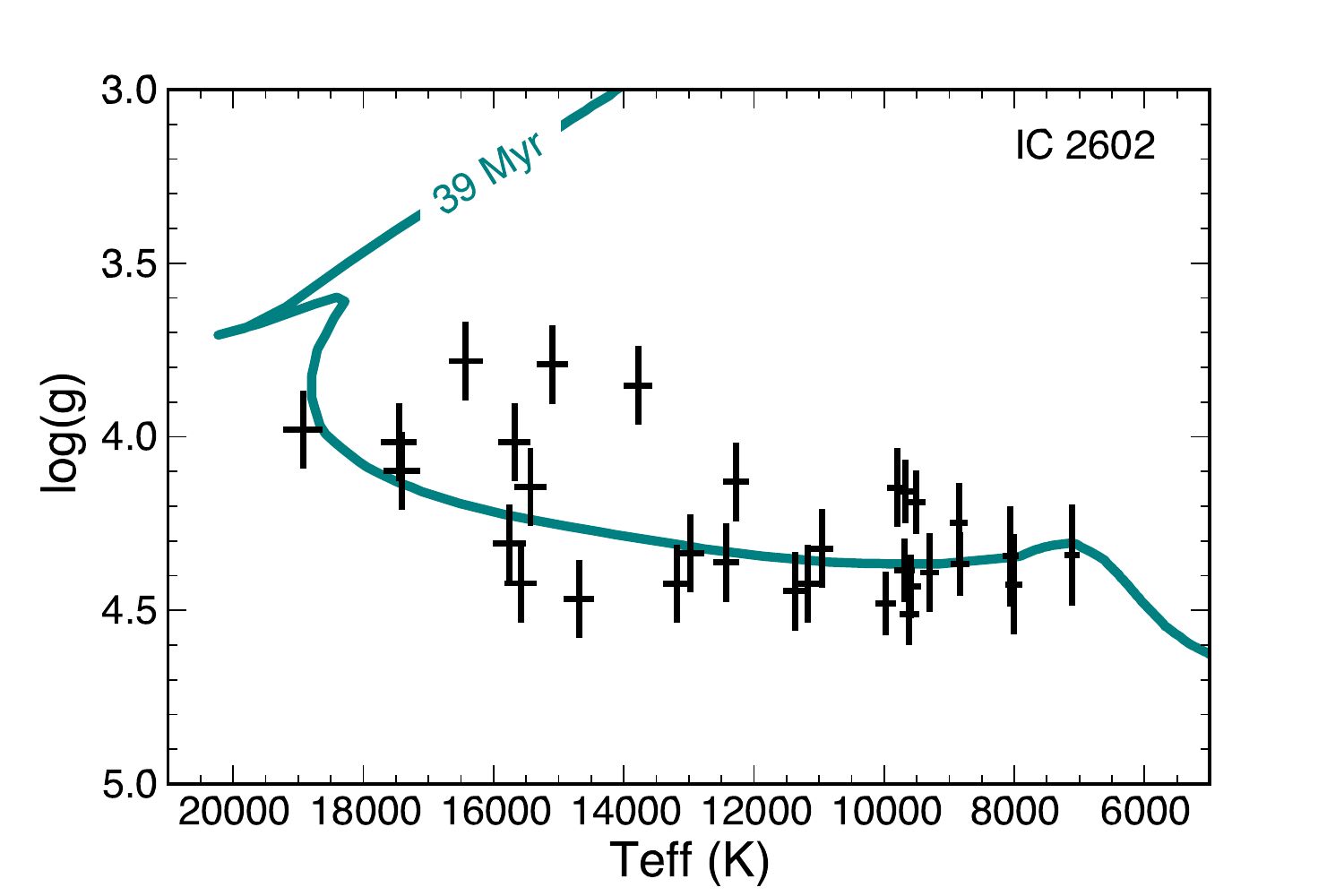}
\includegraphics[width=0.45\textwidth]{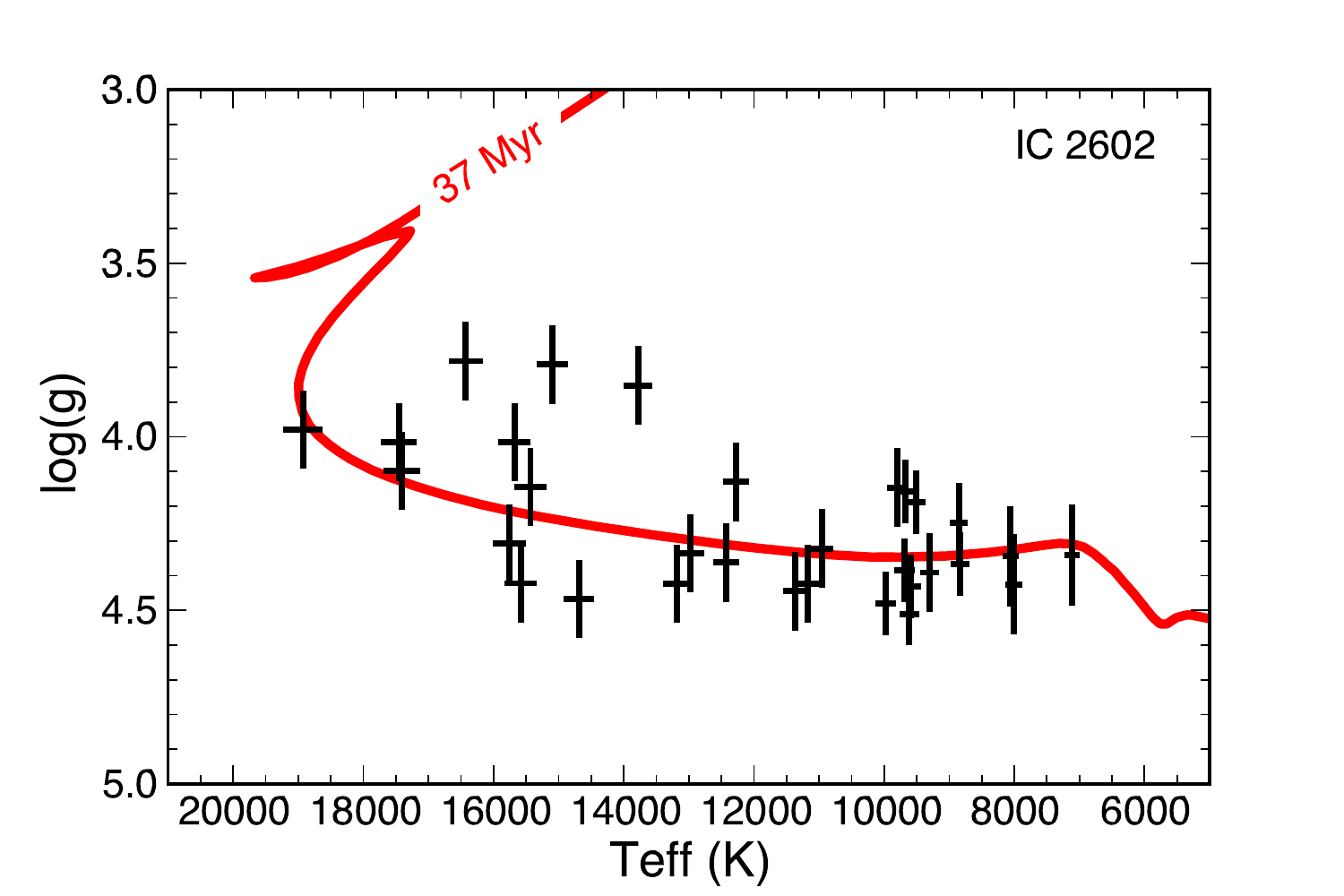}
\includegraphics[width=0.45\textwidth]{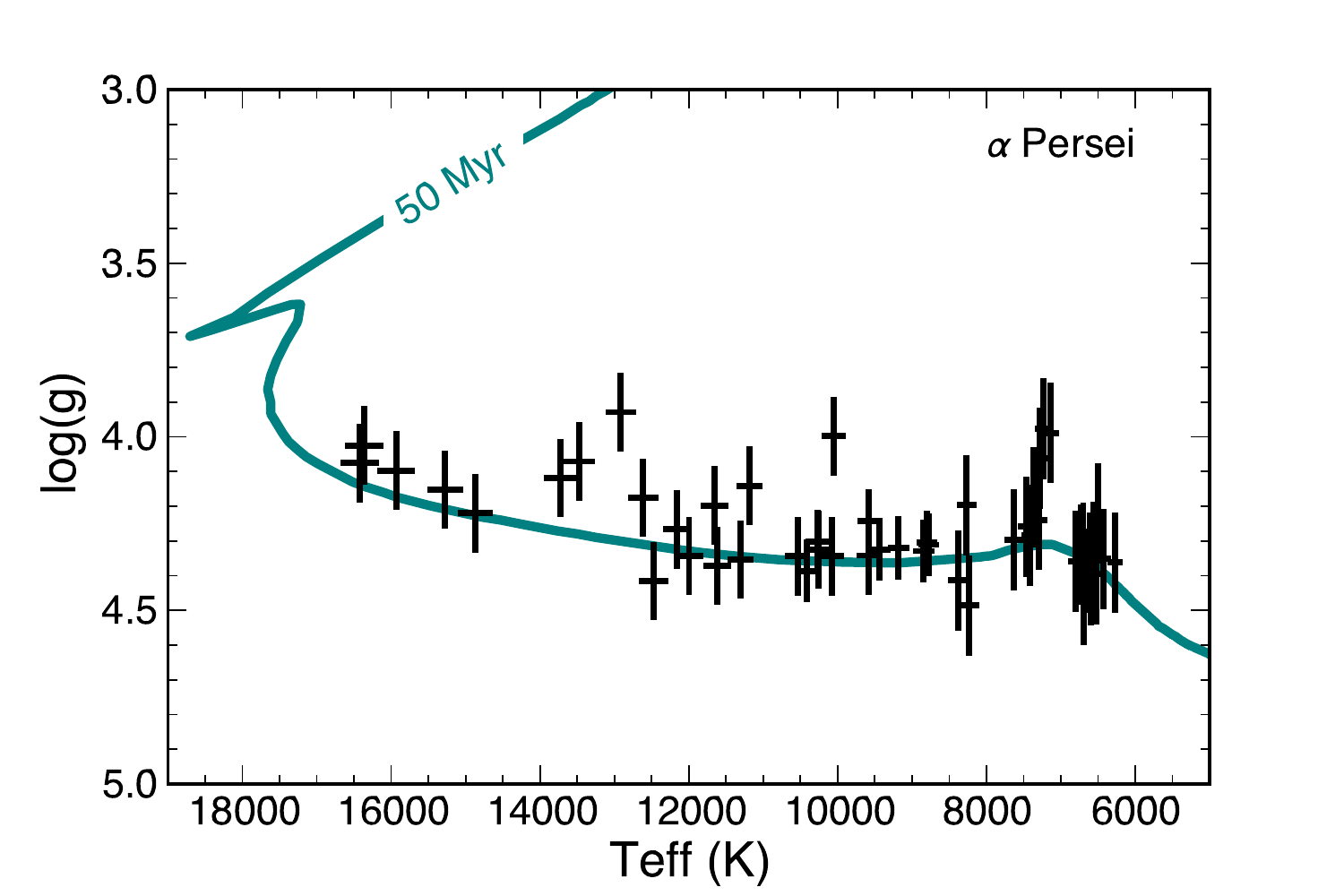}
\includegraphics[width=0.45\textwidth]{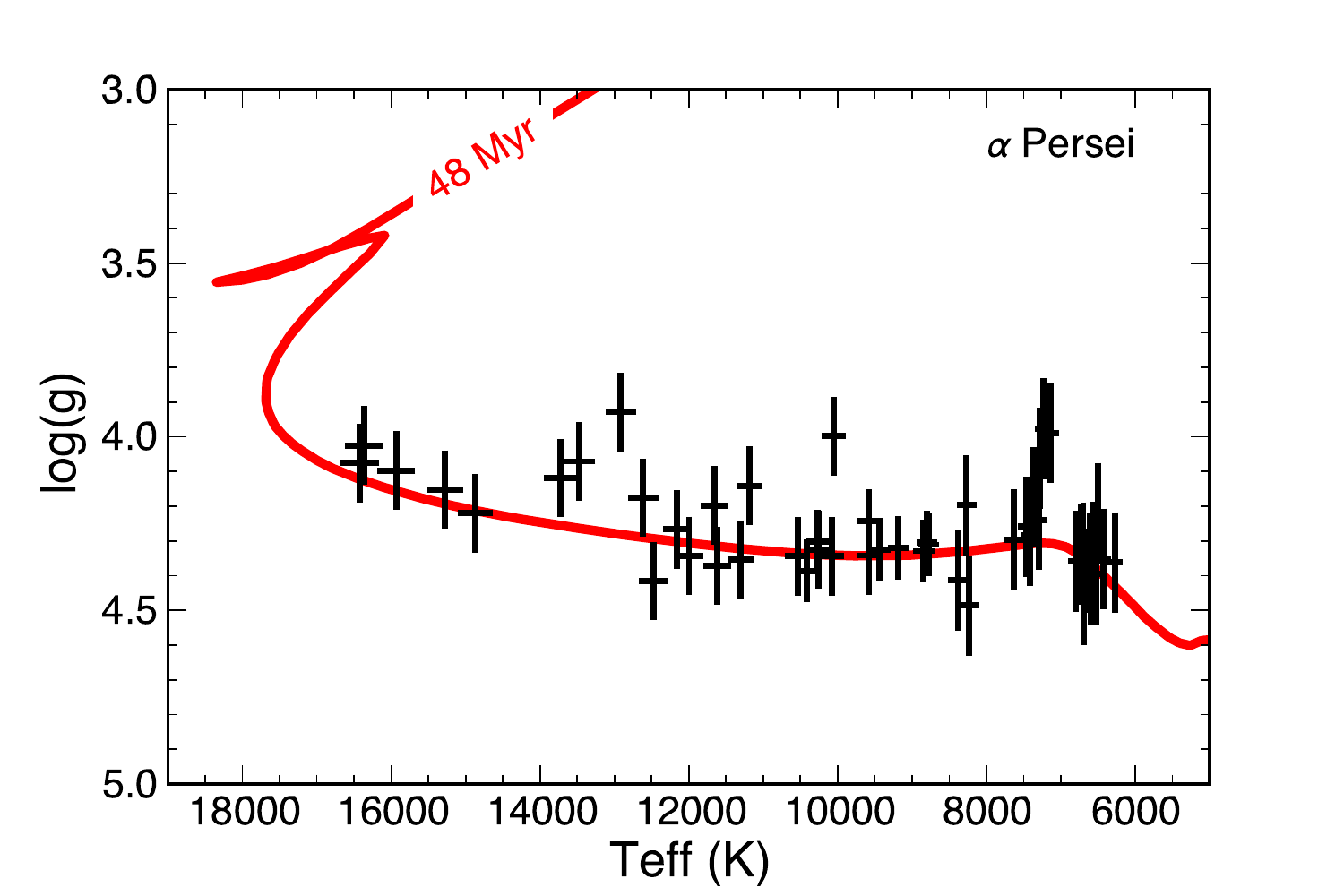}
\includegraphics[width=0.45\textwidth]{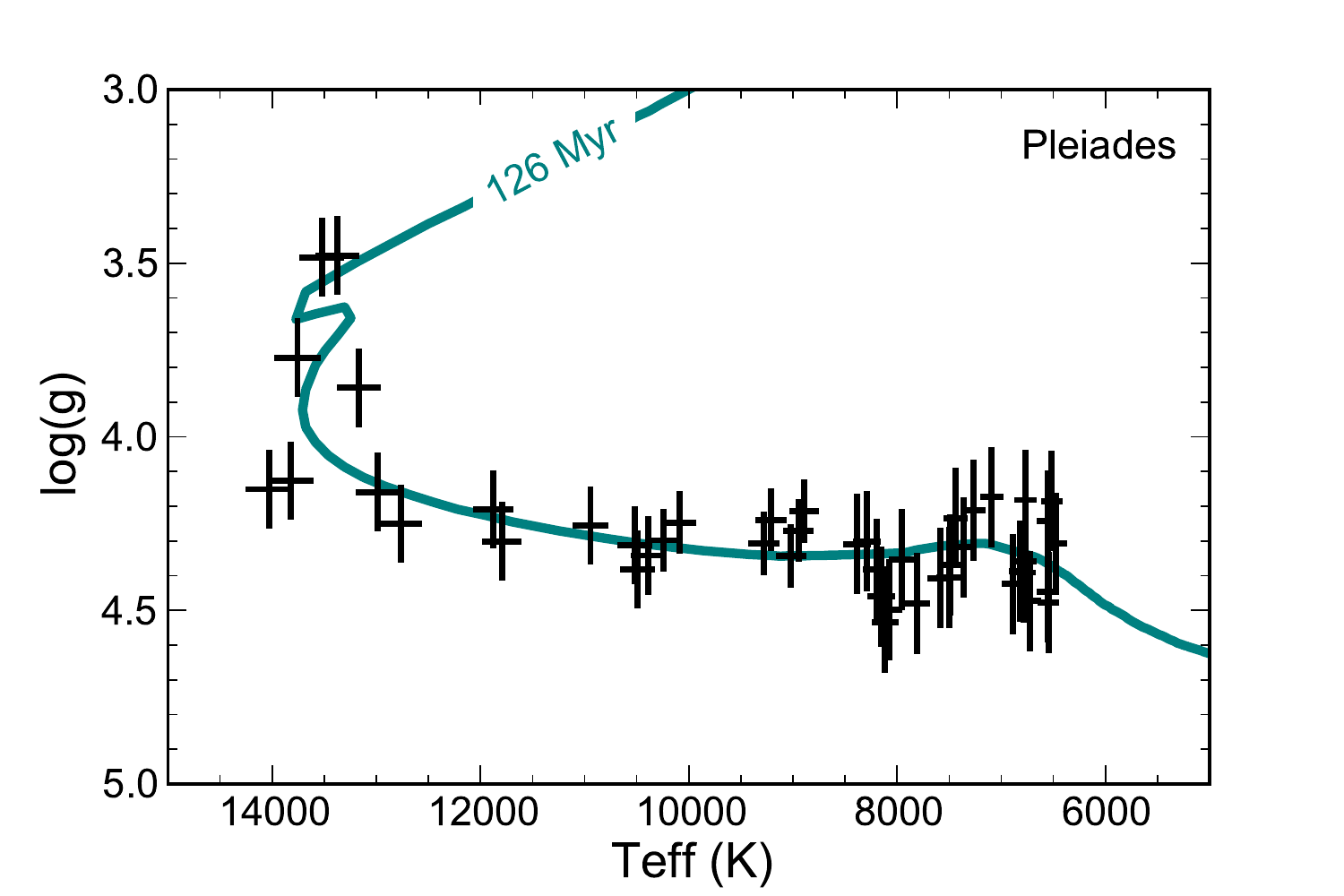}
\includegraphics[width=0.45\textwidth]{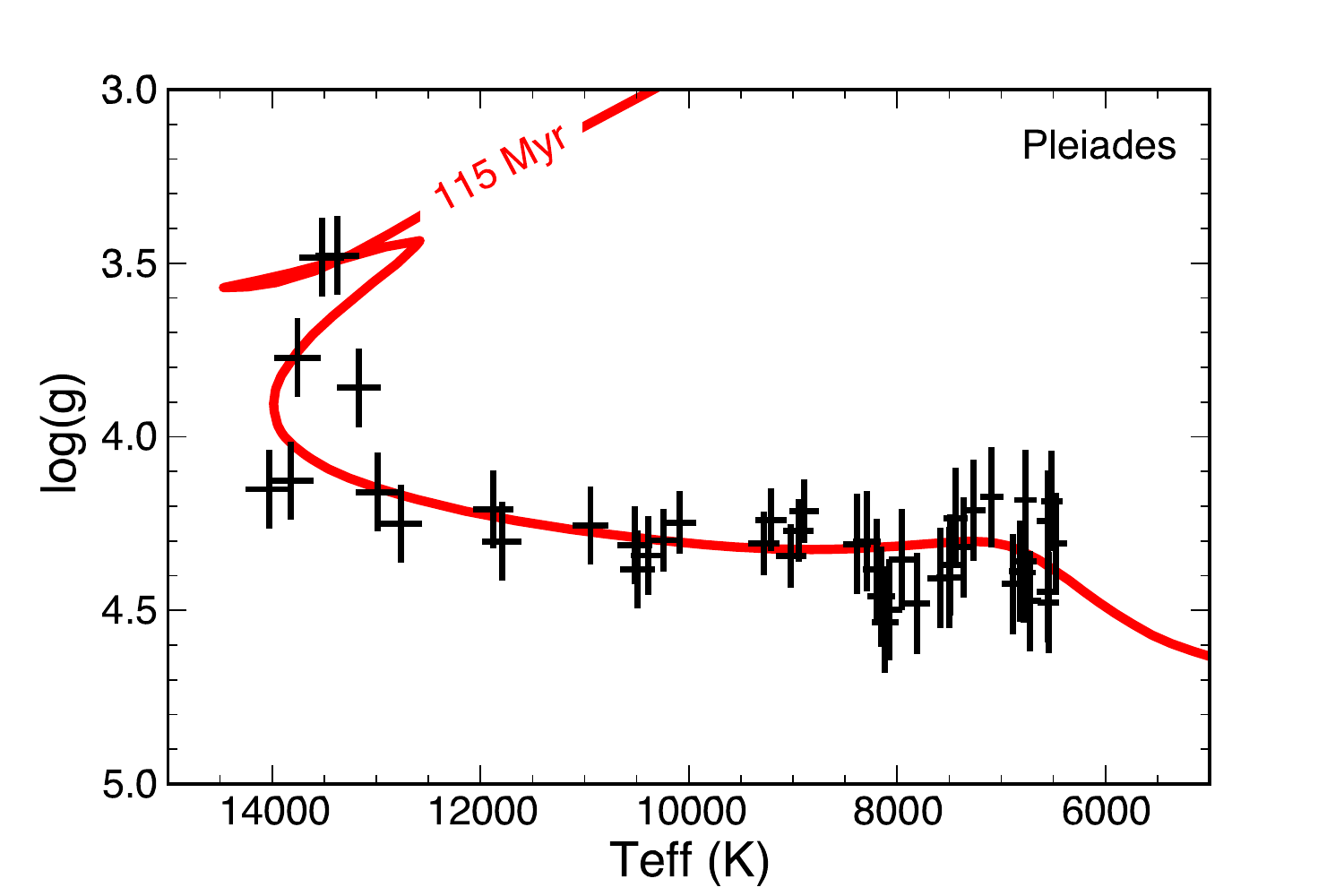}
\includegraphics[width=0.45\textwidth]{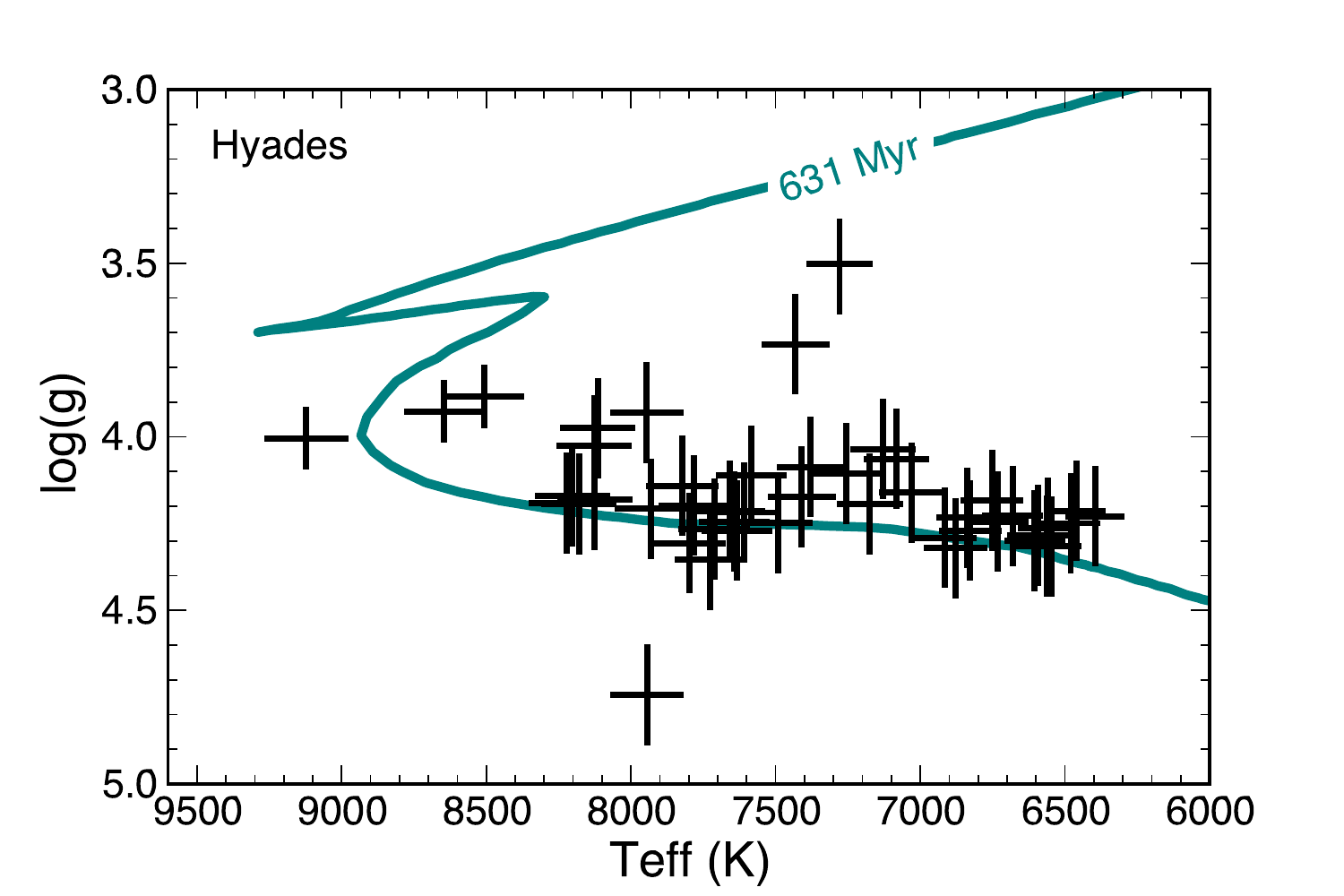}
\includegraphics[width=0.45\textwidth]{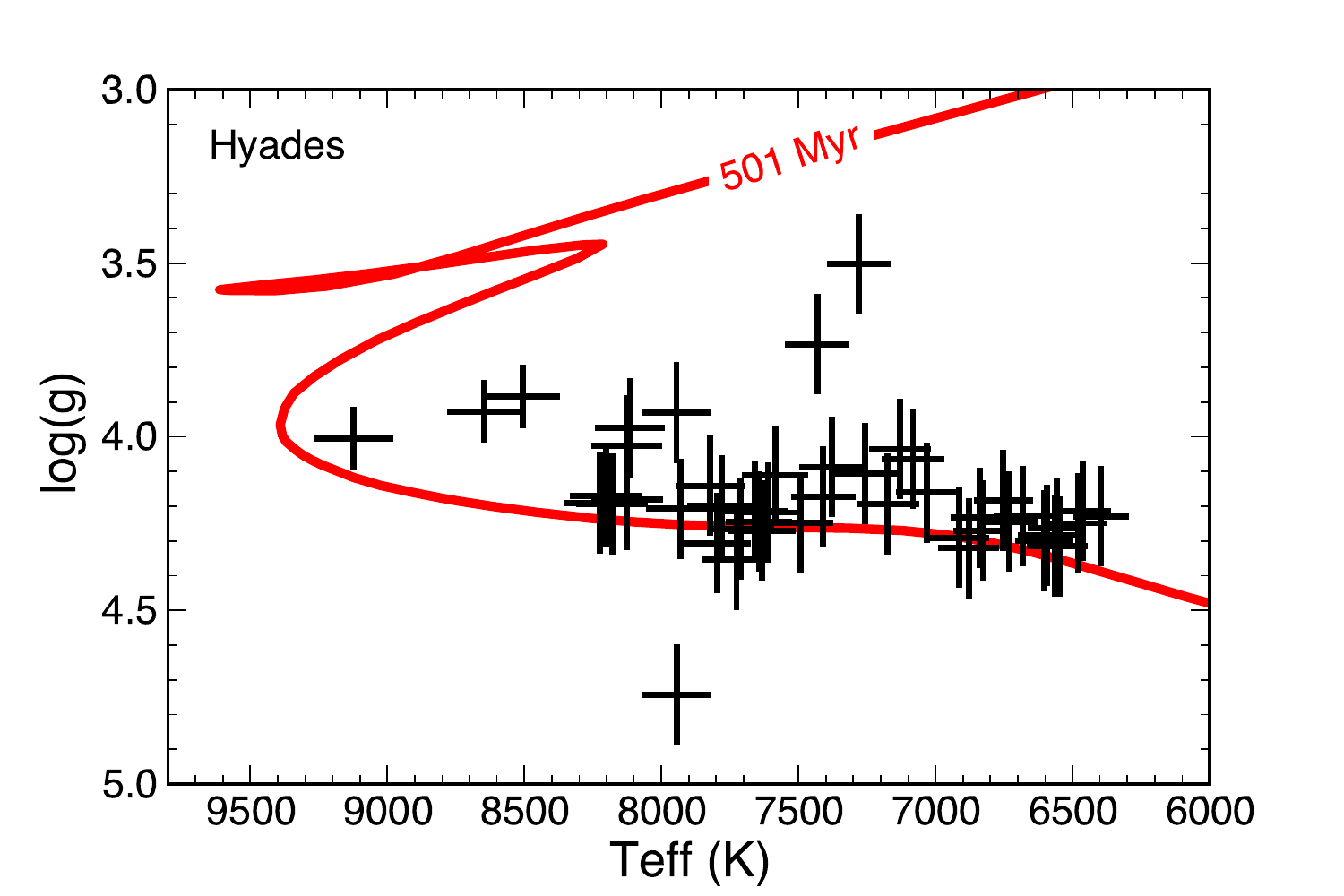}
\caption{Best fitting isochrones found through $\chi^2$-minimization for four open clusters, with atmospheric parameters determined through $uvby\beta$ photometry. Left panels are the fully rotating \cite{ekstrom2012} evolutionary models while right panels are the \cite{bressan2012} models . For the Pleiades, the best fitting isochrone age (126 Myr) from the E12 models is within the currently accepted range of 125$\pm$8 Myr. The B12 models give a best-fit age of 115 Myr, representing a fractional error of $\sim 8\%$ (or 1.25$\sigma$) relative to the accepted age. In the case of the Hyades (lower panels), the low and far left outliers are a spectroscopic binary and a blue straggler, respectively. Excluding these stars yields no change in the best-fitting isochrone for the E12 models and only moderately increases the best-fitting B12 model to 530 Myr.}
\label{fig:bestfit-isochrones}
\end{figure*}

As a final test of the two sets of evolutionary models, we used $\chi^2$-minimization to find the best-fitting isochrone for each cluster. By fitting all members of a cluster simultaneously, we are able to assign a single age to all stars, test the accuracy of the isochrones for stellar ensembles, and test the ability of our $uvby\beta$ method to reproduce the shapes of coeval stellar populations in $T_\mathrm{eff}-\log{g}$ space. For this exercise, we did not interpolate between isochrones, choosing instead to use the default spacing for each set of models (0.1 dex and 0.0125 dex in log(age/yr) for the E12 and B12 models, respectively). For the best results, we consider only the sections of the isochrones with $\log{g}$ between 3.5 and 5.0 dex. The results of this exercise are shown in Figure~\ref{fig:bestfit-isochrones}. The best-fitting E12 isochrone (including rotation) is consistent with accepted ages to within 1\% for the Pleiades and Hyades, $\sim 15\%$ for IC 2602, and $\sim 44\%$ for $\alpha$-Per. For the B12 models, the best-fit isochrones are consistent with accepted ages to $\sim 8\%$ for the Pleiades, $\sim 20\%$ for the Hyades and IC 2602, and $\sim 47\%$ for $\alpha$-Per.  The B12 models produce systematically younger ages than the E12 models, by a fractional amount that increases with absolute age.

As detailed above, the open cluster tests revealed that our method is able to distinguish between ensembles of differing ages, from tens to hundreds of Myr, at least in a statistical sense. For individual stars, large uncertainties may remain, particularly for the later types, owing almost entirely to the difficulty in determining both precise and accurate surface gravities. The open cluster tests also demonstrate the importance of a $v \sin i$ correction for early (B0-A0) and intermediate (A0-A3) group stars in determining accurate stellar parameters. While the $v \sin i$ correction was not applied to the late group (A3-F5 in this case) stars, it is likely that stars in this group experience non-negligible gravity darkening. The typically unknown inclination angle, $i$, also contributes significant uncertainties in derived stellar parameters and hence ages.

\section{The Methodology Applied to Nearby Field Stars}
\label{subsec:fieldstars}

\begin{figure*}
\centering
\includegraphics[width=0.90\textwidth]{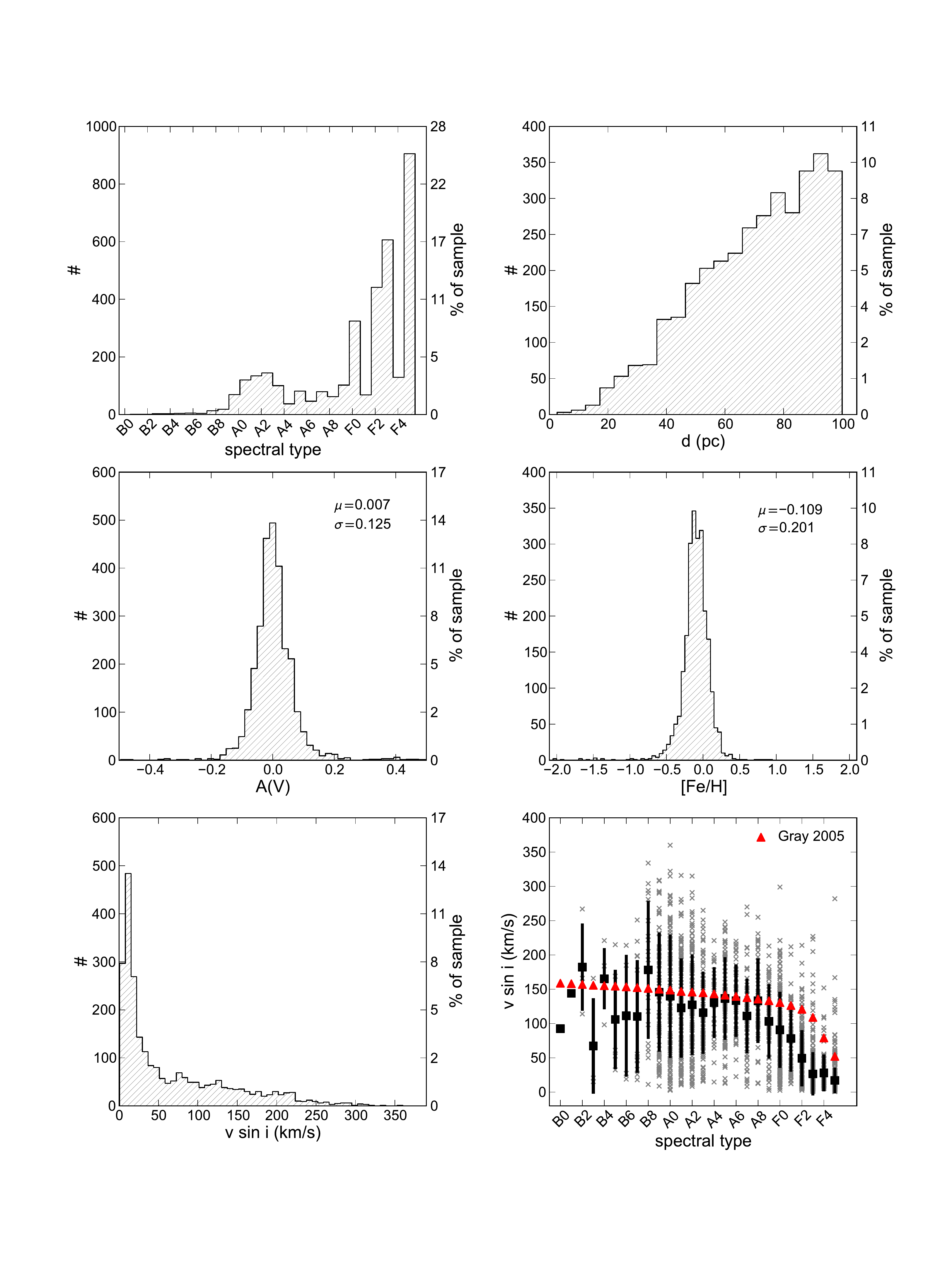}
\caption{Characterization of our sample of 3499 nearby field stars. \emph{Upper panels}: histograms of the spectral types (left) and distances (right) of stars in our sample, taken from \cite{anderson2012}. \emph{Middle panels}: histograms of the V-band extinction in magnitudes (left), as derived by the IDL program described in \S~\ref{subsec:reddening}, and the [Fe/H] values in dex from \cite{anderson2012}. \emph{Lower panels}: histogram of the projected rotational velocities in our sample (left), with data taken from \cite{glebocki2005}, and $v\sin{i}$ as a function of spectral type (right) with grey x's indicating individual stars and black squares representing the mean $v\sin{i}$ in each spectral type bin. The error bars represent the standard deviation in $v\sin{i}$ values for each bin. The red triangles indicate the empirical $T_\mathrm{eff}$-$v\sin{i}$ relation of \cite{gray2005book} using the spectral-type-$T_\mathrm{eff}$ relation of \cite{habets1981}.}
\label{fig:sample-specs}
\end{figure*}

\begin{figure} 
\centering
\includegraphics[width=0.45\textwidth]{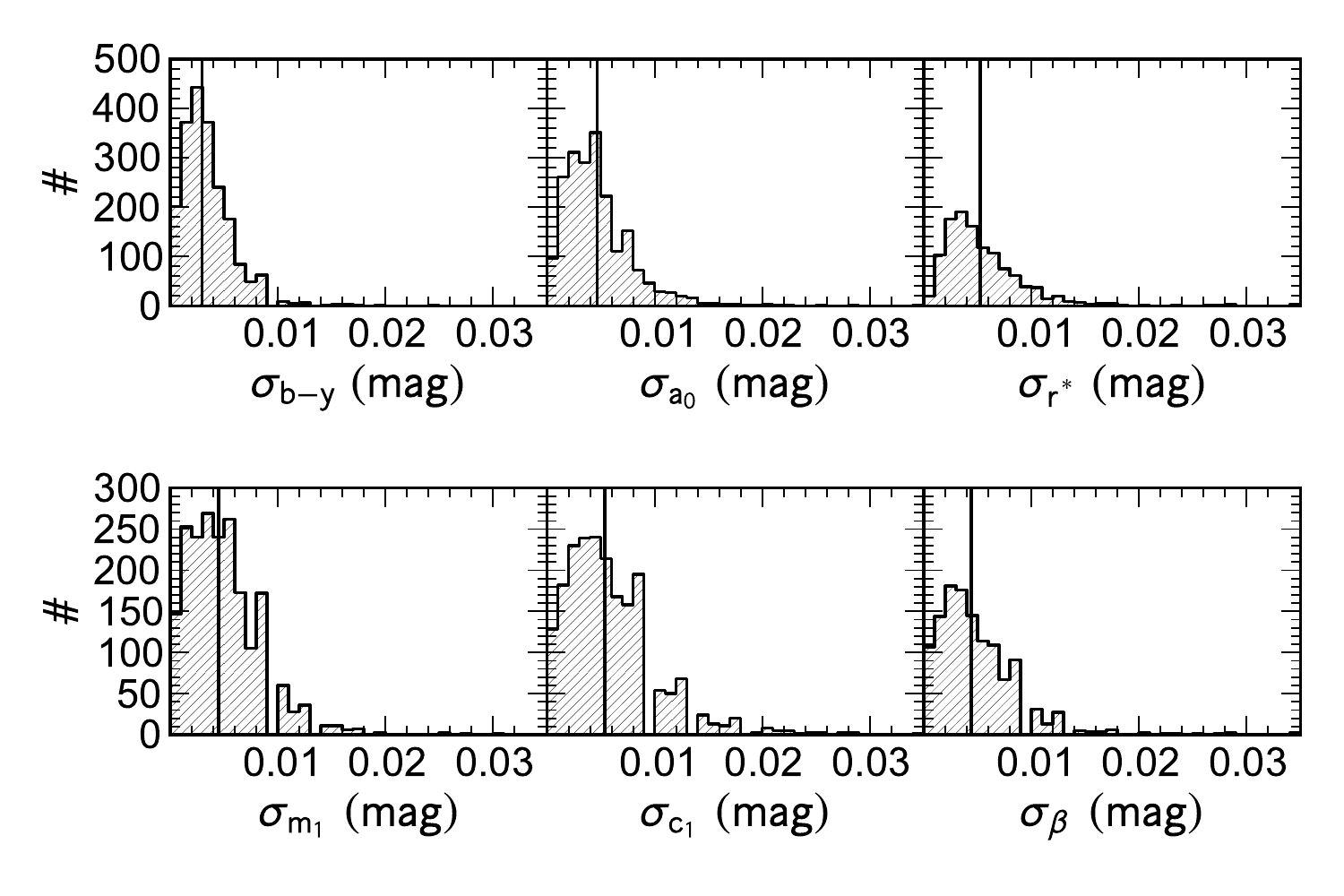}
\caption{Histograms of the uncertainties (in mag) for different $uvby\beta$ indices for the sample of $\sim$ 3500 field stars discussed in \S~\ref{subsec:fieldstars}. The solid lines in each plot indicate the position of the mean uncertainty in that parameter. Uncertainties in $a_0$ and $r^*$ are calculated according to Eqns. (13) \& (14).}
\label{fig:hm98err}
\end{figure}

\begin{figure*}
\centering
\includegraphics[width=0.99\textwidth]{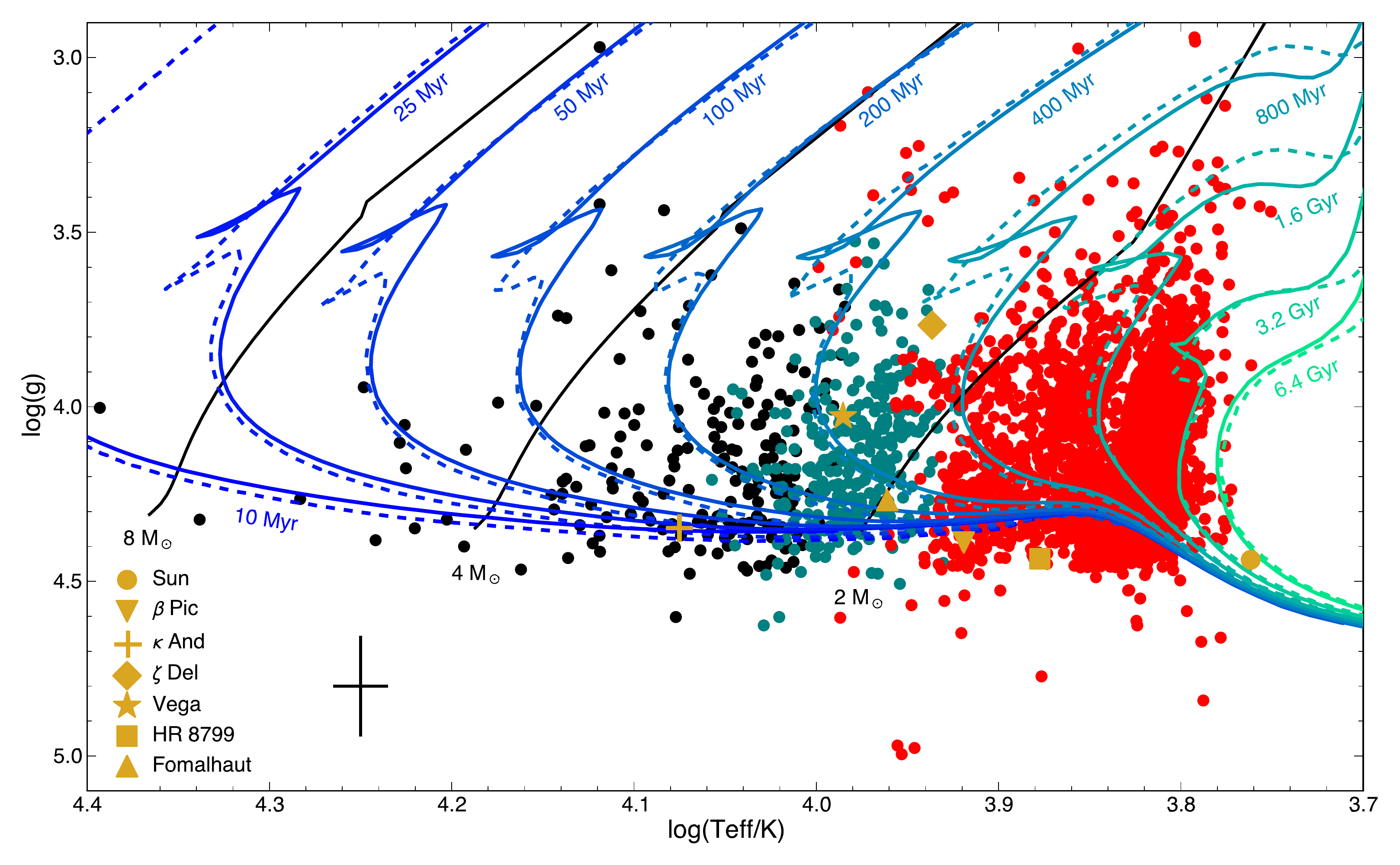}
\caption{H-R diagram for our sample of B0-F5 field stars within 100 pc. Thirteen stars with $\log{g}<2.9$ are excluded in this figure. Several stars of interest are plotted in gold. As before, red, teal, and black scatter points correspond to late, intermediate, and early group stars, respectively. Values for the Sun are also plotted for reference. Of note, $\sim$ 770 of the stars plotted are subgiants according to their XHIP luminosity classes, while only $\sim$ 250 stars have $\log{g} < 3.8$, suggesting some spectral types are in error.}
\label{fig:hrd}
\end{figure*}

As an application of our developed, calibrated, validated, and tested methodology, we consider the complete HM98 photometric catalog of 63,313 stars. We are interested only in nearby stars that are potential targets for high contrast imaging campaigns, and for which interstellar extinction is negligible. We thus perform a distance cut at 100 pc, using distances from the XHIP catalog \citep{anderson2012}. We perform an additional cut in spectral type (using information from XHIP), considering only B0-F5 stars belonging to luminosity classes IV,V, because this is the range for which our method has been shown to work with high fidelity and additionally these are the primary stars of interest to near-term high-contrast imaging surveys. In total, we are left with 3499 stars. Figure~\ref{fig:sample-specs} shows the distribution of our field star sample in spectral type, distance, $A_V$, [Fe/H], and $v\sin{i}$.  The distributions of photometric errors in given $uvby\beta$ indices are shown in Figure ~\ref{fig:hm98err}, and the mean errors in each index are summarized as follows: $\left  \langle \sigma_{b-y}  \right \rangle, \left  \langle \sigma_{m_1}  \right \rangle, \left  \langle \sigma_{c_1}  \right \rangle, \left  \langle \sigma_{\beta}  \right \rangle, \left  \langle \sigma_{a_0}  \right \rangle, \left  \langle \sigma_{r^*}  \right \rangle = 0.003, 0.004, 0.005, 0.004, 0.005, 0.005$ mag.

Projected rotational velocities for the sample of nearby field stars are sourced from the \cite{glebocki2005} compilation, which contains $v\sin{i}$ measurements for 2874 of the stars, or $\sim 82\%$ of the sample. For an additional 8 stars $v\sin{i}$ measurements are collected from \cite{zorec2012}, and for another 5 stars $v\sin{i}$ values come from \cite{schroeder2009}. For the remaining stars without $v\sin{i}$ measurements, a projected rotational velocity is assumed according to the mean $v\sin{i}-T_\mathrm{eff}$ relation from Appendix B of \cite{gray2005book}. Atmospheric parameters are corrected for rotational velocity effects as outlined in \S~\ref{subsec:vsinicorrection}. 

Atmospheric parameter determination was not possible for six stars, due to discrepant positions in the relevant $uvby\beta$ planes: HIP 8016 (a B9V Algol-type eclipsing binary), HIP 12887 (a poorly studied F3V star), HIP 36850 (a well-studied A1V+A2Vm double star system), HIP 85792 (a well-studied Be star, spectral type B2Vne), HIP 97962 (a moderately studied B9V star), and HIP 109745 (an A0III star, classified in XHIP as an A1IV star). Consequently, ages and masses were not computed for these stars.

An H-R diagram of the entire sample is shown in Figure ~\ref{fig:hrd}, with the evolutionary models of \cite{bressan2012} overlaid. Equipped with atmospheric parameters for the remaining 3493 stars, and assuming uniform uncertainties of 3.4\% and 0.14 dex in $T_\mathrm{eff}$ and $\log{g}$, respectively, ages and masses were computed via the process outlined in \S ~\ref{sec:ageestimation}. Posterior probabilities were calculated on a uniform 321$\times$321 grid of the \cite{bressan2012} models, gridded from 1 Myr-10 Gyr in steps of 0.0125 dex in log(age), and from 1-10$M_\odot$ in steps of 0.028$M_\odot$. As the \cite{bressan2012} models exist for high resolution timesteps, no interpolation between isochrones was required.

From the 2D joint posterior PDF, we obtain the marginalized 1D PDFs in age and mass, from which we compute the mean (expected value), median, mode (most probable value), as well as 68\% and 95\% confidence intervals. Interpolated ages and masses are also included, and these values may be preferred, particularly for objects with an interpolated age $\lesssim 10^8$ yr and a $\log{g}$ placing it near the ZAMS (see \S ~\ref{subsec:belowzams} for more detail). The table of ages and masses for all 3943 stars, including our newly derived atmospheric parameters, are available as an electronic table and a portion (sorted in ascending age) is presented here in Table ~\ref{table:fieldstars}. In rare instances (for $\sim 5\%$ of the sample), true 68\% and 95\% confidence intervals were not obtained due to numerical precision, the star's location near the edge of the computational grid, or some combination of the two effects. In these cases the actual confidence interval quoted is noted as a flag in the electronic table.

\begin{deluxetable*}{ccccccccccccccc}
\tabletypesize{\footnotesize}
\tablecolumns{9}
\tablewidth{0.99\textwidth}
\tablecaption{Ages, Masses, and Atmospheric Parameters of Nearby B0-F5 Field Stars \label{table:fieldstars}}
\tablehead{
\colhead{HIP} & 
\colhead{$T_\mathrm{eff}$} & 
\colhead{$\log{g}$} & 
\colhead{Mean} & 
\colhead{Median} &
\colhead{Mode} &
\colhead{68\%} &
\colhead{95\%} &
\colhead{Interp.}&
\colhead{Mean} & 
\colhead{Median} &
\colhead{Mode} &
\colhead{68\%} &
\colhead{95\%} &
\colhead{Interp.} 
\\
 & & & Age & Age & Age & Age & Age & Age & Mass & Mass & Mass & Mass & Mass & Mass
\\
 & (K) & (dex) & (Myr) & (Myr) & (Myr) & (Myr) & (Myr) & (Myr) & (M$_\odot$) & (M$_\odot$) & (M$_\odot$) & (M$_\odot$) & (M$_\odot$) & (M$_\odot$)   
}
\startdata

65474 & 24718 & 4.00 & 6 & 7 & 9 & 5-12 & 2-14 & 13 & 9.59 & 9.61 & 9.62 & 9.4-9.9 & 9.2-10.0 & 10.26 \\
61585 & 21790 & 4.32 & 6 & 7 & 11 & 4-18 & 1-21 & 1 & 7.53 & 7.52 & 7.48 & 7.3-7.7 & 7.1-8.0 & 7.34 \\
61199 & 16792 & 4.18 & 18 & 22 & 33 & 13-53 & 3-60 & 36 & 4.84 & 4.83 & 4.78 & 4.6-5.0 & 4.5-5.2 & 4.95 \\
60718 & 16605 & 4.35 & 19 & 23 & 35 & 13-55 & 3-61 & 1 & 4.75 & 4.74 & 4.70 & 4.6-4.9 & 4.4-5.1 & 4.58 \\
60000 & 15567 & 4.12 & 21 & 26 & 40 & 14-65 & 4-77 & 60 & 4.27 & 4.26 & 4.22 & 4.1-4.4 & 4.0-4.6 & 4.48 \\
100751 & 17711 & 3.94 & 23 & 29 & 43 & 20-50 & 5-52 & 48 & 5.41 & 5.42 & 5.35 & 5.1-5.6 & 5.0-5.9 & 5.91 \\
23767 & 16924 & 4.10 & 23 & 30 & 44 & 18-56 & 4-61 & 46 & 4.96 & 4.95 & 4.92 & 4.7-5.2 & 4.5-5.4 & 5.14 \\
92855 & 19192 & 4.26 & 24 & 29 & 34 & 23-38 & 8-40 & 8 & 6.39 & 6.37 & 6.25 & 6.0-6.6 & 5.8-7.1 & 5.95 \\
79992 & 14947 & 3.99 & 26 & 31 & 48 & 18-78 & 4-89 & 88 & 4.01 & 4.00 & 3.97 & 3.8-4.1 & 3.7-4.3 & 4.45

\enddata
\tablenotetext{}{The fractional uncertainty in our $T_\mathrm{eff}$ determinations is 3.4\% and the uncertainty in our $\log{g}$ determinations is 0.14 dex. All ages and masses are computed from the \cite{bressan2012} models. Statistical measures are quoted for marginalized PDFs in log(age) rather than age, e.g. column 4 is $10^{\left \langle \log(\tau) \right \rangle}$ rather than $\left \langle \tau \right \rangle$. Confidence intervals are computed via the HPD method. The full table containing ages, masses, and atmospheric parameters for all 3493 stars is available electronically. Table \ref{table:fieldstars} is published in its entirety in the electronic edition of \emph{ApJ}, A portion is shown here for guidance regarding its form and content.}
\end{deluxetable*}

\begin{figure}
\centering
\includegraphics[width=0.45\textwidth]{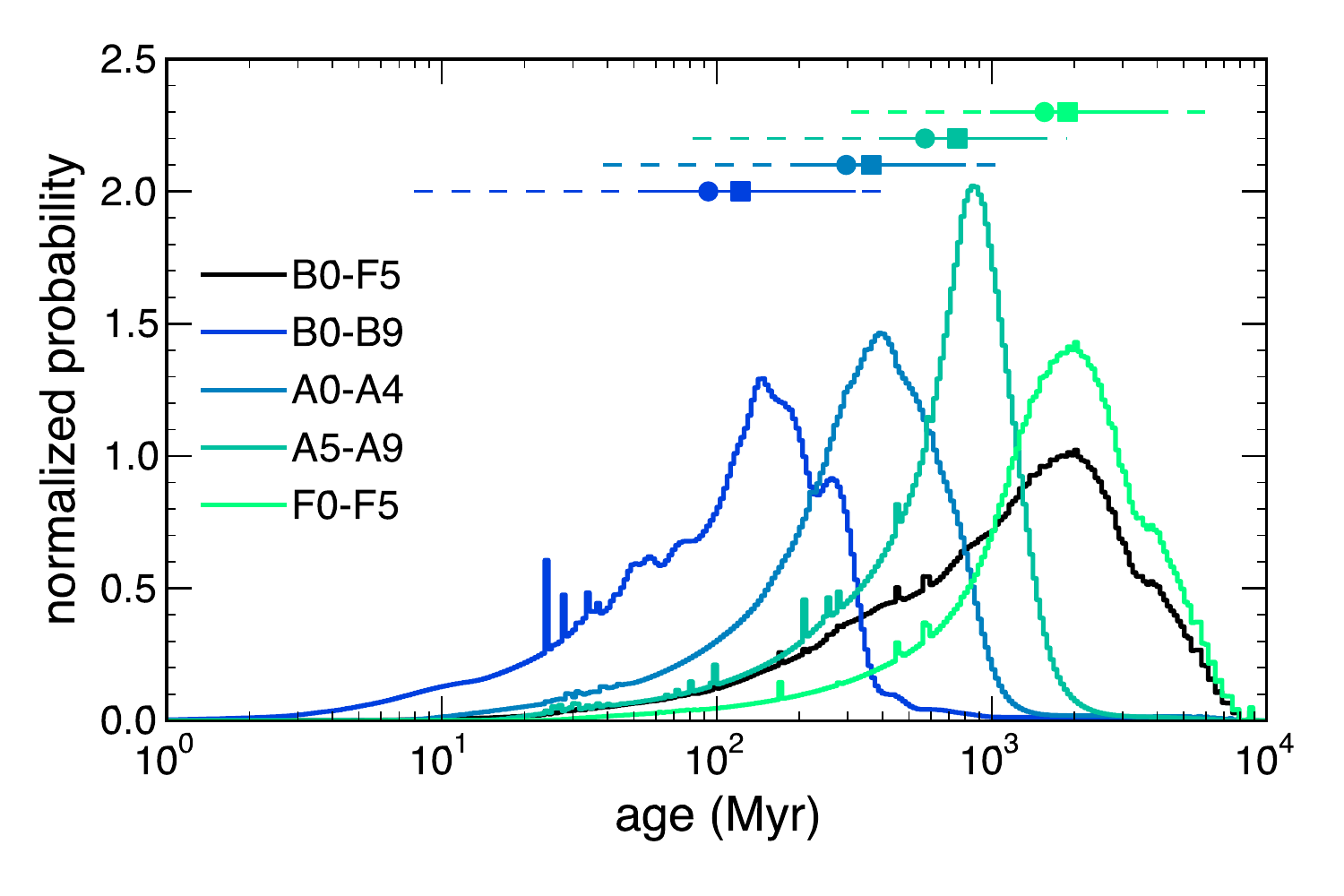}
\caption{Normalized composite age PDFs for our sample of field B0-F5 stars within 100 pc. The normalized composite PDFs are created by summing the normalized, 1D marginalized age PDFs of individual stars in a given spectral type grouping. The black curve represents the composite pdf for all spectral types, while the colored curves represent the composite PDFs for the spectral type groups B0-B9, A0-A4, A5-A9, F0-F5 (see legend). Circles represent the expectation values of the composite PDFs, while squares represent the medians. The solid and dashed lines represent the 68\% and 95\% confidence intervals, respectively, of the composite PDFs. The statistical measures for these composite PDFs are also presented in Table~\ref{table:compositeagepdfs}.}
\label{fig:compositeage}
\end{figure}

As with the open clusters, we can sum the individual, normalized PDFs in age to produce composite PDFs for various subsets of our sample. Figure ~\ref{fig:compositeage} depicts the composite age PDF for our entire sample, as well as age PDFs for the subsets of B0-B9, A0-A4, A5-A9, and F0-F5 stars. From these PDFs we can ascertain the statistical properties of these subsets of solar neighborhood stars, which are presented in Table ~\ref{table:compositeagepdfs}.

\begin{deluxetable*}{cccccc}
\tabletypesize{\footnotesize}
\tablewidth{0pc}
\tablecaption{Statistics of Composite Age PDFs}
\tablehead{
\colhead{Sp. Types} & 
\colhead{Mean Age} & 
\colhead{Median Age} &
\colhead{Mode Age} &
\colhead{68\% C.I.} &
\colhead{95\% C.I.}
\\
 & (Myr) & (Myr) & (Myr) & (Myr) & (Myr)
}
\startdata
B0-B9 & 93 & 122 & 147 & 56-316 & 8-410 \\
A0-A4 & 296 & 365 & 392 & 200-794 & 39-1090 \\
A5-A9 & 572 & 750 & 854 & 434-1372 & 82-1884 \\
F0-F5 & 1554 & 1884 & 2024 & 1000-4217 & 307-6879 \\
\enddata
\label{table:compositeagepdfs}
\end{deluxetable*}

\subsection{Empirical Mass-Age Relation}
\label{subsec:massagerelation}

From our newly derived set of ages and masses of solar-neighborhood B0-F5 stars, we can determine an empirical mass-age relation. Using the mean ages and masses for all stars in our sample, we performed a linear least squares fit using the NumPy polyfit routine, yielding the following relation, valid for stars $1.04<M/M_\odot<9.6$:

\begin{equation}
\log(\mathrm{age/yr}) = 9.532 - 2.929 \log\left (  \frac{M}{M_\odot}\right ).
\end{equation}

The RMS error between the data and this relation is a fairly constant 0.225 dex as a function of stellar mass.

\begin{figure*}
\centering
\includegraphics[width=0.49\textwidth]{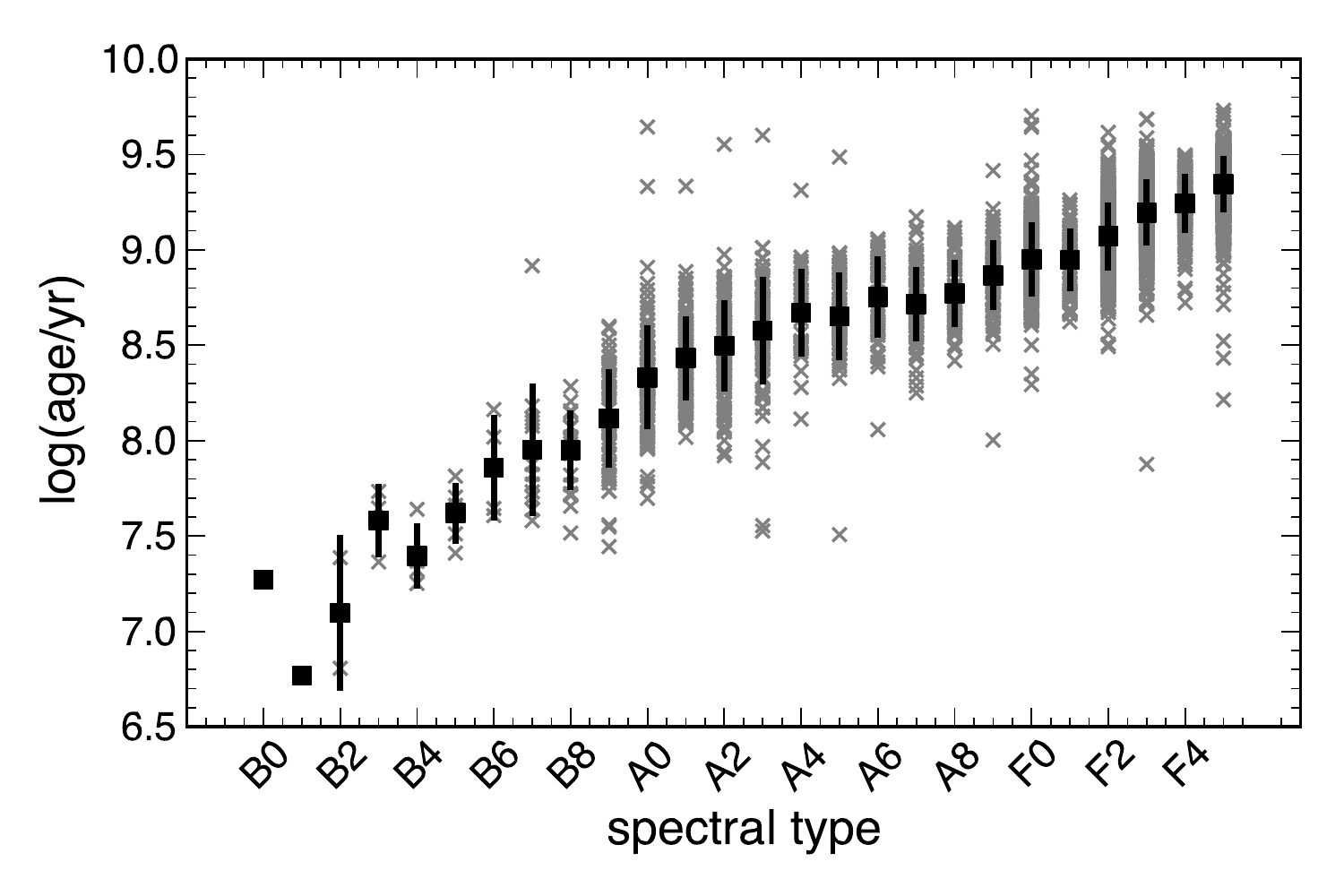}
\includegraphics[width=0.49\textwidth]{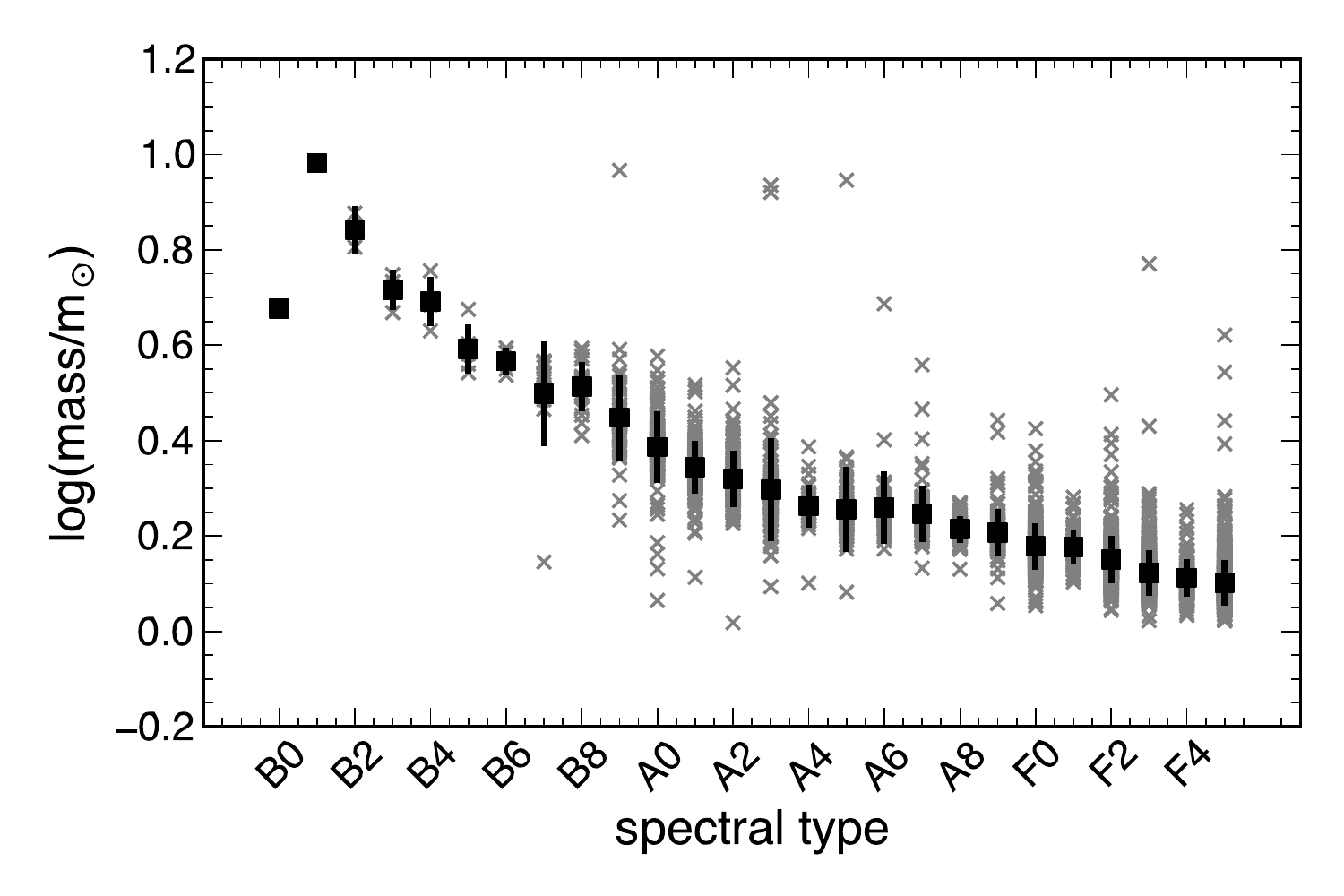}
\caption{Empirical spectral-type-age relation (left) and spectral-type-mass relation (right) for solar neighborhood B0-F5 stars. Grey x's represent individual stars, while the black scatter points represent the mean value in a given spectral type bin and the error bars represent the scatter in a that bin.}
\label{fig:sptrelations}
\end{figure*}

\subsection{Empirical Spectral-Type-Age/Mass Relations}
\label{subsec:sptagerelation}
We can also derive empirical spectral-type-age and spectral-type-mass relations for the solar neighborhood, using the mean masses derived from our 1D marginalized posterior PDFs in age, and spectral type information from XHIP. These relations are plotted in Figure ~\ref{fig:sptrelations}, and summarized in Table ~\ref{table:sptagerelation}.

\begin{deluxetable}{cccccc}
\tabletypesize{\footnotesize}
\tablewidth{0pc}
\tablecaption{Empirical Spectral-Type Relations for Main Sequence B0-F5 Stars}
\tablehead{
\colhead{Sp. Type} & 
\colhead{$\left \langle \tau \right \rangle$} & 
\colhead{$\sigma_\tau$} &
\colhead{$\left \langle M \right \rangle$} & 
\colhead{$\sigma_M$} &
\colhead{No. of Stars}
\\
 & (Myr) & (Myr) & ($M_\odot$) & ($M_\odot$) & 
}
\startdata
B0 & 19 & \textemdash & 4.75 & \textemdash & 1 \\
B1 & 6 & \textemdash & 9.59 & \textemdash & 1 \\
B2 & 15 & 13 & 6.96 & 0.81 & 2 \\
B3 & 41 & 16 & 5.22 & 0.50 & 3 \\
B4 & 26 & 12 & 4.94 & 0.59 & 4 \\
B5 & 44 & 16 & 3.94 & 0.49 & 5 \\
B6 & 84 & 51 & 3.69 & 0.23 & 4 \\
B7 & 140 & 209 & 3.23 & 0.60 & 13 \\
B8 & 99 & 43 & 3.28 & 0.38 & 18 \\
B9 & 154 & 86 & 2.88 & 0.88 & 67 \\
A0 & 285 & 437 & 2.47 & 0.40 & 120 \\
A1 & 313 & 217 & 2.23 & 0.29 & 132 \\
A2 & 373 & 320 & 2.11 & 0.30 & 144 \\
A3 & 462 & 412 & 2.07 & 0.96 & 100 \\
A4 & 540 & 333 & 1.84 & 0.19 & 37 \\
A5 & 514 & 350 & 1.86 & 0.81 & 81 \\
A6 & 628 & 265 & 1.85 & 0.49 & 46 \\
A7 & 574 & 262 & 1.78 & 0.30 & 79 \\
A8 & 642 & 272 & 1.64 & 0.11 & 62 \\
A9 & 800 & 339 & 1.62 & 0.21 & 102 \\
F0 & 994 & 544 & 1.52 & 0.19 & 324 \\
F1 & 948 & 352 & 1.51 & 0.13 & 68 \\
F2 & 1280 & 526 & 1.42 & 0.19 & 441 \\
F3 & 1687 & 633 & 1.34 & 0.23 & 605 \\
F4 & 1856 & 600 & 1.30 & 0.12 & 129 \\
F5 & 2326 & 697 & 1.27 & 0.18 & 905 \\
\enddata
\label{table:sptagerelation}
\end{deluxetable}

%%%%% DISCUSSION %%%%%

\section{Discussion}
The precision of the age-dating method described here relies on the use of Str\"{o}mgren $ubvy\beta$ photometry to finely distinguish stellar atmosphere parameters and compare them to isochrones from stellar evolution models. For ages $\leq$ 10 Myr and $\gtrsim$ 100 Myr, in particular, there is rapid evolution of $\log{T_\text{eff}}$ and $\log{g}$ for intermediate-mass stars (see Figure ~\ref{fig:evolutiont}).  This enables greater accuracy in age determination through isochrone placement for stars in this mass and age range.  Fundamentally, our results rely on the accuracy of both the stellar evolution models and the stellar atmosphere models that we have adopted.  Accuracy is further set by the precision of the photometry, the derived atmospheric parameters, the calibration of the isochrones, and the ability to determine whether an individual star is contracting onto the main sequence or expanding off of it.  By using isochrones that include both pre-MS and post-MS evolution in a self-consistent manner \citep{bressan2012}, we can determine pre-ZAMS in addition to post-ZAMS ages for every data point in $T_\mathrm{eff}, \log{g}$).

Above, we have described our methodology in detail, including corrections for reddening and rotation,  and we have presented quality control tests that demonstrate the precision and accuracy of our ages. In the section we describe several aspects of our analysis of specific interest, including the context of previous estimates of stellar ages for early type stars (\S ~\ref{sec:context}), how to treat stars with locations apparently below the ZAMS (\S ~\ref{subsec:belowzams}), and discussion of notable individual objects (\S ~\ref{subsec:specialstars}). We will in the future apply our methods to new spectrophotometry.

\subsection{Methods Previously Employed in Age Determination for Early Type Stars}\label{sec:context}

In this section we place our work on nearby open cluster stars and approximately 3500 nearby field stars in the context of previous age estimation methods for BAF stars.

\cite{song2001} utilized a method quite similar to ours, employing $uvby\beta$ data from the catalogs of \cite{hauck1980, olsen1983, olsen1984}, the color grids of \cite{moon1985} including a temperature-dependent gravity modification suggested by \cite{napiwotzki1993}, and isochrones from \cite{schaller1992}, to determine the ages of 26 Vega-like stars.

For A-type stars, \cite{vican2012} determined ages for \emph{Herschel} DEBRIS survey stars by means of isochrone placement in log($T_\text{eff}$)-log($g$) space using \cite{li2008} and \cite{pinsonneault2004} isochrones, and atmospheric parameters from the literature. \cite{rieke2005} published age estimates for 266 B- \& A-type main sequence stars using cluster/moving group membership, isochrone placement in the H-R diagram, and literature ages (mostly coming from earlier application of $uvby\beta$ photometric methods).

Among later type F dwarfs, previous age estimates come primarily from the Geneva-Copenhagen Survey \citep{casagrande2011}, but their reliability is caveated by the substantially different values published in various iterations of the catalog \citep{nordstrom2004, holmberg2007, holmberg2009, casagrande2011} and the inherent difficulty of isochrone dating these later type dwarfs.

More recently, \cite{nielsen2013} applied a Bayesian inference approach to the age determination of 70 B- \& A-type field stars via $M_V$ vs $B-V$ color-magnitude diagram isochrone placement, assuming a constant star formation rate in the solar neighborhood and a Salpeter IMF. \cite{derosa2014} estimated the ages of 316 A-type stars through placement in a $M_K$ vs $V-K$ color-magnitude diagram.  Both of these broad-band photometric studies used the theoretical isochrones of \cite{siess2000}.

Considering the above sources of ages, the standard deviation among them suggests scatter among authors of only 15\% for some stars up to 145\%, with a typical value of 40\%. The full range (as opposed to the dispersion) of published ages is 3-300\%, with a peak around the 80\% level. The value of the age estimates presented here resides in the large sample of early type stars and the uniform methodology applied to them.

\subsection{Stars Below the Main Sequence}
\label{subsec:belowzams}

In Figure~\ref{fig:hrd} it may be noted that many stars, particularly those with $\log \mathrm{T_{eff}} \leq 3.9$, are located well below the model isochrones. Using rotation-corrected atmospheric parameters, $\sim$ 540 stars or $\sim 15\%$ of the sample, fall below the theoretical ZAMS. 

Prior studies also faced a similarly large fraction of stars falling below the main sequence. \cite{song2001} arbitrarily assigned an age of 50 Myr to any star lying below the 100 Myr isochrone used in that work. \cite{tetzlaff2011} arbitrarily shifted stars towards the ZAMS and treated them as ZAMS stars.

Several possibilities might be invoked to explain the large population of stars below the $\log{g}-\log{T_\mathrm{eff}}$ isochrones: (1) failure of evolutionary models to predict the true width of the MS, (2) spread of metallicities, with the metal-poor MS residing beneath the solar-metallicity MS, (3) overaggressive correction for rotational velocity effects, or (4) systematics involved in surface gravity or luminosity determinations. Of these explanations, we consider (4) the most likely, with (3) also contributing somewhat. \citet{valenti2005} found a 0.4 dex spread in $\log{g}$ among their main sequence FGK stars along with a 0.1 dex shift downward relative to the expected zero metallicity main sequence. 

\begin{figure}
\centering
\includegraphics[width=0.4\textwidth]{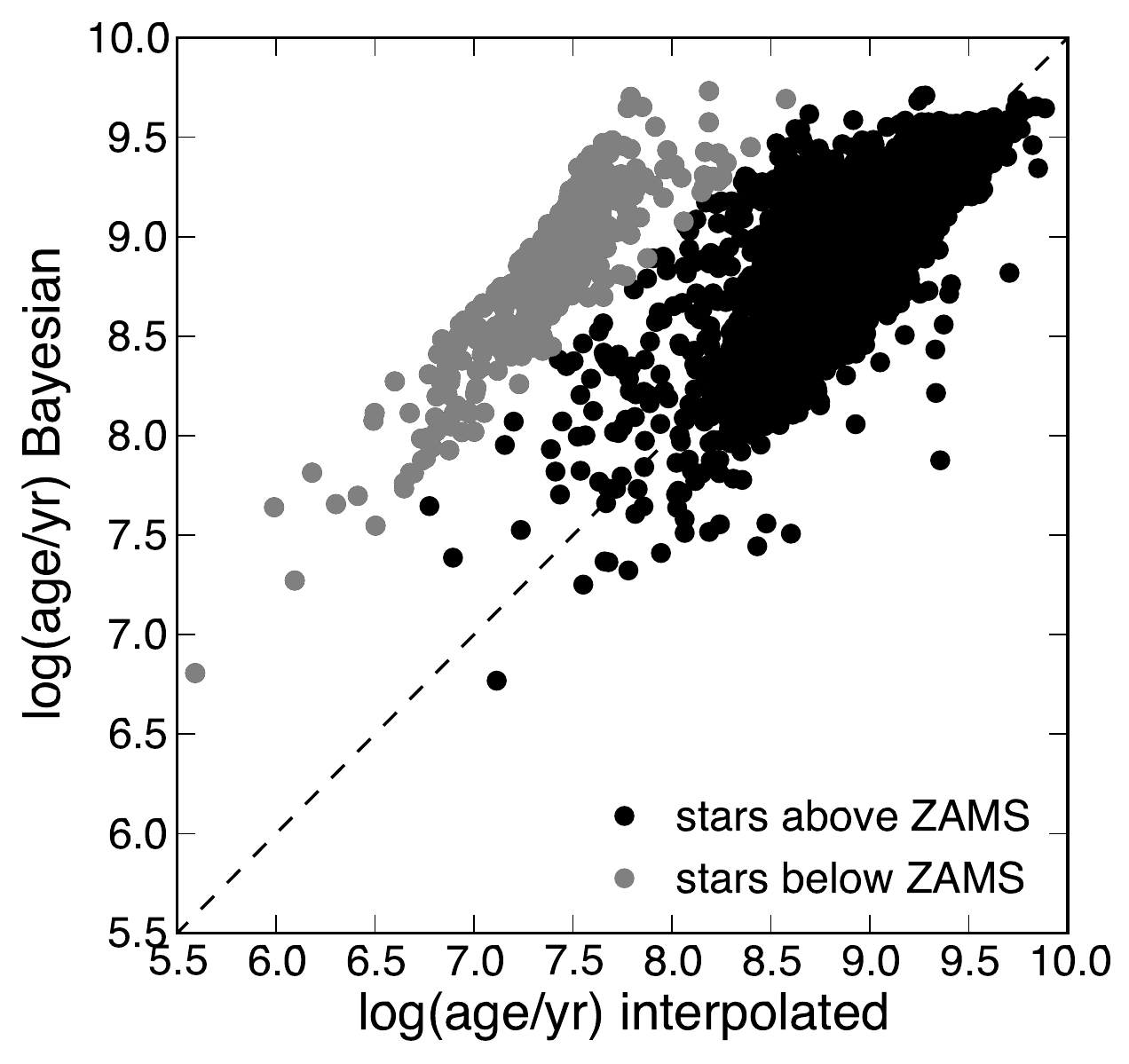}
\caption{Comparison of ages for BAF field stars derived through 2D linear interpolation and Bayesian inference. Grey points represent those stars with $\Delta \log{\mathrm{age/yr}} > 1$ (in the sense of Bayesian minus interpolated), which coincide with the same stars that reside below the MS.}
\label{fig:bayesinterp}
\end{figure}

The Bayesian age estimates for stars below the MS are likely to be unrealistically old, so we compared the ages for these stars with interpolated ages. Using the field star atmospheric parameters and \cite{bressan2012} models, we performed a 2D linear interpolation with the SciPy routine \texttt{griddata}. Stars below the main sequence could be easily identified by selecting objects with $\log{\mathrm{(age/yr)}}_\mathrm{Bayes} - \log{\mathrm{(age/yr)}}_\mathrm{interp} > 1.0$. Notably, for these stars below the MS, the linear interpolation produces more realistic ages than the Bayesian method. A comparison of the Bayesian and interpolated ages for all stars is presented in Figure ~\ref{fig:bayesinterp}. Of note, there is closer agreement between the Bayesian and interpolation methods in regards to estimating masses.

Figure ~\ref{fig:bayesinterp} further serves to illustrate the difference between the Bayesian ages and the interpolated ages, which scatters over an order magnitude from a 1:1 relationship. A number of stars that fall below the MS and have independently constrained ages are examined in detail in n \S ~\ref{subsec:specialstars}. These stars have interpolated ages that are more in line with prior studies, and in light of this, we publish the interpolated ages in addition to the Bayesian ages in the final electronic table.

\subsection{Stars of Special Interest}
\label{subsec:specialstars}

In this section we discuss stars of particular interest given that they have either spatially resolved debris disks, detected possibly planetary mass companions, or both.
As a final test of the \cite{bressan2012} evolutionary models and our Bayesian age and mass estimation method, we performed our analysis on these stars, including the Sun.

\begin{figure*}
\centering
\includegraphics[width=0.45\textwidth]{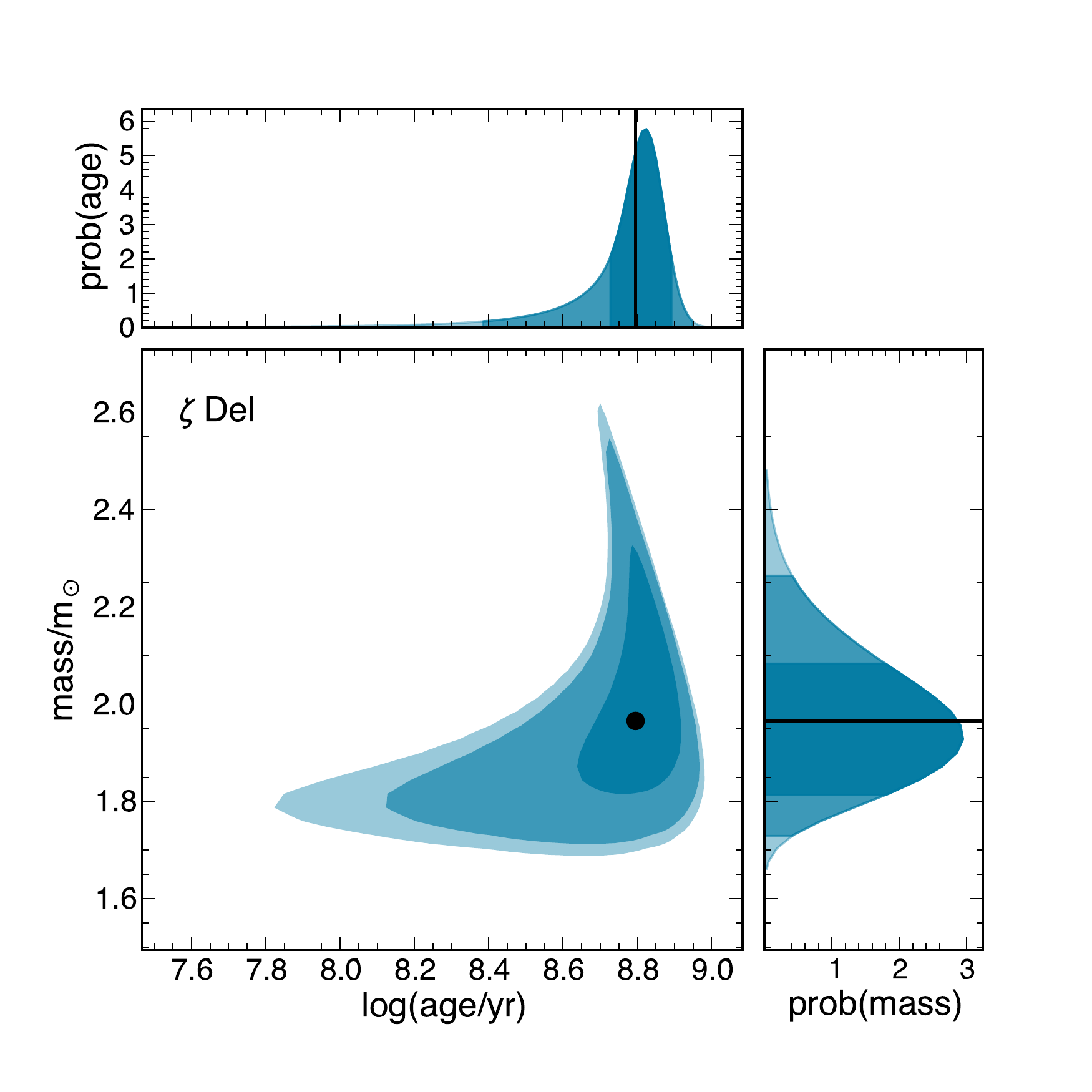}
\includegraphics[width=0.45\textwidth]{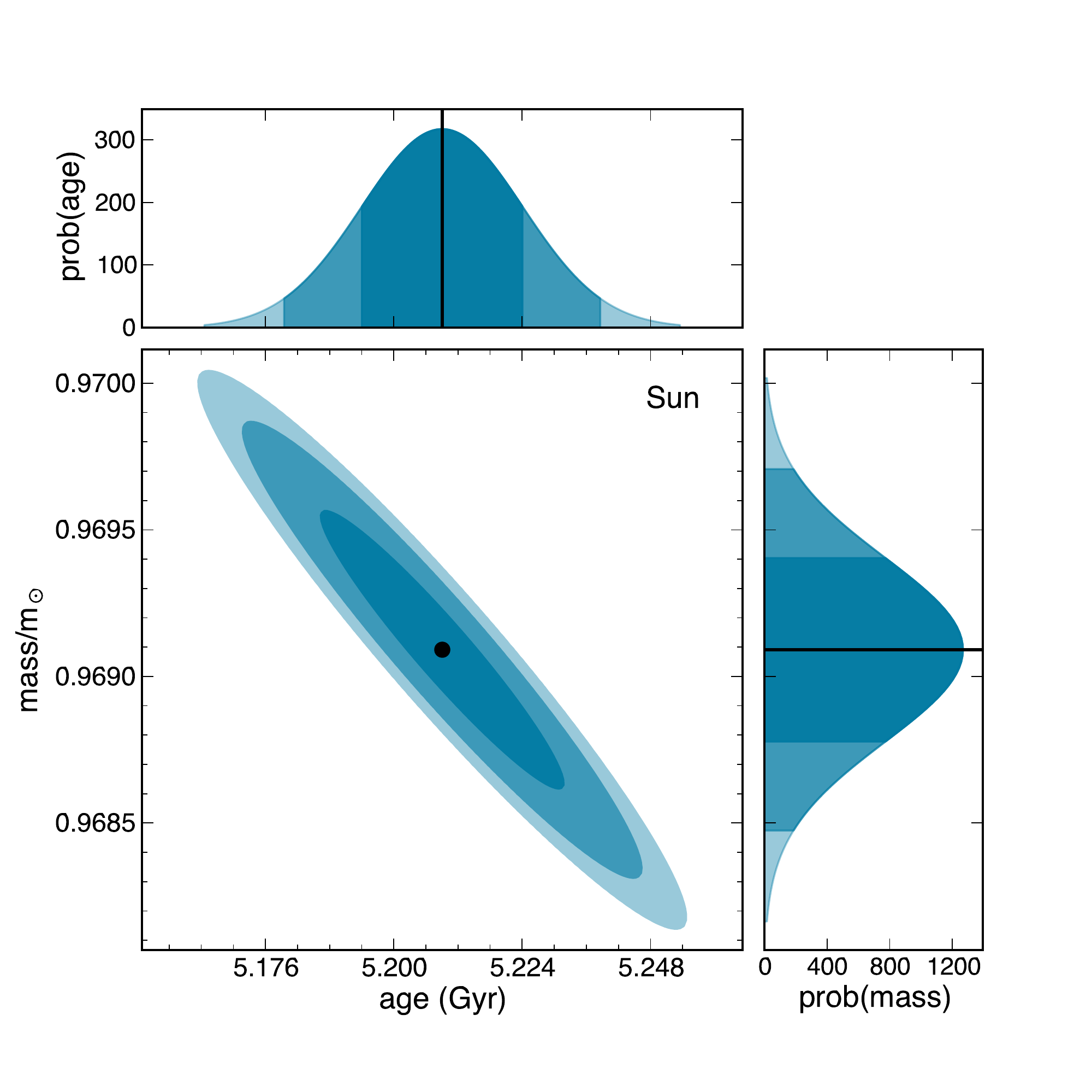}
\caption{2D joint posterior PDFs in age and mass for an early-type star with typical atmospheric parameter uncertainties (\emph{left}) and the Sun (\emph{right}), for which $T_\mathrm{eff}$ and $\log{g}$ are known to high precision. The dark, medium, and light blue shaded regions indicate the 68\%, 95\%, and 99\% confidence contours. Above, 1D marginalized and normalized posterior PDF in age, with the shaded regions representing the same corresponding confidence intervals. Right, the same as above for mass.}
\label{fig:joint-post}
\end{figure*}

\subsubsection{Sun}
The atmospheric parameters of our Sun are known with a precision that is orders of magnitude higher than what is obtainable for nearby field stars. One would thus expect the assumed priors to have a negligible influence on the Bayesian age and mass estimates.

The effective temperature of the Sun is calculated to be $T_\mathrm{eff} = 5771.8 \pm 0.7$ K from the total solar irradiance \citep{kopp2011}, the solar radius \citep{haberreiter2008}, the IAU 2009 definition of the AU, and the CODATA 2010 value of the Stefan-Boltzmann constant. The solar surface gravity is calculated to be $\log{g} = 4.43812 \pm 0.00013$ dex from the IAU 2009 value of $GM_\odot$ and the solar radius \citep{haberreiter2008}. Using these values, our Bayesian analysis yields a median age of 5.209$\pm$0.015 Gyr. The Bayesian estimation also yields a mass estimate of 0.9691$\pm$0.0003 $M_\odot$. Performing a 2D linear interpolation yields a slightly older age of 5.216 Gyr and slightly lower mass of 0.9690 $M_\odot$. As expected, the precise solar values lead to an elliptical joint posterior PDF in age and mass, and symmetric 1D marginalized PDFs. The difference between the Bayesian age estimate and interpolated age is negligible in this regime of extremely small uncertainties. This test also demonstrates that the \cite{bressan2012} evolutionary models may introduce a systematic overestimation of ages and underestimation of masses towards cooler temperatures, though because the Sun is substantially different from our sample stars we do not extrapolate this conclusion to our sample.

\subsubsection{HR 8799}
HR 8799 is located near the ZAMS and is metal-poor with [Fe/H]=$−0.47 \pm 0.10$ dex \citep{gray1999}. However, because HR 8799 is a $\lambda$ Boo peculiar-type star, its photospheric metallicity may not reflect the global stellar metal abundance. The age of HR 8799 is believed to be $30^{+20}_{-10}$ Myr based on its proposed membership to the Columba association \citep{zuckerman2011}.

Figure ~\ref{fig:hrd} shows that HR 8799 lays well below the theoretical ZAMS. This location is well-documented from other spectroscopic and photometric analyses of the star, and is likely due to a combination of its genuine youth and subsolar metallicity. Consistent with the discussion in \S 8.2 and as illustrated in Figure ~\ref{fig:bayesinterp}, our Bayesian age analysis leads to an unrealistically old age for the star, with a median age of 956 Myr and a 68\% confidence interval of 708-1407 Myr. The Bayesian approach also seems to overestimate the mass, with a median mass of $1.59 M_\odot$ and 68\% confidence interval of $1.49-1.68 M_\odot$. Notably, however, 2D linear interpolation leads to more reasonable age estimates: 26 Myr assuming our newly derived atmospheric parameters ($T_\mathrm{eff}$=7540 K, $\log{g}$ = 4.43), or 25 Myr using $T_\mathrm{eff}$=7430 K and $\log{g}$ = 4.35 from \cite{gray1999}.

\subsubsection{$\beta$ Pic}
\cite{zuckerman2001} assigned an age of 12 Myr to $\beta$ Pic based on its proposed membership to the moving group of the same name. Isochronal age estimates for the star have ranged from the ZAMS age to 300 Myr \citep{barrado1999}. \cite{nielsen2013} performed a Bayesian analysis concluding a median age of 109 Myr with a 68\% confidence interval of 82-134 Myr. Although barely below the ZAMS, $\beta$ Pic in our own Bayesian analysis has a much older median age of 524 Myr with a 68\% confidence interval of 349-917 Myr. Prior authors also have noted that $\beta$ Pic falls below the ZAMS on a color-magnitude diagram. As was the case for HR~8799, we conclude that our erroneous age for $\beta$ Pic is due to the dominance of the prior assumption/s in exactly such a scenario. 

However, the interpolated age using our atmospheric parameters of $T_\mathrm{eff}$=8300 K, $\log{g}$=4.389, is 20 Myr. Using the \cite{gray2006} values of $T_\mathrm{eff}$=8052 K (within $1\sigma$ of our determination), $\log{g}$=4.15 ($>1.5\sigma$ away from our surface gravity) we obtain an interpolated age of 308 Myr.

\subsubsection{$\kappa$ And}
$\kappa$ Andromedae is another proposed member of the Columba association \citep{zuckerman2011}. Using the nominal 30 Myr age, \cite{carson2013} suggested a companion  discovered via direct imaging is of planetary mass ($12-13 M_\mathrm{Jup}$). \cite{hinkley2013} refuted this claim, concluding an age of $220\pm100$ Myr from multiple isochronal analyses in \S3.2 of that work. This older age estimate leads to a model-dependent companion mass of $50^{+16}_{-13} M_\mathrm{Jup}$. Our Bayesian analysis allows us to nearly rule out a 30 Myr age with a 95\% confidence interval of 29-237 Myr. The mean, median, mode, and 68\% confidence interval of the 1D marginalized posterior PDF in age for $\kappa$ And are 118, 150, 191, and 106-224 Myr, respectively. Notably, $\kappa$ And has a projected rotational velocity of $v\sin{i} \sim 160$ km s$^{-1}$ \citep{glebocki2005}, and we find its rotation corrected atmospheric parameters ($T_\mathrm{eff} = 11903 \pm 405$ K, $\log{g} = 4.35 \pm 0.14$ dex) produce an interpolated age of 16 Myr. Using uncorrected atmospheric parameters ($T_\mathrm{eff} = 11263 \pm 383$ K, $\log{g} = 4.26 \pm 0.14$ dex) leads to an interpolated age of 25 Myr. 

\subsubsection{$\zeta$ Delphini}
\cite{derosa2014} recently published the discovery of a wide companion to $\zeta$ Delphini (HIP 101589). Those authors estimated the age of the system as 525$\pm$125 Myr, from the star's positions on a color-magnitude and a temperature-luminosity diagram, leading to a model-dependent companion mass of 50$\pm$15 M$_\mathrm{Jup}$. Our method yields a mean age of 552 Myr, with 68\% and 95\% confidence intervals of 531-772 Myr, and 237-866 Myr, respectively. Our revised age is in agreement with the previous estimate of \cite{derosa2014}, although favoring the interpretation of an older system and thus more massive companion. The interpolated ages for $\zeta$ Del are somewhat older: 612 Myr for the rotation-corrected atmospheric parameters $T_\mathrm{eff}$=8305 K, $\log{g}$=3.689, or 649 Myr for the uncorrected parameters $T_\mathrm{eff}$=8639 K, $\log{g}$=3.766. Note, in this case moderate rotation ($v \sin i =$ 99.2 km s$^{-1}$) leads to a discrepancy of only $\approx 6\%$ in the derived ages.

\subsubsection{49 Ceti}
49 Ceti does not have a known companion at present, but does possess a resolved molecular gas disk \citep{hughes2008}. The star is a proposed member of the 40 Myr Argus association, which would require cometary collisions to explain the gaseous disk that should have dissipated by $\sim$ 10 Myr due to radiation pressure \citep{zuckerman2012}. With a mean rotational velocity of $\sim$ 190 km s$^{-1}$ \citep{glebocki2005}, and evidence that the star is highly inclined to our line of sight, rotational effects on photometric H-R diagram placement are prominent. Our $uvby\beta$ atmospheric parameters for 49 Ceti are $T_\mathrm{eff}$ = 10007 $\pm$ 340 K, $\log{g} = 4.37 \pm 0.14$ dex, after rotational effects were accounted for. These parameters place the star essentially on the ZAMS, with an interpolated age of 9 Myr, and calling into question the cometary genesis of its gaseous disk. However, the uncorrected atmospheric parameters ($T_\mathrm{eff} = 9182 \pm 309$ K, $\log{g}=4.22 \pm 0.14$ dex) are more consistent with the A1 spectral type and produces an interpolated age of 57 Myr, which seems to support the cometary collision hypothesis. This case illustrates the importance of high-precision atmospheric parameters.

%%%% CONCLUSIONS %%%%%

\section{Conclusions}

In the absence of finely calibrated empirical age indicators, such as the rotation-activity-age
relation for solar-type stars \citep[e.g.][]{mamajek2008}, ages for early spectral type stars typically 
have come from open cluster and moving group membership, or through association with a late-type
companion that can be age-dated through one of the applicable empirical methods.  Because of their rapid evolution, early type stars are amenable to age dating via isochrones. In this paper we have investigated the use of Str\"{o}mgren photometric techniques for estimating stellar atmospheric parameters, which are then compared to isochrones from modern stellar evolution models.  

Bayesian inference is a particularly useful tool in the estimation of parameters such as age and mass from evolutionary models for large samples that span considerable ranges in temperature, luminosity, mass, and age. The Bayesian approach produces unbiased ages relative to a straightforward interpolation among isochrones which leads to age estimates that are biased towards older ages. However, as noted earlier, stars located beyond the range of the theory (below the theoretical ZAMS in our case) are assigned unreasonably old ages with the Bayesian method. This presumably is due to the clustering of isochrones and the dominance of the prior in inference scenarios in which the prior probability is changing quickly relative to the magnitude of the uncertainty in the atmospheric parameters. Linear interpolation for stars apparently below the MS may produce more reasonable age estimates.

The most important parameter for determining precise stellar ages near the ZAMS is the luminosity or surface gravity indicator. Effective temperatures, or observational proxies for temperature, are currently estimated with suitable precision. However, $\log g$, luminosity, or absolute magnitude (requiring a precise distance as well), are not currently estimated with the precision needed to meaningfully constrain the ages of field stars near the ZAMS. This effect is particularly pronounced for lower temperature stars where, for a given shift in $\log g$, the inferred age can change by many orders of magnitude. Our open cluster tests indicated that the age uncertainties due to the choice of evolutionary models are not significant compared to those introduced by the uncertainties in the surface gravities.

We have derived new atmospheric parameters (taking stellar rotation into account) and model-dependent ages and masses for 3493 BAF stars within 100 pc of the Sun. Our method of atmospheric parameter determination was calibrated and validated to stars with fundamentally determined atmospheric parameters. We further tested and validated our method of age estimation using open clusters with well-known ages. In determining the uncertainties in all of our newly derived parameters we conservatively account for the effects of systematics, metallicity, numerical precision, reddening, photometric errors, and uncertainties in $v\sin{i}$ as well as unknown rotational velocities. 

Field star ages must be considered with caution. At minimum, our homogeneously derived set of stellar ages provides a relative youth ordering. For those stars below the MS we encourage the use of interpolated ages rather than Bayesian ages, unless more precise atmospheric parameters become available. Using the new set of ages, we presented an empirical mass-age relation for solar neighborhood B0-F5 stars. We also presented empirical relations between spectral type and age/mass and we discussed ages in detail for several famous low mass companion and/or debris disk objects. An anticipated use of our catalog is in the prioritization of targets for direct imaging of brown dwarf and planetary mass companions. David \& Hillenbrand (2015b, \emph{in preparation}) will explore how ages derived using this methodology can be applied to investigations such as debris disk evolution.

%%%%% ACKNOWLEDGMENTS %%%%%
The authors wish to thank John Stauffer for his helpful input on sources of $uvby\beta$ data for open clusters and Timothy Brandt for helpful discussions during the proof stage of this work regarding the open cluster analysis, resulting in the appendix material concerning logarithmic versus linear approaches and a modified version of Figure 17. This material is based upon work supported in 2014 and 2015 by the National Science Foundation Graduate Research Fellowship under Grant No. DGE‐1144469. This research has made use of the WEBDA database, operated at the Institute for Astronomy of the University of Vienna, as well as the SIMBAD database and VizieR catalog access tool, operated at CDS, Strasbourg, France.

%%%%% APPENDIX %%%%%
\clearpage
\appendix

%%%%%%%%%%%%%%%%%%%%%%%
% METALLICITY EFFECTS %
%%%%%%%%%%%%%%%%%%%%%%%
\subsection{Metallicity Effects}
\label{metallicity}

We do not account explicitly for metallicity in this study, having assumed solar values in both our atmospheric models and our evolutionary grids. Our analysis in the $T_\mathrm{eff}$ and $\log g$ calibrations found that for stars with fundamentally determined atmospheric parameters and available [Fe/H] measurements, the accuracy with which we can determine atmospheric parameters using $uvby\beta$ photometry does not vary systematically with metallicity.

\begin{figure*}[!h] 
\centering
\includegraphics[width=0.99\textwidth]{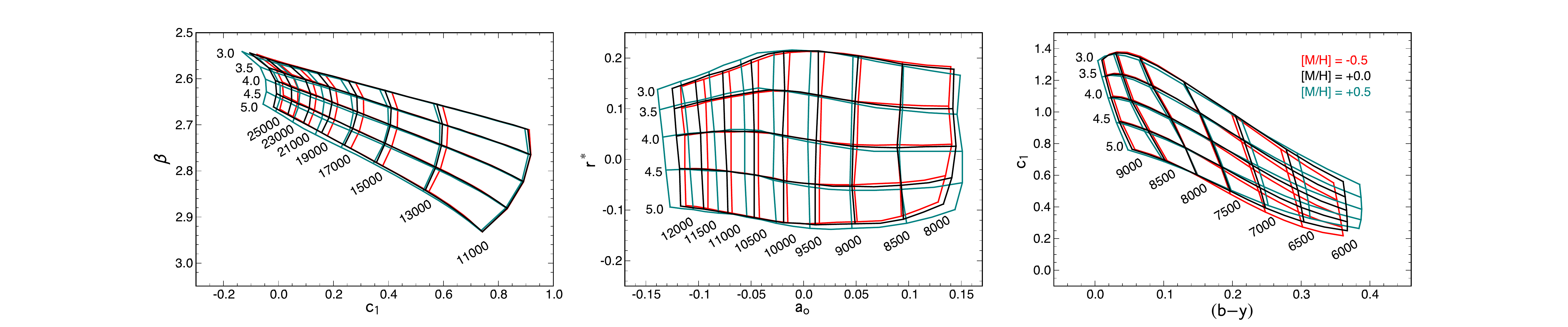}      
\caption{Comparison of ATLAS9 color grids for different metallicities.}
\label{fig:grids_metallicity}
\end{figure*}

The effects of different metallicity assumptions on the Str{\"o}mgren index atmospheric grids is illustrated in Figure~\ref{fig:grids_metallicity}. Moving from the atmospheric grid to the evolutionary grid, Figure 17 of \citet{valenti2005} illustrates that  for the coolest stars under consideration here, which were the focus of their study, variation of metallicity from +0.5 to -0.5 dex in [Fe/H] corresponds to a +0.1 to -0.1 dex shift in $\log g$ of an evolutionary isochrone. Among hotter stars, Figure~\ref{fig:grids_metallicity} shows that metallicity uncertainty affects temperatures only minorly, and gravities not at all or minimally.

\begin{figure*}[!h]
\centering
\includegraphics[width=0.3\textwidth]{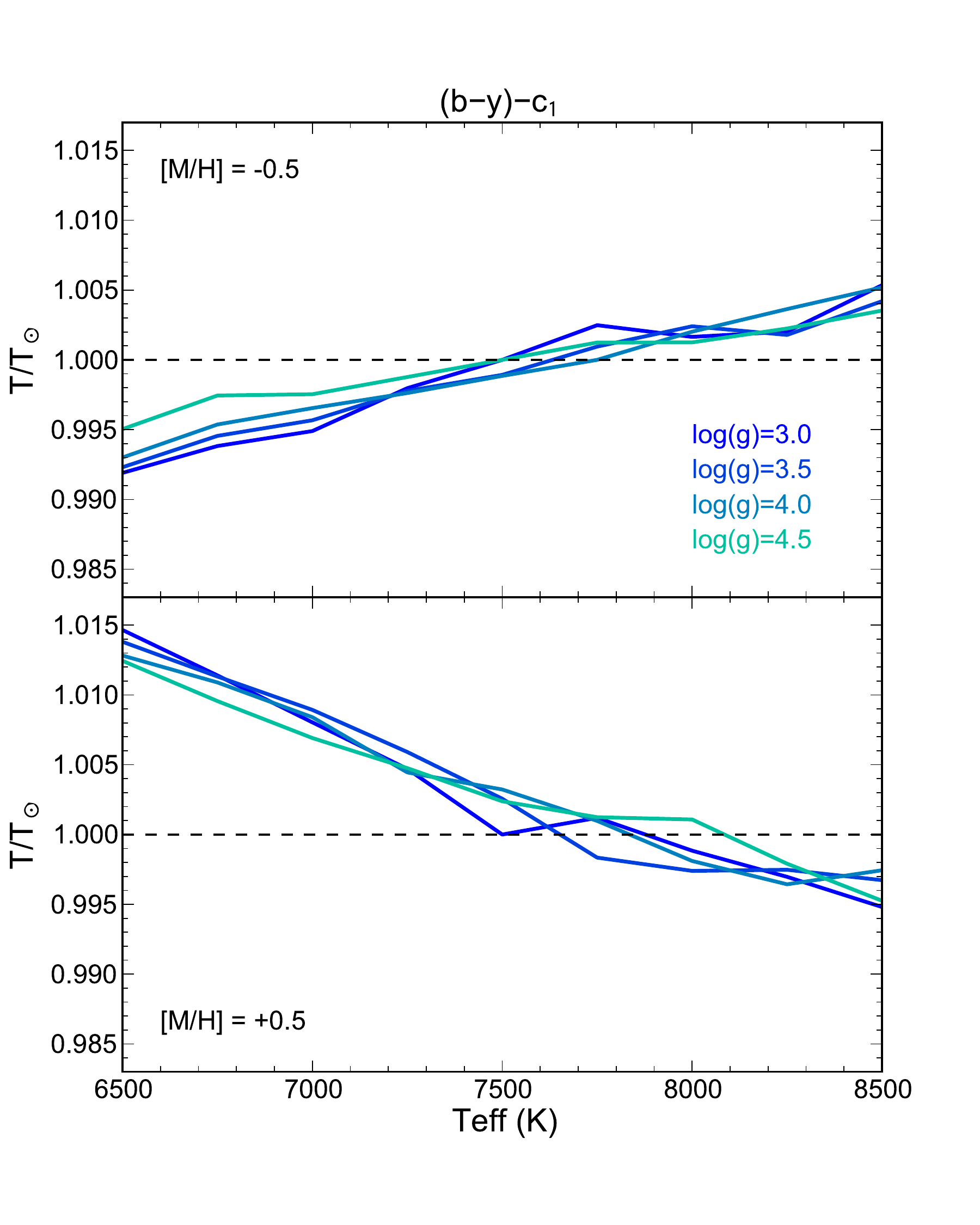}
\includegraphics[width=0.3\textwidth]{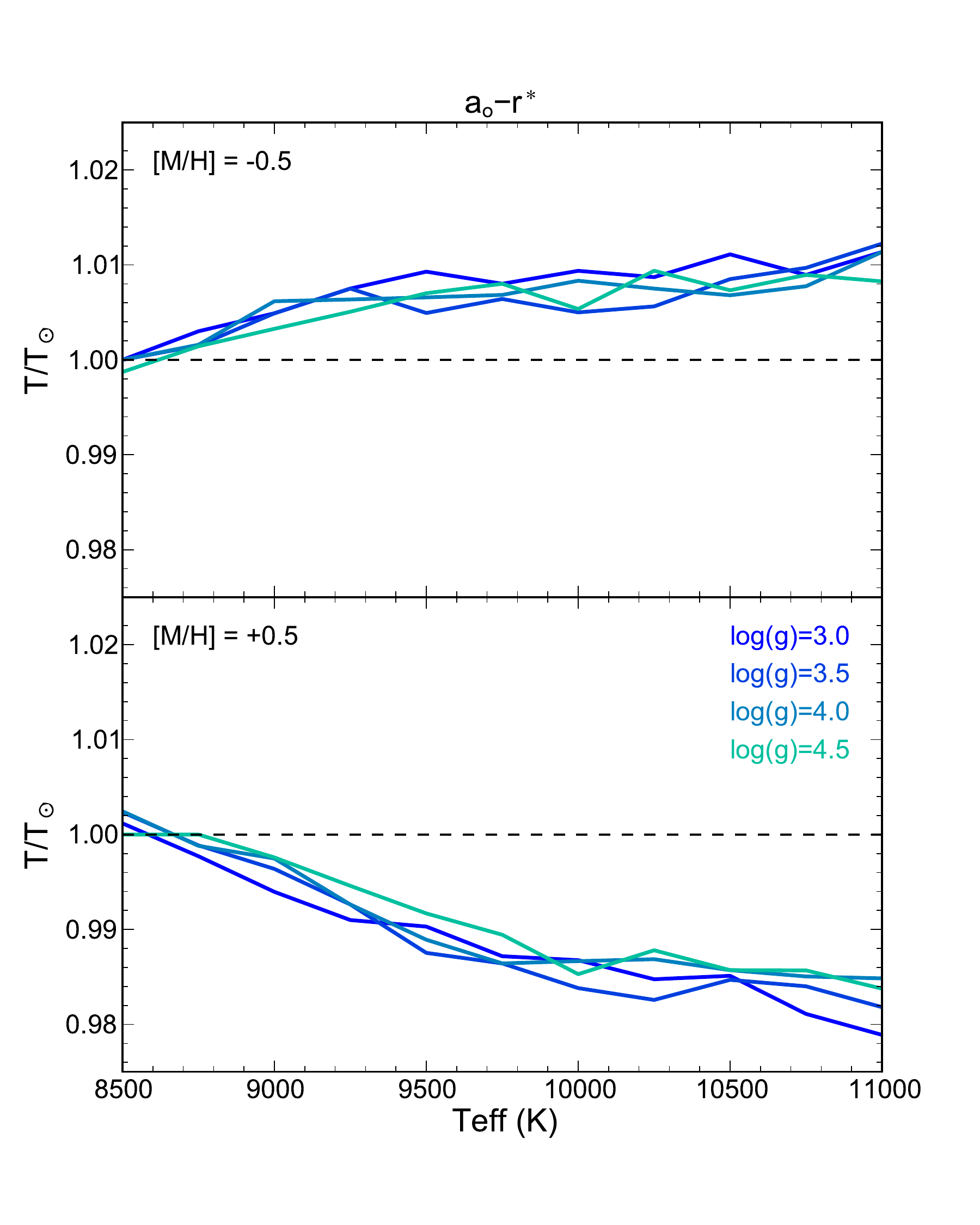}
\includegraphics[width=0.3\textwidth]{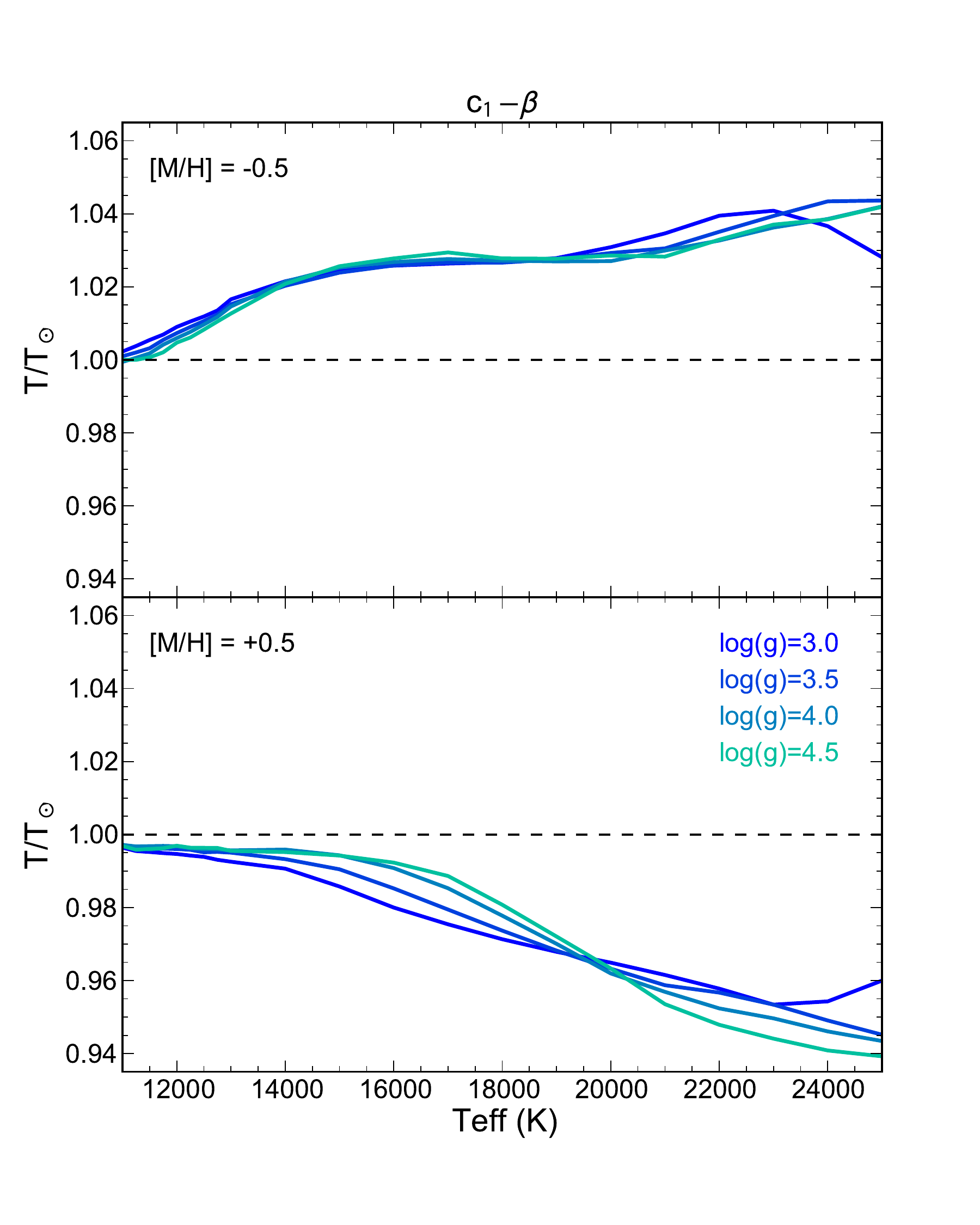}
\caption{The effect of metallicity on $uvby\beta$ determinations of temperature, as predicted by model grids of \cite{castelli2006, castelli2004}. In the left-most figure, for given values of $T_\mathrm{eff}$, or $(b-y)$, the ratio of the temperature given by the grid of metallicity [M/H]=-0.5 to the solar metallicity grid is depicted in the top panel. The bottom panel shows the ratio of the temperature given by a grid of metallicity [M/H]=+0.5 to the temperature given by the solar metallicity grid. In the temperature range of interest ($\approx$ 6500K-8500K, or spectral types F5-A4), a shift of 0.5 dex in [M/H] can produce variations up to $\sim 1\%$ in $T_\mathrm{eff}$, with the smallest discrepancies occurring at approximately the F0-A9 boundary. The middle figure is analogous to the left figure, for the $a_0-r^*$ grids which are used for stars between $\approx$ 8500K-11000K (A3-B9). In this regime, shifts of 0.5 dex in metallicity can produce variations up to $\sim 2\%$ in temperature. Finally, for the hottest stars ($T_\mathrm{eff} > 11000$ K, spectral types B9 and earlier), a 0.5 dex shift in metallicity can produce variations up to $\sim 6\%$ in effective temperature.}
\label{fig:met-teff}
\end{figure*}

\begin{figure*}[!h]
\centering
\includegraphics[width=0.3\textwidth]{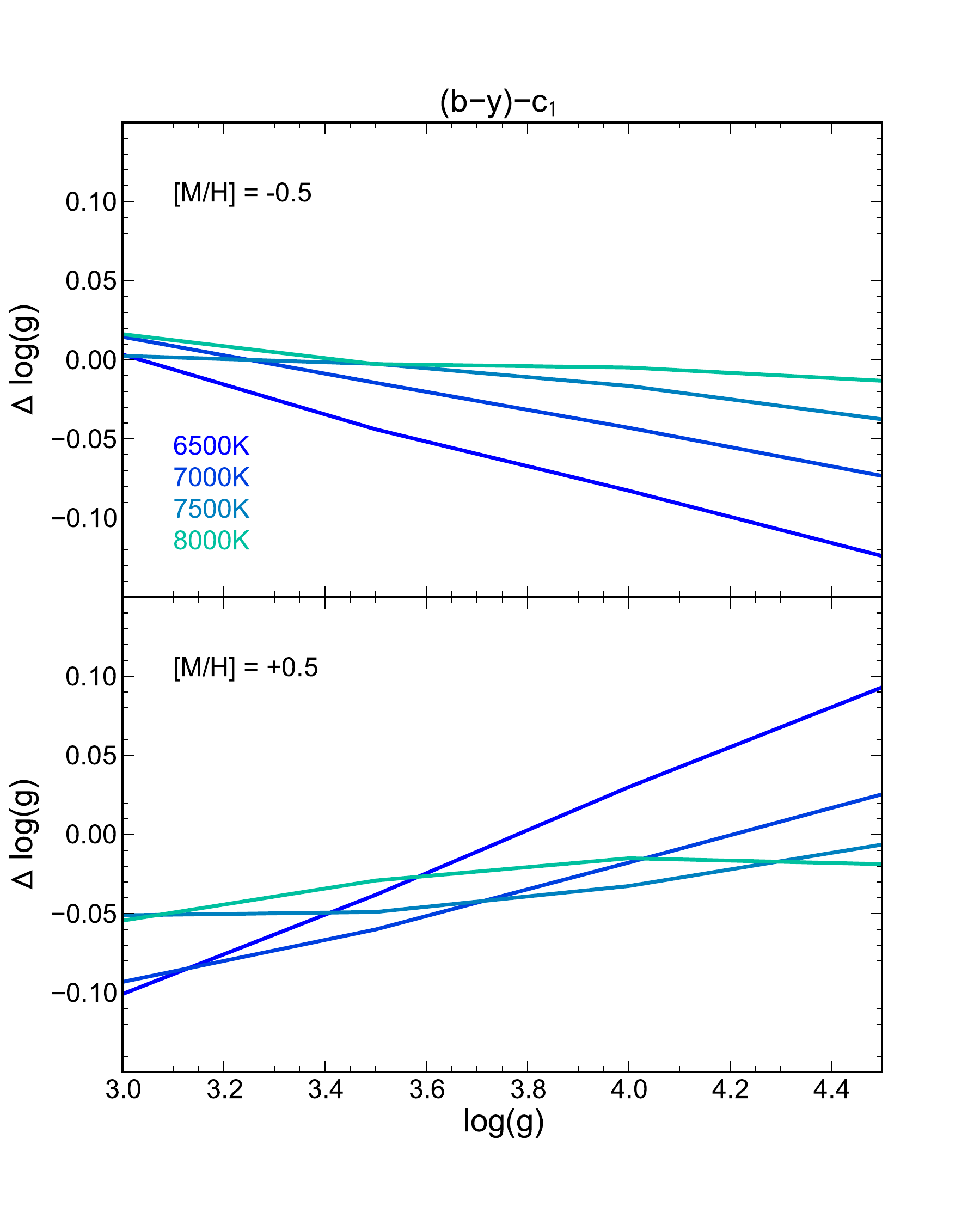}
\includegraphics[width=0.3\textwidth]{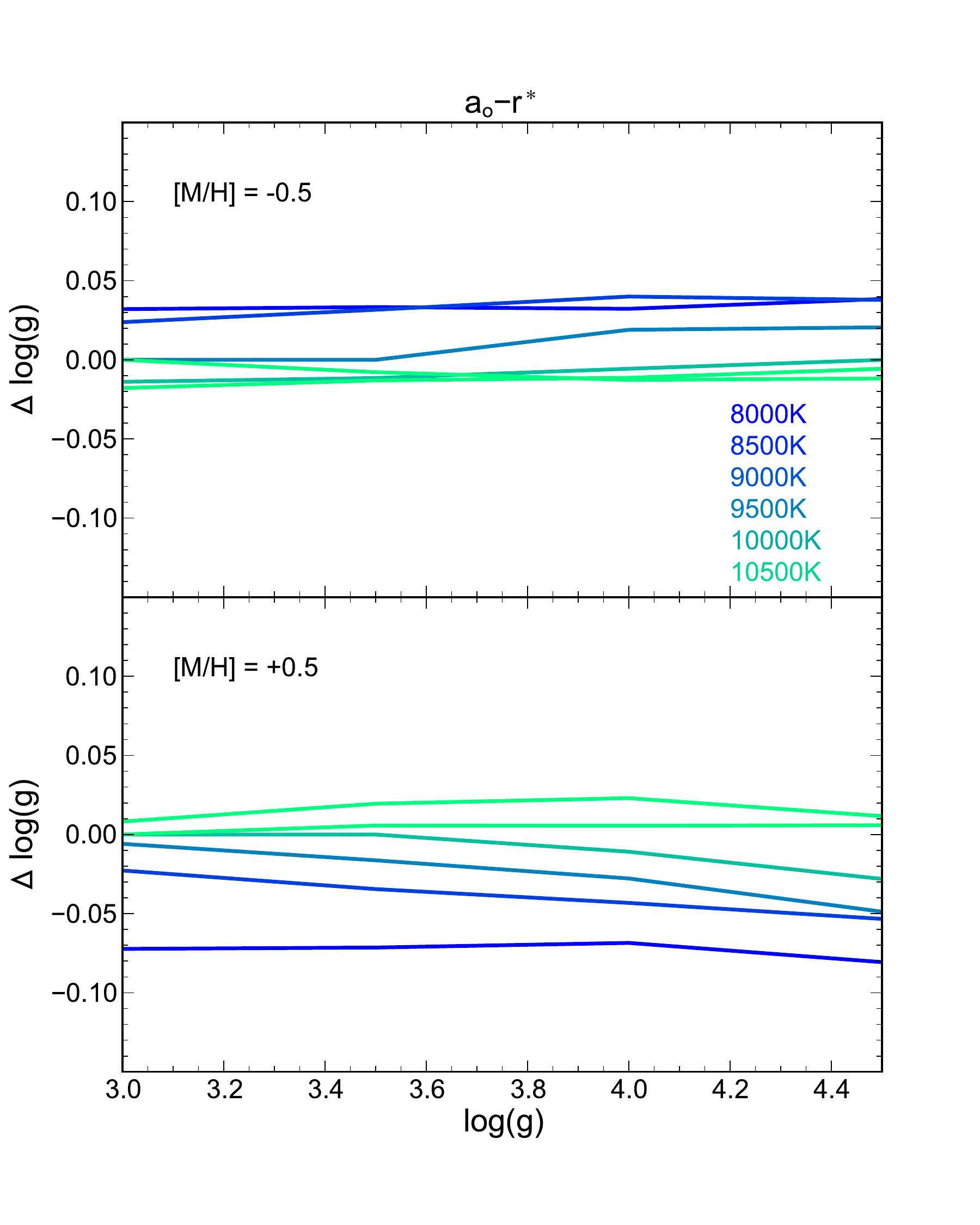}
\includegraphics[width=0.3\textwidth]{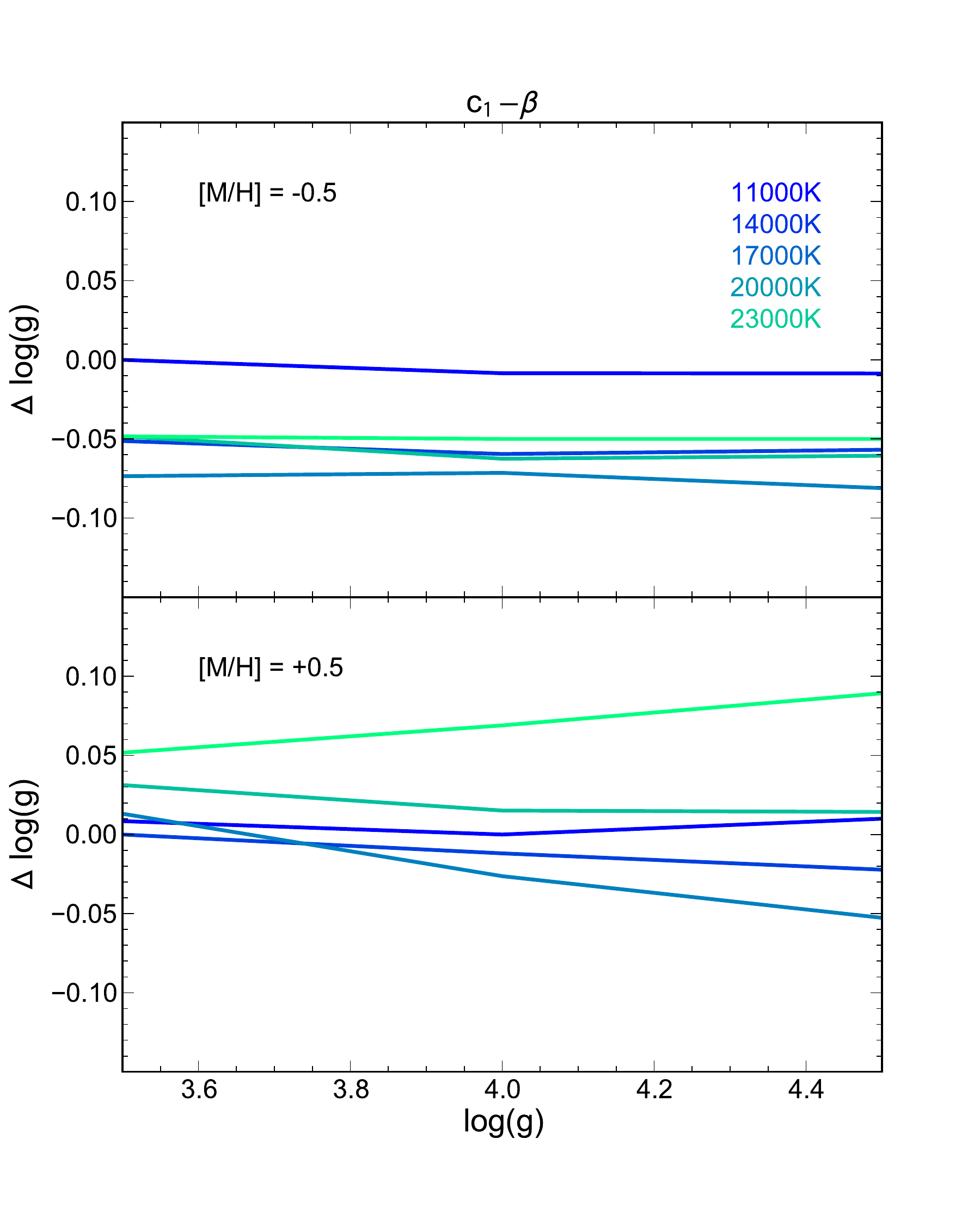}
\caption{The effect of metallicity on $uvby\beta$ determinations of surface gravity, as predicted by model grids of \cite{castelli2006, castelli2004}. In the left-most figure, for given values of $\log{g}$, or $c_1$, the ratio of the temperature given by the grid of metallicity [M/H]=-0.5 to the solar metallicity grid is depicted in the top panel. The bottom panel shows the ratio of the temperature given by a grid of metallicity [M/H]=+0.5 to the temperature given by the solar metallicity grid. In the temperature range of interest ($\approx$ 6500K-8500K, or spectral types F5-A4), a shift of 0.5 dex in [M/H] can produce variations up to $\sim 0.1$ dex in $\log{g}$. The middle figure is analogous to the left figure, for the $a_0-r^*$ grids which are used for stars between $\approx$ 8500K-11000K (A3-B9). In this regime, the gravity indicator $r^*$ is particularly insensitive to metallicity, with shifts of 0.5 dex in metallicity producing variations of only $\sim 0.05$ dex or less in $\log{g}$. Finally, for the hottest stars ($T_\mathrm{eff} > 11000$ K, spectral types B9 and earlier), a 0.5 dex shift in metallicity can produce variations up to $\sim 0.1$ dex in $\log{g}$.}
\label{fig:met-logg}
\end{figure*}

We similarly calculated the effect on atmospheric parameter determination when allowing the model color grids to vary from +0.5 to -0.5 dex in [M/H], which notably represent the extremes of the metallicity range included in our sample (less than 1\% of stars considered here have $\left | [\mathrm{Fe/H}] \right |>0.5$ dex). Figures ~\ref{fig:met-teff} \& ~\ref{fig:met-logg} examine in detail the effects of metallicity on $T_\mathrm{eff}, \log{g}$ determinations in the relevant $uvby\beta$ planes. In summary, $T_\mathrm{eff}$ variations of up to $\sim 1\%$ in the $(b-y)-c_1$ plane, $\sim 2\%$ in the $a_0-r^*$ plane, and $6\%$ in the $c_1-\beta$ plane are possible with shifts of $\pm$ 0.5 dex in [M/H]. Notably, however, $T_\mathrm{eff}$ variations above the 2\% level are only expected in the $c_1-\beta$ plane for stars hotter than $\sim 17000$ K, or roughly spectral type B4, of which there are very few in our sample. Similarly metallicity shifts of $\pm$ 0.5 dex can cause variations of $\sim$ 0.1 dex in $\log{g}$ in the $(b-y)-c_1$ and $c_1-\beta$ planes, while the same variation in the $a_0-r^*$ plane produces surface gravity shifts closer to $\sim 0.05$ dex.

By contrast, metallicity effects are more prominent in color-magnitude techniques. Recently, \citet{nielsen2013} executed a Bayesian analysis of the locations in the $M_V$ vs $B-V$ diagram of Gemini/NICI targets to derive their ages including confidence contours for the stellar masses, ages, and metallicities.  The work demonstrates correlation in this particular color-magnitude diagram of increasing mass and decreasing age with higher metallicity. Metal poor stars will have erroneously young ages attributed to them when solar metallicity is assumed.

%%%%%%%%%%%%%%%%%%%%%%%%
% CONFIDENCE INTERVALS %
%%%%%%%%%%%%%%%%%%%%%%%%

\subsection{Confidence Intervals}

All confidence intervals in age and mass quoted in this work are the bounds of the Highest Posterior Density (HPD) Region. For a given posterior probability density, $p(\theta | x)$, the $100(1-\alpha) \%$ HPD region is defined as the subset, $\mathcal{C}$, of $\theta$ values:

\begin{equation}
\mathcal{C} = \left \{  \theta : p(\theta | x) \geq  p^*\right \},
\end{equation}

where $p^*$ is the largest number such that

\begin{equation}
\int_{\theta: p(\theta | x) \geq p^*}  p(\theta | x) \mathrm{d} \theta = 1- \alpha.
\end{equation}

In other words, the HPD region is the set of most probable values (corresponding to the smallest range in $\theta$) that encloses $100(1-\alpha) \%$ of the posterior mass. The HPD method is particularly suited for finding confidence intervals of skewed probability distributions, such as the stellar age posteriors studied in this work. To find the highest posterior density (HPD) region numerically, a function is created that iteratively integrates a normalized posterior PDF above a test value of $p^*$ while the area/volume under the PDF is less than the desired confidence interval.

\subsection{Open Cluster Tables}

\begin{deluxetable*}{ccccccccc}
\tabletypesize{\footnotesize}
\tablecolumns{9}
\tablewidth{0.99\textwidth}
\tablecaption{IC 2602 members dereddened $uvby\beta$ photometry and atmospheric parameters \label{table:ic2602}}
\tablehead{
\colhead{Star} & 
\colhead{Sp. Type} & 
\colhead{$(b-y)_0$} & 
\colhead{$m_0$} & 
\colhead{$c_0$} &
\colhead{$\beta$} &
\colhead{$T_\mathrm{eff}$} &
\colhead{$\log g$} &
\colhead{$v \sin i$}
\\
 &  & (mag) & (mag) & (mag) & (mag) & (K) & (dex) & (km s$^{-1}$) 
}
\startdata
HD 91711 & B8 V    & -0.062 &  0.146 &  0.457 &  2.745 &  14687 $\pm$    235 &  4.467 $\pm$ 0.113 &  153 \\
HD 91839 & A1 V    &  0.025 &  0.178 &  1.033 &  2.904 &   9509 $\pm$    152 &  4.188 $\pm$ 0.091 &  146 \\
HD 91896 & B7 III  & -0.081 &  0.093 &  0.346 &  2.660 &  16427 $\pm$    263 &  3.782 $\pm$ 0.113 &  155 \\
HD 91906 & A0 V    &  0.016 &  0.177 &  1.005 &  2.889 &   9799 $\pm$    157 &  4.146 $\pm$ 0.113 &  149 \\
HD 92275 & B8 III/IV & -0.056 &  0.125 &  0.562 &  2.709 &  13775 $\pm$    220 &  3.852 $\pm$ 0.113 &  153 \\
HD 92467 & B95III  & -0.026 &  0.168 &  0.833 &  2.851 &  11178 $\pm$    179 &  4.423 $\pm$ 0.113 &  110 \\
HD 92478 & A0 V    &  0.010 &  0.183 &  0.978 &  2.925 &   9586 $\pm$    153 &  4.431 $\pm$ 0.091 &   60 \\
HD 92535 & A5 V n  &  0.104 &  0.194 &  0.884 &  2.838 &   8057 $\pm$    129 &  4.344 $\pm$ 0.145 &  140 \\
HD 92536 & B8 V    & -0.043 &  0.131 &  0.705 &  2.795 &  13183 $\pm$    211 &  4.423 $\pm$ 0.113 &  250 \\
HD 92568 & A  M    &  0.209 &  0.237 &  0.625 &  2.748 &   7113 $\pm$    114 &  4.341 $\pm$ 0.145 &  126 \\
HD 92664 & B8 III P & -0.083 &  0.118 &  0.386 &  2.702 &  15434 $\pm$    247 &  4.145 $\pm$ 0.113 &   65 \\
HD 92715 & B9 V nn & -0.027 &  0.136 &  0.882 &  2.836 &  12430 $\pm$    199 &  4.362 $\pm$ 0.113 &  290 \\
HD 92783 & B85V nn & -0.033 &  0.124 &  0.835 &  2.804 &  12278 $\pm$    196 &  4.130 $\pm$ 0.113 &  230 \\
HD 92837 & A0 IV nn & -0.007 &  0.160 &  0.953 &  2.873 &  10957 $\pm$    175 &  4.322 $\pm$ 0.113 &  220 \\
HD 92896 & A3 IV   &  0.114 &  0.193 &  0.838 &  2.831 &   8010 $\pm$    128 &  4.425 $\pm$ 0.145 &  139 \\
HD 92938 & B3 V n  & -0.075 &  0.105 &  0.384 &  2.690 &  15677 $\pm$    251 &  4.015 $\pm$ 0.113 &  120 \\
HD 92966 & B95V nn & -0.019 &  0.158 &  0.930 &  2.878 &  11372 $\pm$    182 &  4.445 $\pm$ 0.113 &  225 \\
HD 92989 & A05Va   &  0.008 &  0.180 &  0.982 &  2.925 &   9979 $\pm$    160 &  4.480 $\pm$ 0.091 &  148 \\
HD 93098 & A1 V s  &  0.017 &  0.180 &  0.993 &  2.915 &   9688 $\pm$    155 &  4.385 $\pm$ 0.091 &  135 \\
HD 93194 & B3 V nn & -0.078 &  0.105 &  0.357 &  2.668 &  17455 $\pm$    279 &  4.015 $\pm$ 0.113 &  310 \\
HD 93424 & A3 Va   &  0.060 &  0.197 &  0.950 &  2.890 &   8852 $\pm$    142 &  4.247 $\pm$ 0.113 &   95 \\
HD 93517 & A1 V    &  0.052 &  0.196 &  0.976 &  2.919 &   9613 $\pm$    154 &  4.510 $\pm$ 0.091 &  220 \\
HD 93540 & B6 V nn & -0.065 &  0.116 &  0.476 &  2.722 &  15753 $\pm$    252 &  4.308 $\pm$ 0.113 &  305 \\
HD 93549 & B6 V    & -0.066 &  0.123 &  0.454 &  2.729 &  15579 $\pm$    249 &  4.422 $\pm$ 0.113 &  265 \\
HD 93607 & B25V n  & -0.084 &  0.102 &  0.292 &  2.675 &  17407 $\pm$    279 &  4.098 $\pm$ 0.113 &  160 \\
HD 93648 & A0 V n  &  0.041 &  0.188 &  1.025 &  2.890 &   9672 $\pm$    155 &  4.157 $\pm$ 0.091 &  215 \\
HD 93714 & B2 IV-V n & -0.092 &  0.100 &  0.201 &  2.647 &  18927 $\pm$    303 &  3.979 $\pm$ 0.113 &   40 \\
HD 93738 & A0 V nn & -0.027 &  0.158 &  0.842 &  2.817 &  12970 $\pm$    208 &  4.336 $\pm$ 0.113 &  315 \\
HD 93874 & A3 IV   &  0.071 &  0.203 &  0.947 &  2.896 &   8831 $\pm$    141 &  4.367 $\pm$ 0.091 &  142 \\
HD 94066 & B5 V n  & -0.068 &  0.117 &  0.439 &  2.680 &  15096 $\pm$    242 &  3.792 $\pm$ 0.113 &  154 \\
HD 94174 & A0 V    &  0.046 &  0.193 &  0.946 &  2.907 &   9305 $\pm$    149 &  4.391 $\pm$ 0.113 &  149 
\enddata
\end{deluxetable*}

\begin{deluxetable*}{ccccccccc}
\tabletypesize{\footnotesize}
\tablecolumns{9}
\tablewidth{0.99\textwidth}
\tablecaption{$\alpha$ Persei members dereddened $uvby\beta$ photometry and atmospheric parameters \label{table:alphaper}}
\tablehead{
\colhead{Star} & 
\colhead{Sp. Type} & 
\colhead{$(b-y)_0$} & 
\colhead{$m_0$} & 
\colhead{$c_0$} &
\colhead{$\beta$} &
\colhead{$T_\mathrm{eff}$} &
\colhead{$\log g$} &
\colhead{$v \sin i$}
\\
 &  & (mag) & (mag) & (mag) & (mag) & (K) & (dex) & (km s$^{-1}$) 
}
\startdata
BD$+$49 868 & F5 V    &  0.261 &  0.165 &  0.459 &  2.683 &   6693 $\pm$    107 &  4.455 $\pm$ 0.145 &   20 \\
HD 19767 & F0 V N  &  0.176 &  0.178 &  0.756 &  2.765 &   7368 $\pm$    118 &  4.174 $\pm$ 0.145 &  140 \\
HD 19805 & A0 Va   & -0.000 &  0.161 &  0.931 &  2.887 &  10073 $\pm$    161 &  4.344 $\pm$ 0.113 &   20 \\
HD 19893 & B9 V    & -0.031 &  0.131 &  0.850 &  2.807 &  12614 $\pm$    202 &  4.176 $\pm$ 0.113 &  280 \\
HD 19954 & A9 IV   &  0.150 &  0.200 &  0.794 &  2.792 &   7632 $\pm$    122 &  4.297 $\pm$ 0.145 &   85 \\
HD 20135 & A0 P    & -0.011 &  0.186 &  0.970 &  2.848 &  10051 $\pm$    161 &  3.998 $\pm$ 0.113 &   35 \\
BD$+$49 889 & F5 V    &  0.292 &  0.156 &  0.418 &  2.656 &   6430 $\pm$    103 &  4.352 $\pm$ 0.145 &   65 \\
BD$+$49 896 & F4 V    &  0.261 &  0.168 &  0.472 &  2.686 &   6686 $\pm$    107 &  4.410 $\pm$ 0.145 &   30 \\
HD 20365 & B3 V    & -0.079 &  0.103 &  0.346 &  2.681 &  16367 $\pm$    262 &  4.025 $\pm$ 0.113 &  145 \\
HD 20391 & A1 Va n &  0.026 &  0.179 &  1.006 &  2.901 &  10415 $\pm$    167 &  4.386 $\pm$ 0.091 &  260 \\
HD 20487 & A0 V N  & -0.016 &  0.151 &  0.976 &  2.856 &  11659 $\pm$    187 &  4.198 $\pm$ 0.113 &  280 \\
BD$+$47 808 & F1 IV N &  0.183 &  0.179 &  0.759 &  2.763 &   7281 $\pm$    116 &  4.062 $\pm$ 0.145 &  180 \\
BD$+$48 892 & F3 IV-V &  0.246 &  0.167 &  0.524 &  2.696 &   6800 $\pm$    109 &  4.359 $\pm$ 0.145 &   20 \\
BD$+$48 894 & F0 IV   &  0.174 &  0.202 &  0.734 &  2.770 &   7416 $\pm$    119 &  4.284 $\pm$ 0.145 &   75 \\
HD 20809 & B5 V    & -0.074 &  0.109 &  0.395 &  2.696 &  15934 $\pm$    255 &  4.097 $\pm$ 0.113 &  200 \\
HD 20842 & A0 Va   & -0.005 &  0.157 &  0.950 &  2.886 &  10258 $\pm$    164 &  4.325 $\pm$ 0.113 &   85 \\
HD 20863 & B9 V    & -0.034 &  0.134 &  0.810 &  2.813 &  12154 $\pm$    194 &  4.267 $\pm$ 0.113 &  200 \\
BD$+$49 914 & F5 V    &  0.281 &  0.170 &  0.431 &  2.664 &   6520 $\pm$    104 &  4.395 $\pm$ 0.145 &  120 \\
HD 20919 & A8 V    &  0.168 &  0.191 &  0.757 &  2.775 &   7463 $\pm$    119 &  4.259 $\pm$ 0.145 &   50 \\
BD$+$49 918 & F1 V N  &  0.186 &  0.183 &  0.770 &  2.755 &   7235 $\pm$    116 &  3.977 $\pm$ 0.145 &  175 \\
HD 20931 & A1 Va   &  0.018 &  0.174 &  0.979 &  2.911 &   9588 $\pm$    153 &  4.342 $\pm$ 0.113 &   85 \\
BD$+$47 816 & F4 V    &  0.271 &  0.155 &  0.452 &  2.672 &   6600 $\pm$    106 &  4.399 $\pm$ 0.145 &   28 \\
HD 20961 & B95V    & -0.019 &  0.163 &  0.920 &  2.875 &  10537 $\pm$    169 &  4.344 $\pm$ 0.113 &   25 \\
BD$+$46 745 & F4 V    &  0.274 &  0.169 &  0.462 &  2.674 &   6566 $\pm$    105 &  4.332 $\pm$ 0.145 &  160 \\
HD 20969 & A8 V    &  0.186 &  0.192 &  0.715 &  2.758 &   7291 $\pm$    117 &  4.239 $\pm$ 0.145 &   20 \\
HD 20986 & A3 V N  &  0.046 &  0.190 &  1.004 &  2.896 &   9584 $\pm$    153 &  4.243 $\pm$ 0.091 &  210 \\
HD 21005 & A5 V N  &  0.074 &  0.189 &  0.987 &  2.862 &   8266 $\pm$    132 &  4.197 $\pm$ 0.145 &  250 \\
HD 21091 & B95IV nn & -0.019 &  0.152 &  0.938 &  2.856 &  12477 $\pm$    200 &  4.416 $\pm$ 0.113 &  340 \\
HD 21092 & A5 V    &  0.054 &  0.218 &  0.938 &  2.893 &   8775 $\pm$    140 &  4.311 $\pm$ 0.091 &   75 \\
TYC 3320-1715-1 & F4 V    &  0.281 &  0.153 &  0.469 &  2.663 &   6495 $\pm$    104 &  4.220 $\pm$ 0.145 &  110 \\
HD 21152 & B9 V    & -0.018 &  0.158 &  0.943 &  2.868 &  11306 $\pm$    181 &  4.353 $\pm$ 0.113 &  225 \\
HD 232793 & F5 V    &  0.311 &  0.172 &  0.377 &  2.645 &   6274 $\pm$    100 &  4.362 $\pm$ 0.145 &   93 \\
HD 21181 & B85V N  & -0.038 &  0.122 &  0.784 &  2.766 &  13726 $\pm$    220 &  4.119 $\pm$ 0.113 &  345 \\
HD 21239 & A3 V N  &  0.045 &  0.190 &  0.997 &  2.910 &   9182 $\pm$    147 &  4.320 $\pm$ 0.091 &  145 \\
HD 21278 & B5 V    & -0.073 &  0.111 &  0.398 &  2.705 &  15274 $\pm$    244 &  4.152 $\pm$ 0.113 &   75 \\
HD 21302 & A1 V N  &  0.022 &  0.177 &  0.989 &  2.888 &  10269 $\pm$    164 &  4.301 $\pm$ 0.091 &  230 \\
BD$+$48 923 & F4 V    &  0.270 &  0.153 &  0.464 &  2.673 &   6603 $\pm$    106 &  4.362 $\pm$ 0.145 &   20 \\
HD 21345 & A5 V N  &  0.051 &  0.208 &  0.969 &  2.893 &   9435 $\pm$    151 &  4.324 $\pm$ 0.091 &  200 \\
HD 21398 & B9 V    & -0.030 &  0.145 &  0.825 &  2.837 &  11615 $\pm$    186 &  4.372 $\pm$ 0.113 &  135 \\
HD 21428 & B3 V    & -0.077 &  0.105 &  0.363 &  2.686 &  16421 $\pm$    263 &  4.076 $\pm$ 0.113 &  200 \\
HD 21481 & A0 V N  & -0.013 &  0.164 &  0.993 &  2.858 &  11187 $\pm$    179 &  4.141 $\pm$ 0.113 &  250 \\
HD 21527 & A7 IV   &  0.093 &  0.231 &  0.855 &  2.856 &   8231 $\pm$    132 &  4.486 $\pm$ 0.145 &   80 \\
HD 21551 & B8 V    & -0.048 &  0.118 &  0.673 &  2.746 &  14869 $\pm$    238 &  4.220 $\pm$ 0.113 &  380 \\
HD 21553 & A6 V N  &  0.072 &  0.206 &  0.921 &  2.872 &   8381 $\pm$    134 &  4.414 $\pm$ 0.145 &  150 \\
HD 21619 & A6 V    &  0.052 &  0.221 &  0.935 &  2.894 &   8843 $\pm$    141 &  4.329 $\pm$ 0.091 &   90 \\
BD$+$49 957 & F3 V    &  0.258 &  0.168 &  0.500 &  2.687 &   6699 $\pm$    107 &  4.334 $\pm$ 0.145 &   56 \\
HD 21641 & B85V    & -0.042 &  0.131 &  0.721 &  2.747 &  12914 $\pm$    207 &  3.929 $\pm$ 0.113 &  215 \\
BD$+$49 958 & F1 V    &  0.198 &  0.188 &  0.732 &  2.739 &   7137 $\pm$    114 &  3.989 $\pm$ 0.145 &  155 \\
HD 21672 & B8 V    & -0.050 &  0.119 &  0.649 &  2.747 &  13473 $\pm$    216 &  4.071 $\pm$ 0.113 &  225 \\
BD$+$48 944 & A5 V    &  0.063 &  0.220 &  0.931 &  2.886 &   8799 $\pm$    141 &  4.305 $\pm$ 0.091 &  120 \\
HD 21931 & B9 V    & -0.029 &  0.147 &  0.835 &  2.829 &  11998 $\pm$    192 &  4.343 $\pm$ 0.113 &  205 \\
\enddata
\end{deluxetable*}

\begin{deluxetable*}{ccccccccc}
\tabletypesize{\footnotesize}
\tablewidth{0.99\textwidth}
\tablecaption{Pleiades members dereddened $uvby\beta$ photometry and atmospheric parameters}
\tablehead{
\colhead{HD} & 
\colhead{Sp. Type} & 
\colhead{$(b-y)_0$} & 
\colhead{$m_0$} & 
\colhead{$c_0$} &
\colhead{$\beta$} &
\colhead{$T_\mathrm{eff}$} &
\colhead{$\log g$} &
\colhead{$v \sin i$}
\\
 &  & (mag) & (mag) & (mag) & (mag) & (K) & (dex) & (km s$^{-1}$) 
}
\startdata
HD 23157 & A9 V    &  0.168 &  0.190 &  0.725 &  2.778 &   7463 $\pm$    121 &  4.369 $\pm$ 0.145 &  100 \\
HD 23156 & A7 V    &  0.111 &  0.215 &  0.815 &  2.837 &   8046 $\pm$    130 &  4.498 $\pm$ 0.145 &   70 \\
HD 23247 & F3 V    &  0.237 &  0.174 &  0.527 &  2.704 &   6863 $\pm$    111 &  4.424 $\pm$ 0.145 &   40 \\
HD 23246 & A8 V    &  0.170 &  0.184 &  0.758 &  2.773 &   7409 $\pm$    120 &  4.234 $\pm$ 0.145 &  200 \\
HD 23288 & B7 V    & -0.051 &  0.120 &  0.636 &  2.747 &  13953 $\pm$    226 &  4.151 $\pm$ 0.113 &  280 \\
HD 23302 & B6 III  & -0.054 &  0.098 &  0.638 &  2.690 &  13308 $\pm$    216 &  3.478 $\pm$ 0.113 &  205 \\
HD 23289 & F3 V    &  0.244 &  0.164 &  0.521 &  2.699 &   6796 $\pm$    110 &  4.387 $\pm$ 0.145 &   40 \\
HD 23326 & F4 V    &  0.250 &  0.164 &  0.514 &  2.691 &   6741 $\pm$    109 &  4.358 $\pm$ 0.145 &   40 \\
HD 23324 & B8 V    & -0.052 &  0.116 &  0.634 &  2.747 &  13748 $\pm$    223 &  4.126 $\pm$ 0.113 &  255 \\
HD 23338 & B6 IV   & -0.061 &  0.104 &  0.553 &  2.702 &  13696 $\pm$    222 &  3.772 $\pm$ 0.113 &  130 \\
HD 23351 & F3 V    &  0.249 &  0.176 &  0.507 &  2.695 &   6755 $\pm$    109 &  4.391 $\pm$ 0.145 &   80 \\
HD 23361 & A25Va n &  0.069 &  0.201 &  0.959 &  2.872 &   8356 $\pm$    135 &  4.309 $\pm$ 0.145 &  235 \\
HD 23375 & A9 V    &  0.180 &  0.187 &  0.710 &  2.765 &   7336 $\pm$    119 &  4.318 $\pm$ 0.145 &   75 \\
HD 23410 & A0 Va   &  0.004 &  0.164 &  0.975 &  2.899 &  10442 $\pm$    169 &  4.382 $\pm$ 0.113 &  200 \\
HD 23409 & A3 V    &  0.070 &  0.202 &  0.980 &  2.892 &   8903 $\pm$    144 &  4.270 $\pm$ 0.091 &  170 \\
HD 23432 & B8 V    & -0.039 &  0.127 &  0.758 &  2.793 &  12695 $\pm$    206 &  4.250 $\pm$ 0.113 &  235 \\
HD 23441 & B9 V N  & -0.029 &  0.135 &  0.858 &  2.822 &  11817 $\pm$    191 &  4.209 $\pm$ 0.113 &  200 \\
HD 23479 & A9 V    &  0.188 &  0.166 &  0.716 &  2.755 &   7239 $\pm$    117 &  4.212 $\pm$ 0.145 &  150 \\
HD 23489 & A2 V    &  0.033 &  0.183 &  1.012 &  2.907 &   9170 $\pm$    149 &  4.239 $\pm$ 0.091 &  110 \\
HD 23512 & A2 V    &  0.057 &  0.196 &  1.035 &  2.909 &   8852 $\pm$    143 &  4.214 $\pm$ 0.091 &  145 \\
HD 23511 & F5 V    &  0.279 &  0.174 &  0.412 &  2.674 &   6521 $\pm$    106 &  4.477 $\pm$ 0.145 &   28 \\
HD 23514 & F5 V    &  0.285 &  0.179 &  0.443 &  2.668 &   6450 $\pm$    104 &  4.307 $\pm$ 0.145 &   40 \\
HD 23513 & F5 V    &  0.278 &  0.170 &  0.423 &  2.673 &   6528 $\pm$    106 &  4.447 $\pm$ 0.145 &   30 \\
HD 23568 & B95Va n & -0.024 &  0.139 &  0.914 &  2.847 &  11731 $\pm$    190 &  4.301 $\pm$ 0.113 &  240 \\
HD 23567 & F0 V    &  0.159 &  0.196 &  0.735 &  2.788 &   7560 $\pm$    122 &  4.407 $\pm$ 0.145 &   50 \\
HD 23585 & F0 V    &  0.168 &  0.185 &  0.713 &  2.780 &   7472 $\pm$    121 &  4.405 $\pm$ 0.145 &  100 \\
HD 23608 & F5 V    &  0.278 &  0.177 &  0.482 &  2.673 &   6492 $\pm$    105 &  4.185 $\pm$ 0.145 &  110 \\
HD 23607 & F0 V    &  0.108 &  0.203 &  0.814 &  2.841 &   8085 $\pm$    131 &  4.534 $\pm$ 0.145 &   12 \\
HD 23629 & A0 V    & -0.001 &  0.163 &  0.986 &  2.899 &  10340 $\pm$    168 &  4.342 $\pm$ 0.113 &  170 \\
HD 23632 & A0 Va   &  0.006 &  0.167 &  1.009 &  2.899 &  10461 $\pm$    169 &  4.312 $\pm$ 0.113 &  225 \\
HD 23628 & A4 V    &  0.090 &  0.189 &  0.904 &  2.853 &   8163 $\pm$    132 &  4.381 $\pm$ 0.145 &  215 \\
HD 23643 & A35V    &  0.079 &  0.194 &  0.943 &  2.862 &   8258 $\pm$    134 &  4.301 $\pm$ 0.145 &  185 \\
HD 23733 & A9 V    &  0.207 &  0.177 &  0.672 &  2.736 &   7066 $\pm$    114 &  4.174 $\pm$ 0.145 &  180 \\
HD 23732 & F5 V    &  0.258 &  0.172 &  0.460 &  2.688 &   6695 $\pm$    108 &  4.473 $\pm$ 0.145 &   50 \\
HD 23753 & B8 V N  & -0.046 &  0.113 &  0.712 &  2.736 &  13096 $\pm$    212 &  3.859 $\pm$ 0.113 &  240 \\
HD 23791 & A9 V+   &  0.139 &  0.214 &  0.758 &  2.811 &   7776 $\pm$    126 &  4.480 $\pm$ 0.145 &   85 \\
HD 23850 & B8 III  & -0.048 &  0.102 &  0.701 &  2.695 &  13446 $\pm$    218 &  3.483 $\pm$ 0.113 &  280 \\
HD 23863 & A8 V    &  0.116 &  0.201 &  0.857 &  2.826 &   7926 $\pm$    128 &  4.354 $\pm$ 0.145 &  160 \\
HD 23872 & A1 Va n &  0.032 &  0.182 &  1.013 &  2.894 &  10028 $\pm$    162 &  4.247 $\pm$ 0.091 &  240 \\
HD 23873 & B95Va   & -0.023 &  0.143 &  0.907 &  2.852 &  10897 $\pm$    177 &  4.255 $\pm$ 0.113 &   90 \\
HD 23886 & A4 V    &  0.068 &  0.214 &  0.915 &  2.880 &   8974 $\pm$    145 &  4.343 $\pm$ 0.091 &  165 \\
HD 23912 & F3 V    &  0.274 &  0.154 &  0.481 &  2.671 &   6531 $\pm$    106 &  4.242 $\pm$ 0.145 &  130 \\
HD 23924 & A7 V    &  0.100 &  0.223 &  0.852 &  2.852 &   8121 $\pm$    132 &  4.460 $\pm$ 0.145 &  100 \\
HD 23923 & B85V N  & -0.033 &  0.124 &  0.839 &  2.794 &  12911 $\pm$    209 &  4.159 $\pm$ 0.113 &  310 \\
HD 23948 & A1 Va   &  0.033 &  0.191 &  0.984 &  2.905 &   9237 $\pm$    150 &  4.307 $\pm$ 0.091 &  120 \\
HD 24076 & A2 V    &  0.008 &  0.168 &  0.923 &  2.867 &  10196 $\pm$    165 &  4.298 $\pm$ 0.091 &  155 \\
HD 24132 & F2 V    &  0.245 &  0.149 &  0.597 &  2.692 &   6744 $\pm$    109 &  4.182 $\pm$ 0.145 &  230
\enddata
\label{table:pleiades}
\end{deluxetable*}

\begin{deluxetable*}{ccccccccc}
\tabletypesize{\footnotesize}
\tablewidth{0.99\textwidth}
\tablecaption{Hyades members dereddened $uvby\beta$ photometry and atmospheric parameters}
\tablehead{
\colhead{HD} & 
\colhead{Sp. Type} & 
\colhead{$(b-y)_0$} & 
\colhead{$m_0$} & 
\colhead{$c_0$} &
\colhead{$\beta$} &
\colhead{$T_\mathrm{eff}$} &
\colhead{$\log g$} &
\colhead{$v \sin i$}
\\
 &  & (mag) & (mag) & (mag) & (mag) & (K) & (dex) & (km s$^{-1}$) 
}
\startdata
HD 26015 & F3 V    &  0.252 &  0.174 &  0.537 &  2.693 &   6732 $\pm$    109 &  4.244 $\pm$ 0.145 &   25 \\
HD 26462 & F1 IV-V &  0.230 &  0.165 &  0.596 &  2.710 &   6916 $\pm$    112 &  4.291 $\pm$ 0.145 &   30 \\
HD 26737 & F5 V    &  0.274 &  0.168 &  0.477 &  2.674 &   6558 $\pm$    106 &  4.263 $\pm$ 0.145 &   60 \\
HD 26911 & F3 V    &  0.258 &  0.176 &  0.525 &  2.690 &   6682 $\pm$    108 &  4.228 $\pm$ 0.145 &   30 \\
HD 27176 & A7 m    &  0.172 &  0.187 &  0.785 &  2.767 &   7380 $\pm$    120 &  4.087 $\pm$ 0.145 &  125 \\
HD 27397 & F0 IV   &  0.171 &  0.194 &  0.770 &  2.766 &   7410 $\pm$    120 &  4.173 $\pm$ 0.145 &  100 \\
HD 27429 & F2 VN   &  0.240 &  0.171 &  0.588 &  2.693 &   6828 $\pm$    111 &  4.270 $\pm$ 0.145 &  150 \\
HD 27459 & F0 IV   &  0.129 &  0.204 &  0.871 &  2.812 &   7782 $\pm$    126 &  4.198 $\pm$ 0.145 &   35 \\
HD 27524 & F5 V    &  0.285 &  0.161 &  0.461 &  2.656 &   6461 $\pm$    105 &  4.213 $\pm$ 0.145 &  110 \\
HD 27561 & F4 V    &  0.270 &  0.162 &  0.482 &  2.677 &   6594 $\pm$    107 &  4.284 $\pm$ 0.145 &   30 \\
HD 27628 & A2 M    &  0.133 &  0.225 &  0.707 &  2.756 &   7944 $\pm$    129 &  4.743 $\pm$ 0.145 &   30 \\
HD 27819 & A7 IV   &  0.080 &  0.209 &  0.982 &  2.857 &   8203 $\pm$    133 &  4.170 $\pm$ 0.145 &   35 \\
HD 27901 & F4 V N  &  0.238 &  0.178 &  0.597 &  2.704 &   6837 $\pm$    111 &  4.233 $\pm$ 0.145 &  110 \\
HD 27934 & A5 IV-V &  0.064 &  0.201 &  1.053 &  2.867 &   8506 $\pm$    138 &  3.884 $\pm$ 0.091 &   90 \\
HD 27946 & A7 V    &  0.149 &  0.192 &  0.840 &  2.783 &   7584 $\pm$    123 &  4.112 $\pm$ 0.145 &  210 \\
HD 27962 & A3 V    &  0.020 &  0.193 &  1.046 &  2.889 &   9123 $\pm$    148 &  4.004 $\pm$ 0.091 &   30 \\
HD 28024 & A9 IV- N &  0.165 &  0.175 &  0.947 &  2.753 &   7279 $\pm$    118 &  3.503 $\pm$ 0.145 &  215 \\
HD 28226 & A  M    &  0.164 &  0.213 &  0.771 &  2.775 &   7493 $\pm$    121 &  4.248 $\pm$ 0.145 &  130 \\
HD 28294 & F0 IV   &  0.198 &  0.173 &  0.694 &  2.745 &   7174 $\pm$    116 &  4.194 $\pm$ 0.145 &  135 \\
HD 28319 & A7 III  &  0.097 &  0.198 &  1.011 &  2.831 &   7945 $\pm$    129 &  3.930 $\pm$ 0.145 &  130 \\
HD 28355 & A7 m    &  0.112 &  0.226 &  0.908 &  2.832 &   7930 $\pm$    128 &  4.207 $\pm$ 0.145 &  140 \\
HD 28485 & F0 V+ N &  0.200 &  0.192 &  0.717 &  2.740 &   7129 $\pm$    115 &  4.035 $\pm$ 0.145 &  150 \\
HD 28527 & A5 m    &  0.085 &  0.218 &  0.964 &  2.856 &   8180 $\pm$    133 &  4.194 $\pm$ 0.145 &  100 \\
HD 28546 & A7 m    &  0.142 &  0.234 &  0.796 &  2.809 &   7726 $\pm$    125 &  4.354 $\pm$ 0.145 &   30 \\
HD 28556 & F0 IV   &  0.147 &  0.202 &  0.814 &  2.795 &   7645 $\pm$    124 &  4.244 $\pm$ 0.145 &  140 \\
HD 28568 & F5 V    &  0.274 &  0.168 &  0.466 &  2.676 &   6564 $\pm$    106 &  4.315 $\pm$ 0.145 &   55 \\
HD 28677 & F2 V    &  0.214 &  0.176 &  0.654 &  2.725 &   7032 $\pm$    114 &  4.161 $\pm$ 0.145 &  100 \\
HD 28911 & F5 V    &  0.283 &  0.163 &  0.459 &  2.663 &   6481 $\pm$    105 &  4.249 $\pm$ 0.145 &   40 \\
HD 28910 & A9 V    &  0.144 &  0.200 &  0.830 &  2.796 &   7659 $\pm$    124 &  4.213 $\pm$ 0.145 &   95 \\
HD 29169 & F2 V    &  0.236 &  0.183 &  0.567 &  2.708 &   6880 $\pm$    111 &  4.321 $\pm$ 0.145 &   80 \\
HD 29225 & F5 V    &  0.276 &  0.171 &  0.461 &  2.675 &   6547 $\pm$    106 &  4.316 $\pm$ 0.145 &   45 \\
HD 29375 & F0 IV-V &  0.187 &  0.187 &  0.740 &  2.754 &   7257 $\pm$    118 &  4.106 $\pm$ 0.145 &  155 \\
HD 29388 & A5 IV-V &  0.062 &  0.199 &  1.047 &  2.870 &   8645 $\pm$    140 &  3.927 $\pm$ 0.091 &  115 \\
HD 29499 & A  M    &  0.140 &  0.231 &  0.826 &  2.810 &   7713 $\pm$    125 &  4.266 $\pm$ 0.145 &   70 \\
HD 29488 & A5 IV-V &  0.080 &  0.196 &  1.017 &  2.852 &   8127 $\pm$    132 &  4.025 $\pm$ 0.145 &  160 \\
HD 30034 & A9 IV-  &  0.150 &  0.195 &  0.813 &  2.791 &   7610 $\pm$    123 &  4.218 $\pm$ 0.145 &   75 \\
HD 30210 & A5 m    &  0.091 &  0.252 &  0.955 &  2.845 &   8126 $\pm$    132 &  4.181 $\pm$ 0.145 &   30 \\
HD 30780 & A9 V+   &  0.122 &  0.207 &  0.900 &  2.813 &   7823 $\pm$    127 &  4.141 $\pm$ 0.145 &  155 \\
HD 31845 & F5 V    &  0.294 &  0.165 &  0.439 &  2.658 &   6396 $\pm$    104 &  4.229 $\pm$ 0.145 &   25 \\
HD 32301 & A7 IV   &  0.079 &  0.202 &  1.034 &  2.847 &   8116 $\pm$    131 &  3.975 $\pm$ 0.145 &  115 \\
HD 33254 & A7 m    &  0.132 &  0.251 &  0.835 &  2.824 &   7797 $\pm$    126 &  4.306 $\pm$ 0.145 &   30 \\
HD 33204 & A7 m    &  0.149 &  0.245 &  0.803 &  2.796 &   7634 $\pm$    124 &  4.270 $\pm$ 0.145 &   30 \\
HD 25202 & F4 V    &  0.206 &  0.172 &  0.695 &  2.724 &   7082 $\pm$    115 &  4.064 $\pm$ 0.145 &  160 \\
HD 28052 & F0 IV-V N &  0.153 &  0.183 &  0.934 &  2.767 &   7431 $\pm$    120 &  3.733 $\pm$ 0.145 &  170 \\
HD 18404 & F5 IV   &  0.269 &  0.169 &  0.481 &  2.680 &   6605 $\pm$    107 &  4.299 $\pm$ 0.145 &    0 \\
HD 25570 & F4 V    &  0.249 &  0.147 &  0.557 &  2.688 &   6752 $\pm$    109 &  4.183 $\pm$ 0.145 &   34 \\
HD 40932 & A2 M    &  0.079 &  0.205 &  0.978 &  2.853 &   8224 $\pm$    133 &  4.191 $\pm$ 0.145 &   18
\enddata
\label{table:hyades}
\end{deluxetable*}

\subsection{Alternative Treatment of Open Clusters}

As described in \S ~\ref{subsubsec:numericalmethods} The 1-D marginalized PDF in age for an individual star is computed on a model grid that is uniformly spaced in log(age). As such, the prior probability of each bin is also encoded in log(age) (see \S ~\ref{subsubsec:priors}). Thus, the resultant PDF is naturally in the units of $d$ p($\log{\tau}$)/$d\log{\tau}$, where $p$ is probability and $\tau$ is age.

In order to transform p($\log{\tau}$) to p($\tau$) one uses the conversion p($\tau$) = p($\log{\tau}$)/$\tau$. Statistical measures \emph{other than the median}, such as the mean, mode, confidence intervals, etc. will be different depending on whether the PDF being quantified is p($\log{\tau}$) or p($\tau$). For example, $10^{\left \langle \log{\tau} \right \rangle} \neq \left \langle \tau \right \rangle$. Strictly speaking, however, both values are meaningful and authors frequently choose to report one or the other in the literature. In the case at hand, p($\log{\tau}$) for an individual star is more symmetric than the linear counterpart, p($\tau$). As such, one could reasonably argue that $10^{\left \langle \log{\tau} \right \rangle}$ is a more meaningful metric than 
$\left \langle \tau \right \rangle$.

In either case, because the PDFs in age or log(age) are both skewed, the median (which, again, is equal regardless of whether p($\tau$) or p($\log{\tau}$) is under consideration), is actually the most meaningful quantification of the PDF since it is less susceptible to extreme values than either the mean or mode.

With respect to the open clusters, regardless of whether our analyses are performed in logarthmic or linear space, our results favor ages that are younger and older than accepted values for $\alpha$ Per and the Hyades, respectively.

\begin{figure*}
\centering
\includegraphics[width=0.9\textwidth]{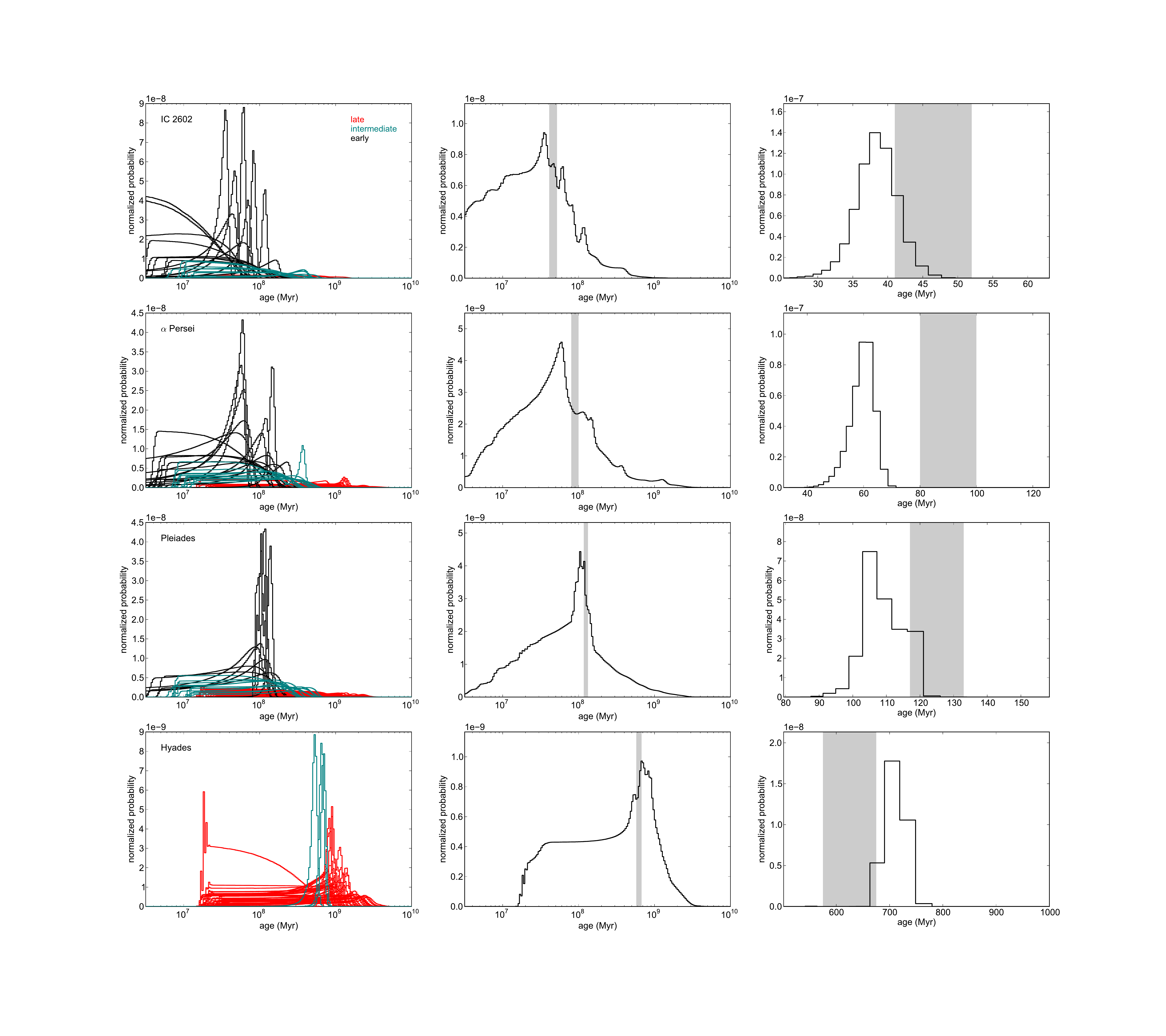}
\caption{\emph{Left panels}: 1D marginalized, normalized posterior PDFs in age, calculated from \cite{bressan2012} evolutionary models, for individual open cluster members. Black, teal, and red histograms represent early, intermediate, and late group stars, respectively. \emph{Middle panels}: Sums of the individual PDFs depicted on the left. This figure shows the total probability associated with the 200 age bins between log(age/yr)=6.5 to 10. The grey shaded regions indicate the currently accepted ages of IC 2602 (46$^{+6}_{-5}$ Myr), $\alpha$ Per (90$\pm$10 Myr), the Pleiades (125$\pm$8 Myr), and the Hyades (625$\pm$50 Myr). \emph{Right panels}: Products of the individual PDFs depicted in the left panels. The grey shaded regions again depict the accepted literature age ranges of each cluster.}
\label{fig:cluster-hist-lin}
\end{figure*}

\begin{deluxetable*}{cccccccccccc}
\tabletypesize{\footnotesize}
\tablecolumns{8}
\tablewidth{0.99\textwidth}
\tablecaption{ Open Cluster Ages: Analysis in Linear Age \label{table:clusters-appendix}}
\tablehead{
&
& 
&
\colhead{Summed PDF} & 
\colhead{Summed PDF} & 
\colhead{Multiplied PDF} & 
\colhead{Multiplied PDF}
\\
\colhead{Cluster}  &
\colhead{Lit. Age}  & 
\colhead{Models} &
\colhead{Median}  &
\colhead{68$\%$ C.I.} &
\colhead{Median}  &
\colhead{68$\%$ C.I.}
\\
&
\colhead{(Myr)} & 
&
\colhead{(Myr)} & 
\colhead{(Myr)} & 
\colhead{(Myr)} & 
\colhead{(Myr)}
}

\startdata

IC 2602 & 46$^{+6}_{-5}$ & \cite{ekstrom2012} & 22 & 3-39 & 41 & 41-42 \\
 &  & \cite{bressan2012} & 24 & 3-40 & 40 & 37-43 \\

$\alpha$ Persei & 90$^{+10}_{-10}$ & \cite{ekstrom2012} & 41 & 3-68 & 63 & 61-68 \\
 &  & \cite{bressan2012} & 45 & 3-71 & 62 & 58-66 \\

Pleiades & 125$^{+8}_{-8}$ & \cite{ekstrom2012} & 61 & 3-113 & 125 & 122-131 \\
 &  & \cite{bressan2012} & 77 & 3-117 & 112 & 107-120 \\

Hyades & 625$^{+50}_{-50}$ & \cite{ekstrom2012} & 118 & 3-403 & 677 & 671-690 \\
 &  & \cite{bressan2012} & 288 & 17-593 & 738 & 719-765 

\enddata
\tablenotetext{}{Literature ages (column 2) come from the sources referenced in \S~\ref{subsec:openclustertests}. For each set of evolutionary models, the median and 68\% confidence interval are computed for both the summed PDF (columns 4,5) and multiplied PDF (columns 6,7). Note, the Hyades analysis includes the blue straggler HD 27962 and the spectroscopic binary HD 27268. Excluding these outliers results in a median and 68\% confidence interval of 322 Myr [17-650 Myr] of the summed PDF or 784 Myr [749-802 Myr] of the multiplied PDF, using the B12 models.}
\end{deluxetable*}

%%%%% REFERENCES %%%%%
\newpage
\bibliography{main}

\begin{thebibliography}{150}
\expandafter\ifx\csname natexlab\endcsname\relax\def\natexlab#1{#1}\fi

\bibitem[{{Abt}(1985)}]{abt1985}
{Abt}, H.~A. 1985, \apjl, 294, L103

\bibitem[{{Allende Prieto} \& {Lambert}(1999)}]{allende1999}
{Allende Prieto}, C., \& {Lambert}, D.~L. 1999, \aap, 352, 555

\bibitem[{{Ammons} {et~al.}(2006){Ammons}, {Robinson}, {Strader}, {Laughlin},
  {Fischer}, \& {Wolf}}]{ammons2006}
{Ammons}, S.~M., {Robinson}, S.~E., {Strader}, J., {Laughlin}, G., {Fischer},
  D., \& {Wolf}, A. 2006, \apj, 638, 1004

\bibitem[{{Anderson} \& {Francis}(2011)}]{anderson2011}
{Anderson}, E., \& {Francis}, C. 2011, VizieR Online Data Catalog, 5137, 0

\bibitem[{{Anderson} \& {Francis}(2012)}]{anderson2012}
---. 2012, Astronomy Letters, 38, 331

\bibitem[{{Asiain} {et~al.}(1997){Asiain}, {Torra}, \& {Figueras}}]{asiain1997}
{Asiain}, R., {Torra}, J., \& {Figueras}, F. 1997, \aap, 322, 147

\bibitem[{{Baines} {et~al.}(2008){Baines}, {McAlister}, {ten Brummelaar},
  {Turner}, {Sturmann}, {Sturmann}, {Goldfinger}, \& {Ridgway}}]{baines2008}
{Baines}, E.~K., {McAlister}, H.~A., {ten Brummelaar}, T.~A., {Turner}, N.~H.,
  {Sturmann}, J., {Sturmann}, L., {Goldfinger}, P.~J., \& {Ridgway}, S.~T.
  2008, \apj, 680, 728

\bibitem[{{Baines} {et~al.}(2012){Baines}, {White}, {Huber}, {Jones},
  {Boyajian}, {McAlister}, {ten Brummelaar}, {Turner}, {Sturmann}, {Sturmann},
  {Goldfinger}, {Farrington}, {Riedel}, {Ireland}, {von Braun}, \&
  {Ridgway}}]{baines2012}
{Baines}, E.~K., {et~al.} 2012, \apj, 761, 57

\bibitem[{{Balona}(1984)}]{balona1984}
{Balona}, L.~A. 1984, \mnras, 211, 973

\bibitem[{{Balona}(1994)}]{balona1994}
---. 1994, \mnras, 268, 119

\bibitem[{{Barrado y Navascues} {et~al.}(1997){Barrado y Navascues},
  {Stauffer}, {Hartmann}, \& {Balachandran}}]{barrado1997}
{Barrado y Navascues}, D., {Stauffer}, J.~R., {Hartmann}, L., \&
  {Balachandran}, S.~C. 1997, \apj, 475, 313

\bibitem[{{Barrado y Navascu{\'e}s} {et~al.}(1999){Barrado y Navascu{\'e}s},
  {Stauffer}, {Song}, \& {Caillault}}]{barrado1999}
{Barrado y Navascu{\'e}s}, D., {Stauffer}, J.~R., {Song}, I., \& {Caillault},
  J.-P. 1999, \apjl, 520, L123

\bibitem[{{Beeckmans}(1977)}]{beeckmans1977}
{Beeckmans}, F. 1977, \aap, 60, 1

\bibitem[{{Bertone} {et~al.}(2004){Bertone}, {Buzzoni}, {Ch{\'a}vez}, \&
  {Rodr{\'{\i}}guez-Merino}}]{bertone2004}
{Bertone}, E., {Buzzoni}, A., {Ch{\'a}vez}, M., \& {Rodr{\'{\i}}guez-Merino},
  L.~H. 2004, \aj, 128, 829

\bibitem[{{Bessell}(2011)}]{bessell2011}
{Bessell}, M.~S. 2011, \pasp, 123, 1442

\bibitem[{{Bigot} {et~al.}(2011){Bigot}, {Mourard}, {Berio}, {Th{\'e}venin},
  {Ligi}, {Tallon-Bosc}, {Chesneau}, {Delaa}, {Nardetto}, {Perraut}, {Stee},
  {Boyajian}, {Morel}, {Pichon}, {Kervella}, {Schmider}, {McAlister}, {ten
  Brummelaar}, {Ridgway}, {Sturmann}, {Sturmann}, {Turner}, {Farrington}, \&
  {Goldfinger}}]{bigot2011}
{Bigot}, L., {et~al.} 2011, \aap, 534, L3

\bibitem[{{Boyajian} {et~al.}(2012){Boyajian}, {McAlister}, {van Belle},
  {Gies}, {ten Brummelaar}, {von Braun}, {Farrington}, {Goldfinger}, {O'Brien},
  {Parks}, {Richardson}, {Ridgway}, {Schaefer}, {Sturmann}, {Sturmann},
  {Touhami}, {Turner}, \& {White}}]{boyajian2012}
{Boyajian}, T.~S., {et~al.} 2012, \apj, 746, 101

\bibitem[{{Boyajian} {et~al.}(2013){Boyajian}, {von Braun}, {van Belle},
  {Farrington}, {Schaefer}, {Jones}, {White}, {McAlister}, {ten Brummelaar},
  {Ridgway}, {Gies}, {Sturmann}, {Sturmann}, {Turner}, {Goldfinger}, \&
  {Vargas}}]{boyajian2013}
---. 2013, \apj, 771, 40

\bibitem[{{Brandt} \& {Huang}(2015)}]{brandt2015}
{Brandt}, T.~D., \& {Huang}, C.~X. 2015, ArXiv e-prints

\bibitem[{{Bressan} {et~al.}(2012){Bressan}, {Marigo}, {Girardi}, {Salasnich},
  {Dal Cero}, {Rubele}, \& {Nanni}}]{bressan2012}
{Bressan}, A., {Marigo}, P., {Girardi}, L., {Salasnich}, B., {Dal Cero}, C.,
  {Rubele}, S., \& {Nanni}, A. 2012, \mnras, 427, 127

\bibitem[{{Burrows} {et~al.}(2004){Burrows}, {Sudarsky}, \&
  {Hubeny}}]{burrows2004}
{Burrows}, A., {Sudarsky}, D., \& {Hubeny}, I. 2004, \apj, 609, 407

\bibitem[{{Caffau} {et~al.}(2011){Caffau}, {Ludwig}, {Steffen}, {Freytag}, \&
  {Bonifacio}}]{caffau2011}
{Caffau}, E., {Ludwig}, H.-G., {Steffen}, M., {Freytag}, B., \& {Bonifacio}, P.
  2011, \solphys, 268, 255

\bibitem[{{Carpenter} {et~al.}(2006){Carpenter}, {Mamajek}, {Hillenbrand}, \&
  {Meyer}}]{carpenter2006}
{Carpenter}, J.~M., {Mamajek}, E.~E., {Hillenbrand}, L.~A., \& {Meyer}, M.~R.
  2006, \apjl, 651, L49

\bibitem[{{Carson} {et~al.}(2013){Carson}, {Thalmann}, {Janson}, {Kozakis},
  {Bonnefoy}, {Biller}, {Schlieder}, {Currie}, {McElwain}, {Goto}, {Henning},
  {Brandner}, {Feldt}, {Kandori}, {Kuzuhara}, {Stevens}, {Wong}, {Gainey},
  {Fukagawa}, {Kuwada}, {Brandt}, {Kwon}, {Abe}, {Egner}, {Grady}, {Guyon},
  {Hashimoto}, {Hayano}, {Hayashi}, {Hayashi}, {Hodapp}, {Ishii}, {Iye},
  {Knapp}, {Kudo}, {Kusakabe}, {Matsuo}, {Miyama}, {Morino}, {Moro-Martin},
  {Nishimura}, {Pyo}, {Serabyn}, {Suto}, {Suzuki}, {Takami}, {Takato},
  {Terada}, {Tomono}, {Turner}, {Watanabe}, {Wisniewski}, {Yamada}, {Takami},
  {Usuda}, \& {Tamura}}]{carson2013}
{Carson}, J., {et~al.} 2013, \apjl, 763, L32

\bibitem[{{Casagrande} {et~al.}(2011){Casagrande}, {Sch{\"o}nrich}, {Asplund},
  {Cassisi}, {Ram{\'{\i}}rez}, {Mel{\'e}ndez}, {Bensby}, \&
  {Feltzing}}]{casagrande2011}
{Casagrande}, L., {Sch{\"o}nrich}, R., {Asplund}, M., {Cassisi}, S.,
  {Ram{\'{\i}}rez}, I., {Mel{\'e}ndez}, J., {Bensby}, T., \& {Feltzing}, S.
  2011, \aap, 530, A138

\bibitem[{{Castelli} \& {Kurucz}(2004)}]{castelli2004}
{Castelli}, F., \& {Kurucz}, R.~L. 2004, ArXiv Astrophysics e-prints

\bibitem[{{Castelli} \& {Kurucz}(2006)}]{castelli2006}
---. 2006, \aap, 454, 333

\bibitem[{{Chauvin} {et~al.}(2004){Chauvin}, {Lagrange}, {Dumas}, {Zuckerman},
  {Mouillet}, {Song}, {Beuzit}, \& {Lowrance}}]{chauvin2004}
{Chauvin}, G., {Lagrange}, A.-M., {Dumas}, C., {Zuckerman}, B., {Mouillet}, D.,
  {Song}, I., {Beuzit}, J.-L., \& {Lowrance}, P. 2004, \aap, 425, L29

\bibitem[{{Chiavassa} {et~al.}(2012){Chiavassa}, {Bigot}, {Kervella}, {Matter},
  {Lopez}, {Collet}, {Magic}, \& {Asplund}}]{chiavassa2012}
{Chiavassa}, A., {Bigot}, L., {Kervella}, P., {Matter}, A., {Lopez}, B.,
  {Collet}, R., {Magic}, Z., \& {Asplund}, M. 2012, \aap, 540, A5

\bibitem[{{Cignoni} {et~al.}(2006){Cignoni}, {Degl'Innocenti}, {Prada Moroni},
  \& {Shore}}]{cignoni2006}
{Cignoni}, M., {Degl'Innocenti}, S., {Prada Moroni}, P.~G., \& {Shore}, S.~N.
  2006, \aap, 459, 783

\bibitem[{{Clem} {et~al.}(2004){Clem}, {VandenBerg}, {Grundahl}, \&
  {Bell}}]{clem2004}
{Clem}, J.~L., {VandenBerg}, D.~A., {Grundahl}, F., \& {Bell}, R.~A. 2004, \aj,
  127, 1227

\bibitem[{{Code} {et~al.}(1976){Code}, {Bless}, {Davis}, \& {Brown}}]{code1976}
{Code}, A.~D., {Bless}, R.~C., {Davis}, J., \& {Brown}, R.~H. 1976, \apj, 203,
  417

\bibitem[{{Collins} \& {Smith}(1985)}]{collinssmith1985}
{Collins}, II, G.~W., \& {Smith}, R.~C. 1985, \mnras, 213, 519

\bibitem[{{Crawford}(1958)}]{crawford1958}
{Crawford}, D.~L. 1958, \apj, 128, 185

\bibitem[{{Crawford}(1966)}]{crawford1966}
{Crawford}, D.~L. 1966, in IAU Symposium, Vol.~24, Spectral Classification and
  Multicolour Photometry, ed. K.~{Loden}, L.~O. {Loden}, \& U.~{Sinnerstad},
  170

\bibitem[{{Crawford}(1979)}]{crawford1979}
---. 1979, \aj, 84, 1858

\bibitem[{{Crawford} \& {Mandwewala}(1976)}]{crawfordmandwewala1976}
{Crawford}, D.~L., \& {Mandwewala}, N. 1976, \pasp, 88, 917

\bibitem[{{Dalle Mese} {et~al.}(2012){Dalle Mese}, {L{\'o}pez-Cruz},
  {Schuster}, {Chavarr{\'{\i}}a}, \& {Garc{\'{\i}}a}}]{dallemese2012}
{Dalle Mese}, G., {L{\'o}pez-Cruz}, O., {Schuster}, W.~J., {Chavarr{\'{\i}}a},
  C., \& {Garc{\'{\i}}a}, J.~G. 2012, ArXiv e-prints

\bibitem[{{Davis} {et~al.}(2011){Davis}, {Ireland}, {North}, {Robertson},
  {Tango}, \& {Tuthill}}]{davis2011}
{Davis}, J., {Ireland}, M.~J., {North}, J.~R., {Robertson}, J.~G., {Tango},
  W.~J., \& {Tuthill}, P.~G. 2011, PASA, 28, 58

\bibitem[{{Davis} \& {Tango}(1986)}]{davis1986}
{Davis}, J., \& {Tango}, W.~J. 1986, \nat, 323, 234

\bibitem[{{De Rosa} {et~al.}(2014){De Rosa}, {Patience}, {Wilson}, {Schneider},
  {Wiktorowicz}, {Vigan}, {Marois}, {Song}, {Macintosh}, {Graham}, {Doyon},
  {Bessell}, {Thomas}, \& {Lai}}]{derosa2014}
{De Rosa}, R.~J., {et~al.} 2014, \mnras, 437, 1216

\bibitem[{{Debernardi} {et~al.}(2000){Debernardi}, {Mermilliod}, {Carquillat},
  \& {Ginestet}}]{debernardi2000}
{Debernardi}, Y., {Mermilliod}, J.-C., {Carquillat}, J.-M., \& {Ginestet}, N.
  2000, \aap, 354, 881

\bibitem[{{Di Folco} {et~al.}(2004){Di Folco}, {Th{\'e}venin}, {Kervella},
  {Domiciano de Souza}, {Coud{\'e} du Foresto}, {S{\'e}gransan}, \&
  {Morel}}]{difolco2004}
{Di Folco}, E., {Th{\'e}venin}, F., {Kervella}, P., {Domiciano de Souza}, A.,
  {Coud{\'e} du Foresto}, V., {S{\'e}gransan}, D., \& {Morel}, P. 2004, \aap,
  426, 601

\bibitem[{{Dobbie} {et~al.}(2010){Dobbie}, {Lodieu}, \& {Sharp}}]{dobbie2010}
{Dobbie}, P.~D., {Lodieu}, N., \& {Sharp}, R.~G. 2010, \mnras, 409, 1002

\bibitem[{{Eggen}(1995)}]{eggen1995}
{Eggen}, O.~J. 1995, \aj, 110, 823

\bibitem[{{Ekstr{\"o}m} {et~al.}(2012){Ekstr{\"o}m}, {Georgy}, {Eggenberger},
  {Meynet}, {Mowlavi}, {Wyttenbach}, {Granada}, {Decressin}, {Hirschi},
  {Frischknecht}, {Charbonnel}, \& {Maeder}}]{ekstrom2012}
{Ekstr{\"o}m}, S., {et~al.} 2012, \aap, 537, A146

\bibitem[{{Figueras} \& {Blasi}(1998)}]{figueras1998}
{Figueras}, F., \& {Blasi}, F. 1998, \aap, 329, 957

\bibitem[{{Figueras} {et~al.}(1991){Figueras}, {Torra}, \&
  {Jordi}}]{figueras1991}
{Figueras}, F., {Torra}, J., \& {Jordi}, C. 1991, \aaps, 87, 319

\bibitem[{{Fischer} \& {Valenti}(2005)}]{fischer2005}
{Fischer}, D.~A., \& {Valenti}, J. 2005, \apj, 622, 1102

\bibitem[{{Fitzpatrick} \& {Massa}(2005)}]{fitzpatrick2005}
{Fitzpatrick}, E.~L., \& {Massa}, D. 2005, \aj, 129, 1642

\bibitem[{{Fortney} {et~al.}(2008){Fortney}, {Marley}, {Saumon}, \&
  {Lodders}}]{fortney2008}
{Fortney}, J.~J., {Marley}, M.~S., {Saumon}, D., \& {Lodders}, K. 2008, \apj,
  683, 1104

\bibitem[{{Gaidos} {et~al.}(2013){Gaidos}, {Fischer}, {Mann}, \&
  {Howard}}]{gaidos2013}
{Gaidos}, E., {Fischer}, D.~A., {Mann}, A.~W., \& {Howard}, A.~W. 2013, \apj,
  771, 18

\bibitem[{{Glebocki} \& {Gnacinski}(2005)}]{glebocki2005}
{Glebocki}, R., \& {Gnacinski}, P. 2005, VizieR Online Data Catalog, 3244, 0

\bibitem[{{Gray}(2005)}]{gray2005book}
{Gray}, D.~F. 2005, {The Observation and Analysis of Stellar Photospheres}

\bibitem[{{Gray} {et~al.}(2006){Gray}, {Corbally}, {Garrison}, {McFadden},
  {Bubar}, {McGahee}, {O'Donoghue}, \& {Knox}}]{gray2006}
{Gray}, R.~O., {Corbally}, C.~J., {Garrison}, R.~F., {McFadden}, M.~T.,
  {Bubar}, E.~J., {McGahee}, C.~E., {O'Donoghue}, A.~A., \& {Knox}, E.~R. 2006,
  \aj, 132, 161

\bibitem[{{Gray} \& {Kaye}(1999)}]{gray1999}
{Gray}, R.~O., \& {Kaye}, A.~B. 1999, \aj, 118, 2993

\bibitem[{{Haberreiter} {et~al.}(2008){Haberreiter}, {Schmutz}, \&
  {Kosovichev}}]{haberreiter2008}
{Haberreiter}, M., {Schmutz}, W., \& {Kosovichev}, A.~G. 2008, \apjl, 675, L53

\bibitem[{{Habets} \& {Heintze}(1981)}]{habets1981}
{Habets}, G.~M.~H.~J., \& {Heintze}, J.~R.~W. 1981, \aaps, 46, 193

\bibitem[{{Hanbury Brown} {et~al.}(1974){Hanbury Brown}, {Davis}, \&
  {Allen}}]{hanburybrown1974}
{Hanbury Brown}, R., {Davis}, J., \& {Allen}, L.~R. 1974, \mnras, 167, 121

\bibitem[{{Hauck} \& {Mermilliod}(1980)}]{hauck1980}
{Hauck}, B., \& {Mermilliod}, M. 1980, \aaps, 40, 1

\bibitem[{{Hauck} \& {Mermilliod}(1998)}]{hauck1998}
---. 1998, \aaps, 129, 431

\bibitem[{{Hinkley} {et~al.}(2013){Hinkley}, {Pueyo}, {Faherty}, {Oppenheimer},
  {Mamajek}, {Kraus}, {Rice}, {Ireland}, {David}, {Hillenbrand}, {Vasisht},
  {Cady}, {Brenner}, {Veicht}, {Nilsson}, {Zimmerman}, {Parry}, {Beichman},
  {Dekany}, {Roberts}, {Roberts}, {Baranec}, {Crepp}, {Burruss}, {Wallace},
  {King}, {Zhai}, {Lockhart}, {Shao}, {Soummer}, {Sivaramakrishnan}, \&
  {Wilson}}]{hinkley2013}
{Hinkley}, S., {et~al.} 2013, \apj, 779, 153

\bibitem[{{Holmberg} {et~al.}(2007){Holmberg}, {Nordstr{\"o}m}, \&
  {Andersen}}]{holmberg2007}
{Holmberg}, J., {Nordstr{\"o}m}, B., \& {Andersen}, J. 2007, \aap, 475, 519

\bibitem[{{Holmberg} {et~al.}(2009){Holmberg}, {Nordstr{\"o}m}, \&
  {Andersen}}]{holmberg2009}
---. 2009, \aap, 501, 941

\bibitem[{{Huber} {et~al.}(2012){Huber}, {Ireland}, {Bedding}, {Brand{\~a}o},
  {Piau}, {Maestro}, {White}, {Bruntt}, {Casagrande}, {Molenda-{\.Z}akowicz},
  {Silva Aguirre}, {Sousa}, {Barclay}, {Burke}, {Chaplin},
  {Christensen-Dalsgaard}, {Cunha}, {De Ridder}, {Farrington}, {Frasca},
  {Garc{\'{\i}}a}, {Gilliland}, {Goldfinger}, {Hekker}, {Kawaler}, {Kjeldsen},
  {McAlister}, {Metcalfe}, {Miglio}, {Monteiro}, {Pinsonneault}, {Schaefer},
  {Stello}, {Stumpe}, {Sturmann}, {Sturmann}, {ten Brummelaar}, {Thompson},
  {Turner}, \& {Uytterhoeven}}]{huber2012}
{Huber}, D., {et~al.} 2012, \apj, 760, 32

\bibitem[{{Hughes} {et~al.}(2008){Hughes}, {Wilner}, {Kamp}, \&
  {Hogerheijde}}]{hughes2008}
{Hughes}, A.~M., {Wilner}, D.~J., {Kamp}, I., \& {Hogerheijde}, M.~R. 2008,
  \apj, 681, 626

\bibitem[{{Ireland} {et~al.}(2011){Ireland}, {Kraus}, {Martinache}, {Law}, \&
  {Hillenbrand}}]{ireland2011}
{Ireland}, M.~J., {Kraus}, A., {Martinache}, F., {Law}, N., \& {Hillenbrand},
  L.~A. 2011, \apj, 726, 113

\bibitem[{{Johnson} {et~al.}(2010){Johnson}, {Aller}, {Howard}, \&
  {Crepp}}]{johnson2010}
{Johnson}, J.~A., {Aller}, K.~M., {Howard}, A.~W., \& {Crepp}, J.~R. 2010,
  \pasp, 122, 905

\bibitem[{{J{\o}rgensen} \& {Lindegren}(2005)}]{jorgensen2005}
{J{\o}rgensen}, B.~R., \& {Lindegren}, L. 2005, \aap, 436, 127

\bibitem[{{Kains} {et~al.}(2011){Kains}, {Wyatt}, \& {Greaves}}]{kains2011}
{Kains}, N., {Wyatt}, M.~C., \& {Greaves}, J.~S. 2011, \mnras, 414, 2486

\bibitem[{{Kalas} {et~al.}(2008){Kalas}, {Graham}, {Chiang}, {Fitzgerald},
  {Clampin}, {Kite}, {Stapelfeldt}, {Marois}, \& {Krist}}]{kalas2008}
{Kalas}, P., {et~al.} 2008, Science, 322, 1345

\bibitem[{{Kervella} {et~al.}(2004){Kervella}, {Th{\'e}venin}, {Morel},
  {Berthomieu}, {Bord{\'e}}, \& {Provost}}]{kervella2004}
{Kervella}, P., {Th{\'e}venin}, F., {Morel}, P., {Berthomieu}, G., {Bord{\'e}},
  P., \& {Provost}, J. 2004, \aap, 413, 251

\bibitem[{{Kervella} {et~al.}(2003){Kervella}, {Th{\'e}venin}, {S{\'e}gransan},
  {Berthomieu}, {Lopez}, {Morel}, \& {Provost}}]{kervella2003a}
{Kervella}, P., {Th{\'e}venin}, F., {S{\'e}gransan}, D., {Berthomieu}, G.,
  {Lopez}, B., {Morel}, P., \& {Provost}, J. 2003, \aap, 404, 1087

\bibitem[{{Kobi} \& {North}(1990)}]{kobinorth1990}
{Kobi}, D., \& {North}, P. 1990, \aaps, 85, 999

\bibitem[{{Kopp} \& {Lean}(2011)}]{kopp2011}
{Kopp}, G., \& {Lean}, J.~L. 2011, \grl, 38, 1706

\bibitem[{{Kraft} \& {Wrubel}(1965)}]{kraftwrubel1965}
{Kraft}, R.~P., \& {Wrubel}, M.~H. 1965, \apj, 142, 703

\bibitem[{{Kunzli} {et~al.}(1997){Kunzli}, {North}, {Kurucz}, \&
  {Nicolet}}]{kunzli1997}
{Kunzli}, M., {North}, P., {Kurucz}, R.~L., \& {Nicolet}, B. 1997, \aaps, 122,
  51

\bibitem[{{Kurucz}(1979)}]{kurucz1979}
{Kurucz}, R.~L. 1979, \apjs, 40, 1

\bibitem[{{Kuzuhara} {et~al.}(2013){Kuzuhara}, {Tamura}, {Kudo}, {Janson},
  {Kandori}, {Brandt}, {Thalmann}, {Spiegel}, {Biller}, {Carson}, {Hori},
  {Suzuki}, {Burrows}, {Henning}, {Turner}, {McElwain}, {Moro-Mart{\'{\i}}n},
  {Suenaga}, {Takahashi}, {Kwon}, {Lucas}, {Abe}, {Brandner}, {Egner}, {Feldt},
  {Fujiwara}, {Goto}, {Grady}, {Guyon}, {Hashimoto}, {Hayano}, {Hayashi},
  {Hayashi}, {Hodapp}, {Ishii}, {Iye}, {Knapp}, {Matsuo}, {Mayama}, {Miyama},
  {Morino}, {Nishikawa}, {Nishimura}, {Kotani}, {Kusakabe}, {Pyo}, {Serabyn},
  {Suto}, {Takami}, {Takato}, {Terada}, {Tomono}, {Watanabe}, {Wisniewski},
  {Yamada}, {Takami}, \& {Usuda}}]{kuzuhara2013}
{Kuzuhara}, M., {et~al.} 2013, \apj, 774, 11

\bibitem[{{Lafreni{\`e}re} {et~al.}(2008){Lafreni{\`e}re}, {Jayawardhana}, \&
  {van Kerkwijk}}]{lafreniere2008}
{Lafreni{\`e}re}, D., {Jayawardhana}, R., \& {van Kerkwijk}, M.~H. 2008, \apjl,
  689, L153

\bibitem[{{Lagrange} {et~al.}(2009){Lagrange}, {Gratadour}, {Chauvin}, {Fusco},
  {Ehrenreich}, {Mouillet}, {Rousset}, {Rouan}, {Allard}, {Gendron}, {Charton},
  {Mugnier}, {Rabou}, {Montri}, \& {Lacombe}}]{lagrange2009}
{Lagrange}, A.-M., {et~al.} 2009, \aap, 493, L21

\bibitem[{{Lagrange} {et~al.}(2010){Lagrange}, {Bonnefoy}, {Chauvin}, {Apai},
  {Ehrenreich}, {Boccaletti}, {Gratadour}, {Rouan}, {Mouillet}, {Lacour}, \&
  {Kasper}}]{lagrange2010}
---. 2010, Science, 329, 57

\bibitem[{{Lejeune} {et~al.}(1999){Lejeune}, {Lastennet}, {Westera}, \&
  {Buser}}]{lejeune1999}
{Lejeune}, T., {Lastennet}, E., {Westera}, P., \& {Buser}, R. 1999, in
  Astronomical Society of the Pacific Conference Series, Vol. 192,
  Spectrophotometric Dating of Stars and Galaxies, ed. I.~{Hubeny}, S.~{Heap},
  \& R.~{Cornett}, 207

\bibitem[{{Lester} {et~al.}(1986){Lester}, {Gray}, \& {Kurucz}}]{lester1986}
{Lester}, J.~B., {Gray}, R.~O., \& {Kurucz}, R.~L. 1986, \apjs, 61, 509

\bibitem[{{Li} \& {Han}(2008)}]{li2008}
{Li}, Z., \& {Han}, Z. 2008, \mnras, 387, 105

\bibitem[{{Ligi} {et~al.}(2012){Ligi}, {Mourard}, {Lagrange}, {Perraut},
  {Boyajian}, {B{\'e}rio}, {Nardetto}, {Tallon-Bosc}, {McAlister}, {ten
  Brummelaar}, {Ridgway}, {Sturmann}, {Sturmann}, {Turner}, {Farrington}, \&
  {Goldfinger}}]{ligi2012}
{Ligi}, R., {et~al.} 2012, \aap, 545, A5

\bibitem[{{Lloyd}(2013)}]{lloyd2013}
{Lloyd}, J.~P. 2013, \apjl, 774, L2

\bibitem[{{Lowrance} {et~al.}(2000){Lowrance}, {Schneider}, {Kirkpatrick},
  {Becklin}, {Weinberger}, {Zuckerman}, {Plait}, {Malmuth}, {Heap}, {Schultz},
  {Smith}, {Terrile}, \& {Hines}}]{lowrance2000}
{Lowrance}, P.~J., {et~al.} 2000, \apj, 541, 390

\bibitem[{{Maeder} \& {Peytremann}(1970)}]{maeder1970}
{Maeder}, A., \& {Peytremann}, E. 1970, \aap, 7, 120

\bibitem[{{Malagnini} {et~al.}(1986){Malagnini}, {Morossi}, {Rossi}, \&
  {Kurucz}}]{malagnini1986}
{Malagnini}, M.~L., {Morossi}, C., {Rossi}, L., \& {Kurucz}, R.~L. 1986, \aap,
  162, 140

\bibitem[{{Mamajek} \& {Hillenbrand}(2008)}]{mamajek2008}
{Mamajek}, E.~E., \& {Hillenbrand}, L.~A. 2008, \apj, 687, 1264

\bibitem[{{Marois} {et~al.}(2008){Marois}, {Macintosh}, {Barman}, {Zuckerman},
  {Song}, {Patience}, {Lafreni{\`e}re}, \& {Doyon}}]{marois2008}
{Marois}, C., {Macintosh}, B., {Barman}, T., {Zuckerman}, B., {Song}, I.,
  {Patience}, J., {Lafreni{\`e}re}, D., \& {Doyon}, R. 2008, Science, 322, 1348

\bibitem[{{Marois} {et~al.}(2010){Marois}, {Zuckerman}, {Konopacky},
  {Macintosh}, \& {Barman}}]{marois2010}
{Marois}, C., {Zuckerman}, B., {Konopacky}, Q.~M., {Macintosh}, B., \&
  {Barman}, T. 2010, \nat, 468, 1080

\bibitem[{{Mawet} {et~al.}(2012){Mawet}, {Pueyo}, {Lawson}, {Mugnier}, {Traub},
  {Boccaletti}, {Trauger}, {Gladysz}, {Serabyn}, {Milli}, {Belikov}, {Kasper},
  {Baudoz}, {Macintosh}, {Marois}, {Oppenheimer}, {Barrett}, {Beuzit},
  {Devaney}, {Girard}, {Guyon}, {Krist}, {Mennesson}, {Mouillet}, {Murakami},
  {Poyneer}, {Savransky}, {V{\'e}rinaud}, \& {Wallace}}]{mawet2012}
{Mawet}, D., {et~al.} 2012, ArXiv e-prints

\bibitem[{{Mermilliod}(1981)}]{mermilliod1981}
{Mermilliod}, J.~C. 1981, \aap, 97, 235

\bibitem[{{Meynet} \& {Maeder}(2000)}]{meynet2000}
{Meynet}, G., \& {Maeder}, A. 2000, \aap, 361, 101

\bibitem[{{Mongui{\'o}} {et~al.}(2014){Mongui{\'o}}, {Figueras}, \&
  {Grosb{\o}l}}]{monguio2014}
{Mongui{\'o}}, M., {Figueras}, F., \& {Grosb{\o}l}, P. 2014, \aap, 568, A119

\bibitem[{{Moon}(1985)}]{ttmoon1985}
{Moon}, T.~T. 1985, in: Communications of the University of London Observatory,
  No. 78

\bibitem[{{Moon} \& {Dworetsky}(1985)}]{moon1985}
{Moon}, T.~T., \& {Dworetsky}, M.~M. 1985, \mnras, 217, 305

\bibitem[{{Mo{\'o}r} {et~al.}(2006){Mo{\'o}r}, {{\'A}brah{\'a}m}, {Derekas},
  {Kiss}, {Kiss}, {Apai}, {Grady}, \& {Henning}}]{moor2006}
{Mo{\'o}r}, A., {{\'A}brah{\'a}m}, P., {Derekas}, A., {Kiss}, C., {Kiss},
  L.~L., {Apai}, D., {Grady}, C., \& {Henning}, T. 2006, \apj, 644, 525

\bibitem[{{Mozurkewich} {et~al.}(2003){Mozurkewich}, {Armstrong}, {Hindsley},
  {Quirrenbach}, {Hummel}, {Hutter}, {Johnston}, {Hajian}, {Elias}, {Buscher},
  \& {Simon}}]{mozurkewich2003}
{Mozurkewich}, D., {et~al.} 2003, \aj, 126, 2502

\bibitem[{{Munari} {et~al.}(2005){Munari}, {Sordo}, {Castelli}, \&
  {Zwitter}}]{munari2005}
{Munari}, U., {Sordo}, R., {Castelli}, F., \& {Zwitter}, T. 2005, \aap, 442,
  1127

\bibitem[{{Napiwotzki} {et~al.}(1993){Napiwotzki}, {Schoenberner}, \&
  {Wenske}}]{napiwotzki1993}
{Napiwotzki}, R., {Schoenberner}, D., \& {Wenske}, V. 1993, \aap, 268, 653

\bibitem[{{Nielsen} {et~al.}(2013){Nielsen}, {Liu}, {Wahhaj}, {Biller},
  {Hayward}, {Close}, {Males}, {Skemer}, {Chun}, {Ftaclas}, {Alencar},
  {Artymowicz}, {Boss}, {Clarke}, {de Gouveia Dal Pino}, {Gregorio-Hetem},
  {Hartung}, {Ida}, {Kuchner}, {Lin}, {Reid}, {Shkolnik}, {Tecza}, {Thatte}, \&
  {Toomey}}]{nielsen2013}
{Nielsen}, E.~L., {et~al.} 2013, ArXiv e-prints

\bibitem[{{Nieva}(2013)}]{nieva2013}
{Nieva}, M.-F. 2013, \aap, 550, A26

\bibitem[{{Nieva} \& {Sim{\'o}n-D{\'{\i}}az}(2011)}]{nieva2011}
{Nieva}, M.-F., \& {Sim{\'o}n-D{\'{\i}}az}, S. 2011, \aap, 532, A2

\bibitem[{{Nordgren} {et~al.}(2001){Nordgren}, {Sudol}, \&
  {Mozurkewich}}]{nordgren2001}
{Nordgren}, T.~E., {Sudol}, J.~J., \& {Mozurkewich}, D. 2001, \aj, 122, 2707

\bibitem[{{Nordstr{\"o}m} {et~al.}(2004){Nordstr{\"o}m}, {Mayor}, {Andersen},
  {Holmberg}, {Pont}, {J{\o}rgensen}, {Olsen}, {Udry}, \&
  {Mowlavi}}]{nordstrom2004}
{Nordstr{\"o}m}, B., {et~al.} 2004, \aap, 418, 989

\bibitem[{{Olsen}(1983)}]{olsen1983}
{Olsen}, E.~H. 1983, \aaps, 54, 55

\bibitem[{{Olsen}(1988)}]{olsen1988}
---. 1988, \aap, 189, 173

\bibitem[{{Olsen} \& {Perry}(1984)}]{olsen1984}
{Olsen}, E.~H., \& {Perry}, C.~L. 1984, \aaps, 56, 229

\bibitem[{{{\"O}nehag} {et~al.}(2009){{\"O}nehag}, {Gustafsson}, {Eriksson}, \&
  {Edvardsson}}]{onehag2009}
{{\"O}nehag}, A., {Gustafsson}, B., {Eriksson}, K., \& {Edvardsson}, B. 2009,
  \aap, 498, 527

\bibitem[{{Perryman} {et~al.}(1998){Perryman}, {Brown}, {Lebreton}, {Gomez},
  {Turon}, {Cayrel de Strobel}, {Mermilliod}, {Robichon}, {Kovalevsky}, \&
  {Crifo}}]{perryman1998}
{Perryman}, M.~A.~C., {et~al.} 1998, \aap, 331, 81

\bibitem[{{Peterson} {et~al.}(2006){Peterson}, {Hummel}, {Pauls}, {Armstrong},
  {Benson}, {Gilbreath}, {Hindsley}, {Hutter}, {Johnston}, {Mozurkewich}, \&
  {Schmitt}}]{peterson2006}
{Peterson}, D.~M., {et~al.} 2006, \nat, 440, 896

\bibitem[{{Pinsonneault} {et~al.}(2004){Pinsonneault}, {Terndrup}, {Hanson}, \&
  {Stauffer}}]{pinsonneault2004}
{Pinsonneault}, M.~H., {Terndrup}, D.~M., {Hanson}, R.~B., \& {Stauffer}, J.~R.
  2004, \apj, 600, 946

\bibitem[{{Pont} \& {Eyer}(2004)}]{pont2004}
{Pont}, F., \& {Eyer}, L. 2004, \mnras, 351, 487

\bibitem[{{Rhee} {et~al.}(2007){Rhee}, {Song}, {Zuckerman}, \&
  {McElwain}}]{rhee2007}
{Rhee}, J.~H., {Song}, I., {Zuckerman}, B., \& {McElwain}, M. 2007, \apj, 660,
  1556

\bibitem[{{Rieke} {et~al.}(2005){Rieke}, {Su}, {Stansberry}, {Trilling},
  {Bryden}, {Muzerolle}, {White}, {Gorlova}, {Young}, {Beichman},
  {Stapelfeldt}, \& {Hines}}]{rieke2005}
{Rieke}, G.~H., {et~al.} 2005, \apj, 620, 1010

\bibitem[{{Salpeter}(1955)}]{salpeter1955}
{Salpeter}, E.~E. 1955, \apj, 121, 161

\bibitem[{{Schaller} {et~al.}(1992){Schaller}, {Schaerer}, {Meynet}, \&
  {Maeder}}]{schaller1992}
{Schaller}, G., {Schaerer}, D., {Meynet}, G., \& {Maeder}, A. 1992, \aaps, 96,
  269

\bibitem[{{Schlaufman} \& {Winn}(2013)}]{schlaufman2013}
{Schlaufman}, K.~C., \& {Winn}, J.~N. 2013, \apj, 772, 143

\bibitem[{{Schlegel} {et~al.}(1998){Schlegel}, {Finkbeiner}, \&
  {Davis}}]{schlegel1998}
{Schlegel}, D.~J., {Finkbeiner}, D.~P., \& {Davis}, M. 1998, \apj, 500, 525

\bibitem[{{Schr{\"o}der} {et~al.}(2009){Schr{\"o}der}, {Reiners}, \&
  {Schmitt}}]{schroeder2009}
{Schr{\"o}der}, C., {Reiners}, A., \& {Schmitt}, J.~H.~M.~M. 2009, \aap, 493,
  1099

\bibitem[{{Siess} {et~al.}(2000){Siess}, {Dufour}, \& {Forestini}}]{siess2000}
{Siess}, L., {Dufour}, E., \& {Forestini}, M. 2000, \aap, 358, 593

\bibitem[{{Smalley}(1993)}]{smalley1993}
{Smalley}, B. 1993, \aap, 274, 391

\bibitem[{{Smalley} \& {Dworetsky}(1995)}]{smalley1995}
{Smalley}, B., \& {Dworetsky}, M.~M. 1995, \aap, 293, 446

\bibitem[{{Soderblom}(2010)}]{soderblom2010}
{Soderblom}, D.~R. 2010, \araa, 48, 581

\bibitem[{{Song} {et~al.}(2001){Song}, {Caillault}, {Barrado y Navascu{\'e}s},
  \& {Stauffer}}]{song2001}
{Song}, I., {Caillault}, J.-P., {Barrado y Navascu{\'e}s}, D., \& {Stauffer},
  J.~R. 2001, \apj, 546, 352

\bibitem[{{Song} {et~al.}(2000){Song}, {Caillault}, {Barrado y Navascu{\'e}s},
  {Stauffer}, \& {Randich}}]{song2000}
{Song}, I., {Caillault}, J.-P., {Barrado y Navascu{\'e}s}, D., {Stauffer},
  J.~R., \& {Randich}, S. 2000, \apjl, 533, L41

\bibitem[{{Stauffer} {et~al.}(1995){Stauffer}, {Hartmann}, \& {Barrado y
  Navascues}}]{stauffer1995}
{Stauffer}, J.~R., {Hartmann}, L.~W., \& {Barrado y Navascues}, D. 1995, \apj,
  454, 910

\bibitem[{{Stauffer} {et~al.}(1998){Stauffer}, {Schultz}, \&
  {Kirkpatrick}}]{stauffer1998}
{Stauffer}, J.~R., {Schultz}, G., \& {Kirkpatrick}, J.~D. 1998, \apjl, 499,
  L199

\bibitem[{{Stauffer} {et~al.}(1999){Stauffer}, {Barrado y Navascu{\'e}s},
  {Bouvier}, {Morrison}, {Harding}, {Luhman}, {Stanke}, {McCaughrean},
  {Terndrup}, {Allen}, \& {Assouad}}]{stauffer1999}
{Stauffer}, J.~R., {et~al.} 1999, \apj, 527, 219

\bibitem[{{Stauffer} {et~al.}(2007){Stauffer}, {Hartmann}, {Fazio}, {Allen},
  {Patten}, {Lowrance}, {Hurt}, {Rebull}, {Cutri}, {Ramirez}, {Young}, {Rieke},
  {Gorlova}, {Muzerolle}, {Slesnick}, \& {Skrutskie}}]{stauffer2007}
---. 2007, \apjs, 172, 663

\bibitem[{{Str{\"o}mgren}(1951)}]{stromgren1951}
{Str{\"o}mgren}, B. 1951, \aj, 56, 142

\bibitem[{{Str{\"o}mgren}(1966)}]{stromgren1966}
---. 1966, \araa, 4, 433

\bibitem[{{Su} {et~al.}(2006){Su}, {Rieke}, {Stansberry}, {Bryden},
  {Stapelfeldt}, {Trilling}, {Muzerolle}, {Beichman}, {Moro-Martin}, {Hines},
  \& {Werner}}]{su2006}
{Su}, K.~Y.~L., {et~al.} 2006, \apj, 653, 675

\bibitem[{{Sweet} \& {Roy}(1953)}]{sweetroy1953}
{Sweet}, I.~P.~A., \& {Roy}, A.~E. 1953, \mnras, 113, 701

\bibitem[{{Takeda} {et~al.}(2007){Takeda}, {Ford}, {Sills}, {Rasio}, {Fischer},
  \& {Valenti}}]{gtakeda2007}
{Takeda}, G., {Ford}, E.~B., {Sills}, A., {Rasio}, F.~A., {Fischer}, D.~A., \&
  {Valenti}, J.~A. 2007, \apjs, 168, 297

\bibitem[{{Tetzlaff} {et~al.}(2011){Tetzlaff}, {Neuh{\"a}user}, \&
  {Hohle}}]{tetzlaff2011}
{Tetzlaff}, N., {Neuh{\"a}user}, R., \& {Hohle}, M.~M. 2011, \mnras, 410, 190

\bibitem[{{Torres} {et~al.}(2010){Torres}, {Andersen}, \&
  {Gim{\'e}nez}}]{torres2010}
{Torres}, G., {Andersen}, J., \& {Gim{\'e}nez}, A. 2010, \aapr, 18, 67

\bibitem[{{Valenti} \& {Fischer}(2005)}]{valenti2005}
{Valenti}, J.~A., \& {Fischer}, D.~A. 2005, \apjs, 159, 141

\bibitem[{{van Belle} \& {von Braun}(2009)}]{vanbelle2009}
{van Belle}, G.~T., \& {von Braun}, K. 2009, \apj, 694, 1085

\bibitem[{{Vican}(2012)}]{vican2012}
{Vican}, L. 2012, \aj, 143, 135

\bibitem[{{Vigan} {et~al.}(2012){Vigan}, {Patience}, {Marois}, {Bonavita}, {De
  Rosa}, {Macintosh}, {Song}, {Doyon}, {Zuckerman}, {Lafreni{\`e}re}, \&
  {Barman}}]{vigan2012}
{Vigan}, A., {et~al.} 2012, \aap, 544, A9

\bibitem[{{Wu} {et~al.}(2011){Wu}, {Singh}, {Prugniel}, {Gupta}, \&
  {Koleva}}]{wu2011}
{Wu}, Y., {Singh}, H.~P., {Prugniel}, P., {Gupta}, R., \& {Koleva}, M. 2011,
  \aap, 525, A71

\bibitem[{{Wyatt}(2008)}]{wyatt2008}
{Wyatt}, M.~C. 2008, \araa, 46, 339

\bibitem[{{Zorec} \& {Royer}(2012)}]{zorec2012}
{Zorec}, J., \& {Royer}, F. 2012, \aap, 537, A120

\bibitem[{{Zuckerman}(2001)}]{zuckerman2001}
{Zuckerman}, B. 2001, \araa, 39, 549

\bibitem[{{Zuckerman} {et~al.}(2012){Zuckerman}, {Melis}, {Rhee}, {Schneider},
  \& {Song}}]{zuckerman2012}
{Zuckerman}, B., {Melis}, C., {Rhee}, J.~H., {Schneider}, A., \& {Song}, I.
  2012, \apj, 752, 58

\bibitem[{{Zuckerman} {et~al.}(2011){Zuckerman}, {Rhee}, {Song}, \&
  {Bessell}}]{zuckerman2011}
{Zuckerman}, B., {Rhee}, J.~H., {Song}, I., \& {Bessell}, M.~S. 2011, \apj,
  732, 61

\end{thebibliography}

\end{document}